\documentclass[11pt,a4paper]{article}

\usepackage{graphicx}
\usepackage{afterpage}
\usepackage{epsfig,cite}
\usepackage{amssymb}
\usepackage{amsmath}
\usepackage{dsfont}
\usepackage{multirow}
\usepackage{url,hyperref}

\usepackage{url}
\usepackage{hyperref}

\def\smallfrac#1#2{\hbox{${{#1}\over {#2}}$}}
\newcommand{\be}{\begin{equation}}
\newcommand{\ee}{\end{equation}}
\newcommand{\bea}{\begin{eqnarray}}
\newcommand{\eea}{\end{eqnarray}}
\newcommand{\bi}{\begin{itemize}}
\newcommand{\ei}{\end{itemize}}
\newcommand{\ben}{\begin{enumerate}}
\newcommand{\een}{\end{enumerate}}
\newcommand{\la}{\left\langle}
\newcommand{\ra}{\right\rangle}
\newcommand{\lc}{\left[}
\newcommand{\rc}{\right]}
\newcommand{\lp}{\left(}
\newcommand{\rp}{\right)}

\def\frac#1#2{{{#1}\over {#2}}}
\def\gsim{\mathrel{\rlap{\lower4pt\hbox{\hskip1pt$\sim$}}
    \raise1pt\hbox{$>$}}}         
\def\lsim{\mathrel{\rlap{\lower4pt\hbox{\hskip1pt$\sim$}}
    \raise1pt\hbox{$<$}}}         

\newcommand{\dat}{\mathrm{dat}}

\newcommand{\rep}{\mathrm{rep}}
\newcommand{\net}{\mathrm{net}}

\newcommand{\tot}{\mathrm{tot}}

\newcommand{\draft}[1]{}

\def\beq{\begin{equation}}  
\def\eeq{\end{equation}}  

\def\nn{\nonumber}

\def \n0{N_j^{(0)}}

\def\lapprox{\lower .7ex\hbox{$\;\stackrel{\textstyle <}{\sim}\;$}}
\def\gapprox{\lower .7ex\hbox{$\;\stackrel{\textstyle >}{\sim}\;$}}
\def\half{\smallfrac{1}{2}}

\newcommand{\tmop}[1]{\ensuremath{\operatorname{#1}}}

\begin{document}
\begin{flushright}
Edinburgh 2011/01\\
IFUM-969-FT\\
FR-PHENO-2010-043\\
RWTH TTK-11-02\\
\end{flushright}
\begin{center}
{\Large \bf Impact of Heavy Quark Masses\\ on Parton Distributions and
  LHC Phenomenology}
\vspace{0.8cm}

{\bf  The NNPDF Collaboration:}\\
Richard~D.~Ball$^{1}$, Valerio~Bertone$^3$, Francesco~Cerutti$^4$,
 Luigi~Del~Debbio$^1$,\\ Stefano~Forte$^2$, Alberto~Guffanti$^3$, 
Jos\'e~I.~Latorre$^4$, Juan~Rojo$^2$ and Maria~Ubiali$^{5}$.

\vspace{1.cm}
{\it ~$^1$ School of Physics and Astronomy, University of Edinburgh,\\
JCMB, KB, Mayfield Rd, Edinburgh EH9 3JZ, Scotland\\
~$^2$ Dipartimento di Fisica, Universit\`a di Milano and
INFN, Sezione di Milano,\\ Via Celoria 16, I-20133 Milano, Italy\\
~$^3$  Physikalisches Institut, Albert-Ludwigs-Universit\"at Freiburg,\\ 
Hermann-Herder-Stra\ss e 3, D-79104 Freiburg i. B., Germany  \\
~$^4$ Departament d'Estructura i Constituents de la Mat\`eria, 
Universitat de Barcelona,\\ Diagonal 647, E-08028 Barcelona, Spain\\
~$^5$ Institut f\"ur Theoretische Teilchenphysik und Kosmologie, RWTH Aachen University,\\ 
D-52056 Aachen, Germany\\}
\end{center}

\vspace{0.8cm}

\begin{center}
{\bf \large Abstract:}
\end{center}

We present a determination of the parton distributions of
the nucleon from a global set of 
hard scattering data using the NNPDF 
methodology including heavy quark mass effects: 
NNPDF2.1. In comparison to the previous NNPDF2.0 parton
determination,  the dataset is enlarged to include deep--inelastic 
charm structure function data.
We implement the FONLL-A general-mass scheme
in the FastKernel framework and assess its accuracy by comparison
to the Les Houches heavy quark benchmarks. 
We discuss the impact on parton distributions of the treatment of
the heavy quark masses, and we provide a determination of the 
uncertainty in the parton distributions due to uncertainty in the 
masses. We assess the impact of these uncertainties on LHC
observables by providing parton 
sets with different values of the charm and bottom quark masses.
Finally, we construct and 
discuss parton sets with a fixed number of flavours.

\clearpage

\tableofcontents

\clearpage

\section{Introduction}

\label{sec-intro}

The inclusion of effects related to heavy quark masses in the
determination of parton distributions (PDFs) has received an
increasing amount of attention over the last few years, driven by the
increase in accuracy and reliability in the determination of the PDFs 
required for phenomenology at the LHC (see
e.g. Ref.~\cite{Forte:2010dt} and references therein). Only a
few years ago, PDFs in common use (such as e.g. 
CTEQ6.1~\cite{Stump:2003yu}) were based on the
so-called zero-mass variable-flavor
number scheme (ZM-VFN), in which heavy quarks decouple at scales below
their mass, $Q^2<m_h^2$, but  are otherwise treated as massless
partons, which amounts to neglecting all contributions of order
${m_h^2}/{Q^2}$. While this approximation only applies to heavy
quark distributions in the vicinity of their respective thresholds,
the ensuing modification of the initial conditions to
perturbative evolution for the heavy quark distributions also affects light
quark PDFs (the momentum sum rule means that a change in any quark's momentum
fraction must be accompanied by corresponding changes in the momentum fractions
carried by all other partons). The high-energy
behaviour of light quark PDFs may then be affected at
the level of several percent, and the ensuing shift in 
predictions for precise high-energy standard candles such as the $W$ and $Z$
cross-sections may be quite significant~\cite{Tung:2006tb}. 
Furthermore, observables which depend on the heavy
quark distributions (such as the single-top production cross-section,
which probes the $b$ distribution)
are substantially affected~\cite{Guffanti:2010yu}.

The purpose of this paper is threefold: first, to present a
determination of parton distributions based on the NNPDF
methodology~\cite{Forte:2002fg,DelDebbio:2004qj,DelDebbio:2007ee,Ball:2008by,Ball:2009mk,Ball:2010de}
with heavy quark mass effects included. Second, to provide tools to
study uncertainties related to heavy quark masses and more general
heavy quark effects. Third, to assess the impact of these
uncertainties on phenomenology.

The first goal will be achieved by repeating a next-to-leading order
global PDF determination based
on exactly the same methodology used in the construction of
the NNPDF2.0 PDF set~\cite{Ball:2010de}, but now with heavy quark mass
effects included up to order $\alpha_s$ through the so-called
FONLL-A scheme~\cite{Forte:2010ta}. The FONLL
method (first suggested in Ref.~\cite{Cacciari:1998it} and generalized to
deep-inelastic scattering in Ref.~\cite{Forte:2010ta}) is especially
convenient in that it allows the inclusion of heavy quark mass effects
to any desired order in $\alpha_s$ and any desired
logarithmic order. The FONLL-A version corresponds to the combination
of $O(\alpha_s)$ mass effects with NLO evolution equations and
coefficient functions: at this NLO-$O(\alpha_s)$  order, 
the FONLL method coincides with
the so-called S-ACOT (simplified~\cite{Kramer:2000hn} ACOT~\cite{acot2}) method,
adopted for instance in the CTEQ6.6~\cite{Nadolsky:2008zw} and
CT10~\cite{Lai:2010vv} NLO PDF determinations. The dataset used here
also coincides with that of  NNPDF2.0, but supplemented by charm
deep-inelastic $F_2^c$ structure function data.

The second goal will be achieved by providing sets of parton
distributions which correspond to different values of the heavy quark
masses: the uncertainty related to the choice of the quark mass can
then be determined simply by variation of the mass value, while
combined PDF+$m_h$ uncertainties  can be determined by constructing
Monte Carlo sets of replicas in which the mass is varied according to
a probability distribution (typically gaussian) with a suitable
width, in analogy to what was done in Refs.~\cite{LHas,Demartin:2010er} to
determine combined PDF+$\alpha_s$ uncertainties. We also provide a
determination of the correlation between the heavy quark masses and
individual PDFs. Finally, we will  provide
PDF sets with various fixed number of flavours.

The third goal will be achieved by computing PDF and heavy quark mass
uncertainties for various LHC standard candles: $W$, $Z$, Higgs and
top production. We will also present a preliminary estimate of
theoretical uncertainties related to higher order heavy quark mass
corrections.

The outline of this paper is the following: in Sect.~\ref{sec:expdata}
we discuss the features of the datasets included in the
NNPDF2.1 analysis, with emphasis on the ZEUS and H1 data on the charm
structure function. Then in Sect.~\ref{sec:evolution} we review 
the FONLL scheme of
Ref.~\cite{Forte:2010ta} for the inclusion of heavy  quark
mass effects in  neutral current structure functions and present its
generalization to charged current deep-inelastic scattering.
In Sect.~\ref{sec:results} we present the  NNPDF2.1 PDF
set and compare it with previous NNPDF releases and with the other
global PDF sets, while in Sect.~\ref{sec:pheno} we perform the same
comparisons for LHC standard candles,
 thus elucidating the impact of the inclusion of 
heavy quark mass
effects in the NNPDF framework. 
In Sect.~\ref{sec:hqmasses} we explore the impact of the uncertainty
on the values of the heavy quark masses both on PDFs themselves and  
on LHC processes using 
NNPDF2.1 sets with varying $m_h$. Finally, in Sect.~\ref{sec:ffnpdfs}
we present NNPDF2.1 sets with various fixed number of flavours.
 Technical details on the
 implementation and benchmarking of FONLL neutral
and charged current structure functions in the FastKernel
computational framework of Ref.~\cite{Ball:2010de}
are collected in Appendices~\ref{sec:massive-nc} and~\ref{sec:massive-cc}.

 
\section{Experimental data}
\label{sec:expdata}

In this Section we discuss the experimental data used for
the NNPDF2.1 analysis.
First of all we motivate the kinematical
cuts that are applied to our dataset. Then we present the details
and kinematical coverage of the NNPDF2.1 dataset, with special
emphasis on the new charm structure function data. Finally,
we discuss the implementation of positivity constraints.
These data have been used to generate Monte Carlo replicas, which have
been checked to reproduce the statistical features of the original
dataset. The replica generation and its testing has been performed in
the same way as in previous NNPDF analyses~\cite{Ball:2008by,Ball:2010de}
and will not be discussed further here.

\subsection{Kinematical cuts}
\label{sec:kincuts}
The NNPDF2.1 dataset has been subjected to some kinematic cuts:
specifically, 
the cut in $W^2$ is the same as in
 previous NNPDF fits, $W^2_{\rm min}=12.5$ GeV$^2$, but the cut in
 $Q^2$ is slightly higher. While in
NNPDF2.0 the cut in $Q^2$
for the DIS data was set to be $Q^2_{\rm min}=2$ GeV$^2$,
in the NNPDF2.1 analysis we use a somewhat more restrictive
kinematical cut in $Q^2$, namely $Q^2_{\rm min}=3$ GeV$^2$. There
are two main motivations for this modification which we now discuss.
First,  very close to the heavy quark 
threshold the predictions for $F_2^c$ from the GM scheme might
suffer from instabilities due to the threshold behaviour. One
would like to avoid having data crossing the charm mass threshold when varying
the heavy quark mass in various fits. 
This suggest to use a value of $Q^2_{\rm min}$
at least as large as the maximum value of the charm mass than
can be considered acceptable. $Q^2_{\rm min}=3$ GeV$^2$ is then a
reasonable choice since then $m_c^{\rm max}\sim 1.7$ GeV.
Furthermore, there is now an
indication of possible deviations from NLO DGLAP in the small-$x$
and $Q^2$ HERA data~\cite{Caola:2010cy,Caola:2009iy}. 
These deviations are mostly relevant
in the smaller $Q^2$ bins of HERA data. The theoretical
uncertainty in the PDFs and LHC observables related to their inclusion in 
the global fit is
moderate as compared to the PDF errors and other
uncertainties, but removing the HERA points below $Q^2_{\rm min}$
reduces these theoretical uncertainties even further. The price to pay
for this reduced theoretical uncertainty is an increase in statistical
uncertainty: indeed, we will see 
in Sect.~\ref{sec:dataset} that removing the data below 
$Q^2_{\rm min}=3$ GeV$^2$ results in an increase of PDF
uncertainty in the small-$x$
gluon PDF due to the reduced experimental information.

On top of the previous general kinematical 
cuts, applied to all the DIS experiments,
we will also perform additional cuts on the HERA $F_2^c$ data. The motivation
for these is that in this work we will use
the FONLL-A general-mass scheme for heavy quarks, and 
as discussed in~\cite{Forte:2010ta},  
FONLL-A\footnote{Note that this is true for any heavy quark
scheme that does not include the $\mathcal{O}\lp \alpha_s^2\rp$ 
corrections, like for example the S-ACOT-$\chi$ used in the
CTEQ/CT family of PDF sets.} provides
a poor description of the data in the smallest $x$ and $Q^2$ bins
due to  missing large $\mathcal{O}\lp \alpha_s^2\rp$ corrections. 
Only the FONLL-B scheme can cure this problem since it includes
consistently $\mathcal{O}\lp \alpha_s^2\rp$ corrections in $F_2^c$
into a NLO fit,
as can be seen 
 in~\cite{Forte:2010ta} and we will review in Sect.~\ref{sec:evolution}.
We will thus remove from the fit HERA $F_{2}^c$ data with $Q^2 \le 4$
GeV$^2$ and data with $Q^2 \le 10$
GeV$^2$ for $x\le 10^{-3}$. These cuts ensure that all $F_2^c$ experimental
data included in the fit are well described by $\mathcal{O}\lp \alpha_s\rp$
theory.

In Table~\ref{tab:kincuts} we summarize the choices
for the initial evolution scale and
kinematical cuts applied in this work,
compared to the choices in other recent PDF determinations.
Note that HERAPDF does not perform a cut in $W^2$ since they only
include HERA data which do not extend to the low $W^2$ region.

\begin{table}
\centering
\begin{tabular}{|c|c|c|c|}
\hline
 & $Q_0^2$ [GeV$^2$]& $Q^2_{\rm min}$ [GeV$^2$] & $W^2_{\rm min}$ [GeV$^2$]\\
\hline
\hline
NNPDF2.1 & 2.0 & 3.0 & 12.5 \\
\hline
NNPDF2.0~\cite{Ball:2010de} & 2.0 & 2.0 & 12.5  \\
CT10~\cite{Lai:2010vv} & 1.69 & 4.0 & 12.25 \\
MSTW08~\cite{Martin:2009iq} & 1  & 2.0 & 15.0\\
ABKM09~\cite{Alekhin:2009ni} & 9  & 2.5 & 3.24 \\
HERAPDF1.0~\cite{H1:2009wt} & 1.9 & 3.5 & 155.75 \\
\hline
\end{tabular}
\caption{\small The values of the initial evolution scale where the
PDFs are parametrized, $Q^2_0$, and the kinematical cuts in $Q^2$
and $W^2$ applied to the fitted DIS dataset, $Q^2_{\rm min}$
and $W^2_{\rm min}$, in the present work and in other recent
PDF determinations. As discussed in the text, further
cuts are applied to $F_2^c$ data in the NNPDF2.1 case. For HERAPDF
the value of $W^2_{\rm min}$ given is the minimum of the HERA dataset
and no cut is performed.\label{tab:kincuts}}
\end{table}

\subsection{NNPDF2.1 dataset}

\begin{figure}[ht]
\begin{center}
\epsfig{width=\textwidth,figure=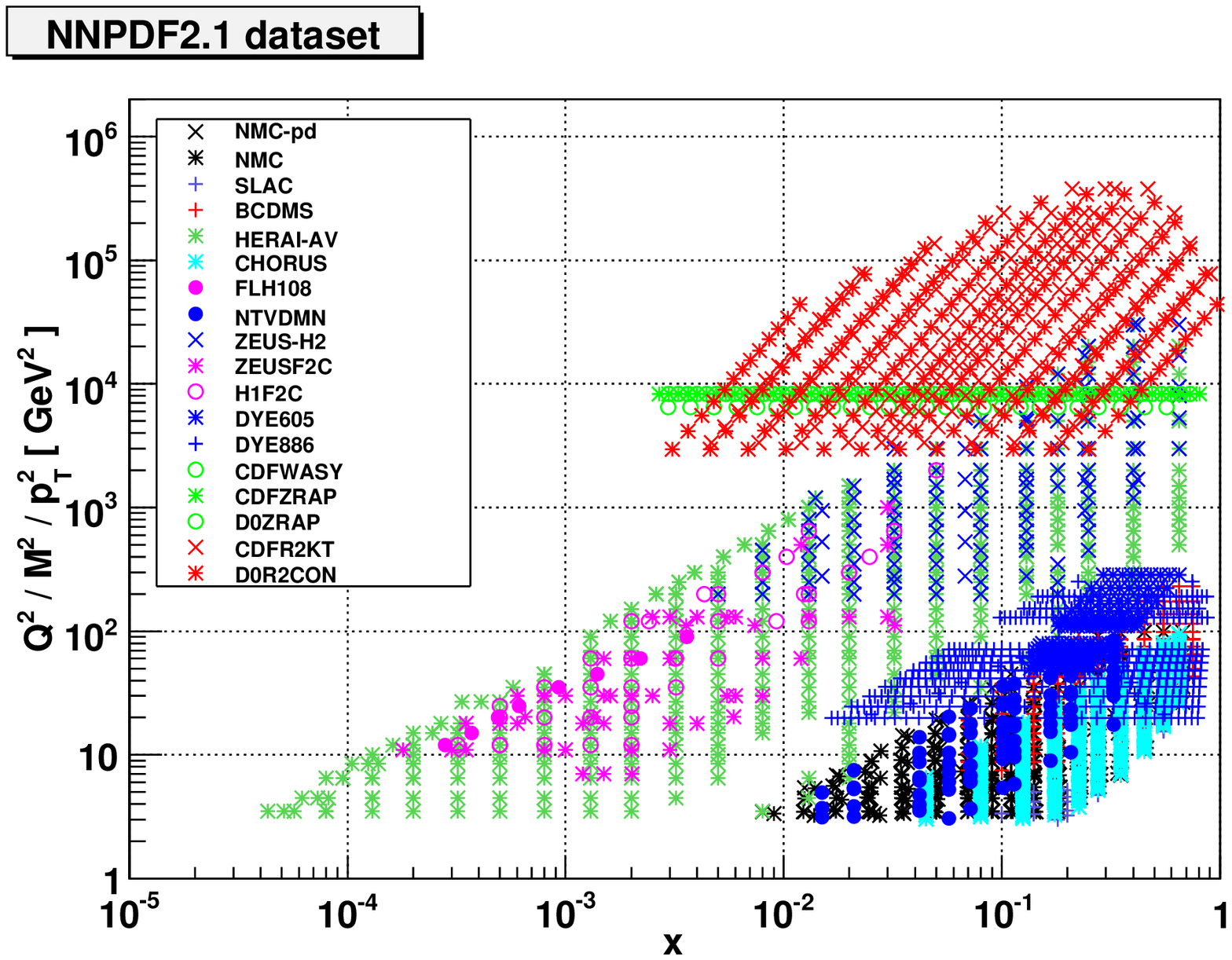}
\caption{ \small Experimental datasets 
which enter the NNPDF2.1 analysis. The kinematical coverage
of each dataset is summarized in
Table~\ref{tab:exp-sets}. 
\label{fig:dataplottot}} 
\end{center}
\end{figure}

Now we discuss the datasets that are included in the
present analysis.
As compared to the NNPDF2.0 analysis~\cite{Ball:2010de}, 
on top of all relevant data from DIS, Drell-Yan and weak vector
boson production\footnote{The impact of the leptonic W asymmetry data
from the Tevatron, not included in NNPDF2.0, as been studied in 
Ref.\cite{reweighting} using
the Bayesian reweighting technique. } and inclusive jet production
we include here all the relevant
charm structure function $F_2^c(x,Q^2)$ data from the H1 and ZEUS
experiments at HERA~\cite{Breitweg:1999ad,Chekanov:2003rb,Chekanov:2008yd,Chekanov:2009kj,Adloff:2001zj,Collaboration:2009jy,H1F2c10:2009ut}. 
These datasets provide a handle on the
small-$x$ gluon, and are sensitive also to the value
of the charm mass $m_c$.
On the other hand, HERA $F_2^b$ has
much larger uncertainties, and is thus not included in the
present analysis.
The kinematical coverage of
all the datasets included in NNPDF2.1 
is summarized in Table~\ref{tab:exp-sets}
and in Fig.~\ref{fig:dataplottot}. Note that 
the only differences with respect
the NNPDF2.0 dataset are the addition of HERA $F_{2}^c$ data and the
new kinematical cut $Q^2_{\rm min}=3$ GeV$^2$.

 \begin{table}
 \tiny
 \centering
 \begin{tabular}{|c|c|c|c|c|c|c|c|}
  \hline
\multicolumn{8}{|c|}{\bf Deep-Inelastic scattering}\\
\hline
  Experiment & Set & Ref. &$N_{\rm dat}$ & $x_{\rm min}$ & $x_{\rm max}$
 & $Q^2_{\rm min}$ [GeV$^2$]  & $Q^2_{\rm max}$ [GeV$^2$]   \\ \hline
\hline
 NMC-pd          & & &          260 (132)  & & & & \\ \hline
 &  NMC-pd          & \cite{Arneodo:1996kd}& 260 (132) &      0.0015 (0.008) &      0.68 &  0.2 (3.5) &  99.0 \\
 \hline
 NMC             & &  &        288 (221)  & & & &  \\ \hline
 &  NMC             &  \cite{Arneodo:1996qe} & 288 (221) &      0.0035 (0.009) &      0.47 &  0.8 (3.2) &  61.2  \\
 \hline
 SLAC            & &  &        422 (74)  & & & & \\ \hline
 &  SLACp           & \cite{Whitlow:1991uw} & 211 (37) &      0.07 (0.1) &      0.85 (0.55) &        0.58 (3.0) &   29.2  \\
 \hline
 &  SLACd           & \cite{Whitlow:1991uw} & 211 (37) &      0.07 (0.1) &      0.85 (0.55) &        0.58  (3.2)&   29.1  \\
 \hline
 BCDMS           & &   &       605 (581)  & & & &\\ \hline
 &  BCDMSp          &\cite{bcdms1}  &351 (333) &      0.07 &      0.75 &        7.5 &      230.0     \\
 \hline
 &  BCDMSd          &  \cite{bcdms2}  &254 (248) &      0.07 &      0.75 &        8.8 &      230.0     \\
 \hline
 HERAI-AV        & & &          741 (592) & & & & \\ \hline
 &  HERA1-NCep      & \cite{H1:2009wt} & 528 (379) &      $6.2\,10^{-7}$ ($4.3\,10^{-5}$) &      0.65 &        0.045 (3.5) &    30000  \\
 \hline
 &  HERA1-NCem      & \cite{H1:2009wt} & 145 &      $1.3\,10^{-3}$ &      0.65 &       90.000 &    30000  \\
 \hline
 &  HERA1-CCep      & \cite{H1:2009wt} &  34 &      0.008 &      0.4 &      300.0 &    15000 \\
 \hline
 &  HERA1-CCem      & \cite{H1:2009wt} &  34 &      0.013 &      0.4 &      300.0 &    30000  \\
 \hline
 CHORUS          & &  &       1214 (862) & & & & \\ \hline
 &  CHORUSnu        & \cite{Onengut:2005kv} & 607 (431) &      0.02 (0.045) &      0.65 &        0.3 (3.0) &       95.2  \\
 \hline
 &  CHORUSnb        & \cite{Onengut:2005kv} & 607 (431)&      0.02 (0.045) &      0.65 &        0.3 (3.0) &       95.2 \\
 \hline
 FLH108          & &  &          8  & & & & \\ \hline
 &  FLH108          & \cite{h1fl} &  8 &      0.00028 &      0.0036 &       12.0 &       90.000 \\
 \hline
 NTVDMN          & &   &        90 (79) & & & &\\ \hline
 &  NTVnuDMN        & \cite{Goncharov:2001qe,MasonPhD} & 45 (41) &      0.027 &      0.36 &        1.1 (3.1)&      116.5 \\
 \hline
 &  NTVnbDMN        & \cite{Goncharov:2001qe,MasonPhD} & 45 (38)&      0.021 &      0.25 &        0.8 (3.1) &       68.3 \\
 \hline
 ZEUS-H2         & &   &       127  & & & & \\ \hline
 &  Z06NC           &  \cite{Chekanov:2009gm}& 90 &      $5\,10^{-3}$ &      0.65 &      200 &    $3\,10^5$  \\
 \hline
 &  Z06CC           & \cite{Chekanov:2008aa} & 37 &      0.015 &      0.65 &      280 &    $3\,10^5$ \\
 \hline
 \hline
\multicolumn{8}{|c|}{\bf HERA charm structure function data}\\
\hline
 ZEUSF2C         & &    &       69 (50) & & & & \\ \hline
 &  ZEUSF2C99       & \cite{Breitweg:1999ad} &  21 (14) &      $5\,10^{-5}$ 
($3\,10^{-4}$) &      0.02 &        1.8 (7.0) &      130 \\
 \hline
 &  ZEUSF2C03       &  \cite{Chekanov:2003rb}& 31 (21) &      $3\,10^{-5}$ 
($1.8\,10^{-5}$)&      0.03 &        2.0 (7.0) &      500\\
 \hline
 &  ZEUSF2C08       & \cite{Chekanov:2008yd}  & 9 (7) &      $2.2\,10^{-4}$
( $6.5\,10^{-4}$) &      0.032 &        7.0 &      112  \\
 \hline
 &  ZEUSF2C09       &  \cite{Chekanov:2009kj} & 8 &      $8\,10^{-4}$ &      0.03 &       30 &     1000 \\
 \hline
 H1F2C           &  &     &      47 (38)  & & & & \\ 
 \hline
&  H1F2C01         & \cite{Adloff:2001zj} & 12 (6) &      $5\,10^{-4}$ &      $3.2\,10^{-3}$ &        1.5 (12) &       60   \\
\hline
 &  H1F2C09         & \cite{Collaboration:2009jy}  & 6 &      $2.4\,10^{-4}$ &      0.025 &      120 &      400  \\
\hline
 &  H1F2C10         & \cite{H1F2c10:2009ut}   &26 &      $2\,10^{-4}$ 
($3.2\,10^{-4}$) &      0.05 &        5.0 (12) &     2000  \\
\hline
\hline
\multicolumn{8}{|c|}{\bf Fixed Target Drell-Yan production}\\
\hline
 Experiment & Set & Ref. & $N_{\rm dat}$ & $\lc y/x^F_{\rm min},y/x^F_{\rm max}\rc $ &  $\lc x_{\rm min}, x_{\rm max}\rc$
 & $M^2_{\rm min}$ [GeV$^2$]  & $M^2_{\rm max}$ [GeV$^2$]
   \\ \hline
\hline
 DYE605          & &           &  119 & & &\\ \hline
 &  DYE605          & \cite{Moreno:1990sf} & 119 &     $\lc -0.20,0.40\rc$  & $\lc 0.14,0.65\rc$  &    50.5 &      286 \\
 \hline
 DYE866          & &   &       390  & & & &\\ \hline
 &  DYE866p         & \cite{Webb:2003ps,Webb:2003bj} &184 &     $\lc  0.0, 0.78\rc$ &  $\lc 0.017,0.87\rc$ &     19.8 &      251.2\\
 \hline
 &  DYE866r         & \cite{Towell:2001nh} & 15 &     $\lc 0.05 ,0.53\rc$  & $\lc 0.025,0.56\rc$ &       21.2 &      166.4 \\
 \hline
 \hline
\multicolumn{8}{|c|}{\bf Collider vector boson production}\\
\hline
 Experiment & Set & Ref.& $N_{\rm dat}$ & $\lc y_{\rm min},y_{\rm max}\rc $ &  $\lc x_{\rm min}, x_{\rm max}\rc$
 & $M^2_{\rm min}$ [GeV$^2$] & $M^2_{\rm max}$ [GeV$^2$]
  \\ \hline
 CDFWASY         & &   &        13  & & & & \\ \hline
 &  CDFWASY         & \cite{Aaltonen:2009ta}  & 13 &      $\lc 0.10, 2.63\rc$ &  $\lc 2.9\,10^{-3},0.56\rc$ &    6463 &     6463   \\
 \hline
 CDFZRAP         & &    &       29  & & & & \\ \hline
 &  CDFZRAP         & \cite{Abazov:2007jy} &  29 &     $\lc  0.05 ,2.85\rc$ & $\lc 2.9\,10^{-3},0.80\rc$ &     8315 &     8315 \\
 \hline
 D0ZRAP          & &   &        28  & & & &\\ \hline
 &  D0ZRAP          & \cite{Aaltonen:2009pc} & 28 &     $\lc  0.05, 2.75\rc $ & $\lc 2.9\,10^{-3},0.72\rc$ &    8315 &     8315 \\ 
 \hline
 \hline
\multicolumn{8}{|c|}{\bf Collider inclusive jet production}\\
\hline
 Experiment & Set & Ref. & $N_{\rm dat}$ &  $\lc y_{\rm min},y_{\rm max}\rc $ &  $\lc x_{\rm min}, x_{\rm max}\rc$
 & $p^{2}_{T,\rm min}$ [GeV$^2$] & $p^2_{T,\rm max}$   [GeV$^2$] \\ \hline
 CDFR2KT         & &    &       76  & & & &\\ \hline
 &  CDFR2KT         & ~\cite{Abulencia:2007ez}  & 76 &      $\lc 0.05 ,  1.85\rc$ & 
$\lc 4.6\,10^{-3},0.90\rc$  &     3364 &   $3.7\,10^5$  \\
 \hline
 D0R2CON         & &    &      110  & & & &\\ \hline
 &  D0R2CON         & \cite{D0:2008hua} & 110 &     $\lc  0.20, 2.20\rc $ & $\lc 3.1\,10^{-3},0.97\rc$  & 3000 & $3.4\,10^5$ \\
 \hline
 \hline
\multicolumn{8}{|c|}{\bf Total}\\
\hline
 Experiment  & &&  $N_{\rm dat}$ & $x_{\rm min}$ & $x_{\rm max}$
 & $Q^2_{\rm min}$ [GeV$^2$]  & $Q^2_{\rm max}$ [GeV$^2$]    \\ \hline
 TOTAL           &&& 4520 (3338)   &  $3.1\,10^{-5}$  & 0.97 & 2.0 & $3.7\,10^5$\\ 
\hline
 \end{tabular}
\caption{\small \label{tab:exp-sets} Experimental datasets included in the NNPDF2.1 global analysis. For DIS experiments we provide in each case the number
of data points and the ranges of the kinematical variables
before and after (in parenthesis) kinematical cuts. 
For hadronic
data we  show the ranges of parton $x$ covered for each
set determined using 
leading order parton kinematics. 
Note that  hadronic data are unaffected by
kinematic cuts. The values of
$x_{\rm min}$ and $Q^2_{\rm min}$ for the total dataset hold after
imposing
kinematic cuts.}
 \end{table}

Now we describe in turn the features of the various $F_2^c(x,Q^2)$ datasets
included in the present analysis. For most experimental sets 
the full correlation is
not available and thus one is forced to add in quadrature systematic
and statistical uncertainties. The full correlation matrix for all data points,
including the cross-correlations between datasets and between H1 and ZEUS
will be provided together with the combined HERA $F_2^c$ dataset: this
combination will thus significantly improve the accuracy of the existing
separate datasets.

The $F_2^c$ data which we use in the NNPDF2.1 analysis are the following:

\begin{itemize}

\item The ZEUS 96-97  $D^{*\pm}$ 
analysis~\cite{Breitweg:1999ad}. \\In this analysis
$F_2^c$ is extracted from the measurement of $D^{*\pm}$ mesons
reconstructed via their hadronic decays using data collected in
the 1996 and 1997 running periods.

\item The ZEUS 98-00 $D^*$ analysis~\cite{Chekanov:2003rb}.\\ As in
the previous case, $F_2^c$ is extracted from the
 measurement of $D^{*\pm}$ mesons
reconstructed via their hadronic decays, and uses  data collected in
the  running period between 1998 and 2000.

\item The 04-05 ZEUS $D^{\pm},D^0$ 
 analysis~\cite{Chekanov:2008yd}. \\In this analysis, based on
the HERA-II running period of 2004 and 2005, $D$ mesons are
reconstructed via their hadronic decays. An improved precision
is obtained reducing the combinatorial background to the $D$ meson signals 
 by using the ZEUS micro-vertex detector to reconstruct displaced 
secondary vertices

\item The 2005 ZEUS muon analysis~\cite{Chekanov:2009kj}. \\This dataset
 is based on the measurement of muons that are generated
in charm production from their semileptonic decays. Data was
collected during the 2005 HERA-II running period.

\item The H1 96-97 $D^{*\pm}$ analysis~\cite{Adloff:2001zj}. \\This
analysis, based on the 1996-1997 running period, used similar
reconstruction strategies as the corresponding ZEUS analysis,
namely the reconstruction of $D^{*\pm}\to D^0\pi^+$
using the $D^*-D^0$ mass difference method.

\item The H1 large $Q^2$ 04-07 $D^{*\pm}$ analysis~\cite{Collaboration:2009jy}.\\
  This analysis
determines $F_2^c$ via identified $D$ mesons produced
at large virtualities $Q^2 \ge 100$ GeV$^2$, and  is based
on data collected in the HERA-II running period 2004 and 2007.
 
\item The H1 low-$Q^2$ 06-07 $D^{*\pm}$ analysis~\cite{H1F2c10:2009ut} \\
This is analogous to the previous  measurement, but now covering the
small and medium $Q^2$ region. It is based on data obtained in the
HERA-II
2006-2007 running period. Events containing heavy quarks are 
distinguished from those containing only light quarks 
using variables that are sensitive to the longer lifetimes of heavy 
flavour hadrons, like the transverse displacement of tracks from the primary vertex.

\end{itemize}

There are more published $F_2^c$ datasets from HERA but the ones
that are included here supersede previous obsolete measurements and
are the basis of the combined HERA $F_2^c$ dataset.
In Sect.~\ref{sec:results} we will quantify the 
impact of the HERA $F_2^c$ data onto the PDFs.

A concern with  $F_2^c$ data which has been sometimes used to
motivate their exclusion from PDF determinations is the fact that the
way $F_2^c$  is usually defined experimentally, as the contribution to
$F_2$ with at least one charmed quark in the final state, is affected
by mass singularities (i.e., it is not finite in the limit in which
$m_c\to 0$). Here we will adopt a definition of $F_2^c$ (as the
contribution to $F_2^c$ when only the charm electric charge is
nonzero)  which is free of mass
singularities; the deviation between this definition and that which is
used to define the experimental observable is estimated in 
Ref.~\cite{Forte:2010ta} by means of a suitable resummation method,
and shown to be negligible in the region of the HERA data.
Also, $F_2^c$ is  affected by  theoretical uncertainties related
to the extrapolation from the experimentally accessible region
(restricted in $p_T$ and $\eta$) to the full phase space. This
theoretical uncertainty is estimated using QCD exclusive partonic
calculations and added as an extra source of systematic uncertainty
in the experimental analysis.

\subsection{Positivity constraints}

As discussed in~\cite{Ball:2010de}, within the
NNPDF framework
general theoretical constraints can be imposed 
guaranteeing that the
fitting procedure only explores the subspace of acceptable 
physical solutions.  An important theoretical
constraint is the positivity of physical cross--sections.
As discussed in Ref.~\cite{Altarelli:1998gn}, positivity should 
be imposed on  observable hadronic cross--sections and not on partonic
quantities, which do not necessarily satisfy this constraint
(except at leading order where the probabilistic interpretation
holds). Positivity constraints may be implemented in various
ways; here we will impose them through Lagrange multipliers, i.e. in
practice by adding
pseudo-datasets for physical cross sections with extremely small
uncertainties in such a way that negative cross sections would lead to
a very large contribution to the $\chi^2$.

In NNPDF2.1 we impose positivity of the following
observables:
\begin{itemize}
\item  The longitudinal structure function $F_L(x,Q^2)$, which
constrains the gluon positivity at small--$x$.
\item The charm production cross section in
neutrino DIS, $d^2\sigma^{\nu,c}/dxdy$~\cite{Ball:2009mk},
which
constrains the strange PDFs both at large and at small-$x$,
beyond the reach of existing data.
\item The neutral current DIS 
charm structure function $F_2^c(x,Q^2)$, useful to
impose the positivity of the gluon at very large-$x$, where it
is not constrained by any experimental dataset. 
\end{itemize}

All the positivity constraints are implemented at a low scale 
$Q^2_{\rm pos}$
that we take to be $Q^2_{\rm pos}=2$ GeV$^2$, in the range 
$x\in \lc 10^{-6},x_{\rm max}\rc$, where $x_{\rm max}$ is the
corresponding kinematical boundary, 
$x_{\rm max} \sim 0.1$ for
NC scattering and $x_{\rm max}\sim 0.5$ for CC scattering. 
DGLAP evolution then takes care or
preserving the positivity properties for higher scales.
We note that the physical observables for the pseudo-data that
implement the positivity constraints are computed consistently
at the same perturbative order as all other physical observables,
in the present case next--to--leading order perturbative QCD.

\section{Structure functions with heavy quark mass effects}
\label{sec:evolution}

The FONLL-A general-mass scheme was introduced for neutral current
structure functions in Ref.~\cite{Forte:2010ta}. We begin this
Section with a brief review of this scheme, emphasizing the impact of
heavy quark effects on DIS structure functions. We then discuss the
values of the heavy quark masses and the associated uncertainties
adopted in the present analysis. The corresponding analysis for charged
current structure functions is presented in the last part of this Section.

\subsection{The FONLL-A General Mass scheme for NC structure functions}

The FONLL general--mass scheme, originally proposed in the context of
heavy quark photo- and hadro-production, was generalized in
Ref.~\cite{Forte:2010ta} to deep--inelastic structure functions. We
refer the reader to Ref.~\cite{Forte:2010ta} for a detailed discussion
of the scheme, and for the notation adopted in this Section. The FONLL
approach allows for a consistent combination of terms determined in a
massive, or decoupling, or fixed-flavour number (FFN) scheme, in which
the heavy quark is subtracted at zero momentum (rather in the
$\overline{\rm MS}$ scheme), so it decouples for scales much below its
mass, and it is included in Feynman diagrams up to some fixed order in
$\alpha_s$ above its threshold for its production, with terms
determined in a massless, or zero-mass (ZM), or simply $\overline{\rm
  MS}$ scheme, in which the heavy flavour is treated as another
massless parton, so it is included in the all-order resummation of
collinear logarithms, up to a suitable chosen logarithmic order (LO, NLO, etc).
A significant
feature of the FONLL method is that the fixed perturbative  order of
the FFN computation and the resummed logarithmic order of the ZM
computation which are being combined  can be chosen
independently of each others.

In the present analysis, we combine FFN massive terms up to order $\alpha_s$
with a NLO ZM computation; this is called FONLL-A in Ref.~\cite{Forte:2010ta}.
As shown in Ref.~\cite{LHhq}, this turns out to be identical to the
S-ACOT~\cite{Kramer:2000hn} scheme used in recent CTEQ/CT  PDF
determinations~\cite{Nadolsky:2008zw,Lai:2010vv}.
Once a specific ``general mass'' (GM) scheme for the combination of
FFN and ZM terms has been chosen, there is still a freedom in the
treatment of subleading terms: indeed, it turns out to be
phenomenologically convenient to suppress subleading terms near the
quark threshold (see  Ref.~\cite{Nadolsky:2009ge,Forte:2010ta}). 
In this work we adopt the so-called threshold or damping factor method
of Ref.~\cite{Forte:2010ta} for the treatment of subleading
terms. In Ref.~\cite{LHhq} the damping factor method is 
benchmarked against various implementations of
the alternative, commonly used $\chi$--scaling
method
for the treatment of subleading terms. 

We now present the explicit
expressions for the $F_{2,h}$ heavy quark structure
function~\footnote{See Ref.~\cite{Forte:2010ta} for the discussion on
  the FONLL expressions for the longitudinal structure
  functions.}. The FONLL-A heavy quark structure function is given by
the sum of two terms:
\begin{equation}
  F_{2,h}^{\tmop{FONLL}} (x, Q^2) =F_{2,h}^{(n_l)} (x,
  Q^2)+\theta\lp Q^2-m_h^2\rp\lp 1- \frac{m_h^2}{Q^2}
\rp^2 F^{(d)}_{2,h} (x, Q^2). 
 \label{eq:FONLLnlo}
\end{equation}
The first contribution on the right-hand side of
Eq.~(\ref{eq:FONLLnlo}) is the massive-scheme heavy-quark structure
function at $\mathcal{O}\lp \alpha_s\rp$:
\begin{equation}
  F_{2,h}^{(n_l)} (x, Q^2)  =  x \int_x^1 \frac{dy}{y} C^{(n_l)}_{2,g} \left(
  \frac{x}{y}, \frac{Q^2}{m_h^2},\alpha_s(Q^2) \right) g^{(n_l+1)}
  (y,Q^2)\, .
\label{eq:hqnlo}
\end{equation}
The heavy quark gluon coefficient function is given by
\begin{equation}
  C^{(n_l)}_{2,g} \left( z, \frac{Q^2}{m_h^2}, \alpha_s(Q^2) \right) =
  \frac{\alpha_s(Q^2)}{2\pi} 2 e_h^2 C_{2,g}^{(n_l),1} \left( z, \frac{Q^2}{m_h^2} \right) + 
  \mathcal{O}\lp \alpha_s^2\rp\, . \label{eq:C0nl}
\end{equation}
The $\mathcal{O}\lp \alpha_s\rp$ coefficient is
\begin{eqnarray}
\label{eq:cgnl1}
  C_{2,g}^{(n_l),1} \left( z, \frac{Q^2}{m_h^2} \right) & = & \theta \left( W^2 - 4
  m_h^2 \right) \times T_R [ (z^2 + (1 - z)^2 + 4 \epsilon z (1 - 3 z) - 8
  \epsilon^2 z^2) \log \frac{1 + v}{1 - v} \nonumber\\
  &  & + (8 z (1 - z) - 1 - 4 \epsilon z (1 - z)) v \left. \right], \label{eq:C0nlc}
\end{eqnarray}
where we have defined
\begin{equation}
 \epsilon \equiv m^2_h / Q^2,\quad v
  \equiv \sqrt{1 - 4 m^2_h / W^2} ,\label{eq:kindefns}
\end{equation}
and the partonic center of mass energy $W^2=Q^2(1-z)/z$.

The second term on the right-hand side of Eq.~(\ref{eq:FONLLnlo}) is
the ``difference'' contribution
\begin{eqnarray}
  F^{(d)}_{2,h} (x, Q^2) & = & x \int_x^1 \frac{dy}{y} \left[ C_{2,q}^{(n_l+1)} \left(
  \frac{x}{y}, \alpha_s(Q^2) \right) \lc h^{(n_l+1)} (y, Q^2) + \bar{h}^{(n_l+1)} (y, Q^2)\rc  +
  \right. \phantom{aaaaa} \nonumber\\
  &  & \left. \left( C_{2,g}^{(n_l+1)} \left(
  \frac{x}{y}, \alpha_s(Q^2) \right) -   {B}^{(0)}_{g,\, h} \left( \frac{x}{y},
  \frac{Q^2}{m^2_h},\alpha_s(Q^2)  \right) \right) g^{(n_l+1)} (y, Q^2) \right],  \label{eq:fdnlo}
\end{eqnarray}
where $h,\bar{h}$ are the heavy quark parton distributions; at
first-order in $\alpha_s$, ${B}^{(0)}_{g,\, h}$ is given by
\begin{equation}
  \label{eq:mznlo}
        {B}^{(0),\,1}_{g,\, h} \left( z, \frac{Q^2}{m^2_h} \right) =  2 e_h^2 
        {C}_{2,g}^{(n_l,0), 1} \left( z, \frac{Q^2}{m^2_h} \right), 
\end{equation}
and the massless limit of the massive coefficient function
is 
\begin{equation}
  \label{eq:mznloexp}
        {C}_{2,g}^{(n_l,0), 1}\left( z, \frac{Q^2}{m^2_h} \right) =  T_R \left[
          (z^2 + (1 - z)^2) \log \frac{Q^2 (1 - z)}{m^2_h z} 
          + (8 z (1 - z) - 1) \right]
        \, , 
\end{equation}
which in the limit $Q^2=m^2_h$ reproduces as required the usual 
massless scheme coefficient function. 

Note that in all terms in Eq.~(\ref{eq:FONLLnlo}) PDFs and $\alpha_s$
are expressed in the
same factorization scheme, namely, the decoupling $n_f=3$
scheme. Exploiting this fact, it is easy to check explicitly that the
``difference''  term Eq.~(\ref{eq:fdnlo}) is formally of
higher order near the heavy quark threshold~\cite{Forte:2010ta} (and
thus in particular it can be suppressed using a suitable threshold
prescription). 

\begin{figure}[t!]
\begin{center}
\includegraphics[width=0.99\textwidth]{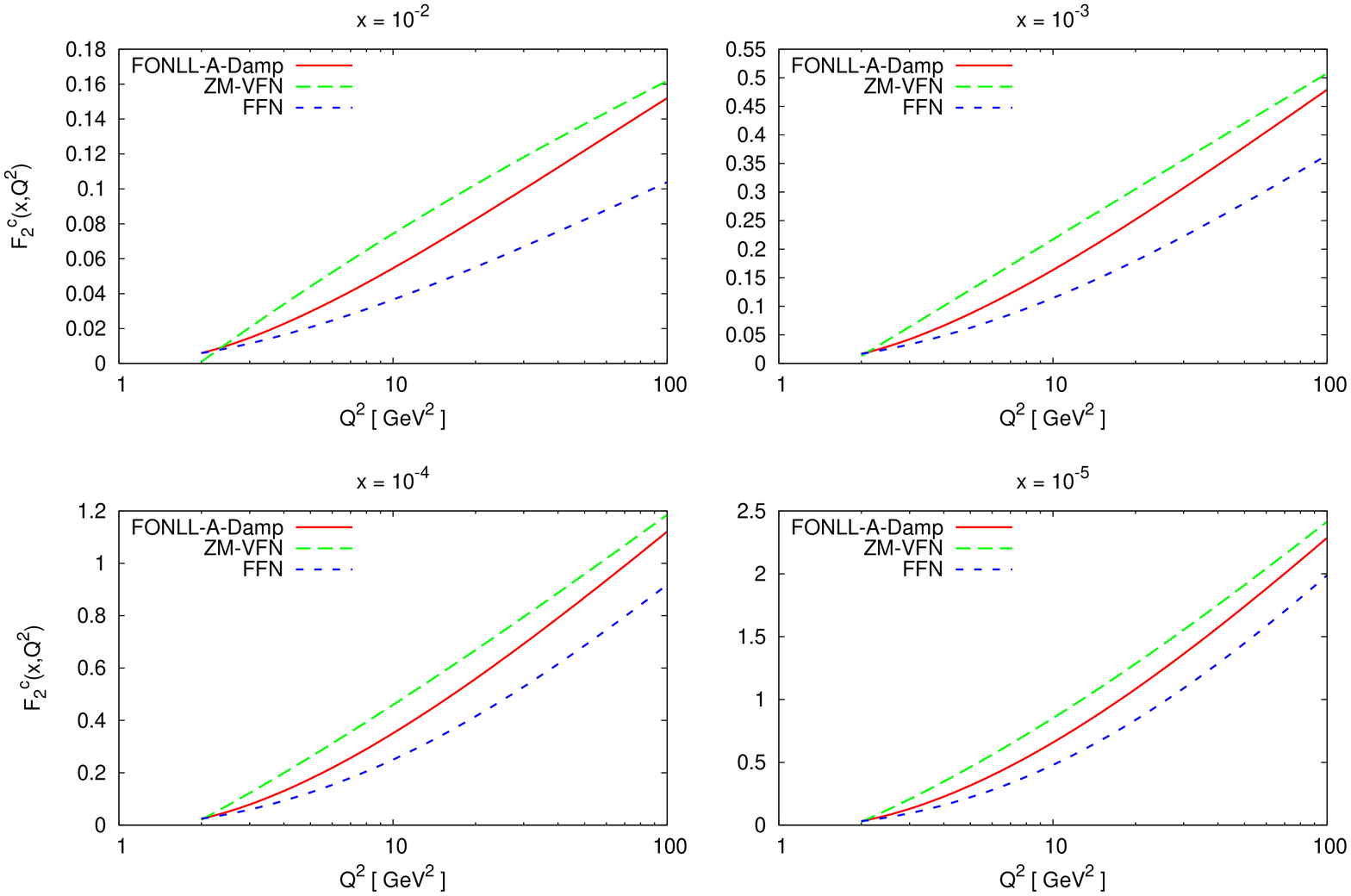}
\includegraphics[width=0.99\textwidth]{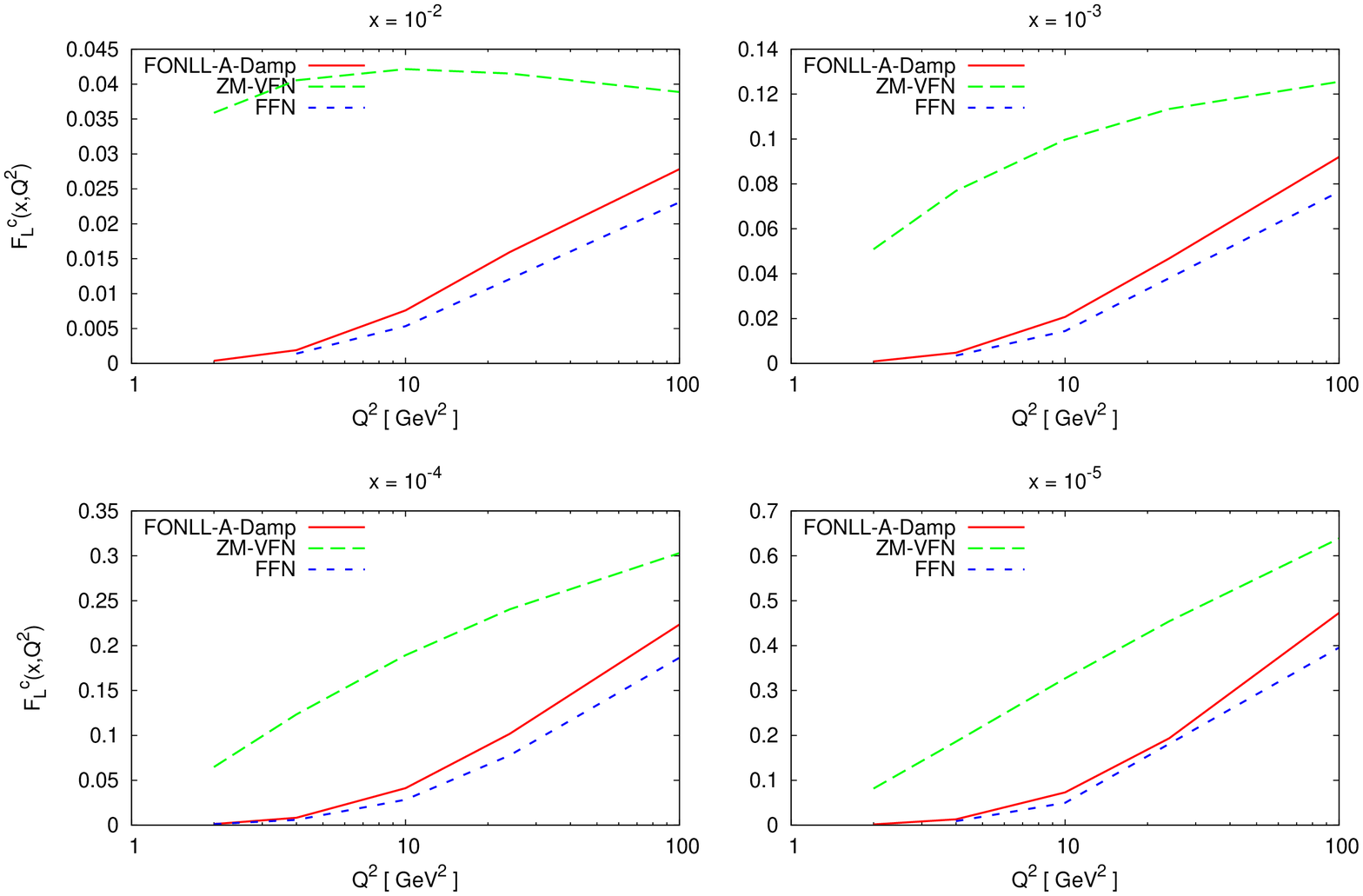}
\end{center}
\caption{\small The charm structure functions $F_{2,c}(x,Q^2)$ and
  $F_{L,c}(x,Q^2)$ as a function of $Q^2$ for different values of $x$
  from $x=10^{-5}$ to $x=10^{-2}$ in various heavy quark schemes,
  computed using the FastKernel method: FONLL-A, ZM-VFN and the FFN
  scheme. The PDFs and settings are identical to those of the Les
  Houches heavy quark benchmark comparison.}
\label{fig:f2c_vs_q2}
\end{figure}

Eq.~(\ref{eq:FONLLnlo}) interpolates smoothly between
the massive scheme at small $Q^2$ and the massless scheme suitable at
large $Q^2$.  As an illustration of the differences between various
schemes for the heavy quark structure functions, in
Fig.~\ref{fig:f2c_vs_q2} we compare the $F_{2,c}$ and the
$F_{L,c}$ charm structure functions for various schemes: ZM, FONLL-A
and the FFN scheme as a function of $Q^2$ for different values of
$x$. It is clear that  FONLL-A interpolates smoothly between
the FFN scheme near threshold and the massless scheme at large
$Q^2$ (also thanks to the use of a damping factor in
Eq.~(\ref{eq:FONLLnlo}).
For this comparison, PDFs and other settings, like the value of $m_c$,
are identical to those of the Les Houches heavy quark benchmark
comparison~\cite{LHhq}. The comparison for the longitudinal structure
function $F_{L,c}$ shows that mass effects are much larger than in
$F_{2,c}$, so the ZM computation is completely unreliable.

\begin{figure}[t!]
\begin{center}
\includegraphics[width=0.49\textwidth]{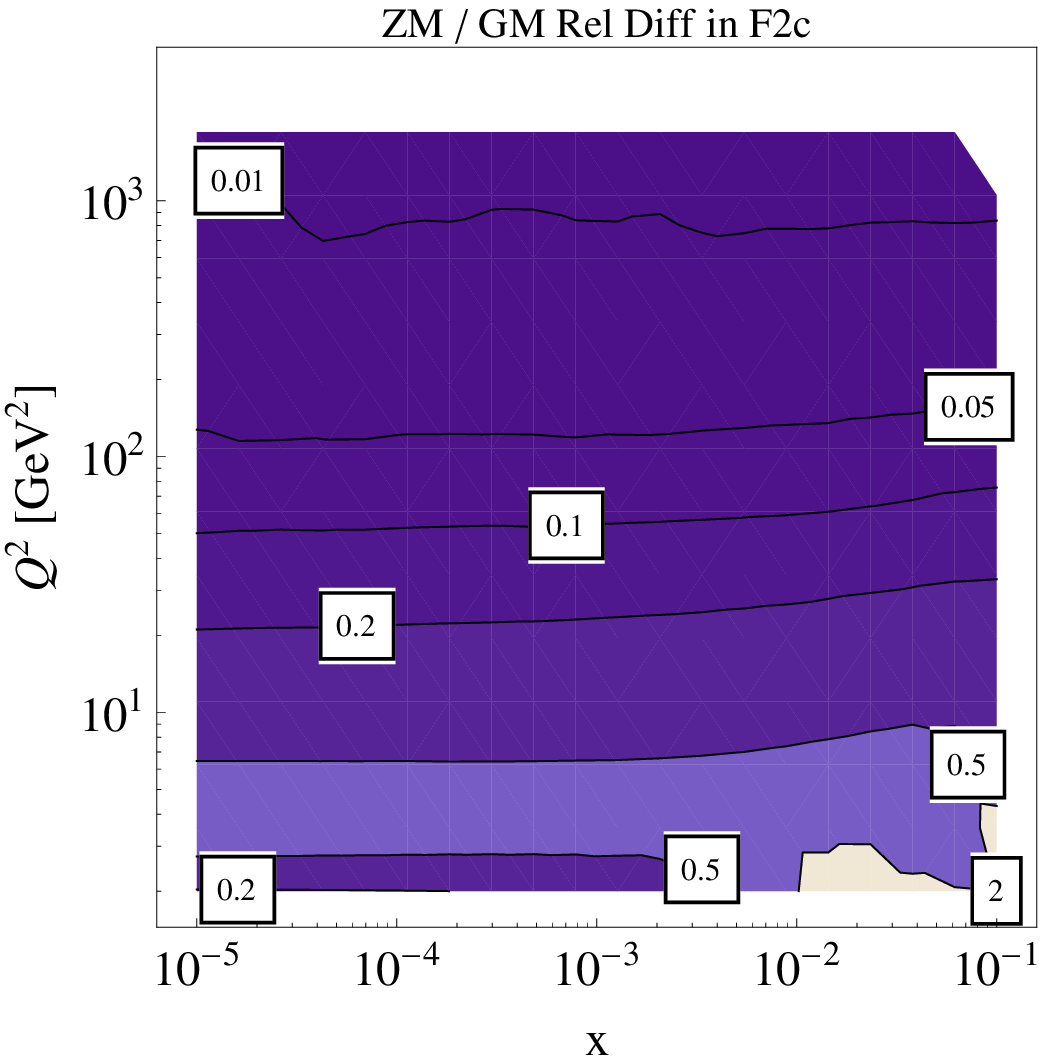}
\includegraphics[width=0.49\textwidth]{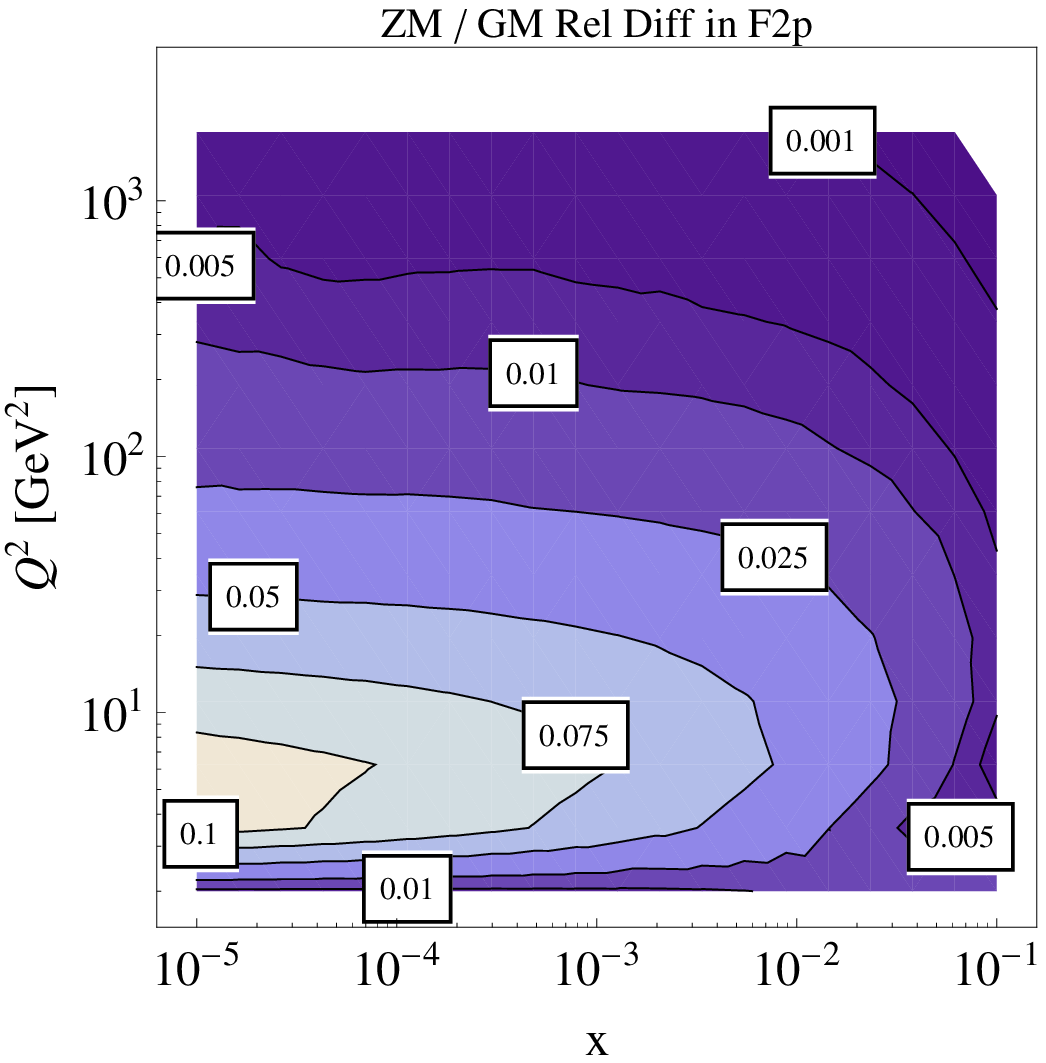}
\includegraphics[width=0.49\textwidth]{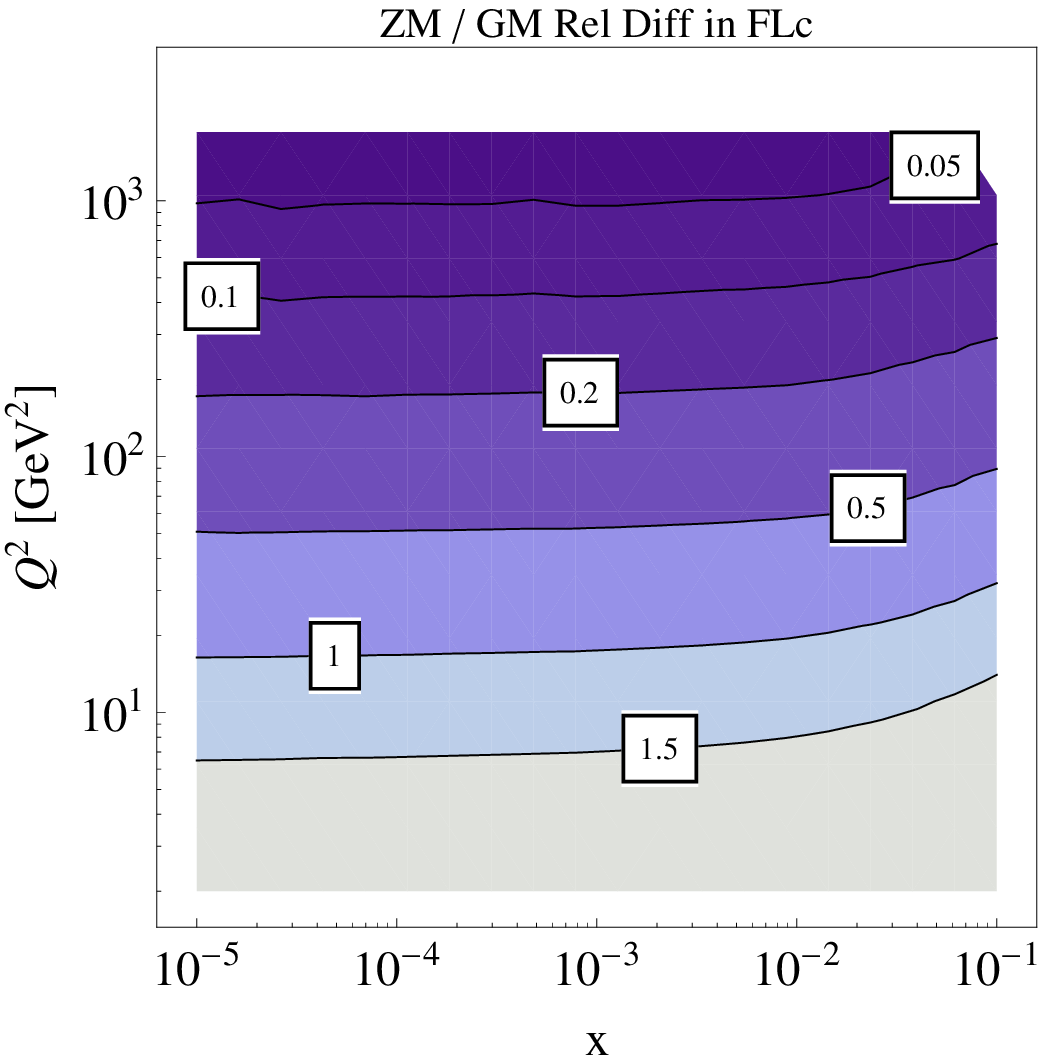}
\includegraphics[width=0.49\textwidth]{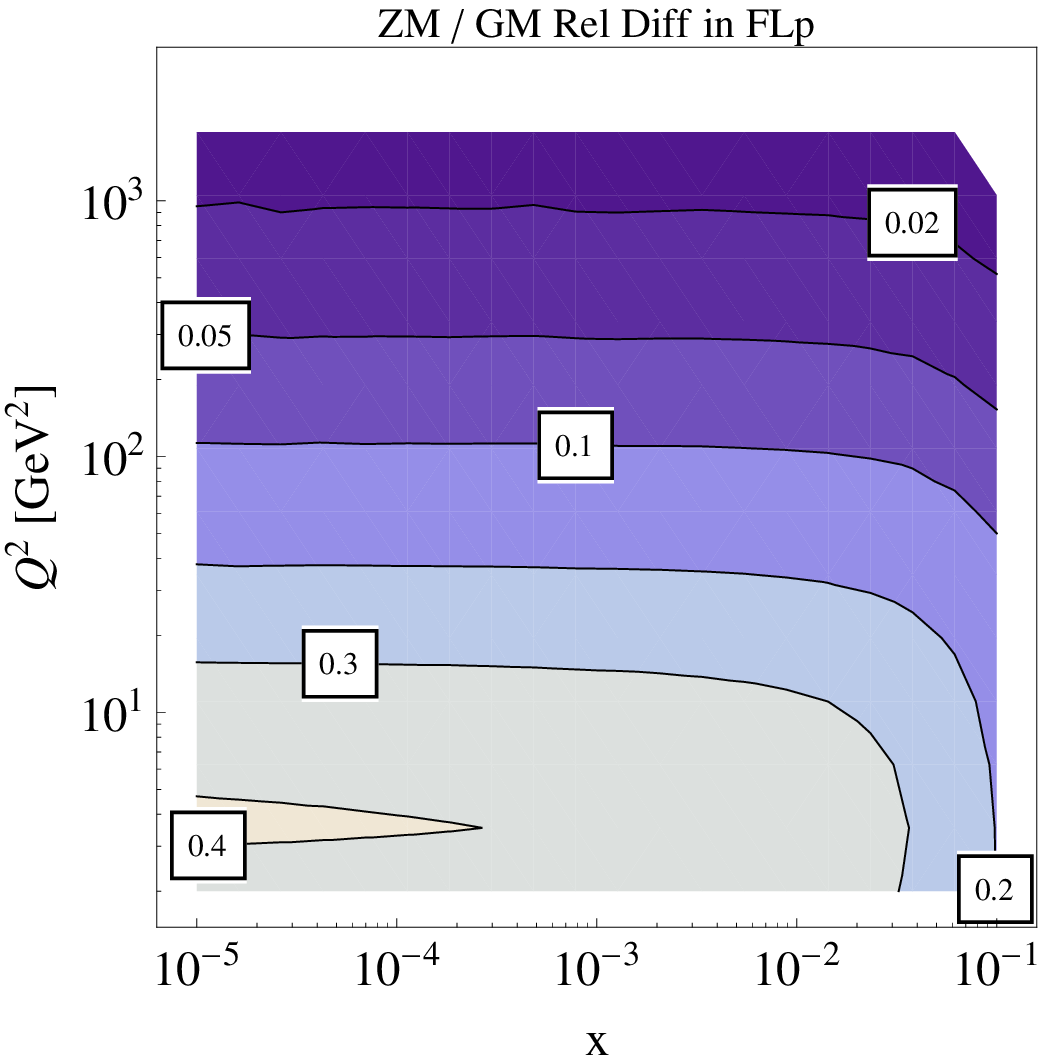}
\end{center}
\caption{\small Left plots: the relative difference for the charm
  structure functions $F_{2,c}(x,Q^2)$ and $F_{L,c}(x,Q^2)$ computed in
  the ZM and FONLL-A schemes as a function of $x$ and $Q^2$. The PDFs
  and settings are identical to those of the Les Houches heavy quark
  benchmark comparison. Right plots: the same but now for the
  inclusive structure functions $F_2^p(x,Q^2)$ and $F_L^p(x,Q^2)$.}
\label{fig:f2c_flc_cont}
\end{figure}

The impact of heavy quark mass effects in DIS structure functions is
further quantified in Fig.~\ref{fig:f2c_flc_cont}, where the relative
difference between the ZM and FONLL-A schemes is computed as a
function of $x$ and $Q^2$, both for the inclusive structure functions
$F_2^p$ and $F_L^p$ and for the charm structure functions $F_{2,c}$
and $F_{L,c}$. For the phenomenologically more relevant case of
$F_2^p$, we see that heavy quark mass effects can be as large as $\sim
10\%$, decreasing fast for increasing $x$ and $Q^2$. As in the case of
Fig.~\ref{fig:f2c_vs_q2} the Les Houches heavy quark benchmark
settings have been used.  Note that while the qualitative features of
Fig.~\ref{fig:f2c_flc_cont} are general, the quantitative detail can
depend on specific features of the general--mass heavy quark scheme,
like for example the prescription to suppress the subleading threshold
terms.

\begin{figure}[t!]
\begin{center}
\includegraphics[width=0.60\textwidth]{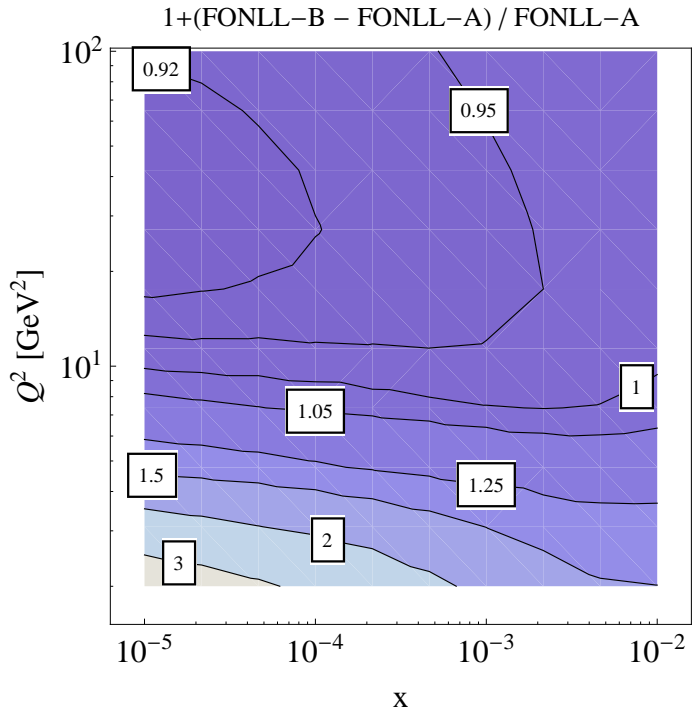}
\end{center}
\vspace{-0.5cm}
\caption{\small Relative difference for the charm structure functions
  $F_{2,c}(x,Q^2)$ between the FONLL-B and FONLL-A general-mass
  schemes, in units of FONLL-A, as a function of $x$ and $Q^2$. The
  PDFs and settings are identical to those of the Les Houches heavy
  quark benchmark comparison. }
\label{fig:f2c_fonll-confr}
\end{figure}

As discussed in Ref.~\cite{Forte:2010ta} it is possible to account
for the phenomenologically relevant $\mathcal{O}\lp \alpha_s^2\rp$
corrections to $F_{2,c}$ into an NLO PDF fit by means of the FONLL-B
scheme. We show in Fig.~\ref{fig:f2c_fonll-confr} the relative
difference between $F_{2,c}$ computed in the FONLL-B and FONLL-A
schemes. As shown in the figure, their difference at small $x$ and
$Q^2$ is rather large. The inadequacy of $\mathcal{O}\lp \alpha_s\rp$
theory to describe the low $x$ and $Q^2$ $F_{2,c}$ data motivates the
cuts to the $F_{2,c}$ HERA datasets discussed in
Sect.~\ref{sec:kincuts}. At larger values of $x$ and $Q^2$ the
differences between the two schemes become of the order of a few
percent, much smaller than the typical experimental uncertainties,
thus validating the inclusion of the $F_{2,c}$ data into the present fit
based on the FONLL-A scheme.

The $\mathcal{O}\lp \alpha_s\rp$ massive scheme heavy quark
coefficient function, Eq.~(\ref{eq:cgnl1}), was first computed in
Refs.~\cite{Witten:1975bh,Shifman:1977yb,Leveille:1978px}, while
its Mellin
transform, hitherto not available\footnote{A numerical parametrization of the Mellin
  space heavy quark coefficient functions up to $\mathcal{O}\lp
  \alpha_s^2\rp$ was provided in Ref.~\cite{Alekhin:2003ev}.} is
presented
  in 
Appendix~\ref{sec:massive-nc}.  
Details of the implementation of the FONLL-A scheme in the FastKernel
framework used in the NNPDF analysis are also given in
Appendix~\ref{sec:massive-nc}.
We have assessed the accuracy FONLL implementation computing the Les
Houches heavy quark benchmark tables~\cite{LHhq}, showing that the
accuracy is sufficient for precision PDF determination.

\subsection{FONLL Charged Current structure functions}
\label{sec:pdfevol-cc}
\label{sec:evolution-cc}

The FONLL method for  charged currents was only mentioned briefly in
Ref.~\cite{Forte:2010ta}: here we provide a detailed explanation
of this general--mass scheme for CC structure functions.  
In the charged current sector, coefficient functions are only known up
to $\mathcal{O}(\alpha_s)$, so the FONLL-A scheme can be constructed,
while the  FONLL-B and C which are based on
unavailable\footnote{Partial 
knowledge  of $\mathcal{O}(\alpha_s^2)$ massive terms  is
available in the form of the asymptotic~\cite{Buza:1997mg} and
threshold limits~\cite{Corcella:2003ib}.}
$\mathcal{O}(\alpha_s^2)$
massive results cannot.

Heavy quark mass effects are required to describe charm production in
neutrino DIS (the dimuon process) and to a lesser extent also the
inclusive neutrino reduced cross sections, since in both cases most of
the data lie close to the charm threshold, $Q^2 \gsim m_c^2$. HERA
charged current data on the other hand are at large $Q^2$ and thus for
practical purposes any general--mass scheme reduces to the ZM-VFN
scheme.

Here we generalize the FONLL-A scheme to
charged current structure functions.  Its implementation in the
FastKernel framework requires the analytic computation of the Mellin
transforms of the $\mathcal{O}\lp \alpha_s\rp$ charged current heavy
quark coefficient functions~\cite{Gluck:1996ve}. The detailed
description of the implementation is given in
Appendix~\ref{sec:massive-cc}.  We then benchmark the FONLL
implementation in FastKernel, using the Les Houches heavy quark
benchmark settings~\cite{LHhq} comparing with a $x$--space code written
for this purpose that implements FONLL-A for CC structure functions.
For simplicity, we will make the assumption
that $|V_{\rm cs}|=1$, and the rest of the CKM matrix elements are
zero. The generalization to realistic CKM elements,
as actually implemented in FastKernel, is straightforward.  We assume
also a single heavy quark, the charm quark with mass $m_c$.  The
factorization scale is set to be equal to $\mu_F^2=Q^2$.  Finally, we
consider only neutrino induced charm production, the anti-neutrino
case is again straightforward.

\begin{figure}[ht]
  \begin{center}
    \includegraphics[width=0.30\textwidth]{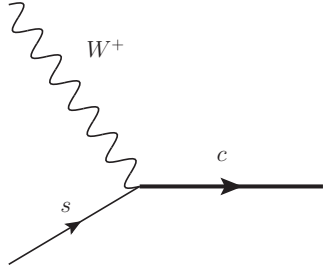}
  \end{center}
  \caption{\small Feynman diagram for the LO contribution to
    $F_{2,c}^{(n_l),\rm CC}$ in the FFNS scheme. Thick solid lines indicate a heavy quark (charm) and thin
    solid lines a light quark (strange). }
  \label{fig:ffns}
\end{figure}

\begin{figure}[ht]
  \begin{center}
    \includegraphics[width=0.60\textwidth]{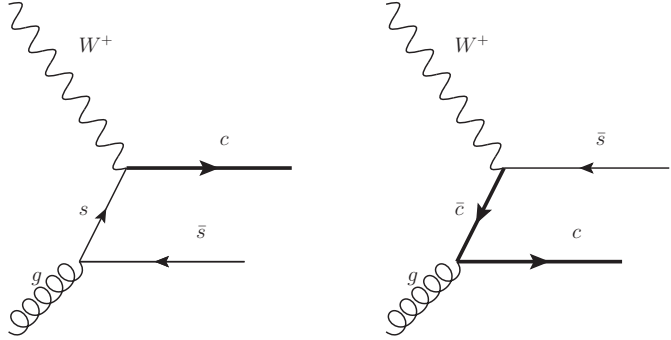}
  \end{center}
  \caption{\small Feynman diagrams for the NLO ($\mathcal{O}\lp
    \alpha_s\rp$) gluon-induced
    contribution 
to
    $F_{2,c}^{(n_l),\rm CC}$ in the FFNS scheme.}
  \label{fig:ffns2}
\end{figure}

In the FFN massive scheme, the charged current charm production
$F_{2,c}^{\rm CC}$ structure function for neutrino induced scattering
has been computed in $x$  space in
Refs.~\cite{Gottschalk:1980rv,Gluck:1996ve}: 
\begin{eqnarray}
  F_{2,c}^{(n_l),\rm
    CC}(x,Q^2)&=&2\xi s\lp\xi,Q^2\rp \nonumber \\ 
  &+& 2\xi
  \frac{\alpha_s\lp Q^2\rp}{2\pi}
  \bigg\{\int\limits_{\xi}^{1}\frac{dz}{z}\bigg[C^{(n_l),1}_{2,h}(z,Q^2,\lambda)
    s\lp \frac{\xi}{z},Q^2\rp \nonumber \\ 
    &+&C^{(n_l),1}_{2,g}(z,Q^2,\lambda)g\lp \frac{\xi}{z},Q^2\rp
    \bigg]\bigg\} \ ,
  \label{eq:f2cffns}
\end{eqnarray}
where 
\begin{equation}
  \xi=x\lp 1+\frac{m_c^2}{Q^2}\rp \ , \qquad \lambda\equiv
  \frac{Q^2}{Q^2+ m_c^2} \, .
\end{equation}
In Eq.~(\ref{eq:f2cffns}),   $C^{(n_l),1}_{2,g}$ includes the
contributions in which the gluon splits into a $s$ and a $\bar{c}$
quark, both of which contribute to
$F_{2,c}^{(n_l),\rm CC}$ at NLO. The Feynman diagrams for the
LO and NLO gluon-induced 
subprocesses are shown in Figs.~\ref{fig:ffns},~\ref{fig:ffns2}.

The $x$--space expressions for the $\mathcal{O}\lp \alpha_s\rp$
charged current coefficient functions in Eq.~(\ref{eq:f2cffns}) are
given in Refs.~\cite{Gottschalk:1980rv,Gluck:1996ve}.  The quark
coefficient function can be separated into a delta function piece, a
regular piece and a singular piece regulated with the usual plus
prescription, 
\begin{eqnarray}
  C^{(n_l),1}_{2,h}(z,Q^2,\lambda) =
  C^{(n_l)}_{h,\delta}\lp \lambda \rp \delta\lp 1-z \rp +
  C^{(n_l)}_{h,{\rm r}}\lp \lambda, z \rp + \lc C^{(n_l)}_{h,{\rm s}}\lp
  \lambda, z, Q^2 \rp\rc_+\, .
  \label{eq:qcoeffnlopieces}
\end{eqnarray}
The explicit expressions for the different pieces are the
following. For the delta term we have
\begin{equation}
 C^{(n_l)}_{h,\delta}\lp \lambda \rp = -C_F \lp 4 + \frac{1}{2\lambda}
 +\frac{\pi^2}{3} + \frac{1+\lambda}{2\lambda}K_A \rp \ , \ee \be K_A
 = \lp 1-\lambda \rp\ln \lp 1-\lambda \rp/\lambda \ .
\end{equation}
The regular piece can be written as
\begin{eqnarray}
  C^{(n_l)}_{h,{\rm r}}\lp \lambda, z  \rp &=& C_F\Bigg[
    -\lp 1+z\rp\lp 2\ln\lp 1-z\rp - \ln\lp 1-\lambda z\rp\rp -\frac{\lp 1+z^2\rp\ln z}{1-z} \nonumber
    \\
    &+& \lp 2z+2-\frac{2}{z}\rp + \lp \frac{2}{z}-1-z\rp\frac{1}{1-\lambda z} 
    \Bigg] \, ,
\end{eqnarray}
and finally the singular piece reads
\begin{eqnarray}
  C^{(n_l)}_{h,{\rm s}}\lp \lambda, z ,Q^2 \rp &=& C_F\Bigg[ -\frac{1+z^2}{1-z}
    \ln \frac{Q^2+m_c^2}{Q^2} + 2\frac{2\ln \lp 1-z\rp-\ln \lp 
      1-\lambda z\rp}{1-z}\nonumber \\ 
    &-&\frac{2}{1-z} + \frac{1}{2}\frac{1-z}{\lp 
      1-\lambda z\rp^2} 
    \Bigg] \, ,
\end{eqnarray}
where the first term is the contribution that depends on the
factorization scale and is proportional to the $qq$ splitting
function. Separating the massive quark coefficient functions into the
various contributions is important to properly evaluate their Mellin
transforms, as will be discussed below.

Finally, we give the expression for the FFN gluon
coefficient function. In this case there are no
singular terms and it reads
\begin{eqnarray}
  C^{(n_l),1}_{2,g}(z,Q^2,\lambda) &=& \Bigg[ T_f\lp z^2+\lp
    1-z\rp^2\rp \lp \ln \frac{1-\lambda z}{(1-\lambda)z} + \ln
    \frac{Q^2+m_c^2}{Q^2}\rp \nonumber \\ 
    &+& T_f\lp z^2+\lp
    1-z\rp^2\rp \lp 2\ln\lp 1-z\rp -\ln\lp 1-\lambda z\rp -\ln z\rp
    \nonumber \\ 
    &+&\lp 8-18\lp 1-\lambda\rp +12\lp 1+\lambda\rp^2\rp
    z(1-z) +\lp \frac{1-\lambda}{1-\lambda
      z}-1\rp \label{eq:gluoncoeffffn} \nonumber \\ 
    &+&\lp
    1-\lambda\rp z \ln \frac{1-\lambda z}{(1-\lambda) z} \lp 6\lambda
    -12\lambda^2 z\rp \Bigg] .
\end{eqnarray}
Again the last term in the first line is the scale-dependent
contribution and is proportional to $P_{qg}^{(0)}$.  Note that both
the diagrams shown in Fig.~\ref{fig:ffns2}
contribute~\cite{Gluck:1996ve}.
Analogous expressions for the charged
current $F_{3,c}$ and $F_{L,c}$ structure functions can be found in
Refs.~\cite{Gottschalk:1980rv,Gluck:1996ve}.

As in the case of neutral currents, the massless limits of the FFN
structure functions is easily obtained. For the
massive $F_{2,c}^{\rm CC}$ structure function it has the structure
\begin{eqnarray} 
  F_{2,c}^{(n_l,0),\rm CC}(x,Q^2)&=&2x s\lp x,Q^2\rp
  \nonumber \\ 
  &+& 2x \frac{\alpha_s\lp Q^2\rp}{2\pi}
  \bigg\{\int\limits_{x}^{1}\frac{dz}{z}\bigg[C^{(n_l,0),1}_{2,h}(z,Q^2,\lambda)
    s\lp \frac{x}{z},Q^2\rp \nonumber \\ 
    &+&C^{(n_l,0),1}_{2,g}(z,Q^2,\lambda)g\lp \frac{x}{z},Q^2\rp
    \bigg]\bigg\} \, ,
\end{eqnarray}
where
\begin{eqnarray}
  C^{(n_l,0),1}_{2,h}(z,Q^2,\lambda) =
  C^{(n_l,0)}_{h,\delta} \delta\lp 1-z \rp + C^{(n_l,0)}_{h,{\rm r}}\lp
  z \rp + \lc C^{(n_l,0)}_{h,{\rm s}}\lp z \rp\rc_+ \, ,
\end{eqnarray}
\begin{equation}
  C^{(n_l,0)}_{h,\delta} = -C_F \lp \frac{9}{2} +\frac{\pi^2}{3} \rp \, ,
\end{equation}
\begin{equation}
  C^{(n_l,0)}_{h,{\rm r}}\lp z \rp = C_F\Bigg[ -\lp 1+z\rp
    \ln\lp 1-z\rp -\frac{\lp 1+z^2\rp\ln z}{1-z} + 3+2z \Bigg] \, , 
\end{equation}
\begin{eqnarray}
  C^{(n_l,0)}_{h,{\rm s}}\lp z \rp = C_F\Bigg[ 2 \lp \frac{\ln \lp
      1-z\rp}{1-z}\rp - \frac{3}{2}\lp \frac{1}{1-z} \rp \Bigg] \, ;
\end{eqnarray}
and for the gluon
\begin{eqnarray}
  C^{(n_l,0),1}_{2,g}(z,Q^2)&=& 2T_f\lc
  \lp z^2+(1-z)^2\rp\ln \frac{1-z}{z} +8z\lp 1-z\rp -1\rc \nonumber
  \\ 
  &+& T_f \lp z^2+(1-z)^2\rp\ln \frac{Q^2}{m_c^2} \ .
  \label{eq:cnl01massless}
\end{eqnarray}

For completeness, we provide also the ZM-VFN quark coefficient
functions for quarks and gluons, 
\begin{eqnarray}
  &&C^{(n_l+1),1}_{2,h}(z) =
  C_F\Bigg[ 2\lp \frac{\ln\lp 1-z\rp}{1-z}\rp_+ -\frac{3}{2}\lp
  \frac{1}{1-z}\rp_+\nonumber \\ 
  &&-(1+z)\ln\lp 1-z\rp
  -\frac{(1+z^2)\ln z}{1-z} +3+2z +\delta\lp 1-z\rp \lp
  -\frac{\pi^2}{3}-\frac{9}{2}\rp \Bigg] \, ,
  \label{eq:quarkcoefzm}
\end{eqnarray}
\begin{equation}
  \label{eq:gluoncoefzm}
  C^{(n_l+1),1}_{2,g}(z) = T_F\Bigg[ \lp z^2+\lp 1-z\rp^2\rp\ln \frac{1-z}{z}
    +(8z(1-z)-1)\Bigg]\, .
\end{equation}
Note that the above gluon coefficient function,
Eq.~(\ref{eq:gluoncoefzm}), is defined according to  the notation of
Ref.~\cite{Zijlstra:1992qd}, that is, it corresponds to the production
of a single quark or antiquark.

Comparing the FFNS0 and ZM-VFN coefficient functions we
find that for the gluon piece the following relation holds
\begin{equation}
  \label{eq:compg}
  C^{(n_l,0),1}_{2,g}(z,Q^2) = 2C^{(n_l+1),1}_{2,g}(z) +
  T_f  \lp z^2+(1-z)^2\rp\ln \frac{Q^2}{m_c^2} \, ,
\end{equation}
where the overall factor 2 is due to the fact that the ZM coefficient
function, Eq.~(\ref{eq:gluoncoefzm}), has been defined for a single
quark, while in Eq.~(\ref{eq:gluoncoeffffn}) the gluon coefficient
function accounts for the production of two quarks ($s$ and
$\bar{c}$).  Note also the presence of the usual collinear logarithm.
For the quark piece we find 
\begin{equation}
  \label{eq:compq}
  C^{(n_l,0),1}_{2,h}(z) = C^{(n_l+1),1}_{2,h}(z) \, ,
\end{equation}
without any collinear logarithm.

\begin{figure}[ht]
  \begin{center}
    \includegraphics[width=0.60\textwidth]{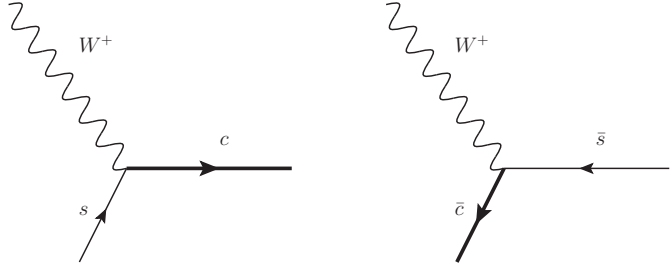}
  \end{center}
  \caption{\small Feynman diagrams that contribute to
    $F_{2,c}^{(n_l+1),\rm CC}$ in the ZM-VFN scheme
    at leading order. The NLO diagrams are the same
    as in the FFNS scheme.}
  \label{fig:zmvfn}
\end{figure}

The definition of the heavy CC structure function in the ZM scheme  is
not unique: here we define it as the contribution to the structure
function which includes all contributions to
the inclusive structure function which survive when all CKM
elements but $|V_{\rm cs}|$ are set to zero. With this definition,
both the leading-order processes $c W^+ \to s$ and   $\bar{c} W^+ \to
\bar{s}$ contribute to it (see  Fig.~\ref{fig:zmvfn}). This definition
coincides with the experimental one because the struck charm antiquark
must be accompanied by an (observed) charm quark, and it is
free of mass singularities. The gluon initiated NLO contributions
remain those shown in Fig.~\ref{fig:ffns2}.
The structure function in the massless
scheme above charm threshold is then given by
\begin{eqnarray}
  \label{eq:f2czm}
  F_{2,c}^{(n_l+1),\rm CC}(x,Q^2)&=&2 x
  \lp s\lp x,Q^2\rp + \bar{c}\lp x,Q^2\rp  \rp \nonumber \\ 
  &+& 2x
  \frac{\alpha_s\lp Q^2\rp}{2\pi}
  \bigg\{\int\limits_{x}^{1}\frac{dz}{z}\bigg[C^{(n_l+1),1}_{2,h}(z,Q^2,\lambda) 
    \lp s\lp \frac{x}{z},Q^2\rp+ \bar{c} \lp \frac{x}{z},Q^2\rp\rp\nonumber \\
    &+&2C^{(n_l+1),1}_{2,g}(z,Q^2,\lambda)g\lp \frac{\xi}{z},Q^2\rp
    \bigg]\bigg\} \, .
\end{eqnarray}
The ZM-VFN massless
coefficient functions have been defined in
Eqs.~(\ref{eq:quarkcoefzm})-(\ref{eq:gluoncoefzm}).  Note the
factor two in front of the gluon coefficient function, to account for the
production of two quarks in the two NLO
subprocesses of Fig.~\ref{fig:ffns2}.

Finally, the various schemes can be combined to construct the FONLL-A
structure functions.  As in the NC case, we define the FONLL structure
function as follows
\begin{equation}
  F_{2,c}^{(\rm FONLL),\rm CC}(x,Q^2) \equiv
  F_{2,c}^{(n_l),\rm CC}(x,Q^2) + \theta\lp Q^2-m_c^2\rp \lp 1-
  \frac{m_c^2}{Q^2}\rp^2 F_{2,c}^{(\rm d),\rm CC}(x,Q^2) 
\end{equation}
\begin{equation}
  F_{2,c}^{(\rm d),\rm CC}(x,Q^2) = F_{2,c}^{(n_l+1),\rm CC}(x,Q^2) -
  F_{2,c}^{(n_l,0),\rm CC}(x,Q^2) \, ,
\end{equation}
where as in the case of neutral currents we use the damping factor
as default threshold prescription.

Using the explicit expressions derived in the previous section for the
difference between the ZM and FFNS0 coefficient functions,
Eqs.~(\ref{eq:compq}) and~(\ref{eq:compg}), we can write the
difference term as  
\begin{equation}
  \label{eq:f2cdiff}
  F_{2,c}^{(\rm d),\rm CC} = 2x\bar{c}\lp x,Q^2\rp -
  2x\frac{\alpha_s}{2\pi} \ln \frac{Q^2}{m_c^2} \int_x^1 \frac{dz}{z}
  T_f \lp z^2+(1-z)^2\rp g\lp \frac{x}{z},Q^2\rp + \mathcal{O}
  (\alpha_s^2)\, ,
\end{equation}
where we have used the fact that the heavy quark distribution is
$\mathcal{O} (\alpha_s)$.  Now, it is easy to see explicitly that, in
the region where $L\equiv \ln Q^2/m_c^2$ is not large, the
``difference'' term is of order $\mathcal{O} (\alpha_s^2)$: to first
order in $\alpha_s$ the FONLL expression coincides with the
massive--scheme one also for charged current scattering. The use of
the leading--order QCD evolution equations immediately leads to
\begin{equation}
  c (x, Q^2)=\bar{c} (x, Q^2) = \frac{\alpha_s(Q^2)}{2 \pi} \ln
  \frac{Q^2}{m_c^2} \int_x^1 \frac{d z}{z} T_f (z^2 + (1 - z)^2) g \lp
  \frac{x}{z}, Q^2\rp + \mathcal{O} (\alpha_s^2)\, .
\end{equation}
Inserting this expansion in Eq.~(\ref{eq:f2cdiff}), it is trivial to
check the explicit cancellation of the $\mathcal{O} (\alpha_s)$ terms,
that is, that near the heavy quark threshold the difference term is of
order $  F_{2,c}^{(\rm d),\rm CC}= \mathcal{O} (\alpha_s^2)$.

The final FONLL-A expressions for the charged current
charm production structure function $F_{2,c}^{\rm CC}$ 
is given by
\begin{eqnarray}
  \label{eq:f2cfonll}
  &&F_{2,c}^{(\rm FONLL),\rm CC}(x,Q^2) = 2\xi s\lp\xi,Q^2\rp +
  \theta\lp Q^2-m_c^2\rp \lp 1- \frac{m_c^2}{Q^2}\rp^2 2x\bar{c}\lp
  x,Q^2\rp \\ 
  \nonumber &+& 2\xi \frac{\alpha_s\lp Q^2\rp}{2\pi}
  \bigg\{\int\limits_{\xi}^{1}\frac{dz}{z}\bigg[C^{(n_l),1}_{2,h}(z,Q^2,\lambda)
    \lp s\lp \frac{\xi}{z},Q^2\rp + \theta\lp Q^2-m_c^2\rp \lp 1- \frac{m_c^2}{Q^2}\rp^2 2x\bar{c}\lp \frac{\xi}{z},Q^2\rp\rp\\
&&\qquad
+ C^{(n_l),1}_{2,g}(z,Q^2,\lambda)g\lp
    \frac{\xi}{z},Q^2\rp \bigg]\bigg\} \\
  &-& \theta\lp Q^2-m_c^2\rp
  \lp 1- \frac{m_c^2}{Q^2}\rp^2 2x \frac{\alpha_s\lp Q^2\rp}{2\pi}
  \int_x^1 \frac{d z}{z} T_f (z^2 + (1 - z)^2) g \lp \frac{x}{z},
  Q^2\rp \nonumber \, .
\end{eqnarray}

It can be easily verified that Eq.~(\ref{eq:f2cfonll}) reduces to the FFN
scheme, Eq.~(\ref{eq:f2cffns}) at the heavy quark threshold
$Q^2=m_c^2$, and to the ZM-VFN expression Eq.~(\ref{eq:f2czm}) in the
asymptotic region $Q^2 \gg m_c^2$. 

The above derivation generalizes straightforwardly to the other relevant
charged current structure functions $xF^{\rm CC}_{3,c}$ and $F^{\rm
  CC}_{L,c}$, as well as to the case with a general CKM quark mixing
matrix. Note that in all the results shown below the standard CKM
mixing has been assumed, with the CKM matrix elements set to their PDG
values~\cite{Nakamura:2010zzi}.

\begin{figure}[ht]
  \begin{center}
    \epsfig{file=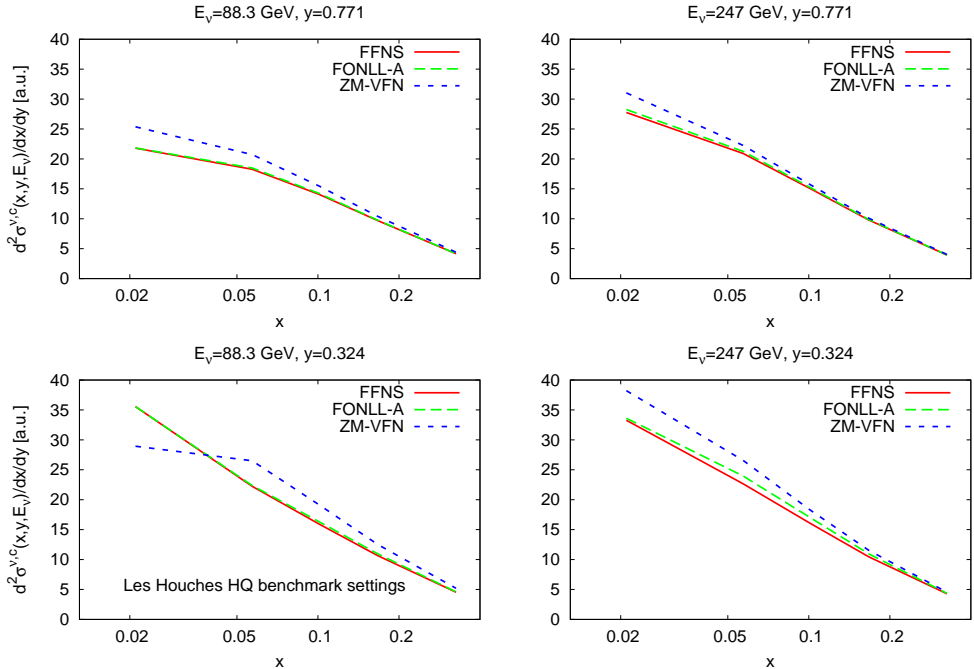,width=0.99\textwidth}
  \end{center}
  \caption{\small Comparison of different schemes for charm production
    in neutrino-induced DIS. The kinematic range is representative of
    the NuTeV dimuon data range. We compare the ZM-VFN, FFN and
    FONLL-A schemes at the level of the neutrino induced charm
    production cross section, Eq.~(\ref{eq:nuxsecdimuon}). 
The settings are the same as those of the Les
    Houches heavy quark benchmark
    comparison\cite{LHhq}. \label{fig:cchqcomp}}
\end{figure}

Now that we have defined the FONLL-A general mass scheme for charged
current structure functions, we can compare the various schemes (ZM,
FFNS, FONLL-A) in the kinematic region that is most relevant in the
global PDF analysis, namely the region covered by the NuTeV dimuon
measurements~\cite{Mason:2007zz} (see Fig.~\ref{fig:dataplottot}).  In
Fig.~\ref{fig:cchqcomp} we show the results of such a comparison
between various schemes for charm production in neutrino-induced
charged current scattering. Results are compared at the level of the
phenomenologically relevant charm production reduced cross section,
defined as~\cite{Ball:2009mk}:
\begin{eqnarray}
  \label{eq:nuxsecdimuon}
  &&\tilde{\sigma}^{\nu (\bar{\nu}),c}(x,y,Q^2)\equiv 
  \frac{1}{E_{\nu}}\frac{d^2\sigma^{\nu(\bar{\nu}),c}}{dx\,dy}
  (x,y,Q^2)\nonumber\\
  &&\qquad =\frac{G_F^2M_N}{2\pi(1+Q^2/M_W^2)^2}
  \Bigg[
    \left( \lp Y_+ - \frac{2M^2_Nx^2y^2}{Q^2} -y^2\rp \lp
    1+ \frac{m_c^2}{Q^2}\rp +y^2\right)
    F_{2,c}^{\nu(\bar{\nu})}(x,Q^2) \nonumber\\ 
    &&\qquad\qquad\qquad\qquad\qquad\qquad\qquad
    -y^2F_{L,c}^{\nu(\bar{\nu})}(x,Q^2)\pm 
    \,Y_-\,xF_{3,c}^{\nu(\bar{\nu})}(x,Q^2)
    \Bigg] \,
\end{eqnarray}
with $Q^2=2M_NE_{\nu}xy$ and $ Y_\pm = 1\pm(1-y)^2$.  In
Fig.~\ref{fig:cchqcomp} we compare the various schemes in some
representative bins of the NuTeV dimuon
kinematics~\cite{Mason:2007zz}.  PDFs and other settings are those of
the Les Houches heavy quark benchmark comparison~\cite{LHhq}.  We
observe that in the kinematic region of 
neutrino data (both inclusive CHORUS data and dimuon NuTeV data),
the FONLL-A result is very close to the FFN scheme computation, and it
only begins to differ from it at the highest energies, where
resummation of charm mass collinear logarithms begins to become
relevant.

\begin{figure}[ht]
  \begin{center}
    \epsfig{file=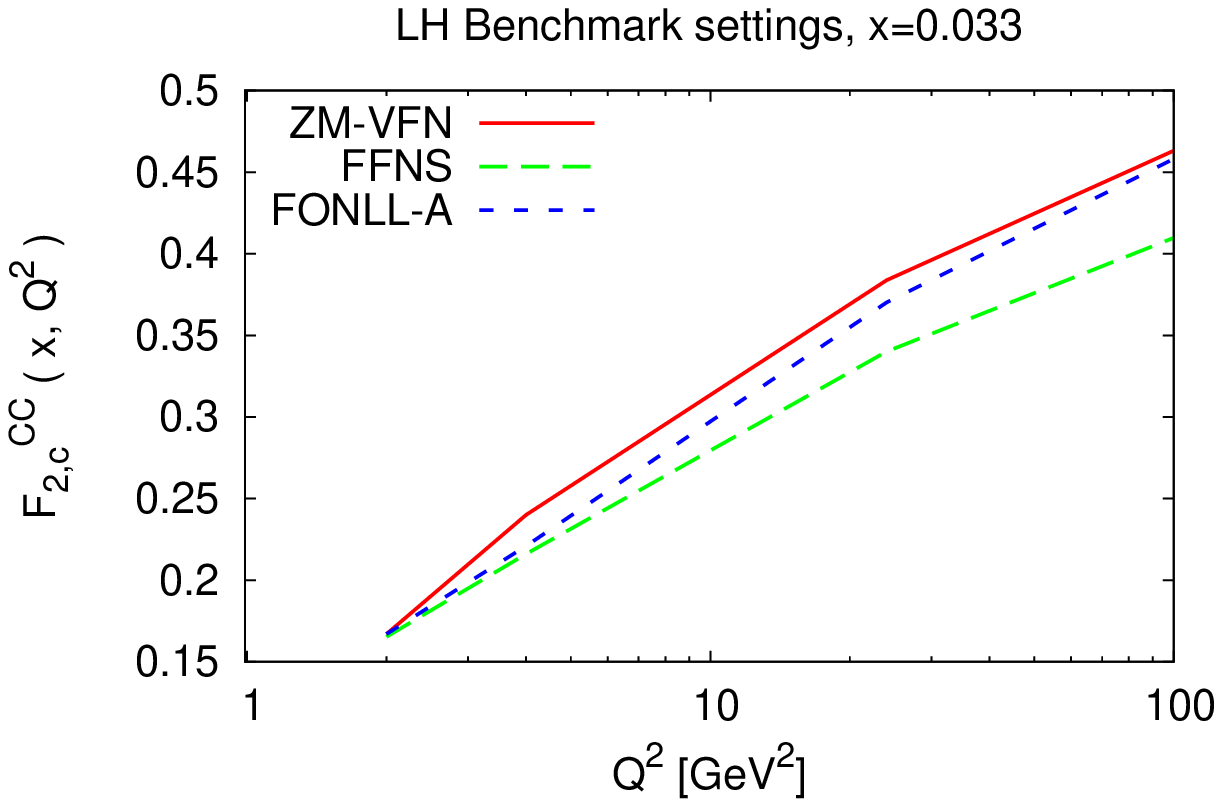,width=0.49\textwidth}
    \epsfig{file=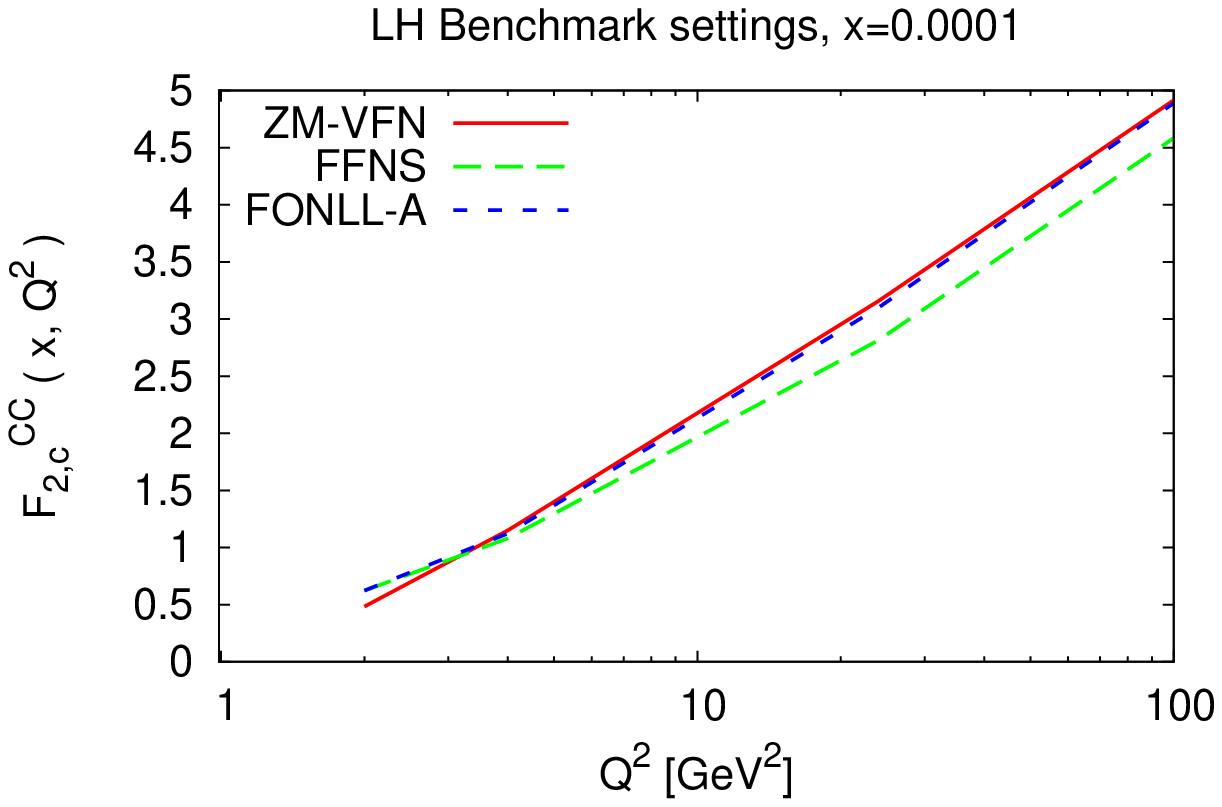,width=0.49\textwidth}
  \end{center}
  \caption{\small Comparison of different schemes for charm production
    in neutrino-induced DIS. We show the $F_{2,c}^{\rm CC}$ structure
    function in the massless, massive and FONLL-A schemes; in this
    case the FONLL-A expression is given by
    Eq.~(\ref{eq:f2cfonll}). The settings
    are the same as those of the Les Houches heavy quark benchmark
    comparison~\cite{LHhq}. \label{fig:cchqcomp2}}
\end{figure}

Even if the differences between the FFN and FONLL-A
schemes for charged current scattering in the NuTeV kinematic region
are  moderate, as shown in Fig.~\ref{fig:cchqcomp}, they become rather
more important at small-$x$ and medium-large $Q^2$, where
the charm and gluon PDFs become larger.
 To illustrate this,
in Fig.~\ref{fig:cchqcomp2} we compare the charged current charm
structure function $F_{2,c}^{\rm CC}$ as a function of $Q^2$ for two
different values of $x$. Notice in particular that at very small $x$
the FONLL-A expression is essentially the massless result.  However,
producing dimuons at $ x\sim 10^{-3} $ and $Q^2\sim 10$ GeV$^2$, where
differences are larger, requires a fixed target neutrino experiment
with a neutrino beam with energy in the multi-TeV range, which is not
foreseen in the near future. Therefore one can conclude that any
reasonable general--mass scheme for charged current scattering will be
very close to the FFNS in the region of experimental data.

The FONLL-A calculation of charged current structure functions has
been implemented in a $x$--space code, FONLLdisCC, that we will use
for benchmarking purposes. This is the analogue of the FONLLdis
code for neutral currents~\cite{FONLLdis}, however is rather simpler
since the unknown $\mathcal{O}\lp \alpha_s^2\rp$ massive coefficient
functions do not have to be implemented.
Our implementation of the FFNS calculations has been
benchmarked with the corresponding results of the MSTW08
code~\cite{wattpriv}, finding perfect agreement.  
We have also compared the FONLL-A and MSTW08 general--mass schemes for
charged currents, finding qualitative agreement but some quantitative
differences.  A detailed comparison between different general--mass
schemes for charged current structure functions, analogous to 
the Les Houches
benchmarks for neutral current structure functions~\cite{LHhq} is
still missing and would be highly desirable.

\section{Results}
\label{sec:results}

In this Section we present the NNPDF2.1 
parton determination. First, we discuss the statistical features
of the fit, then we turn to a comparison of NNPDF2.1 PDFs
and uncertainties with  other
PDF determinations and with
previous NNPDF releases. 
A detailed
comparison between the NNPDF2.1 and 2.0 sets
follows, in which we discuss one by one the
impact of the differences between the two fits,
due both to the choice of dataset and to
the different theoretical framework. The implications
of the NNPDF2.1 set for LHC observables are discussed in the next
sections.

\subsection{Statistical features}
\begin{table}
\centering
\begin{tabular}{|c|c|}
\hline 
$\chi^{2}_{\tot}$ &      1.16 \\
$\la E \ra \pm \sigma_{E} $   &     $2.24\pm 0.09$     \\
$\la E_{\rm tr} \ra \pm \sigma_{E_{\rm tr}}$&   $2.22 \pm 0.11$      \\
$\la E_{\rm val} \ra \pm \sigma_{E_{\rm val}}$&   $2.28 \pm 0.12$        \\
$\la{\rm TL} \ra \pm \sigma_{\rm TL}$   &  $\lp 1.6\pm 0.6 \rp \,10^{4}$     \\
\hline
$\la \chi^{2(k)} \ra \pm \sigma_{\chi^{2}} $  &  $1.25\pm 0.09$     \\
\hline
 $\la \sigma^{(\exp)}
\ra_{\dat}$(\%) & 11.3\% \\
 $\la \sigma^{(\net)}
\ra_{\dat} $(\%)&  4.4\% \\
\hline
 $\la \rho^{(\exp)}
\ra_{\dat}$ & 0.18 \\
 $\la \rho^{(\net)}
\ra_{\dat}$& 0.56 \\
\hline
\end{tabular}
\caption{\small \label{tab:estfit1} Table of statistical estimators
  for NNPDF2.1 with $N_{\rm rep}=
1000$ replicas. The total average uncertainty is given in percentage.
All the $\chi^2$ and $E$ values 
have been computed using the the same $t_0$ covariance 
matrix~\cite{Ball:2009qv} 
used for minimization. }
\end{table}

{
\begin{table}
\centering
\small
\begin{tabular}{|c||c|c||c|c|c|c|c|}
\hline 
Experiment    & $\chi^2 $& $\chi^2_{\rm 2.0}$& $\la E\ra $   & $\la \sigma^{(\exp)}\ra_{\dat}$(\%) & $\la \sigma^{(\net)}\ra_{\dat}$(\%) & $\la \rho^{(\exp)}\ra_{\dat}$ & $\la \rho^{(\net)}\ra_{\dat}$ \\
\hline 
NMC-pd       & 0.97 & 1.04  &2.04 & 1.9\% &  0.5\%& 0.03 & 0.37 \\
\hline
NMC         & 1.73 & 1.73 & 2.79 & 5.0\% & 1.5\% & 0.16 & 0.71 \\
\hline
SLAC        & 1.27 & 1.42 & 2.34 & 4.4\% & 1.6\% & 0.31 & 0.79 \\
\hline
BCDMS    & 1.28 & 1.30 & 2.33 & 5.7\% & 2.3\% & 0.47 & 0.60 \\
\hline
HERAI-AV   & 1.07 & 1.15 & 2.15 & 2.5\% & 1.2\% & 0.06 & 0.35 \\
\hline
CHORUS    & 1.15 & 1.24 &2.23 & 15.1\% & 4.7\% & 0.08& 0.32\\
\hline
FLH108    & 1.37 & 1.50 & 2.36 & 72.0\% & 4.0\% & 0.64 & 0.67 \\
\hline
NTVDMN    & 0.76 & 0.73 & 1.77 & 21.1\% & 14.1\% & 0.04 & 0.62 \\
\hline
ZEUS-H2   & 1.29 & 1.33 & 2.32 & 13.4\% & 1.2\% & 0.27 & 0.51 \\
\hline
ZEUSF2C   & 0.78 & - & 1.80 &   23.3\% &  3.1\%& 0.08 & 0.41 \\
\hline
H1F2C   & 1.50 & - & 2.52 & 17.3\% & 3.0\% & 0.30 & 0.40\\
\hline
DYE605    & 0.84 & 0.87 &1.92 & 22.3\% & 7.9\% & 0.47 & 0.76 \\
\hline
DYE866    & 1.27 & 1.29 &2.37 & 20.1\% & 9.2\% & 0.20 & 0.52 \\
\hline
CDFWASY   &  1.86 & 1.84 & 3.08 & 6.0\% & 4.4\% & 0.51 & 0.75\\
\hline
CDFZRAP   & 1.65 & 1.85 & 2.80 & 11.5\% & 3.6\% & 0.82  & 0.72 \\
\hline
D0ZRAP    & 0.60 & 0.60 &1.62 & 10.2\% & 3.1\% & 0.53 & 0.76 \\
\hline
CDFR2KT   & 0.97 & 1.01  &2.10 & 22.2\% & 4.0\% & 0.78 & 0.57 \\
\hline
D0R2CON   & 0.84 & 0.86 & 1.92 & 16.8\%  & 4.5\% & 0.77 & 0.59 \\
\hline
\end{tabular}
\caption{\small \label{tab:estfit2} Same as Table \ref{tab:estfit1}
  for individual experiments. All estimators
have been obtained with
$N_{\rm rep}= 1000$ replicas. Note that
experimental uncertainties are always given in percentage. In the second
and third column the NNPDF2.1 and NNPDF2.0 set~\cite{Ball:2010de} $\chi^2$ have been
computed with the $t_0$ prescripion. }
\end{table}
}


Statistical estimators for the NNPDF2.1 
fit are shown in Tab.~\ref{tab:estfit1} for the global fit and in
Tab.~\ref{tab:estfit2} for individual experiments. In
Tab.~\ref{tab:estfit2}
the $\chi^2$ values
for NNPDF2.0 are also shown for comparison. As in
Ref.~\cite{Ball:2010de}, $\chi^{2}_{\tot}$ is computed comparing the
central (average) NNPDF2.1 fit  to the original experimental data, 
$\la \chi^{2(k)} \ra$ is computed comparing to the data 
each NNPDF2.1 replica and averaging over replicas, while $\la E \ra$
is the quantity which is minimized, i.e. it coincides with the
$\chi^{2}$ computed comparing each NNPDF2.1 replica to the data
replica it is fitted to, with the three values given corresponding to
the total, training and validation data sets. It is important to
observe that all values of $\chi^2$ shown in
Tabs.~\ref{tab:estfit1}-\ref{tab:estfit2} are obtained using the
covariance matrix with normalization uncertainties included according
to the $t_0$ method of Ref.~\cite{Ball:2009qv} (also given as Eq.~(1)
in Ref.~\cite{Ball:2010de}). In Tabs.~9-10 of
  Ref.~\cite{Ball:2010de} all values of 
$\chi^{2}_{\tot}$ and $\la \chi^{2(k)} \ra$ were instead given with
  $\chi^2$ defined using the ``standard'' covariance matrix (given
  e.g. in Eq.~(52) of that reference), which includes
  normalization uncertainties less accurately than the $t_0$
  covariance matrix. This was done  in order to ease comparison
  between the results of Ref.~\cite{Ball:2010de} and the NNPDF1.x
  PDFs, in which the $t_0$ method was not yet used and normalization
  uncertainties were not fully accounted for. The values of
  $\chi^{2}_{\tot}$ for the NNPDF2.0 shown
  here in Tab.~\ref{tab:estfit1}-\ref{tab:estfit2} has been
  recomputed using the $t_0$ covariance matrix in order to ease
  comparison with NNPDF2.1 and with other PDF determinations which
  also include normalization uncertainties albeit with various other
  methods. The value of $\chi^2_{\rm tot}$ for the NNPDF2.0 global fit
  computed using the $t_0$ method, to be compared to
  the NNPDF2.1 value of Tab.~\ref{tab:estfit1}, is $\chi^2_{\rm tot}=1.23$
  (very close to the the value $\chi^2_{\rm tot}=1.21$ of Tab.~9 in
  Ref.~\cite{Ball:2010de},
 computed with the ``standard'' covariance matrix).

The NNPDF2.1 PDF fit has the following noticeable features:
\begin{itemize}
\item The quality of the global fit as measured by the value
  $\chi^2=1.16$ is rather better than for the NNPDF2.0 fit without
  heavy quark mass effects.
\item As compared to the NNPDF2.0 results, the quality of the fit
to all datasets improves or remains similar. The most noticeable
improvements can be found for the HERA-I average dataset and
for CHORUS. The improvement in the description of HERA data
arises both from the improved heavy flavour treatment and the more
conservative kinematical cuts. 
\item An excellent description of the combined HERA-I inclusive
data, $\chi^2=1.07$, is obtained. Similarly, a reasonable
description of the HERA charm structure function data is achieved.
\item The quality of the fit to hadronic data is  not affected by the use
of the FONLL-A GM scheme for deep--inelastic
observables, as it can be seen by comparing the second and third
column of Table~\ref{tab:estfit2}.
\end{itemize}

\begin{figure}[t]
\begin{center}
\epsfig{width=0.49\textwidth,figure=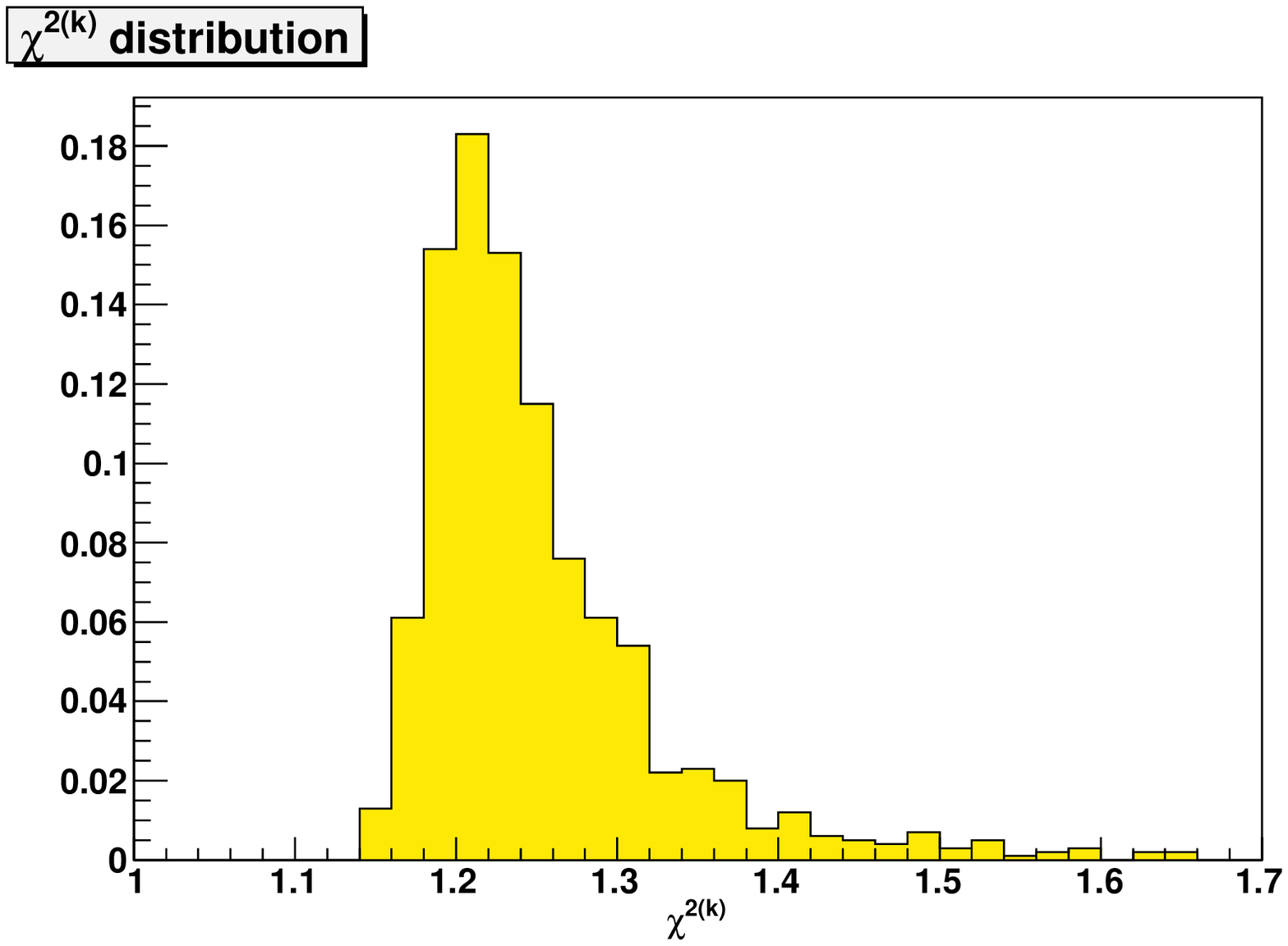}
\epsfig{width=0.49\textwidth,figure=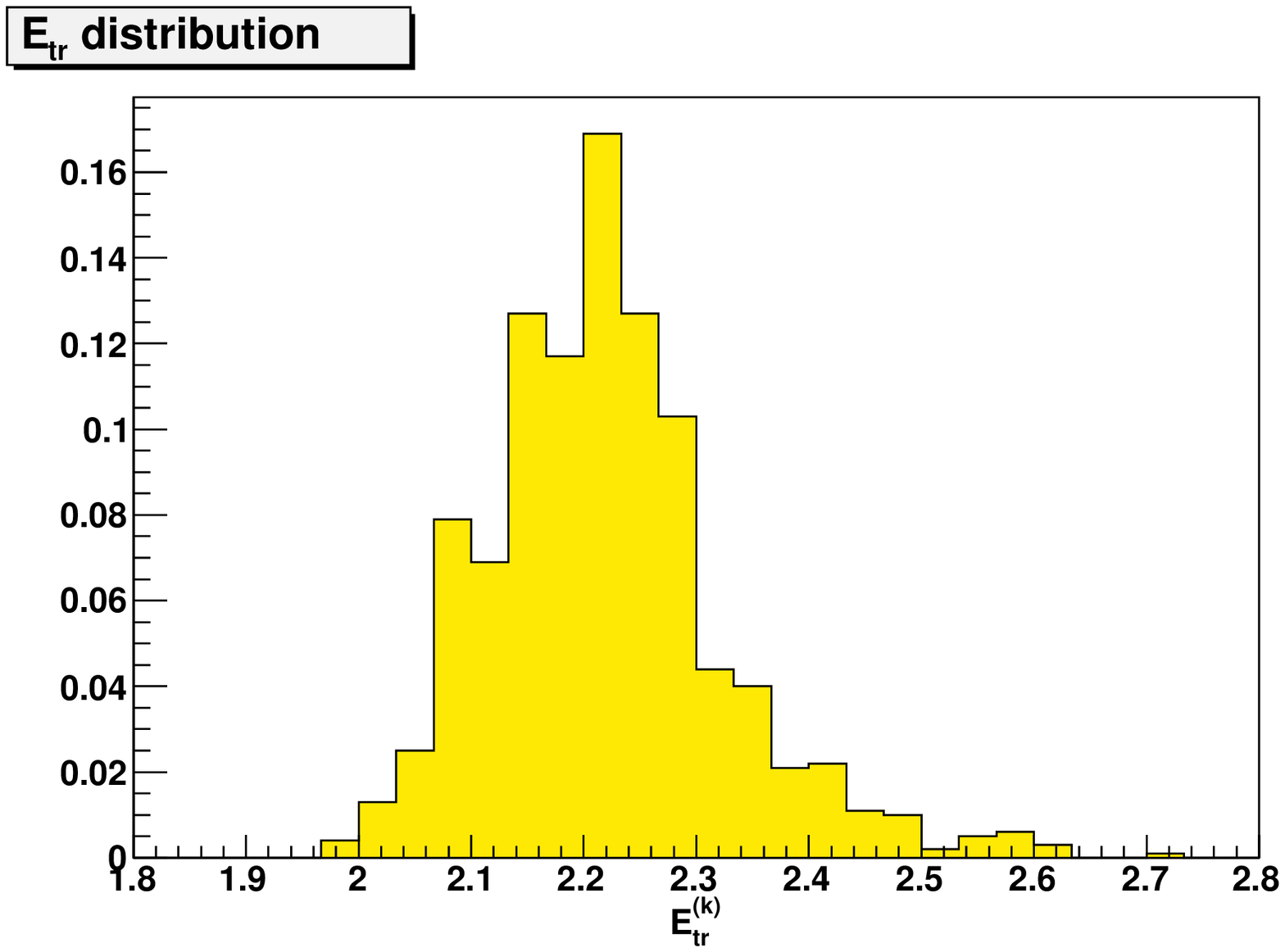}
\caption{\small Distribution of $\chi^{2(k)}$ (left) and  $E^{(k)}_{\rm tr}$ (right),
over the sample of $N_{\mathrm rep}=1000$ replicas. \label{chi2histoplots}} 
\end{center}
\end{figure}

\begin{figure}[ht!]
\begin{center}
\epsfig{width=0.60\textwidth,figure=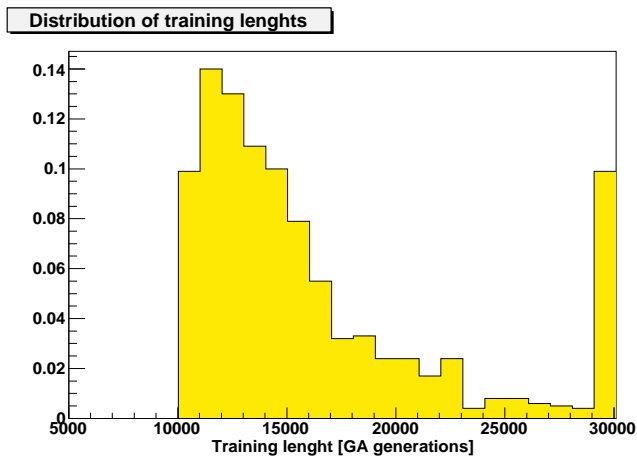}
\caption{\small Distribution of training lengths over the sample of
  $N_{\mathrm rep}=1000$ replicas.  
\label{fig:tl}} 
\end{center}
\end{figure}

The distribution of 
$\chi^{2(k)}$, $E^{(k)}_{\rm tr}$ and
  training lengths among the $N_{\rm
  rep}=1000$ NNPDF2.1 replicas are shown in Fig.~\ref{chi2histoplots} and
Fig.~\ref{fig:tl} respectively.  While
most of the replicas fulfill the stopping criterion, a  fraction
($\sim 12\%$)
of them stops at the maximum training length $N_{\rm gen}^{\rm max}$
which has been introduced in
order to avoid unacceptably long fits. 
This causes some loss of accuracy for outliers fits (i.e. those in the
tail of the distribution): we have
checked that as $N_{\rm gen}^{\rm max}$ is raised more and more of
these replicas stop, and that the loss of accuracy due to this
choice of value of $N_{\rm gen}^{\rm max}$ is reasonably small, in that
the features of the global fit change very little if $N_{\rm gen}^{\rm
  max}$ is raised.

\begin{table}
\begin{center}
\begin{tabular}{|c||c|c||c|c|}
\hline
& \multicolumn{2}{|c|}{NNPDF2.1} &   \multicolumn{2}{|c|}{CT10} \\
Experiment & $N_{\rm dat}$   &        $\chi^2$   &        $N_{\rm dat}$ &  $\chi^2$ \\
\hline
\hline
NMC-pd	   & 132 &   0.97        &  121 &      1.28   \\
NMC	    & 221&   1.73        &  196 &      1.71   \\
BCDMSp	   & 333 &   1.28        &   337 &     1.14   \\
BCDMSd	   & 248 &   1.15        &   250 &     1.12   \\
HERAI-AV     & 592 &   1.07        &  579 &      1.17   \\
\hline
NTVnuDMN	   & 41 &   0.50        &  38 &      0.94   \\
NTVnbDMN	   & 38&   0.42        &  33 &      0.91   \\
\hline
DYE605	          & 119&    0.85        &   119 &     0.81  \\
DYE866p	  & 184&    1.31        &    184 &    1.21  \\
DYE866r             & 15&    0.77        &  15 &      0.64  \\
CDFZRAT	          & 29&    1.62        &   29 &     1.44  \\
D0ZRAP	          & 28&    0.59        &   28 &     0.54  \\
\hline
 CDFR2KT 	  & 76 &    0.97        &  76 &      1.55  \\
 D0R2CON 	  & 110 &    0.84        &  110 &      1.13  \\
\hline
\end{tabular}
\end{center}
\caption{\small Comparison of $\chi^2$ per data point
 for experiments which are
common to the NNPDF2.1 and CT10 PDF determinations. For each
PDF set the number of data points obtained with the
kinematic cuts of Table~\ref{tab:kincuts} is given.  \label{tab:ct10chi2}}
\end{table}

It is instructive to compare the quality of the fit
with the corresponding results obtained
in the recent CT10 analysis.\footnote{We thank Pavel Nadolsky
for providing us with these numbers.} In Table~\ref{tab:ct10chi2} we compare
the $\chi^2$ of the common sets in NNPDF2.1 and CT10, along
with the number of data points in each fit (which differ because of
different kinematic cuts, see Table~\ref{tab:kincuts}). 
It should be borne in mind that 
the $\chi^2$ is defined in a somewhat different way by the CTEQ/CT
group, specifically, but not only, in
what concerns the treatment of normalization errors (see
Ref.~\cite{Lai:2010vv}): hence this
comparison should be taken with care. 
From this comparison, we can see that the two sets
have a comparable fit quality to fixed target DIS,
 CT10 being somewhat better for BCDMS proton and 
NNPDF2.1 rather better for NMC deuteron/proton ratio.
The fit to HERA-I and Tevatron jet data is rather better in NNPDF2.1.
Comparable fit quality to the Drell-Yan and vector boson production data
is obtained in the two cases,
with somewhat smaller $\chi^2$ in the CT10 fit. No comparison
is attempted for the HERA $F_2^c$ data because of the very
different kinematic cuts used in the two fits.
A similar comparison to MSTW08 would be less significant because in
the MSTW08 fit correlated systematics are not included in the
covariance matrix for some datasets.

\subsection{Parton distributions}
\label{sec-pdfcomp}

\begin{figure}[ht!]
\begin{center}
\epsfig{width=0.49\textwidth,figure=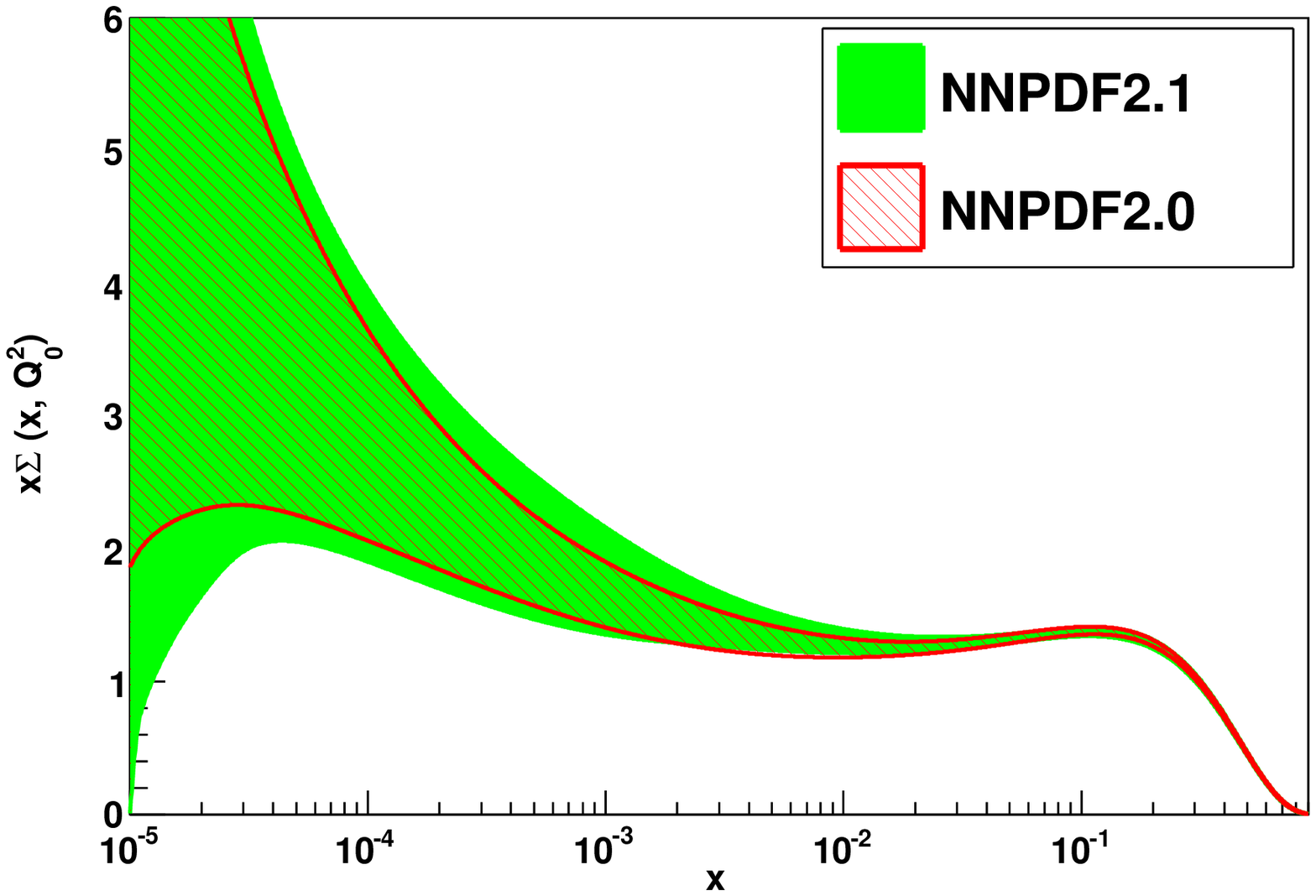}
\epsfig{width=0.49\textwidth,figure=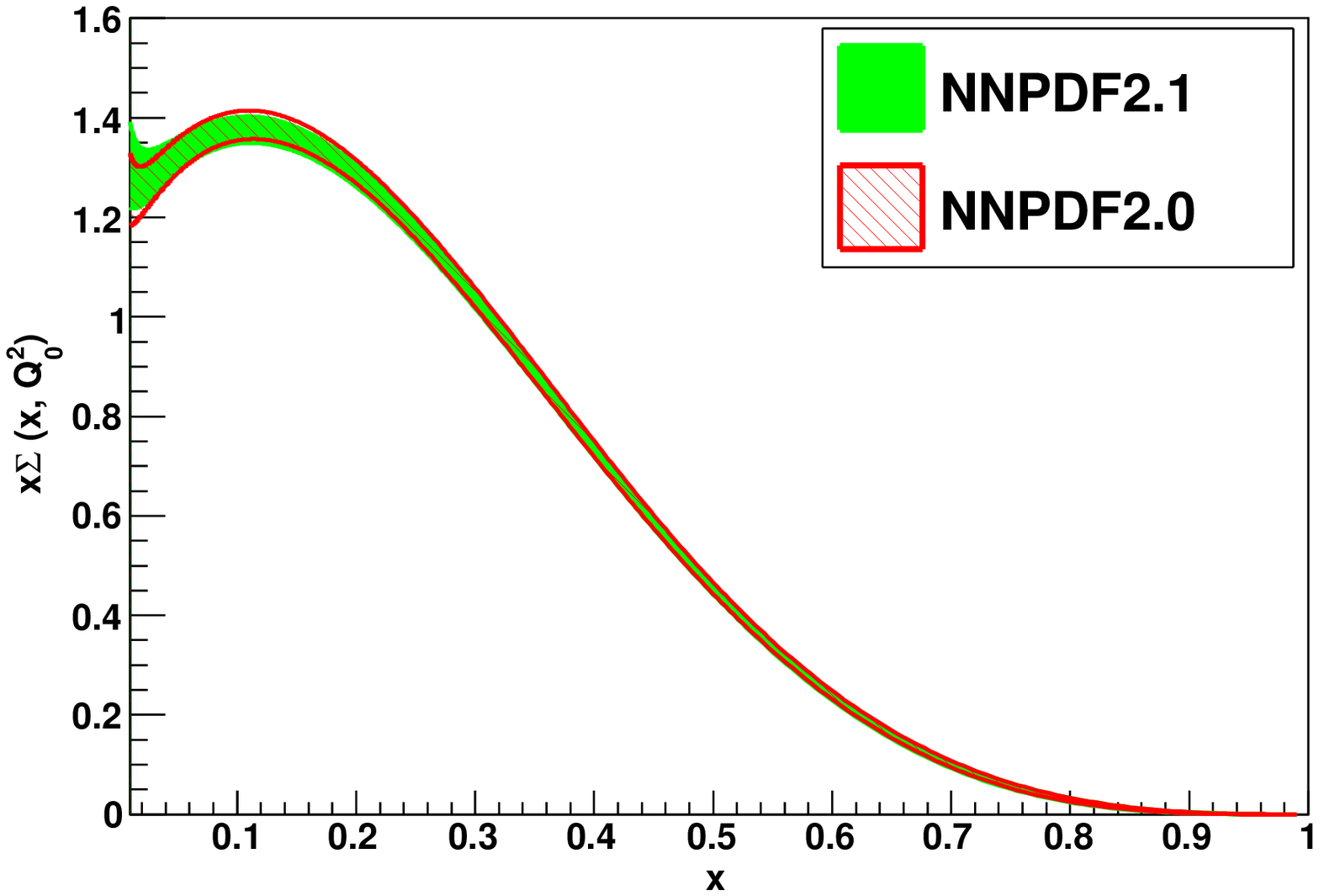}
\epsfig{width=0.49\textwidth,figure=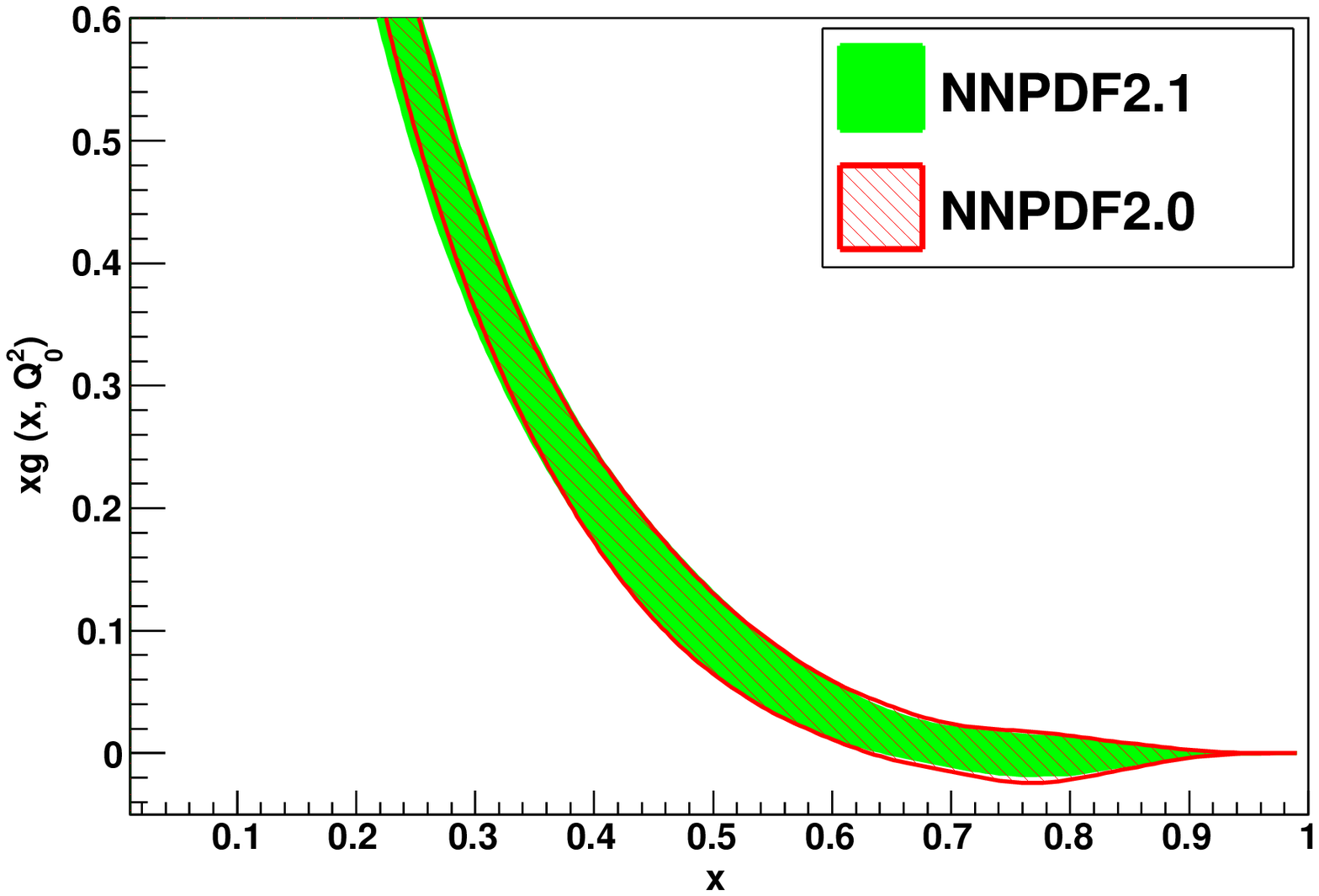}
\epsfig{width=0.49\textwidth,figure=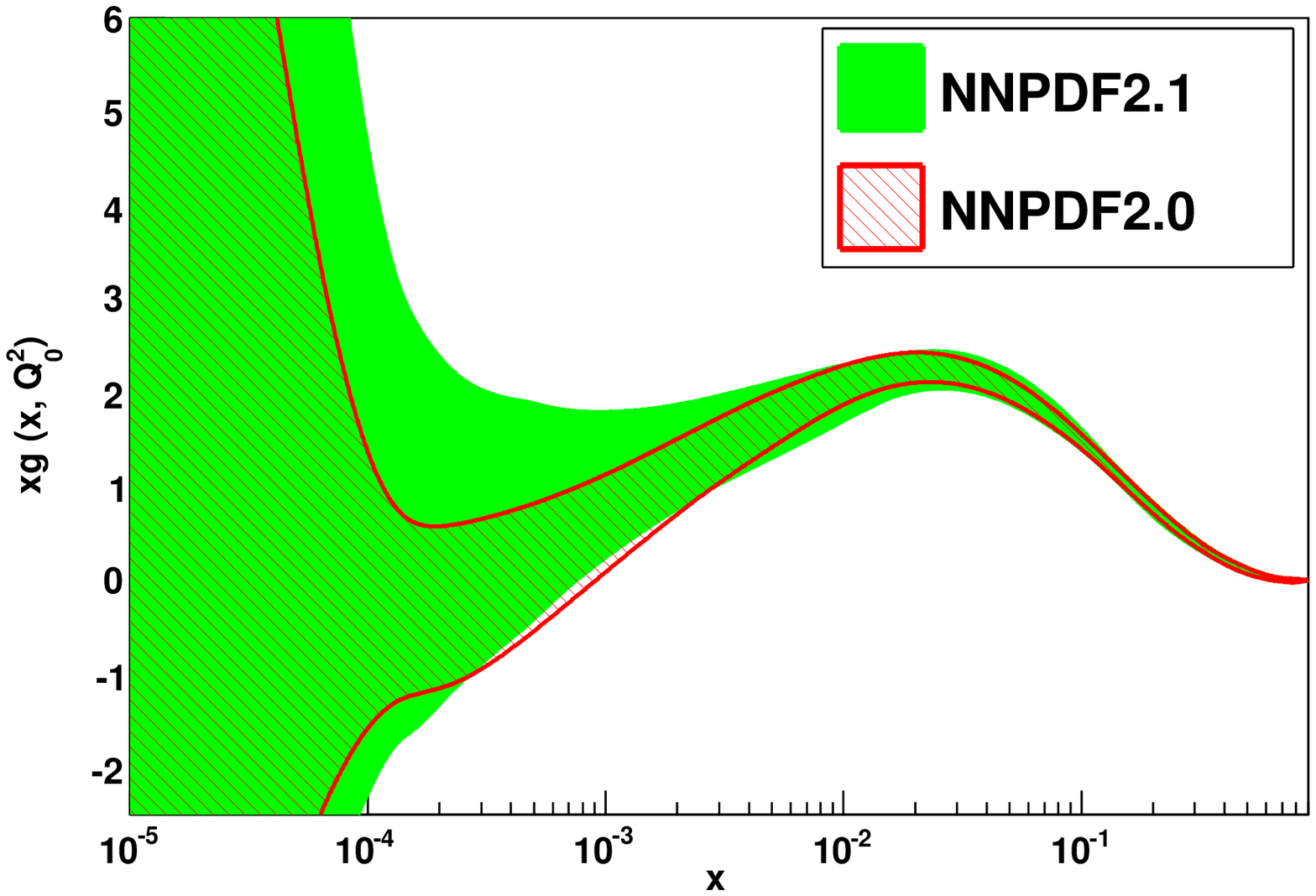}
\caption{\small Comparison of  NNPDF2.1 and NNPDF2.0 singlet sector PDFs, computed using $N_{\rm rep}=1000$ replicas from both sets. All
  error bands shown correspond to one-$\sigma$.
 \label{fig:singletPDFs}} 
\end{center}
\end{figure}

The NNPDF2.1 PDFs are compared to the
previous NNPDF2.0 PDFs in
Figs.~\ref{fig:singletPDFs} (singlet sector) and~\ref{fig:valencePDFs}
(non--singlet sector).

\begin{itemize}

\item The singlet PDF at medium and small-$x$
is rather similar in the two cases, but it is somewhat larger
in the NNPDF2.1 set.
\item Thanks to the new positivity constraint on $F_2^c$ in NNPDF2.1, the gluon
remains always positive even at the largest values of $x$, where
occasionally went very slightly negative in NNPDF2.0.
\item In NNPDF2.1 the small-$x$ gluon is larger than in
NNPDF2.0. We will show that this arises from the use of a GM
scheme as compared to the ZM scheme in NNPDF2.0. 
Also, the medium and small-$x$ gluon has a somewhat larger
uncertainty in NNPDF2.1 as compared to NNPDF2.0. We will show
below that this uncertainty increase is due to the new kinematic
cut. 
\item As expected differences in the large-$x$ valence PDFs are rather
modest. The GM scheme does not affect the large-$x$ PDFs but the
cross-talk induced by the sum rules and other constraints
induces small modifications also in the valence sector, always
well below the one-$\sigma$ level.
\item The strange PDF in NNPDF2.1 is somewhat smaller than in NNPDF2.0.
This may appear surprizing as the main expected effect of the inclusion
of heavy quark mass effects is a suppression of charm which then leads
to an enhancement of all other PDFs. However, in NNPDF2.0 for dimuon data 
(which have
a sizable impact on strangeness) instead of a pure zero-mass scheme, 
the so-called
improved zero-mass (IZM) approximation of Ref.~\cite{Nadolsky:2009ge} was
used to approximate heavy quark mass effects. It turns out that this
IZM method actually overestimates heavy quark mass effects, thus
leading to a slight over-suppression of strangeness in NNPDF2.0. We
will check explicitly below (see Fig.~\ref{fig:cchqcomp}) that when
comparing NNPDF2.1 to a pure zero-mass fit strangeness is somewhat
enhanced as one would expect.
\begin{figure}[ht!]
\begin{center}
\epsfig{width=0.49\textwidth,figure=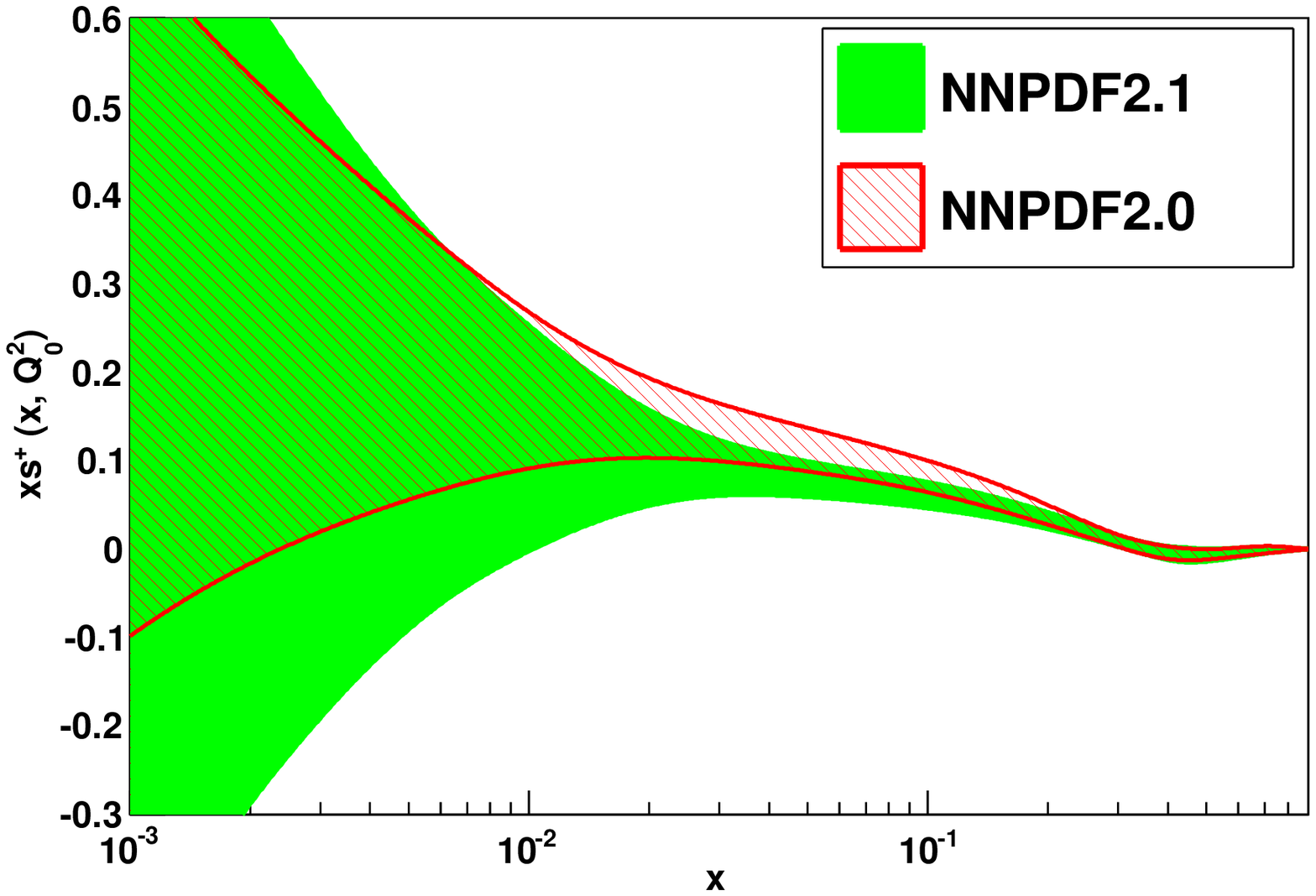}
\epsfig{width=0.49\textwidth,figure=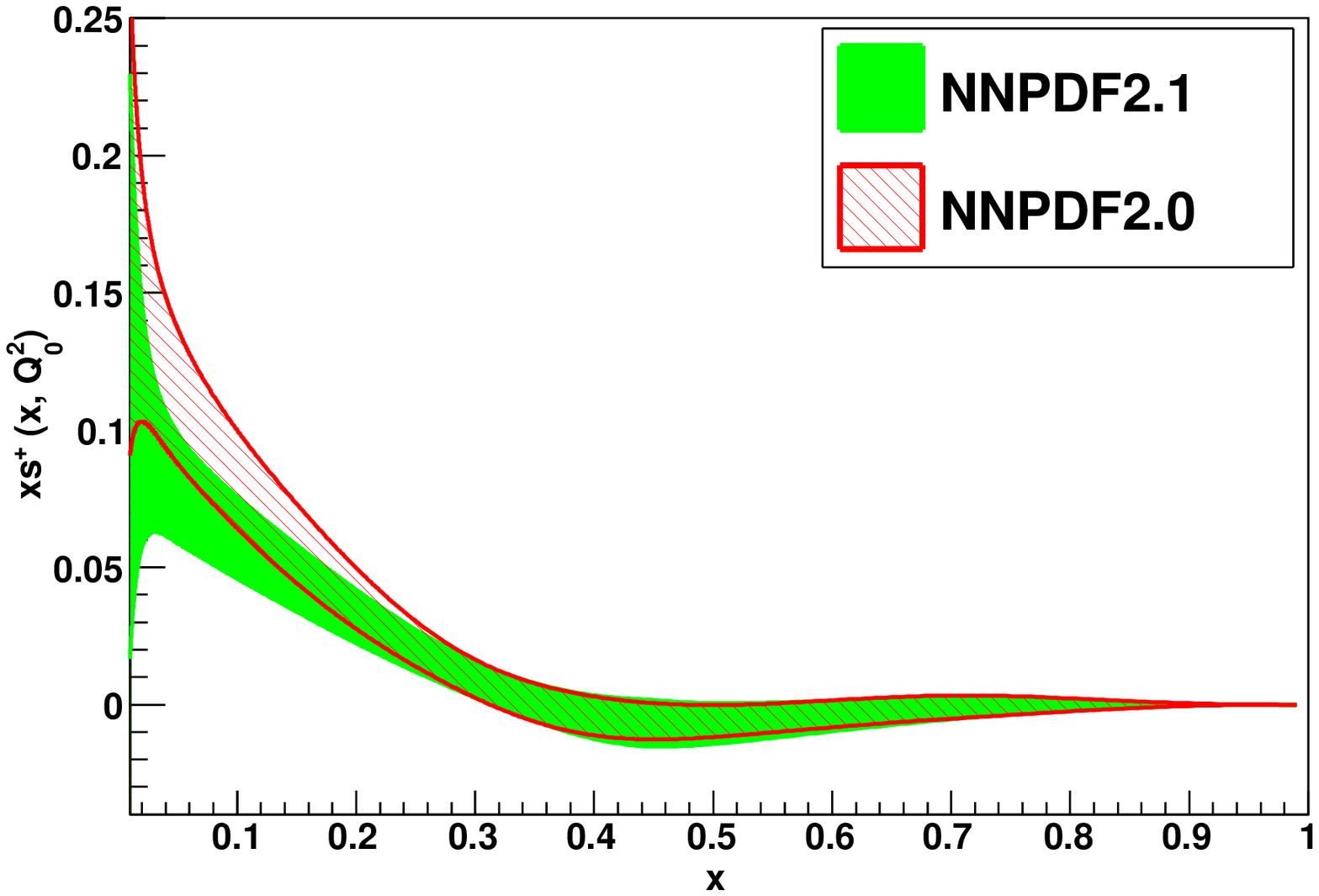}
\epsfig{width=0.49\textwidth,figure=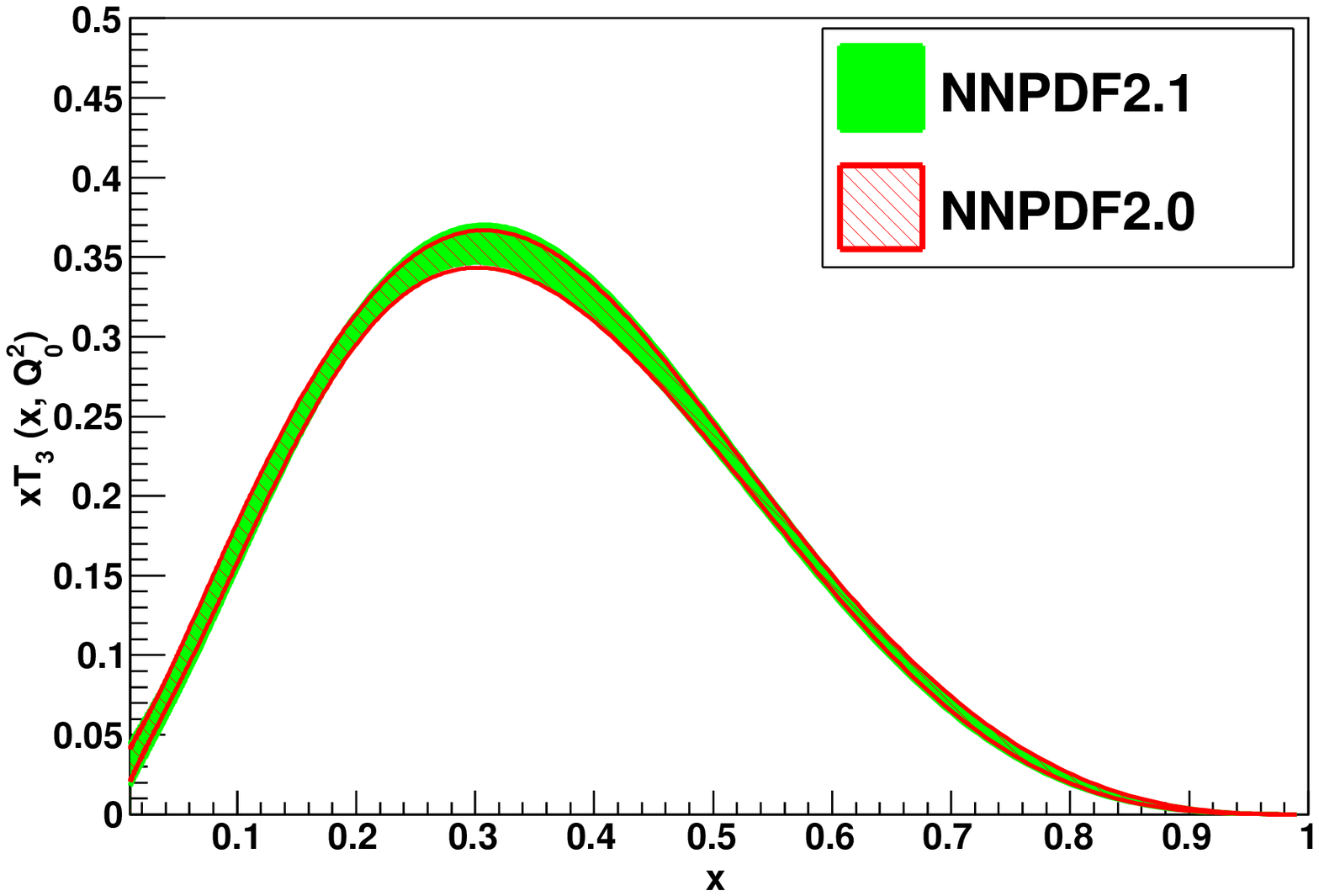}
\epsfig{width=0.49\textwidth,figure=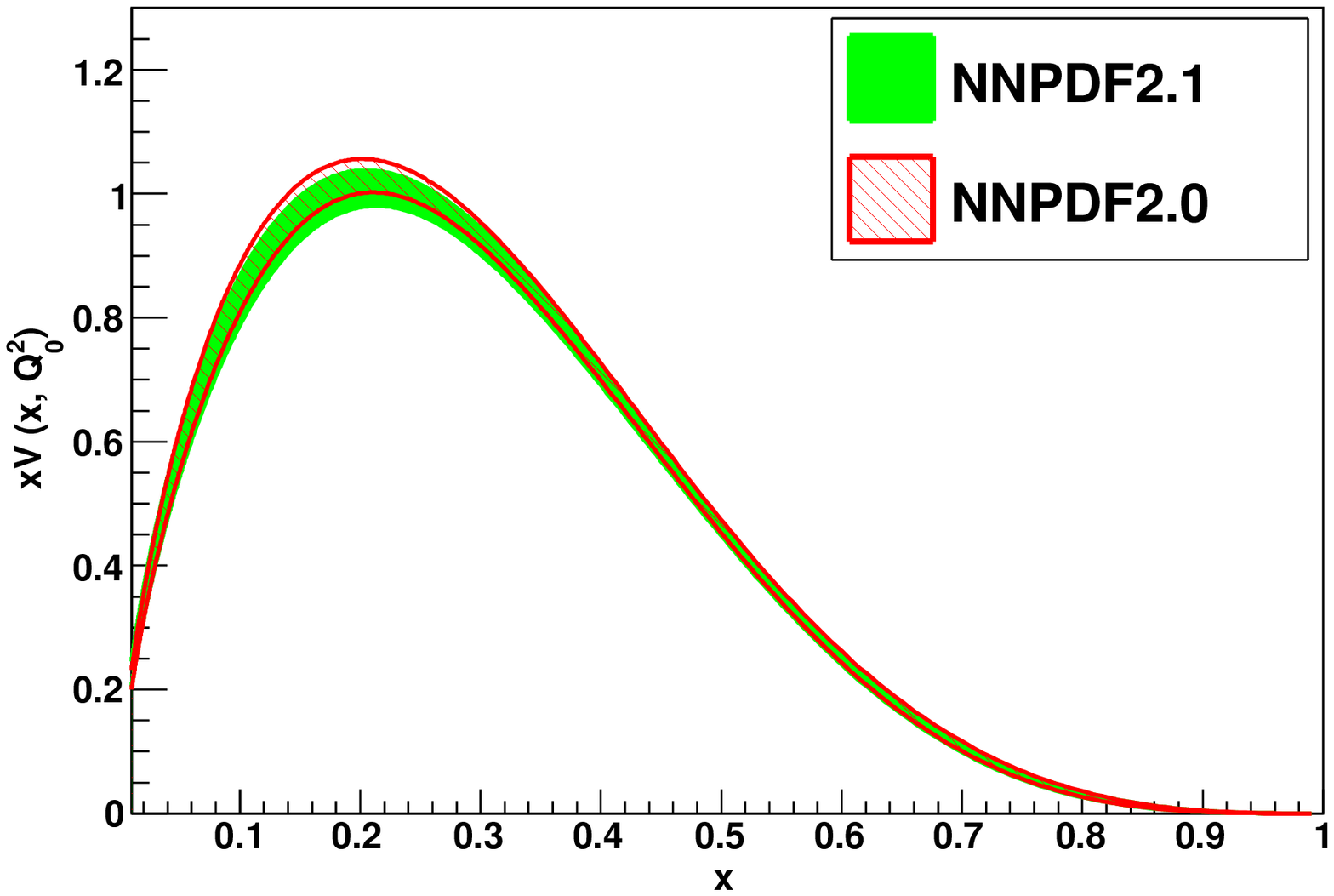}
\epsfig{width=0.49\textwidth,figure=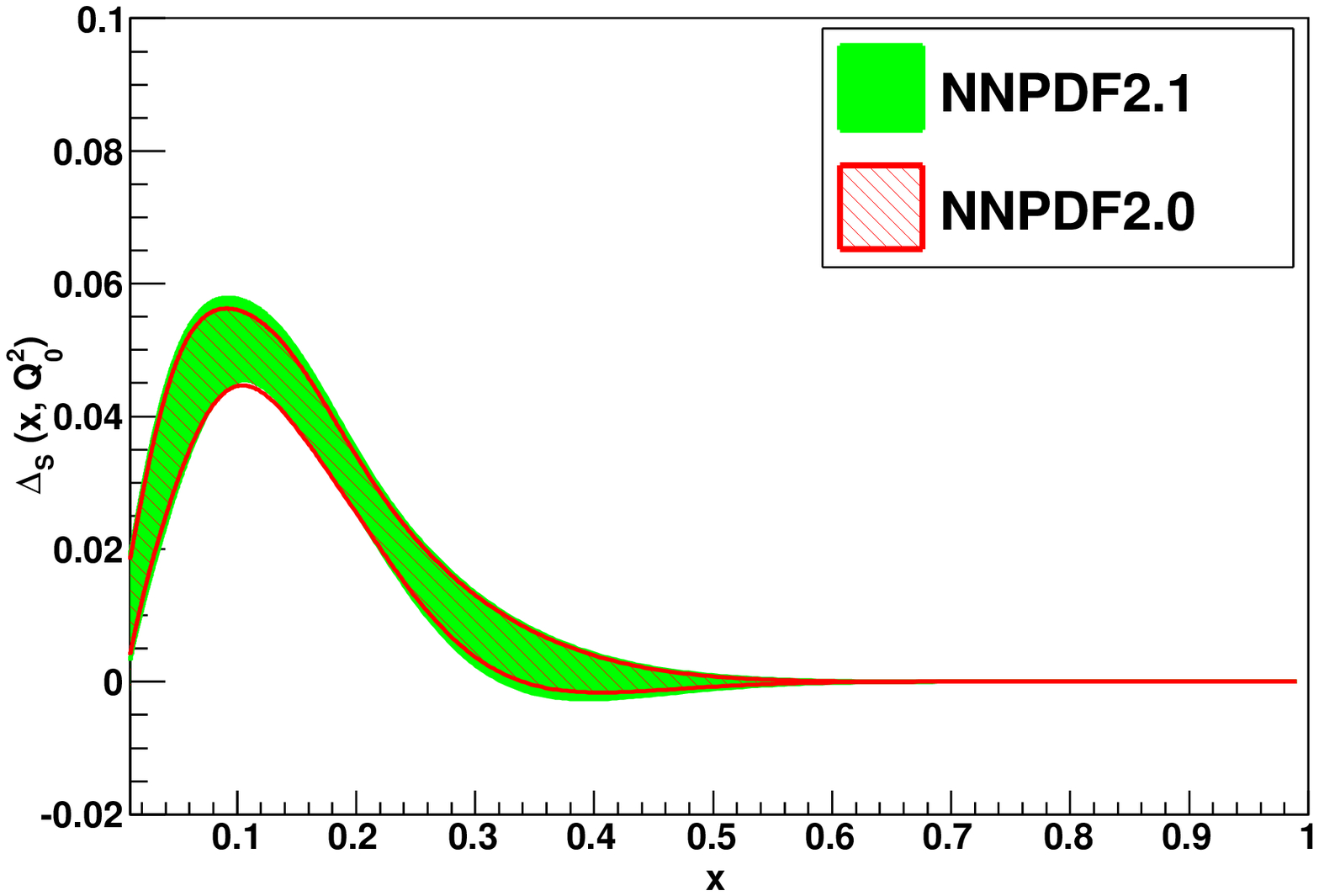}
\epsfig{width=0.49\textwidth,figure=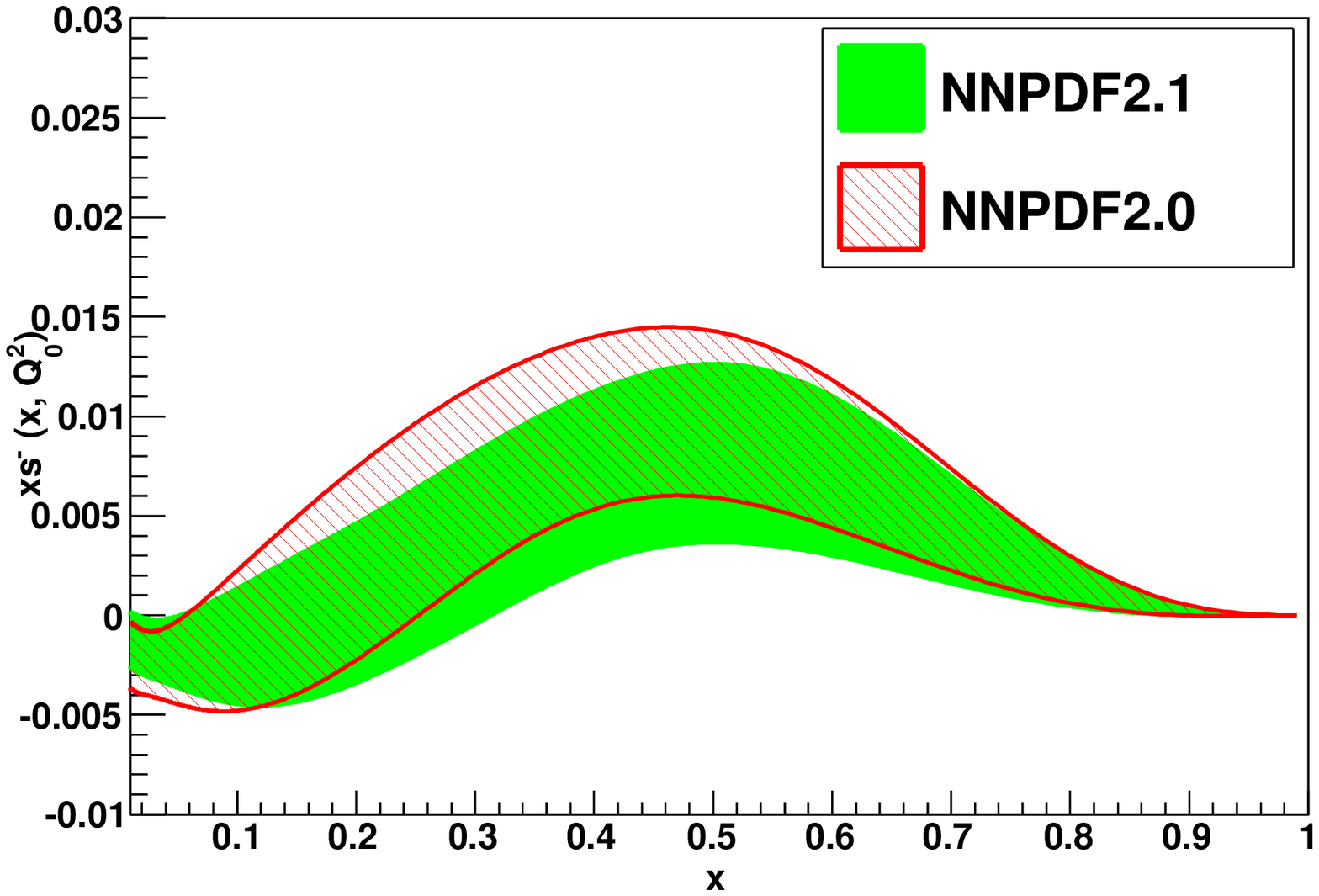}
\caption{\small Same as Fig.~\ref{fig:singletPDFs}
for the non--singlet sector PDFs.
 \label{fig:valencePDFs}} 
\end{center}
\end{figure}
\item We see from Fig.~\ref{fig:valencePDFs} that the strange asymmetry in 
NNPDF2.1 is very close
to that of NNPDF2.0. An important result of the NNPDF2.0 analysis was that the
strange asymmetry $s^-(x,Q^2)$ was of the proper size to completely
cancel the so-called NuTeV anomaly with rather reduced uncertainties.
It is clear that this holds true also with the updated NNPDF2.1 set,
confirming the results of the analysis of~\cite{Ball:2009mk} that showed that
heavy quark mass effects have a very moderate impact in the
determination of the strangeness asymmetry. The implications for the
NuTeV anomaly will be discussed in Sect.~\ref{sec:nutev}.
\end{itemize}

\begin{figure}[t!]
\begin{center}
\epsfig{width=0.99\textwidth,figure=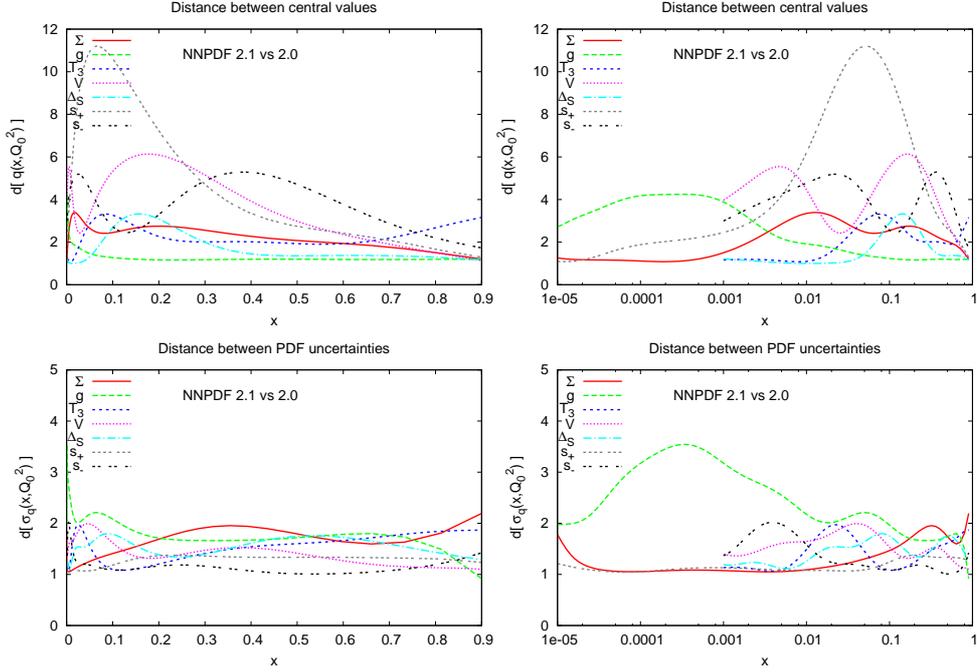}
\caption{\small Distance between the NNPDF2.0 and
NNPDF2.1 parton sets. All distances are computed from
sets of $N_{\rm rep}=100$ replicas. \label{fig:dist_20_21}}
\end{center}
\end{figure}
\begin{figure}[ht!]
\begin{center}
\epsfig{width=0.49\textwidth,figure=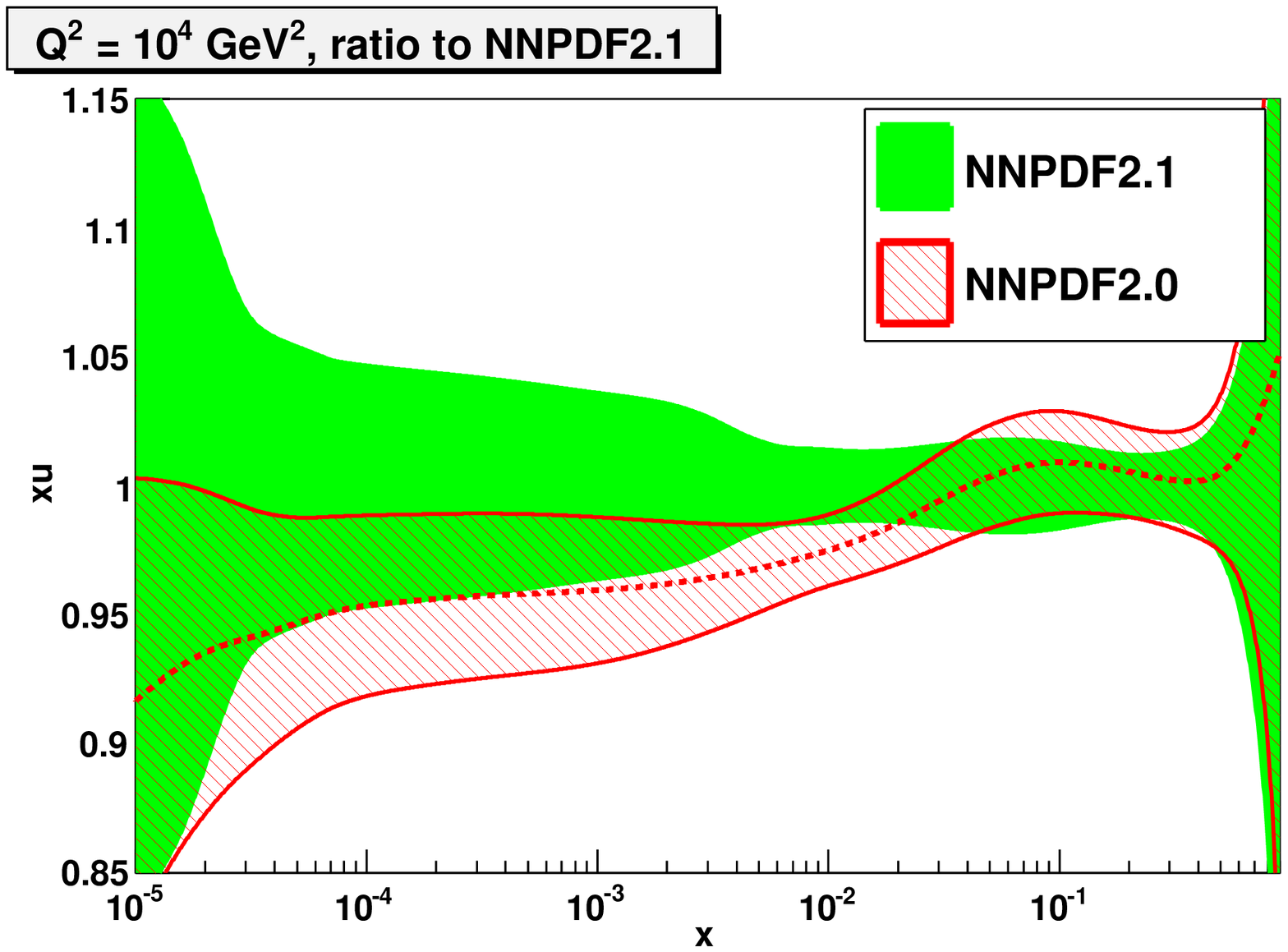}
\epsfig{width=0.49\textwidth,figure=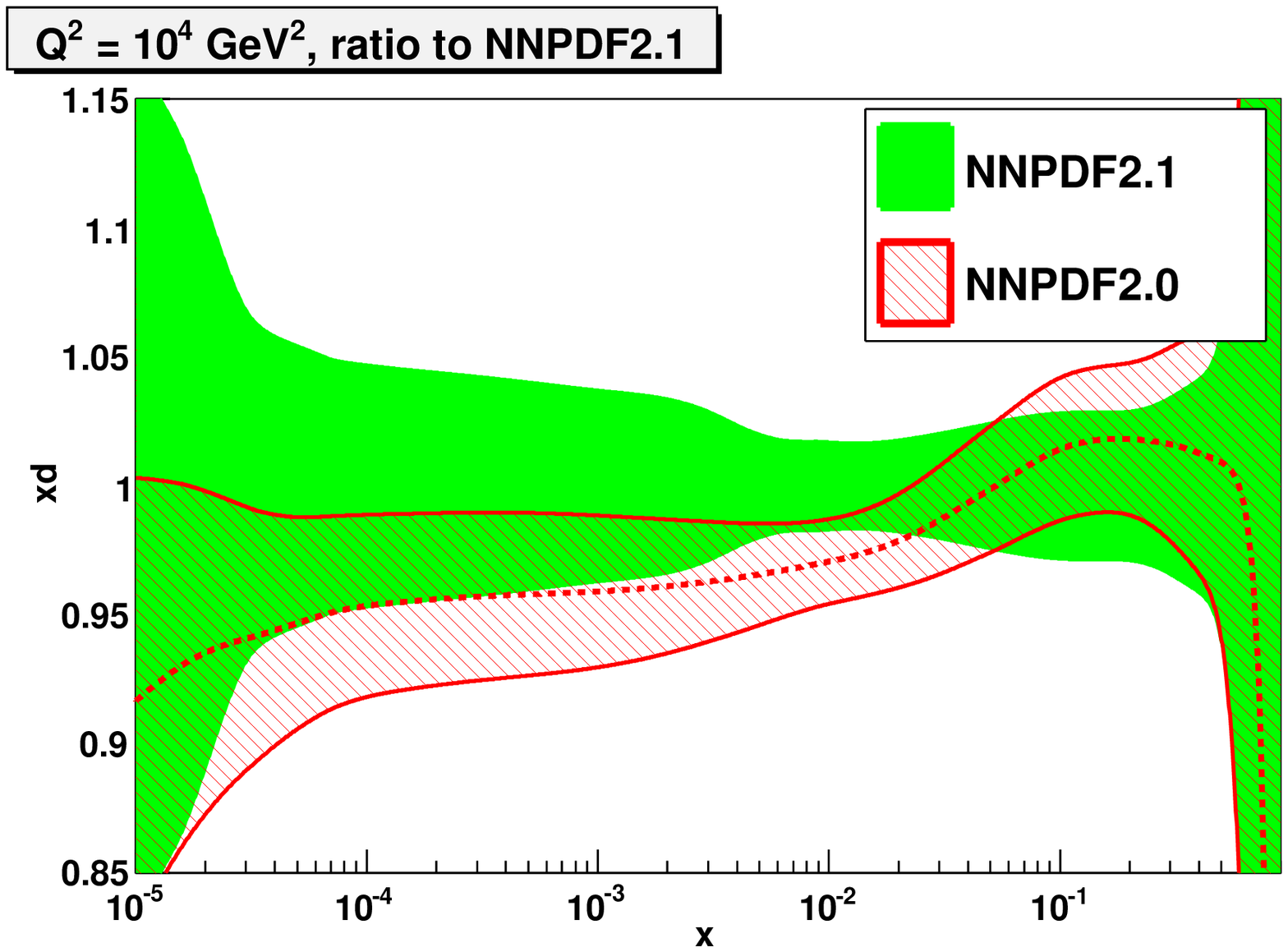}
\epsfig{width=0.49\textwidth,figure=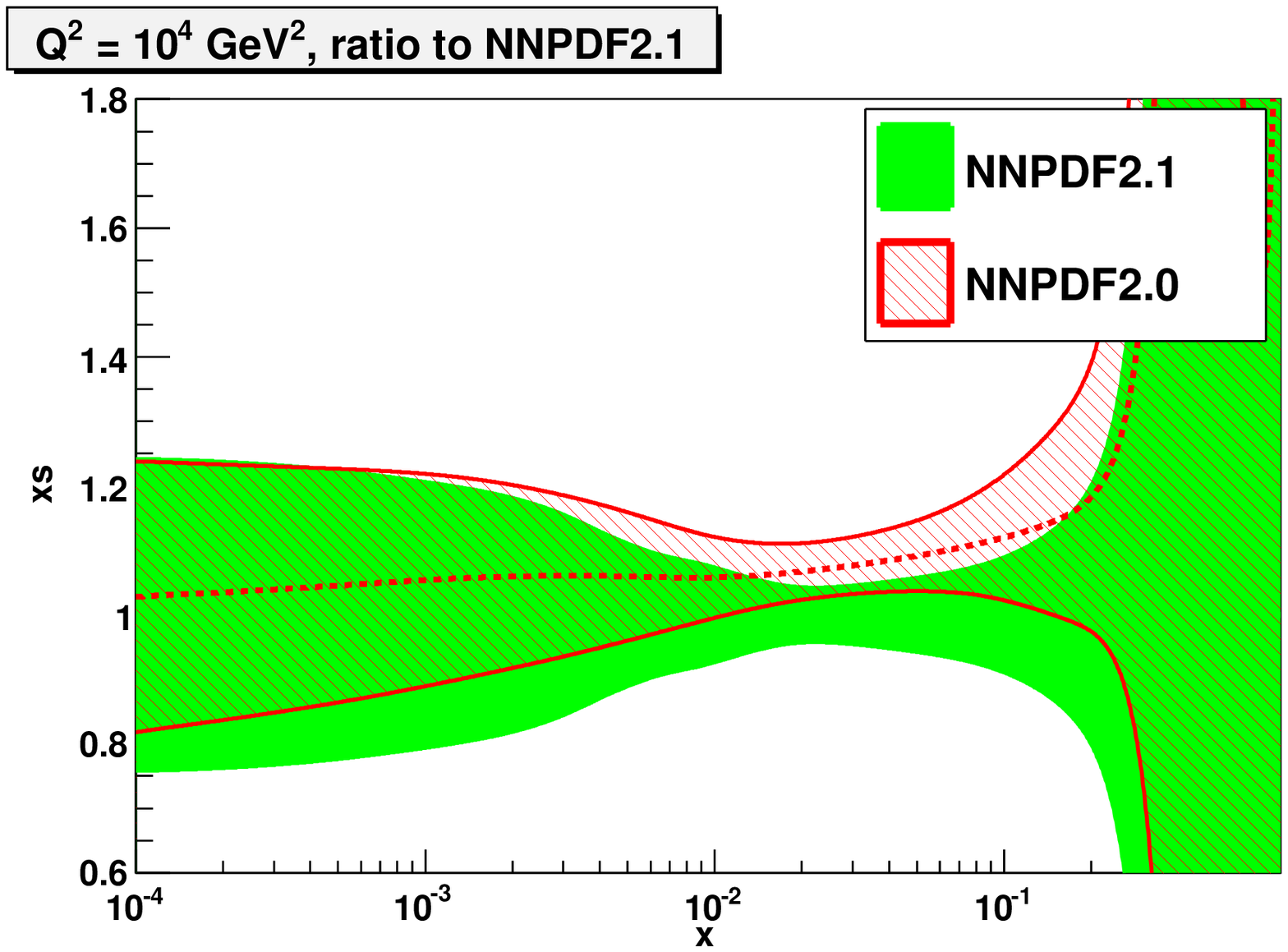}
\epsfig{width=0.49\textwidth,figure=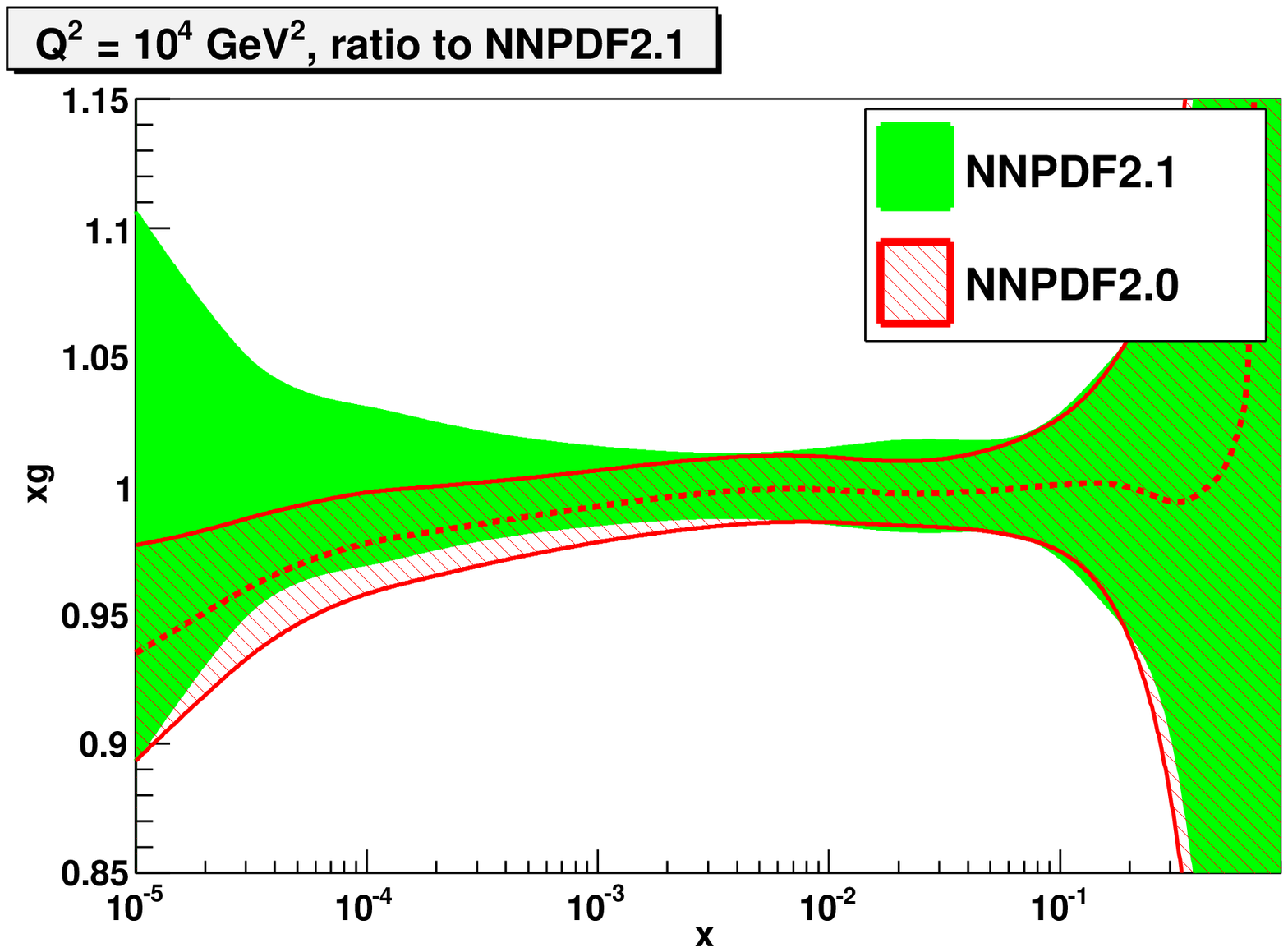}
\caption{\small Comparison between NNPDF2.1 and NNPDF2.0 light quark
  and gluon PDFs at $Q^2=10^4$
  GeV$^2$.
All curves are shown as ratios to the central NNPDF2.1 result.
\label{fig:nnpdf21ratcomp}} 
\end{center}
\end{figure}
\begin{figure}[hb!]
\begin{center}
\epsfig{width=0.49\textwidth,figure=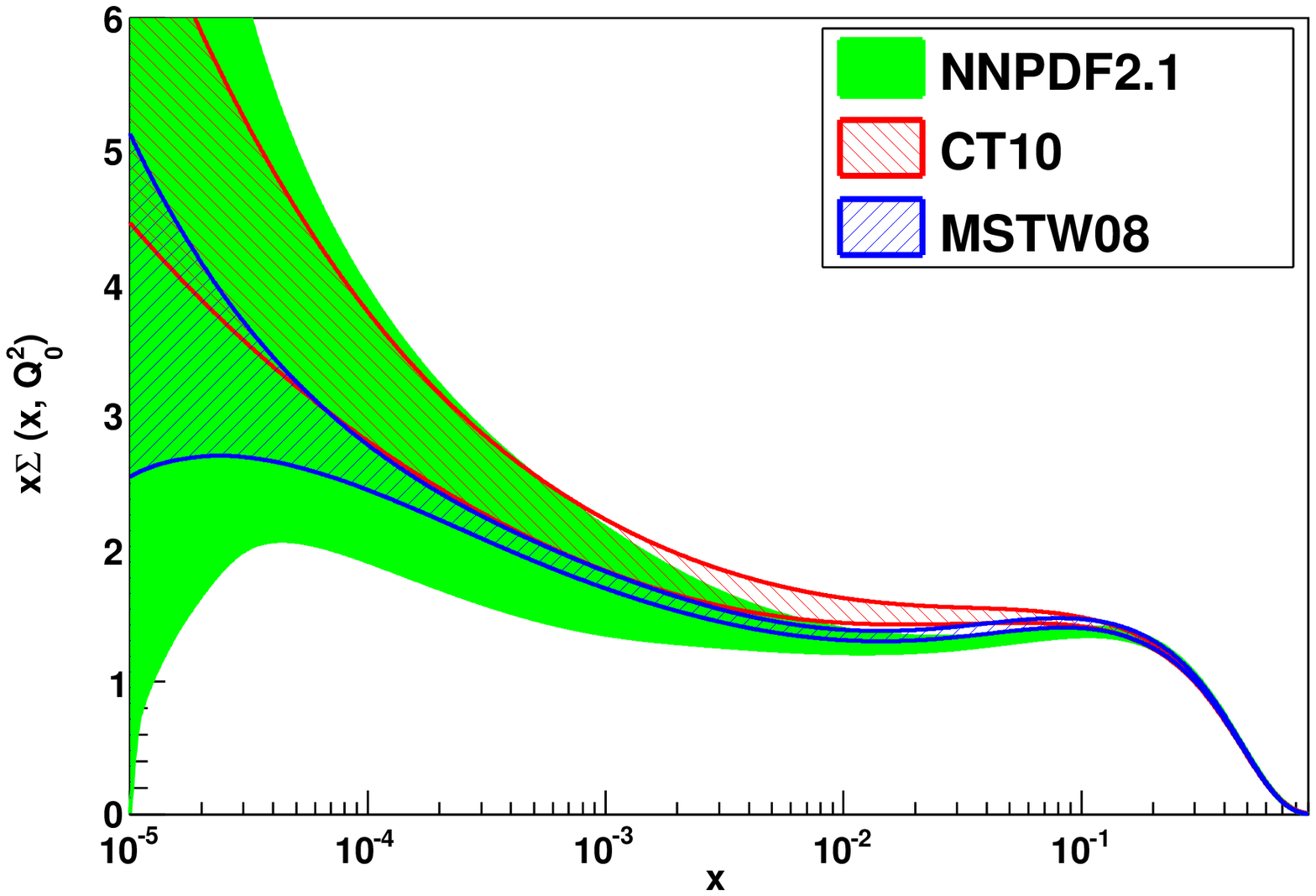}
\epsfig{width=0.49\textwidth,figure=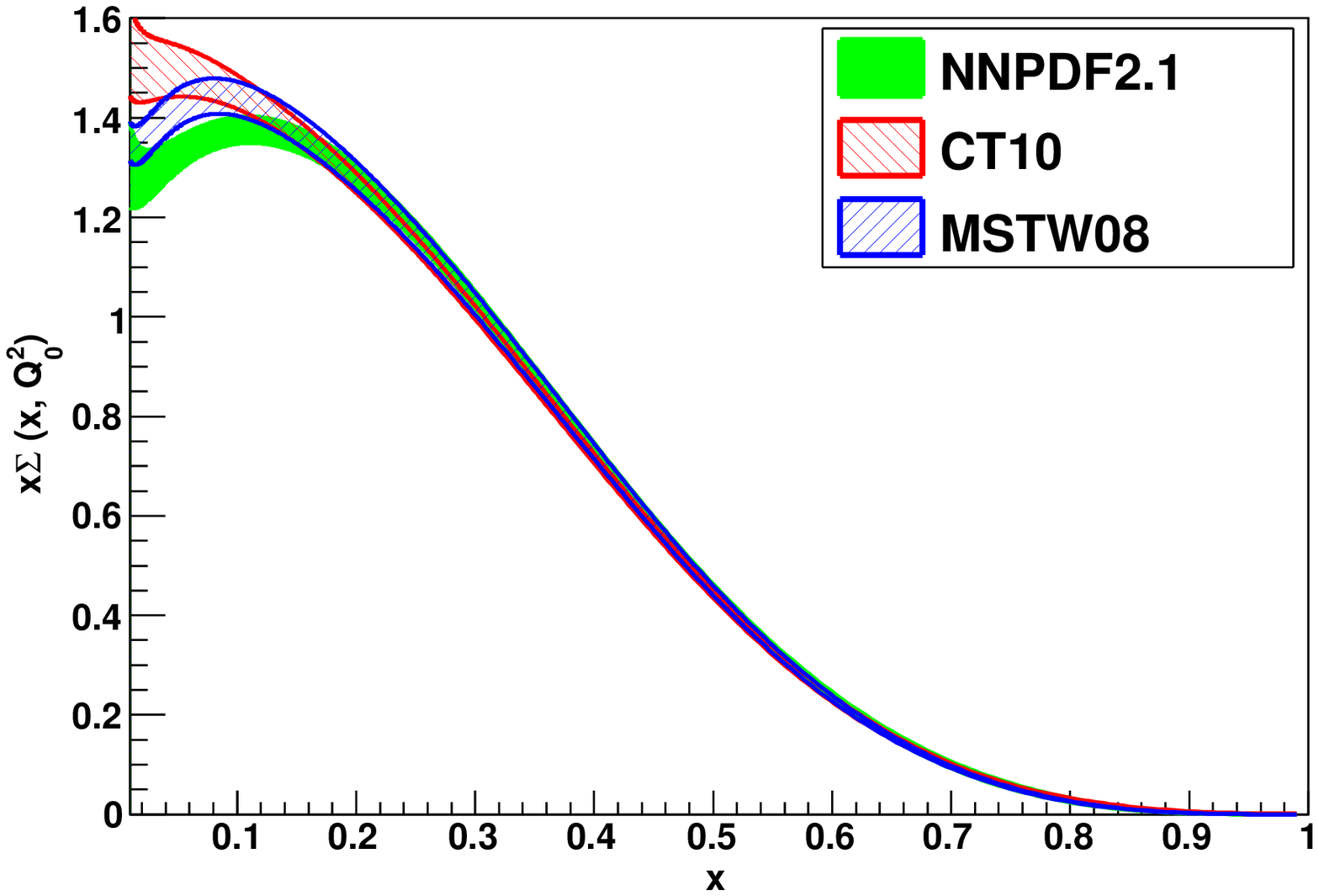}
\epsfig{width=0.49\textwidth,figure=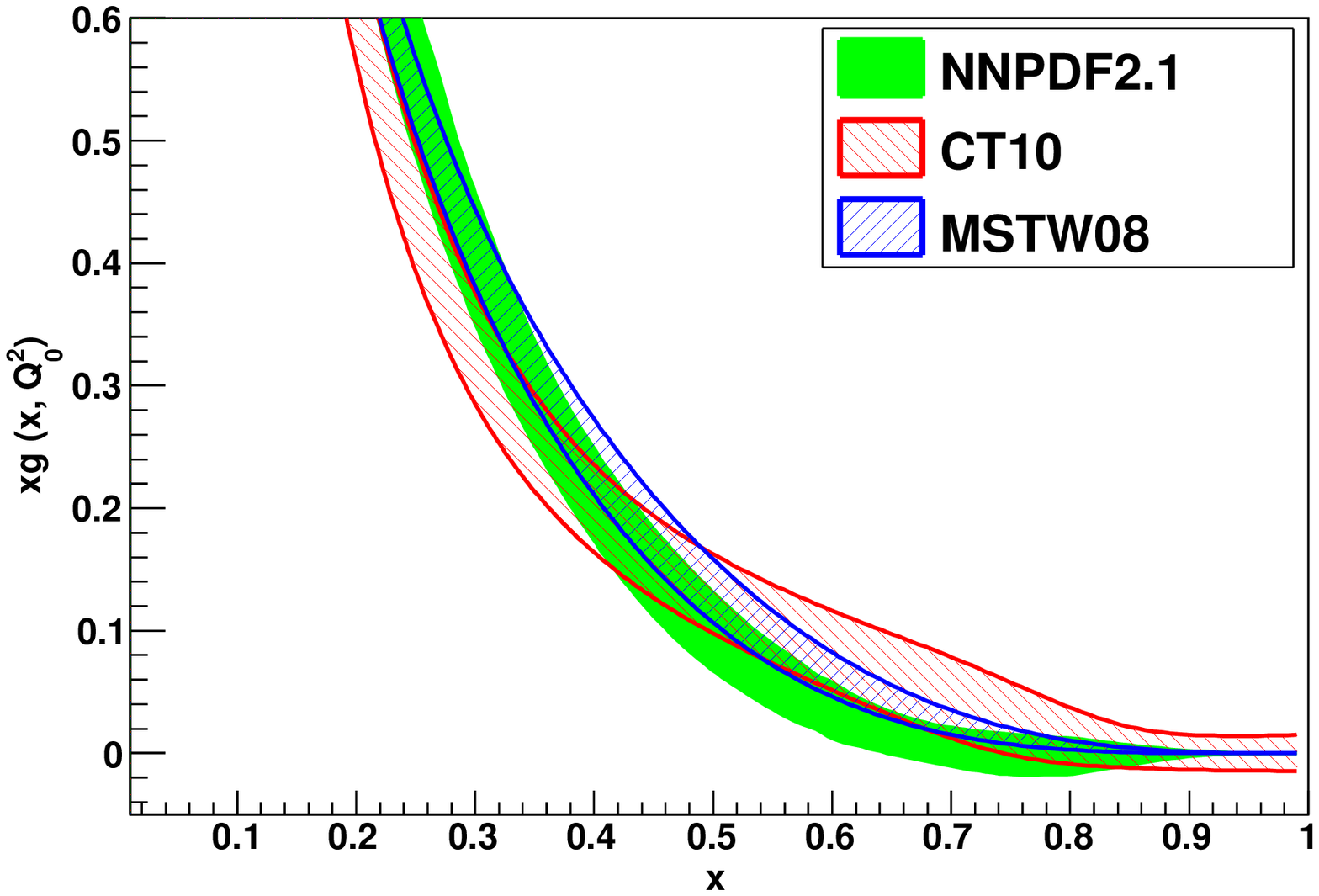}
\epsfig{width=0.49\textwidth,figure=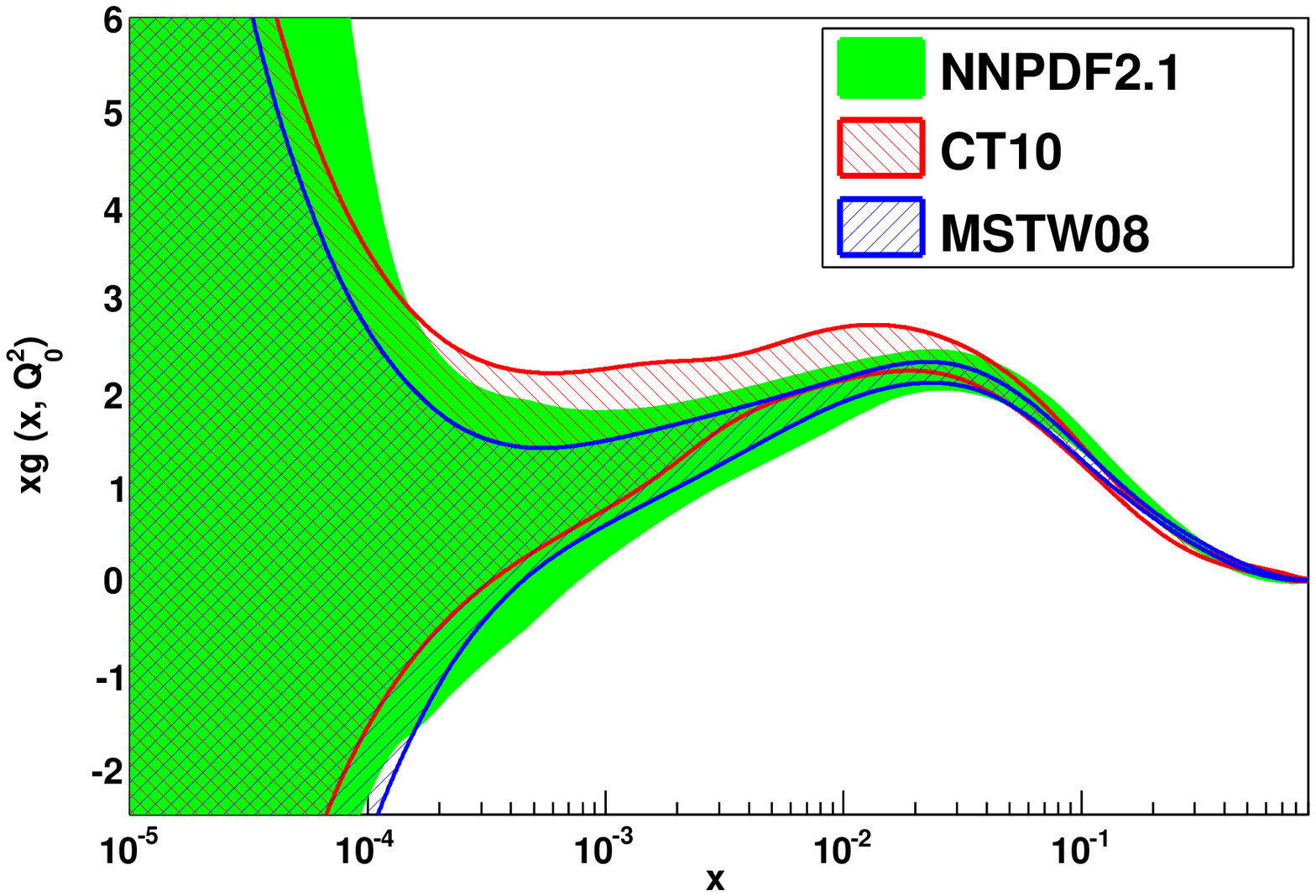}
\caption{\small The NNPDF2.1 singlet sector PDFs, compared
with the CT10 and
MSTW08 PDFs.
The results for  NNPDF2.1  have been obtained with
$N_{\rm rep}=1000$ replicas. All PDF errors are given 
as one-$\sigma$ uncertainties.
 \label{fig:singletPDFs-lhapdf}} 
\end{center}
\end{figure}
\begin{figure}[ht!]
\begin{center}
\epsfig{width=0.49\textwidth,figure=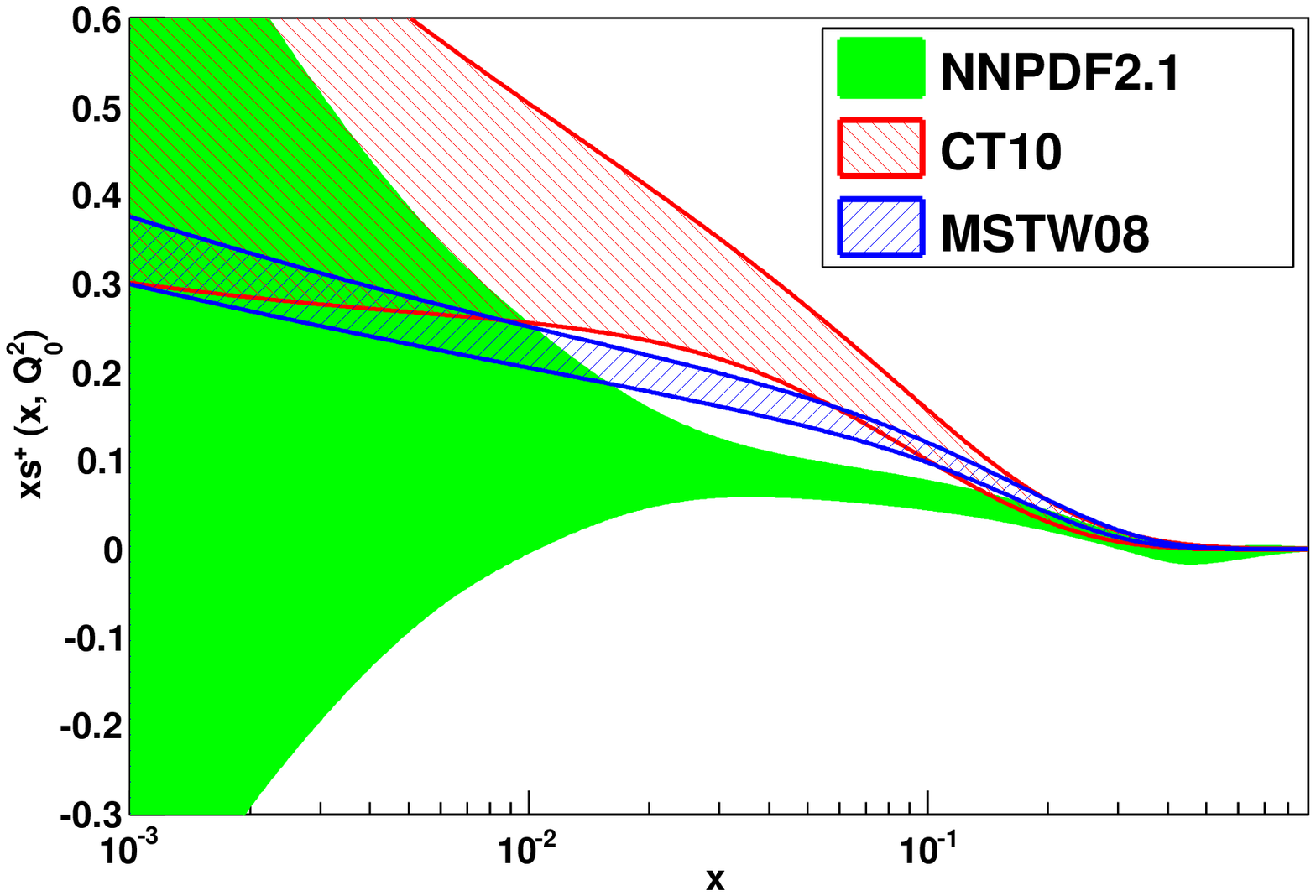}
\epsfig{width=0.49\textwidth,figure=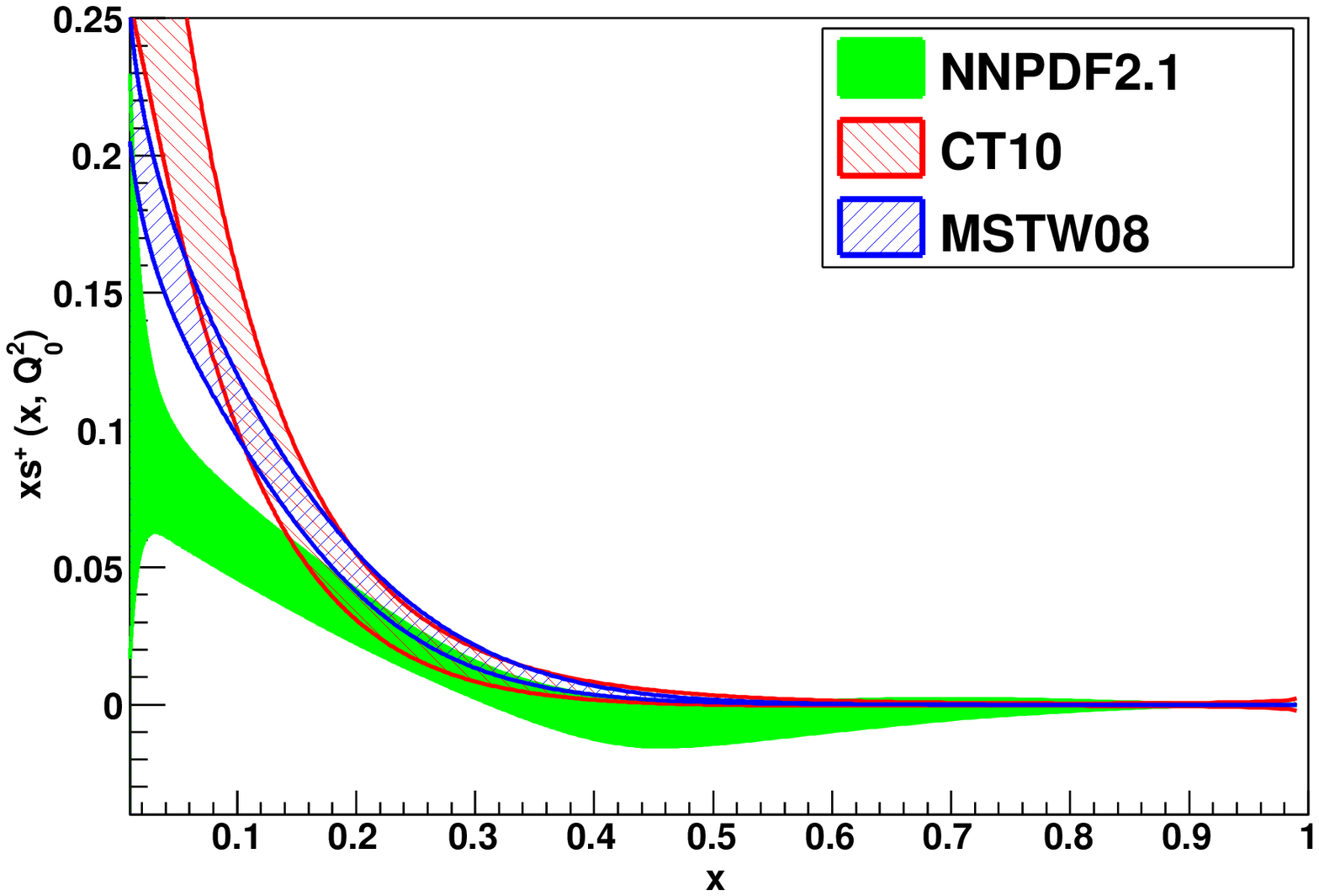}
\epsfig{width=0.49\textwidth,figure=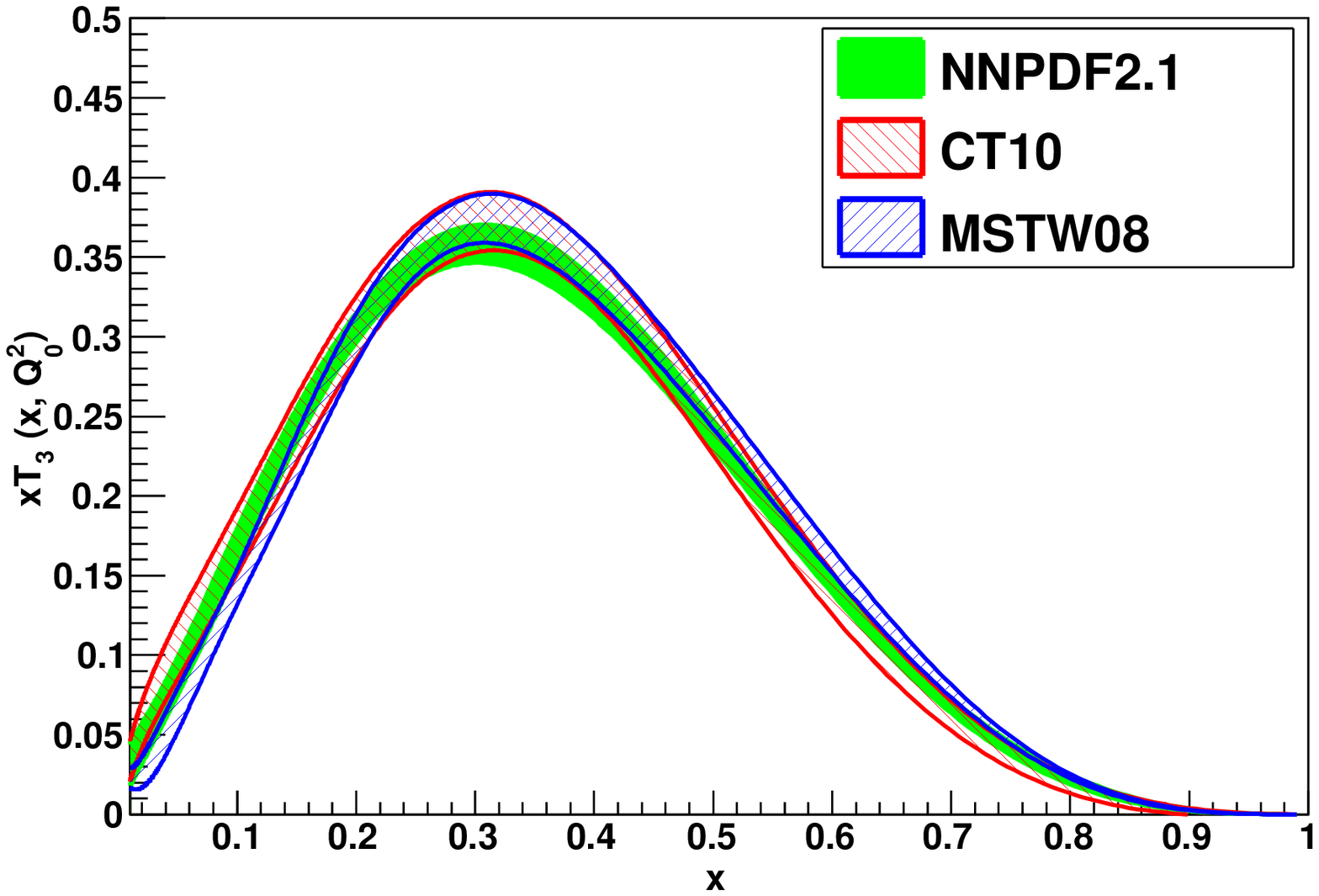}
\epsfig{width=0.49\textwidth,figure=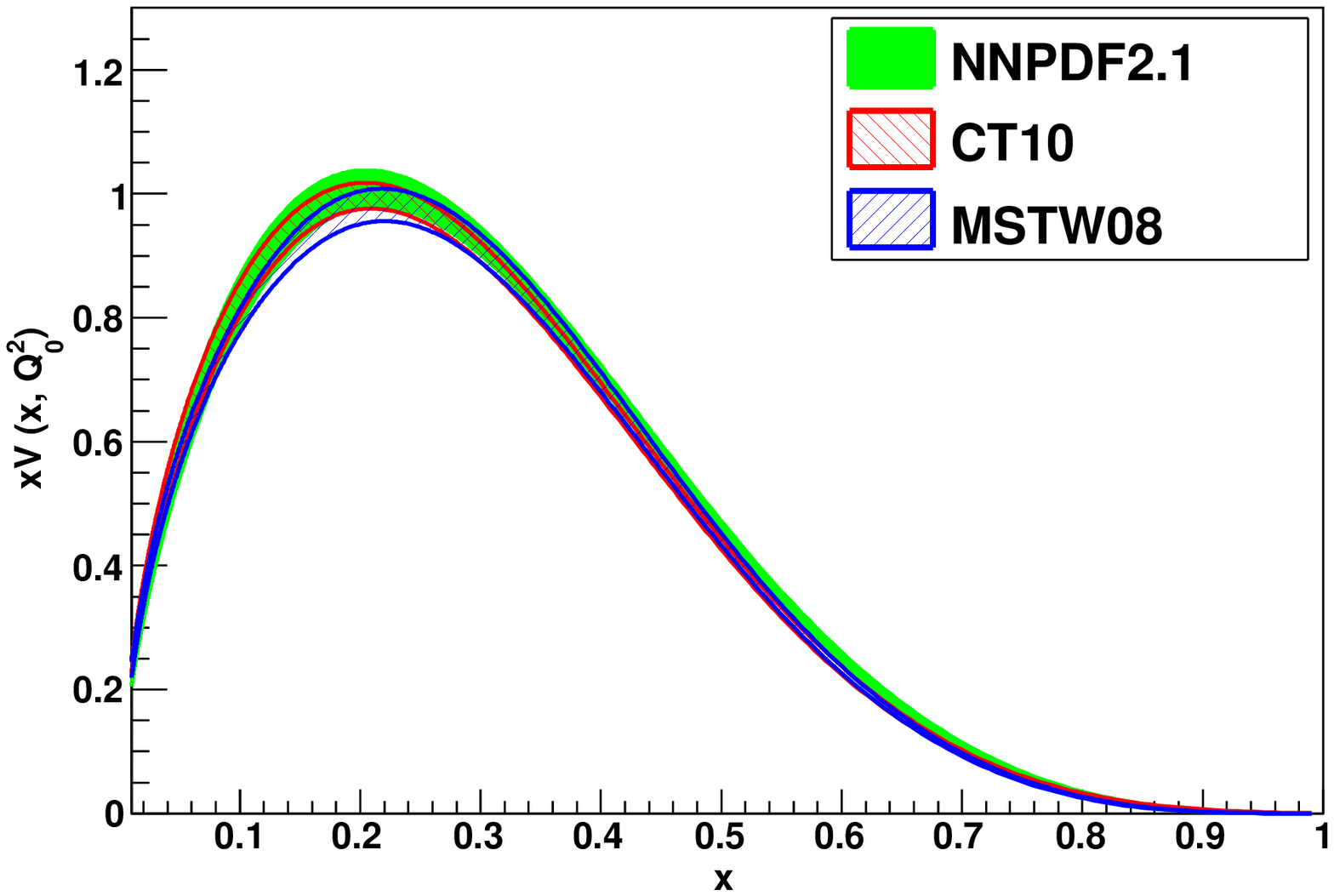}
\epsfig{width=0.49\textwidth,figure=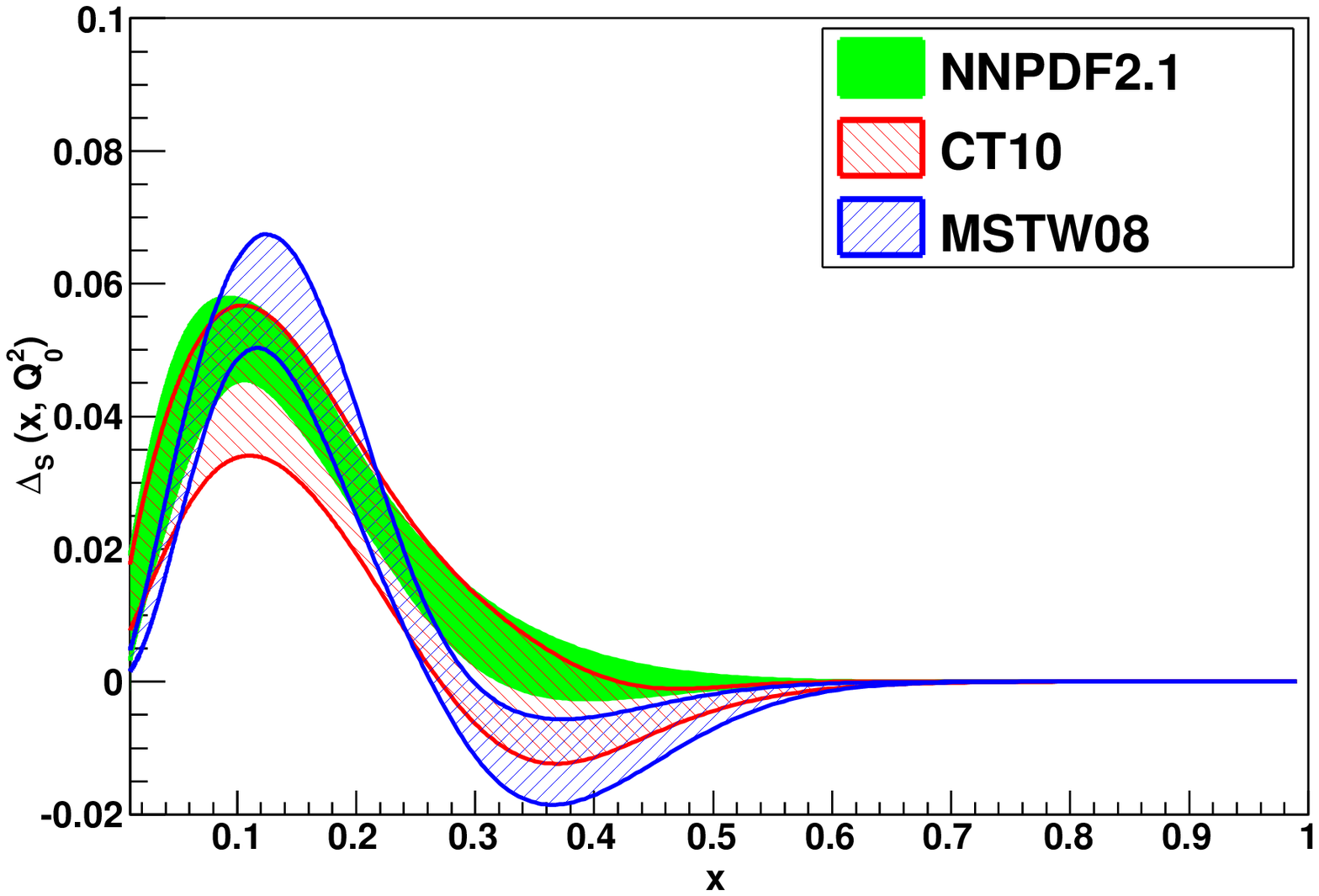}
\epsfig{width=0.49\textwidth,figure=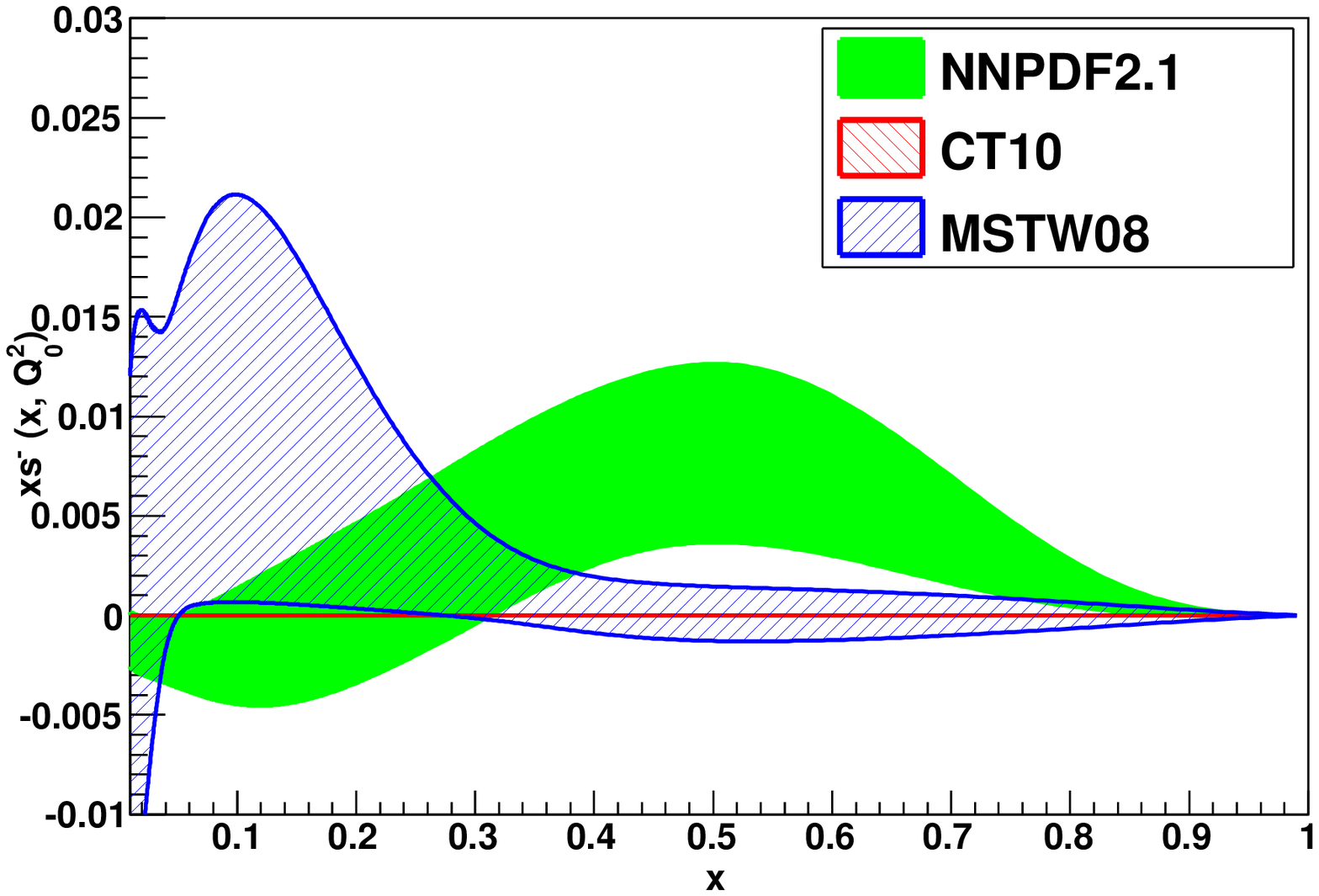}
\caption{\small Same as Fig.~\ref{fig:singletPDFs-lhapdf}
for the non--singlet sector PDFs.
 \label{fig:valencePDFs-lhapdf}} 
\end{center}
\end{figure}
The comparison between NNPDF2.0 and NNPDF2.1 is
further quantified by the computation of the
distance~\cite{Ball:2010de} between the two sets, shown in  
Fig.~\ref{fig:dist_20_21}. Note that $d\sim 1$ corresponds to two sets
of replicas which come from the same underlying probability
distribution, while (using $N_{\rm rep}=100$ replicas)
$d\sim 7$ corresponds to a
one-$\sigma$ difference (see Appendix~A of
Ref.~\cite{Ball:2010de}). 
One concludes that while clearly the two
sets do not come from the same underlying distributions, all PDFs but
the strange are consistent at the one-$\sigma$, and even the
strangeness is consistent at the $90\%$ confidence level. The largest
differences are seen in the medium-$x$ strangeness and to a lesser
extent in the medium and small-$x$ gluon.

The differences between NNPDF2.0 and NNPDF2.1 
PDFs at the initial scale displayed in
Figs.~\ref{fig:singletPDFs}-\ref{fig:dist_20_21}, once evolved up to
the $W$ and $Z$ scale,  are sufficient to
lead to differences between gluon and light sea quark distributions up
to the one-$\sigma$ level at small-$x$, as  shown in
Fig.~\ref{fig:nnpdf21ratcomp} where we plot the NNPDF2.0/NNPDF2.1
ratio for individual light flavours and the gluon at $Q^2=10^4$~GeV$^2$.

The NNPDF2.1 PDFs are compared to the other
global PDF sets CT10~\cite{Lai:2010vv} and
MSTW08~\cite{Martin:2009iq} in 
Figs.~\ref{fig:singletPDFs-lhapdf}-\ref{fig:valencePDFs-lhapdf}, 
which is interesting  to contrast to the
analogous plot which in Ref.~\cite{Ball:2010de} (Figs.~18-19 of that
reference) compared the NNPDF2.0, CTEQ6.6 and MSTW08 PDF sets.

\begin{itemize}
\item  The general 
agreement of the gluon in the
  medium-/small-$x$ region is improved, both because the central value
of NNPDF2.1 is now in
  better agreement with MSTW08 (most likely due to the inclusion of
  heavy quark mass effects) and also because the CT10 central value
  and especially uncertainty are in much better agreement with the
  wider NNPDF and MSTW uncertainties, due to the use of a more
  flexible gluon parametrization in CT10 with respect to CTEQ6.6. 
  The large-$x$ gluon however is in marginal agreement.
\item The small changes in valence and triplet distributions between
  NNPDF2.0 and NNPDF2.1 go anyway in the direction of improving the
  agreement with the other global sets.
\item The strange PDFs are quite different, presumably due to the fact
  that a much less flexible parametrization is adopted by CT/CTEQ and
  MSTW in comparison to NNPDF.
\item The medium-$x$ singlet is in marginal agreement. This may
  be a byproduct of the poor agreement in strangeness. 
\end{itemize}
The effect on LHC observables will be discussed in
Sect.~\ref{sec:pheno}, where we will show that even though there
is  generally a 
reasonable agreement between global sets, there remain  some
significant differences, mostly related to the rather
different large-$x$ gluon in CT10 as shown in
Fig.~\ref{fig:singletPDFs-lhapdf}.


\subsection{Detailed comparison to NNPDF2.0: theoretical framework and dataset}
\label{sec:dataset}

We now assess the separate impact of the inclusion of heavy quark mass
effects and of the charm structure function data on the NNPDF2.1 PDF
determination. 

First, we must  discuss the impact of raising the kinematic cut $Q^2_{\rm min}$ 
within the ZM scheme; then we compare the ZM 
and GM fits; next  we investigate the impact of including the HERA charm 
structure function data into the NNPDF2.1 analysis; finally we estimate the 
impact of ambiguities related to the treatment of heavy quarks.
In each case, we will show the distances between PDFs as well as the 
PDFs that are most affected by each step.

\begin{itemize}

\item New kinematical cut $Q_{\rm min}^2$

In Fig.~\ref{fig:distances-21-olddataset} we show
the distance between NNPDF2.0 PDFs
and a fit with the same dataset but with the new
cut $Q^2_{\rm cut}= 3$ GeV$^2$, denoted
by NNPDF2.0RED (reduced). Also, in order to ease the subsequent
discussion of the impact of heavy quark mass effects,
in NNPDF2.0RED a pure ZM scheme is
used for all observables, rather than the IZM
scheme~\cite{Nadolsky:2009ge} used for dimuon data in
Ref.~\cite{Ball:2010de}. 

\begin{figure}[t!]
\begin{center}
\epsfig{width=0.99\textwidth,figure=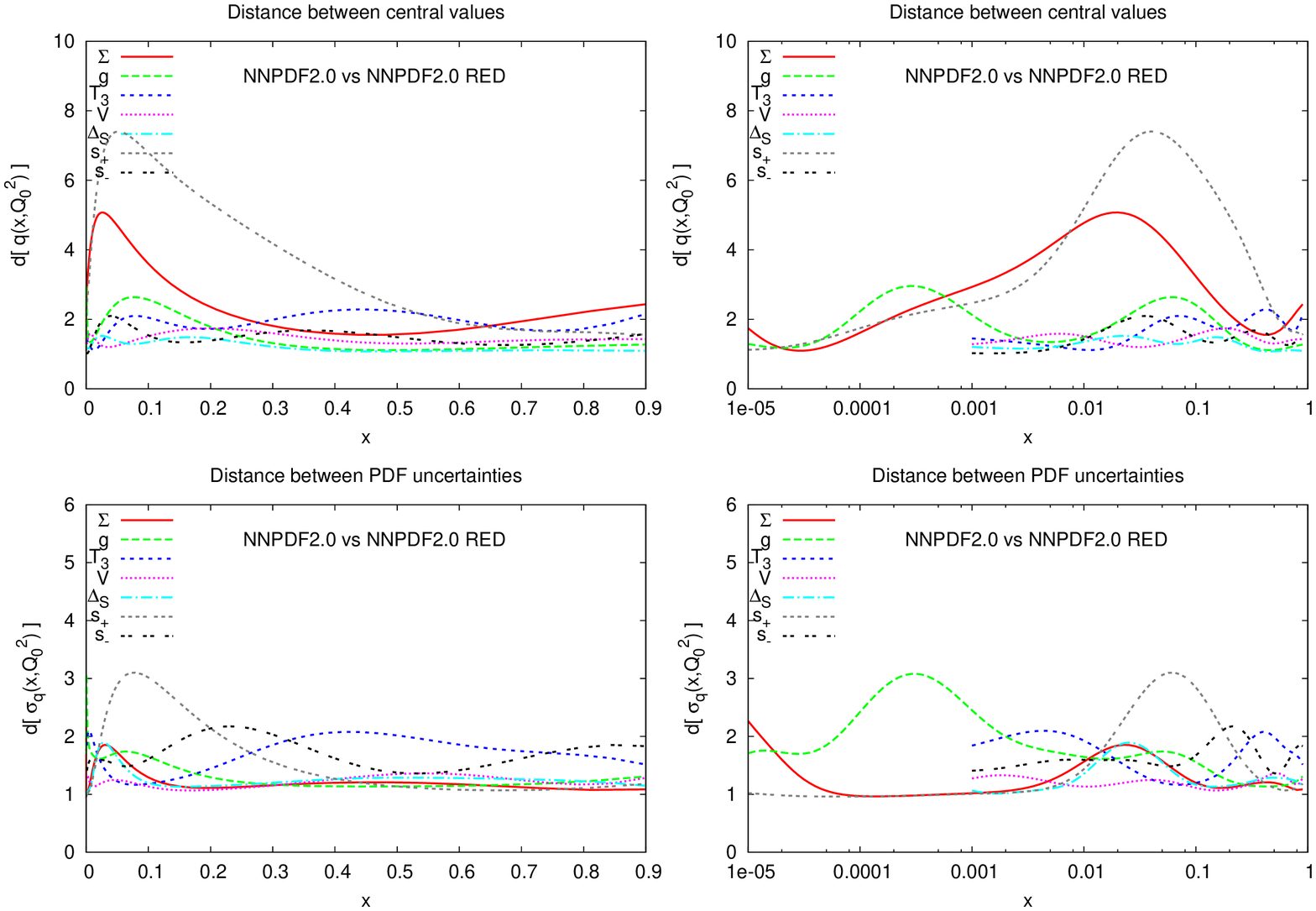}
\caption{\small Distance between the NNPDF2.0 PDF set
and a fit to the same data but with $Q^2_{\rm cut}= 3$ GeV$^2$
and the ZM-VFN scheme for all observables (NNPDF2.0 RED).
All distances are computed from sets of $N_{\rm rep}=100$ replicas.
\label{fig:distances-21-olddataset}} 
\epsfig{width=0.49\textwidth,figure=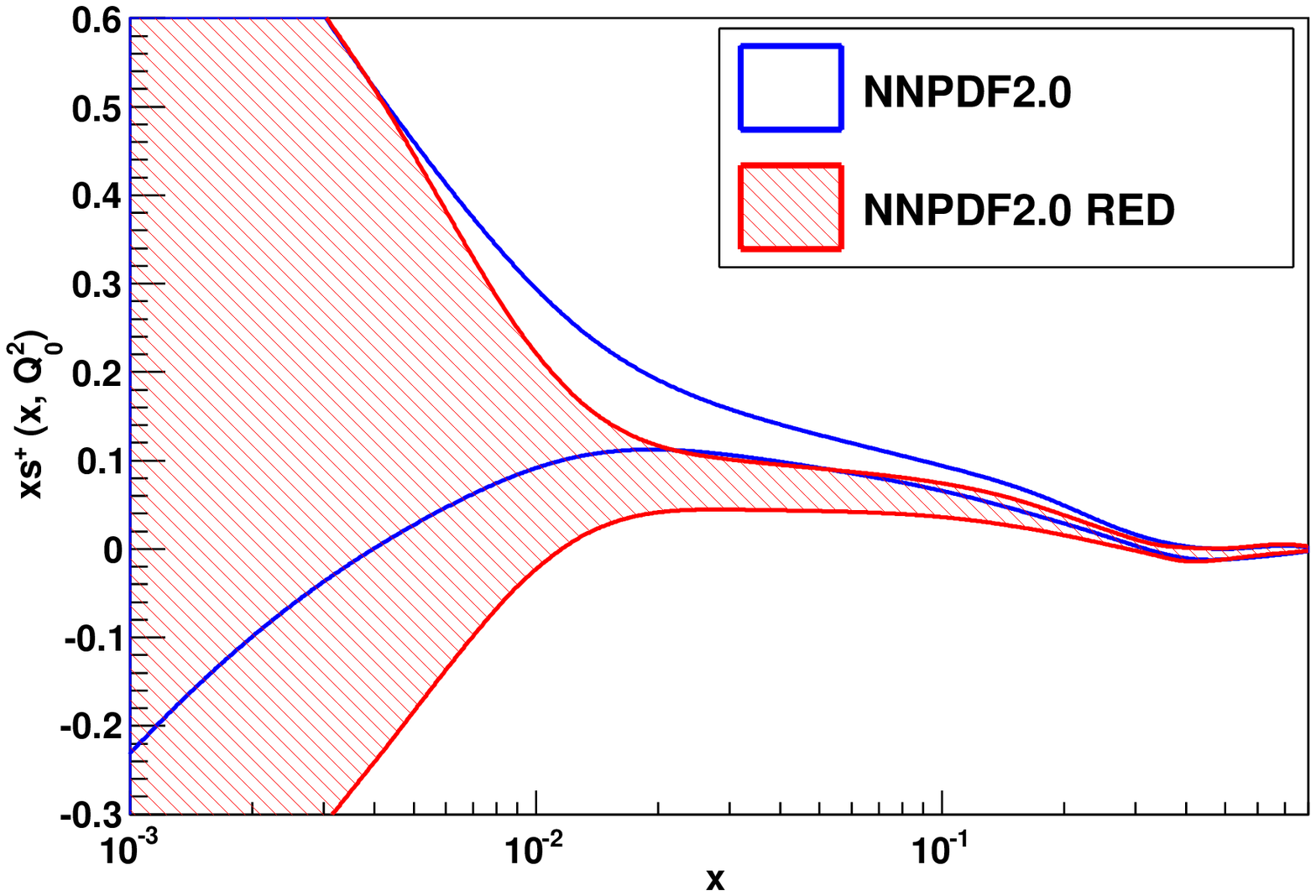}
\epsfig{width=0.49\textwidth,figure=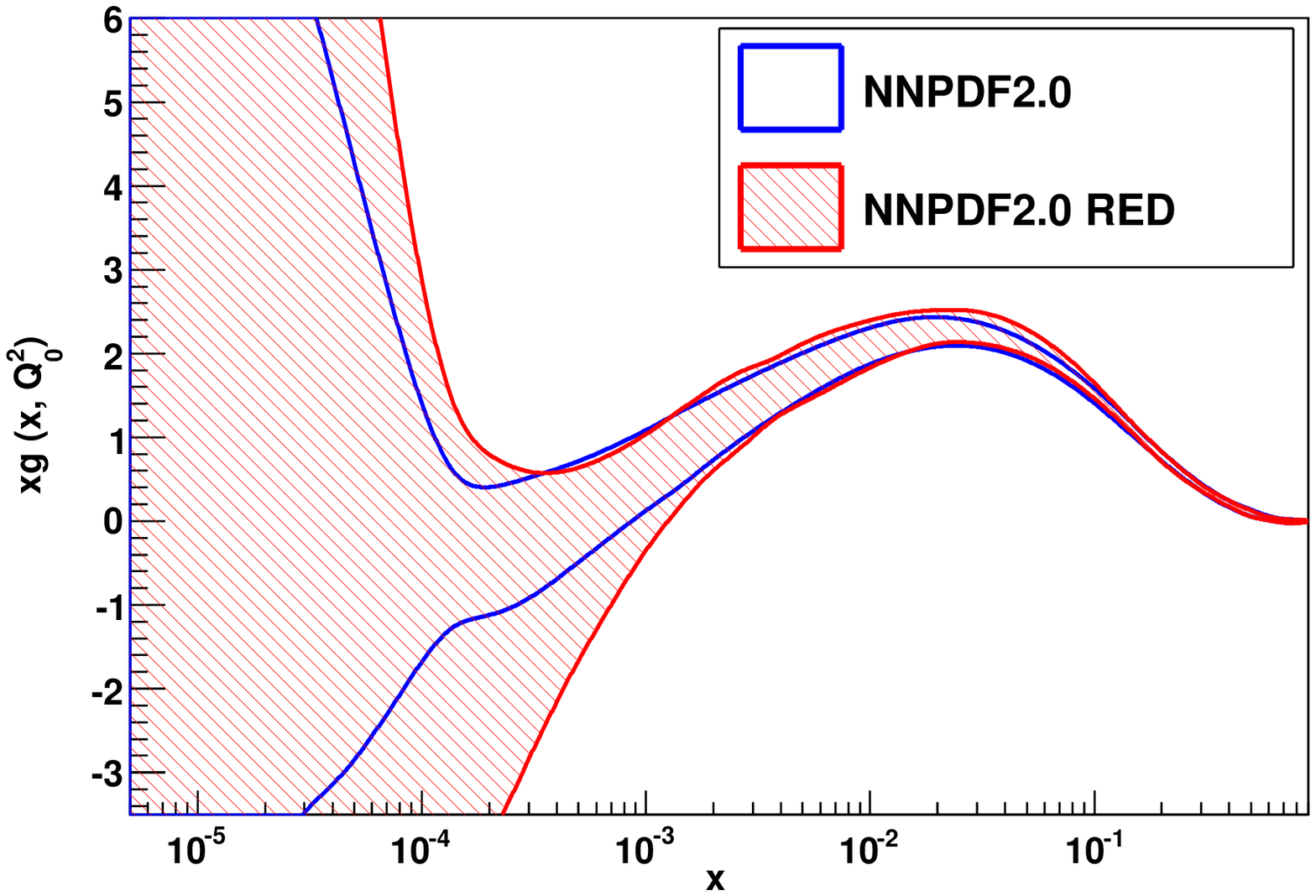}
\caption{\small Comparison of the small--$x$ total strangeness  and
  gluon in NNPDF2.0 and
in NNPDF2.0RED (the distances are shown  in
Fig.~\ref{fig:distances-21-olddataset}).
 \label{fig:PDFs-zm+newcut}} 
\end{center}
\end{figure}

The largest distances correspond to the medium-$x$
strange PDFs and the small-$x$ gluon. These PDFs are shown
in Fig.~\ref{fig:PDFs-zm+newcut}. The strange is rather smaller in the
ZM as compared to the IZM scheme, where it was enhanced due to the
approximate inclusion of charm suppression.
The gluon is somewhat
smaller at small-$x$ and with rather larger uncertainties, due
to  the reduction in dataset at small-$x$ caused by the new kinematic cut.
From the distances we see that the singlet is also modified, but one
can check that this is completely due to the strange contribution to it.

\item Impact of the general-mass scheme

The impact of the FONLL-A GM is assessed by now comparing the NNPDF2.0RED 
fit to a NNPDF2.1 fit without $F_2^c$ data: the dataset is identical,
the only difference is in the treatment of heavy quark masses.
The distances between these two sets are shown in
Fig.~\ref{fig:distances-21-gm-vs-zm}.

\begin{figure}[t!]
\begin{center}
\epsfig{width=0.99\textwidth,figure=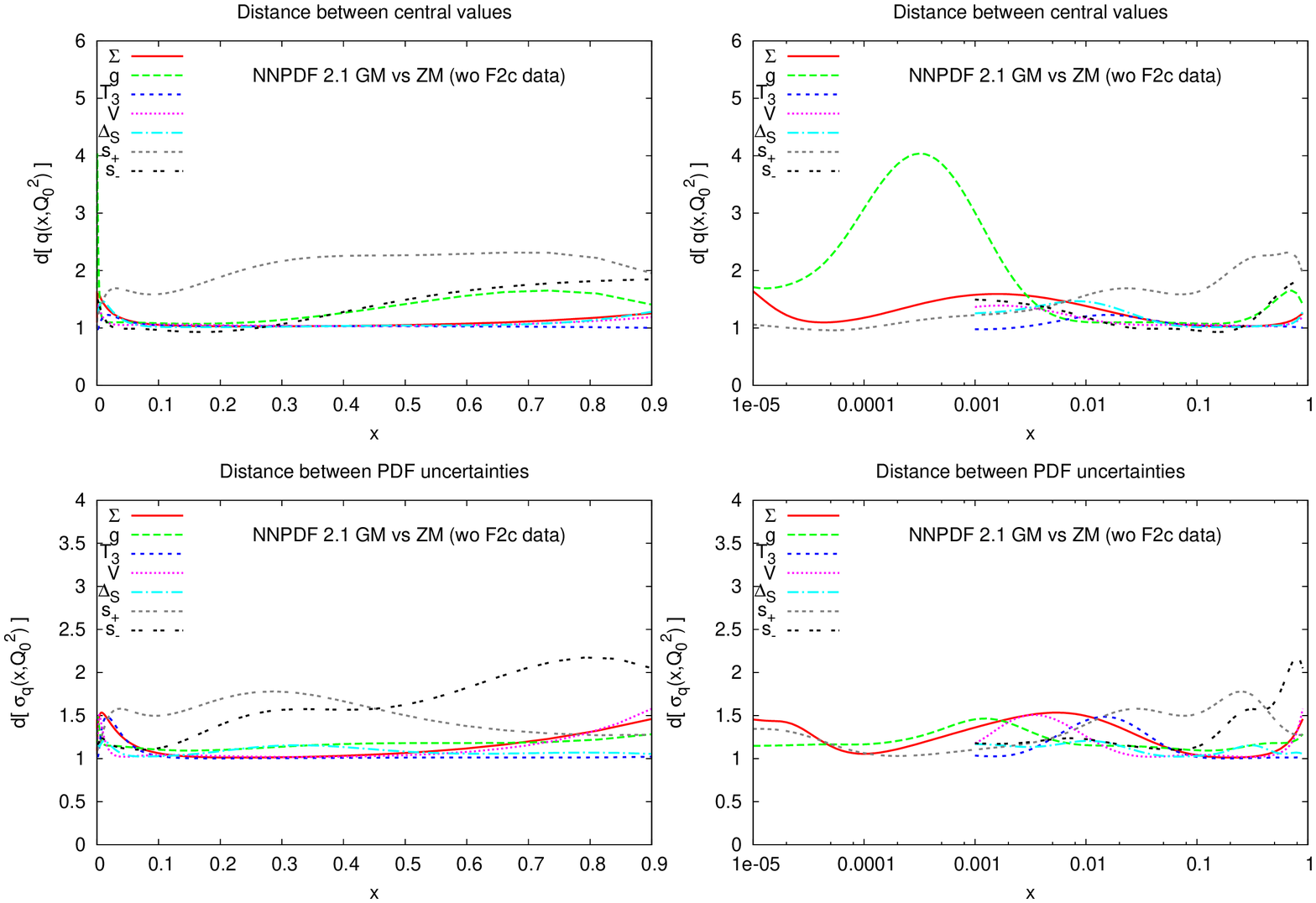}
\caption{\small Distance between the NNPDF2.1 PDF sets
in the GM and in the ZM schemes, in both
cases without HERA $F_2^c$ data.
All distances are computed from sets of $N_{\rm rep}=100$ replicas.
\label{fig:distances-21-gm-vs-zm}} 
\end{center}
\end{figure}
\begin{figure}[h!]
\begin{center}
\epsfig{width=0.49\textwidth,figure=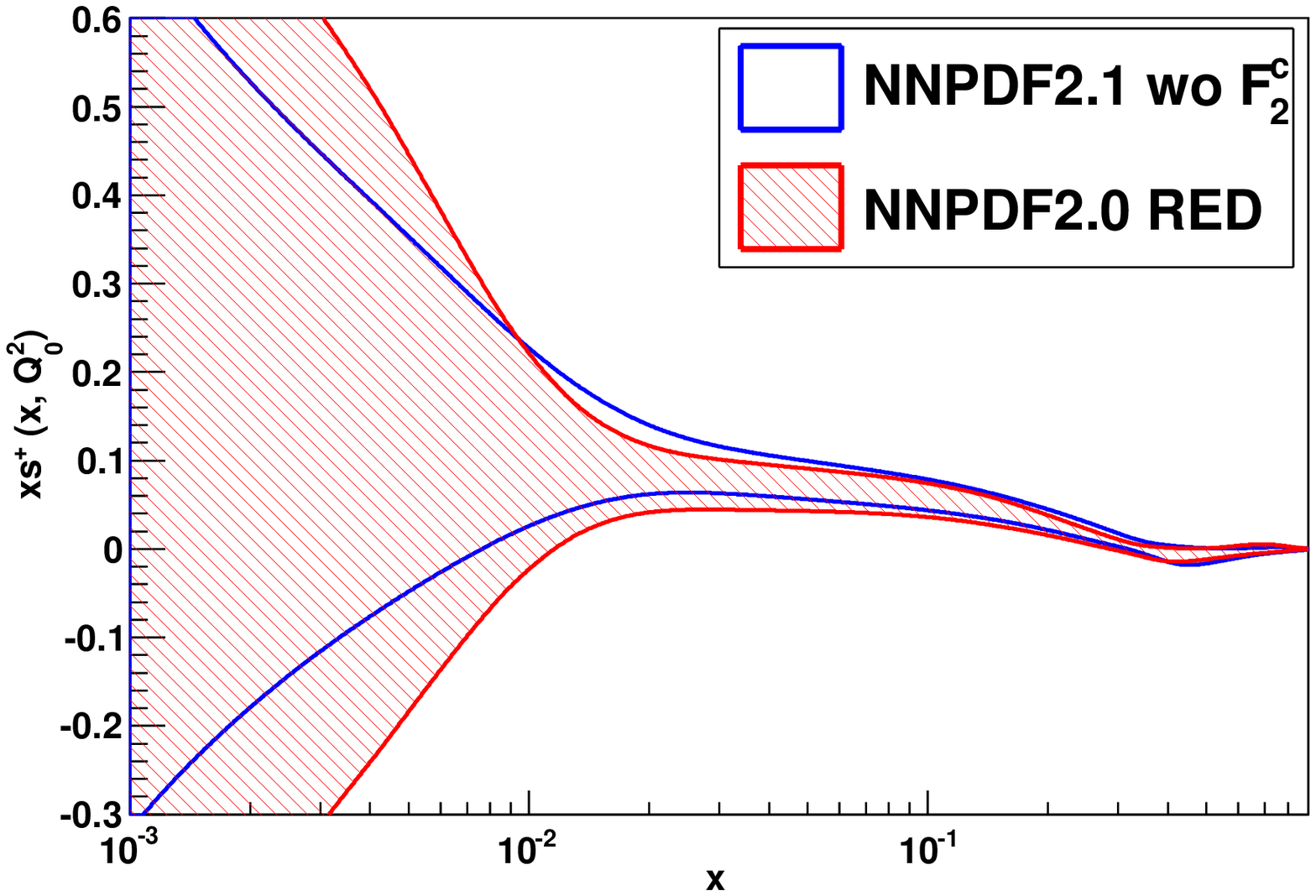}
\epsfig{width=0.49\textwidth,figure=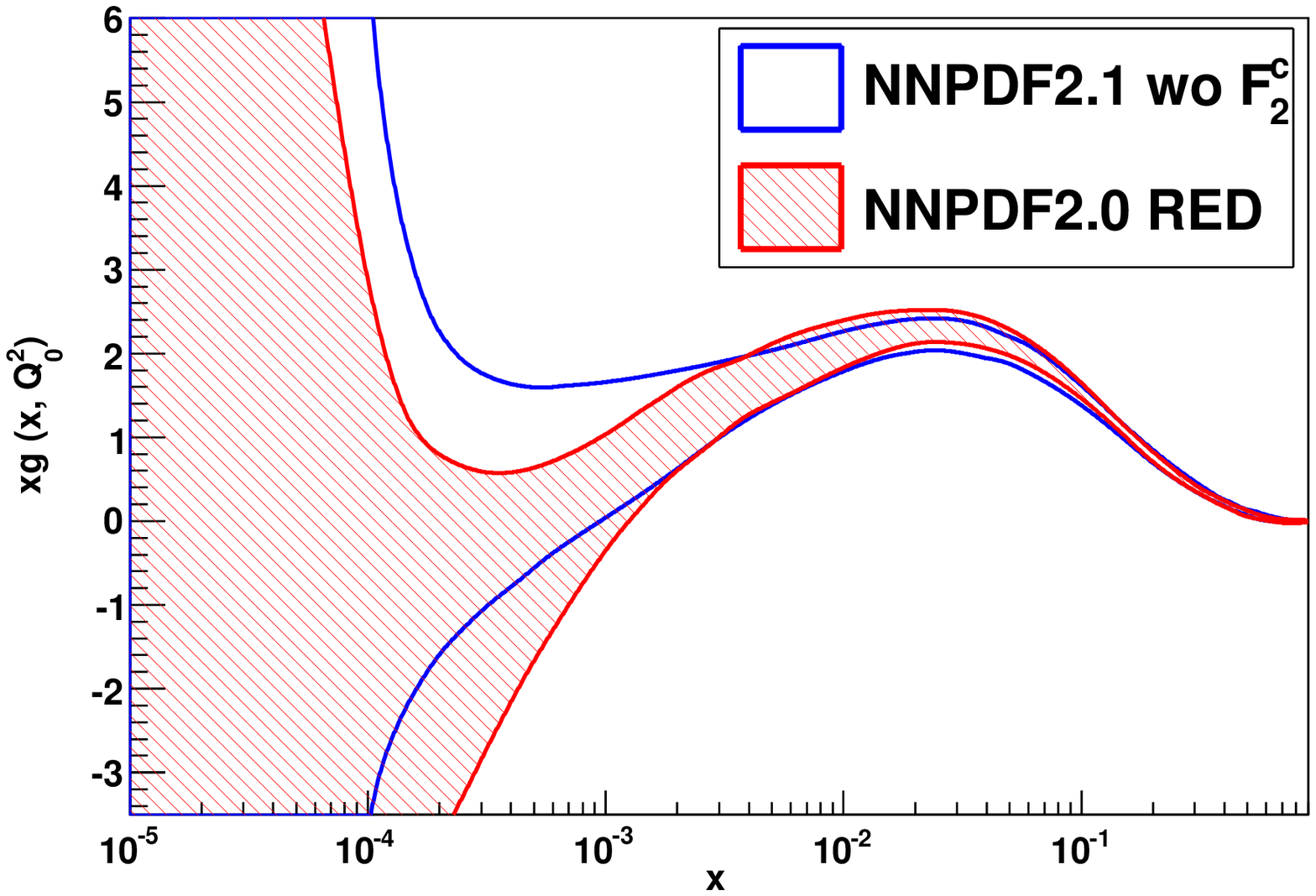}
\caption{\small 
Comparison of the small--$x$ total strangeness  and
  gluon in NNPDF2.0RED and
NNPDF2.1 without $F_2^c$ data (distances are shown  in
Fig.~\ref{fig:distances-21-gm-vs-zm}).
 \label{fig:PDFs-gm-vs-zm}} 
\end{center}
\end{figure}

The impact of the inclusion of heavy quark masses 
is mostly on the small-$x$ gluon and, to a lesser
extent, on  medium-$x$ strangeness. These two PDFs are shown in
Fig.~\ref{fig:PDFs-gm-vs-zm}. The GM scheme leads to a larger
gluon for $x\lsim 2\cdot 10^{-3}$, as well as to a somewhat larger
strangeness, but it leaves the singlet unaffected.
This shows that indeed, as argued in Sect.~\ref{sec-pdfcomp}, the
relatively large total strangeness in NNPDF2.0 was due to the use of 
the IZM approximation for dimuon data, which overestimates charm mass 
effects.
 
\item Impact of  HERA $F_2^c$ data

The impact of HERA $F_2^c$ data is estimated  comparing the
results of the NNPDF2.1 fit with and without these
data. The distances displayed  in Fig.~\ref{fig:distances-21-f2c} 
show that current $F_2^c$
data have little effect: the two fits are almost statistically
equivalent, with most distances of order one. This is partly due to
the relatively large uncertainties on current these $F_2^c$, and also 
to the fact that
low $x$ and $Q^2$ data, which are most sensitive to the gluon PDF, are
excluded by our kinematic cuts. Inclusion of ${\mathcal O}(\alpha_s^2)$ heavy
quark mass effects (e.g. by means of the FONLL-B scheme) is necessary
in order to take advantage of these data.

\begin{figure}[t]
\begin{center}
\epsfig{width=0.99\textwidth,figure=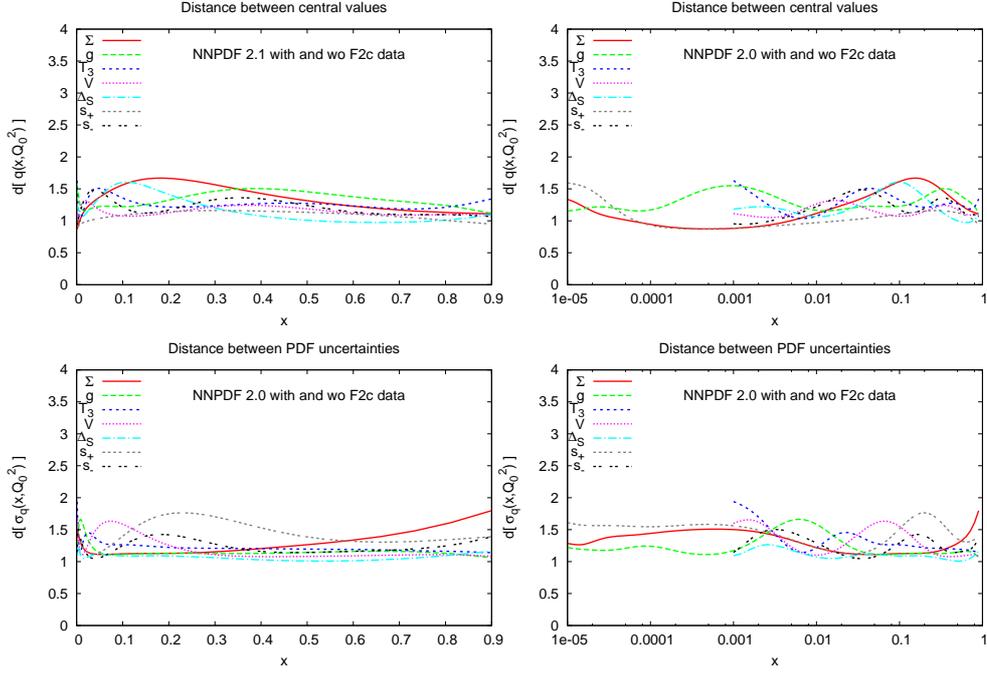}
\caption{\small Distance between  NNPDF2.1 PDF sets
with and without HERA $F_2^c(x,Q^2)$ data.  Distances
have been computed from sets of $N_{\rm rep}=100$ replicas.
\label{fig:distances-21-f2c}} 
\end{center}
\end{figure}

\item Impact of threshold prescription in the GM scheme

Finally, we have repeated the NNPDF2.1 fit with a different treatment
of subleading terms in the inclusion of heavy quark mass effects:
namely, we have used FONLL-A but  without
the threshold damping factor in
Eqs.~(\ref{eq:FONLLnlo},\ref{eq:f2cfonll}). 
Indeed, the benchmarking exercise of Ref.~\cite{LHhq} suggest that the
difference between these cases should provide a reasonable estimate of
the spread of results obtained by including heavy quark masses
according to different prescriptions.
Distances
are shown in Fig.~\ref{fig:dist-21-nodamp}:
the PDFs that are most affected are the singlet and gluon 
PDFs at medium-$x$, shown in 
Fig.~\ref{fig:PDFs-nodamp}. 
Without damping factor, the $F_2^c$ structure
function is closer to the massless result even at moderate
$Q^2$, and  this explains why the singlet PDF is somewhat
smaller  at medium-$x$.

\begin{figure}[t]
\begin{center}
\epsfig{width=0.99\textwidth,figure=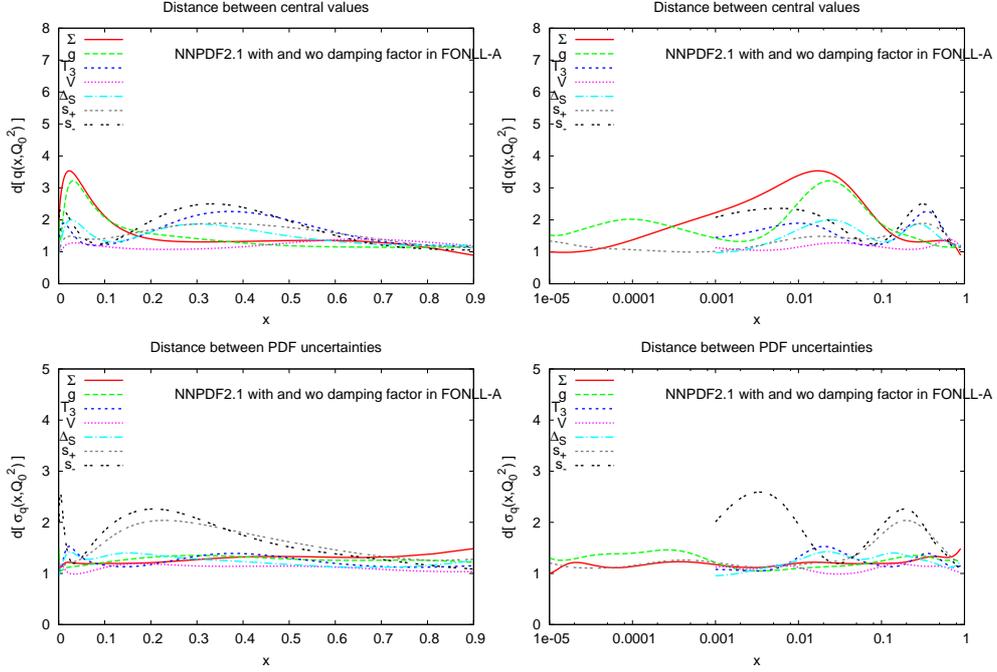}
\caption{\small Distance between the NNPDF2.1 reference
set and the same set obtained without threshold damping factor
in the computation of the FONLL-A structure functions.
\label{fig:dist-21-nodamp}.} 
\end{center}
\end{figure}

\begin{figure}[t]
\begin{center}
\epsfig{width=0.49\textwidth,figure=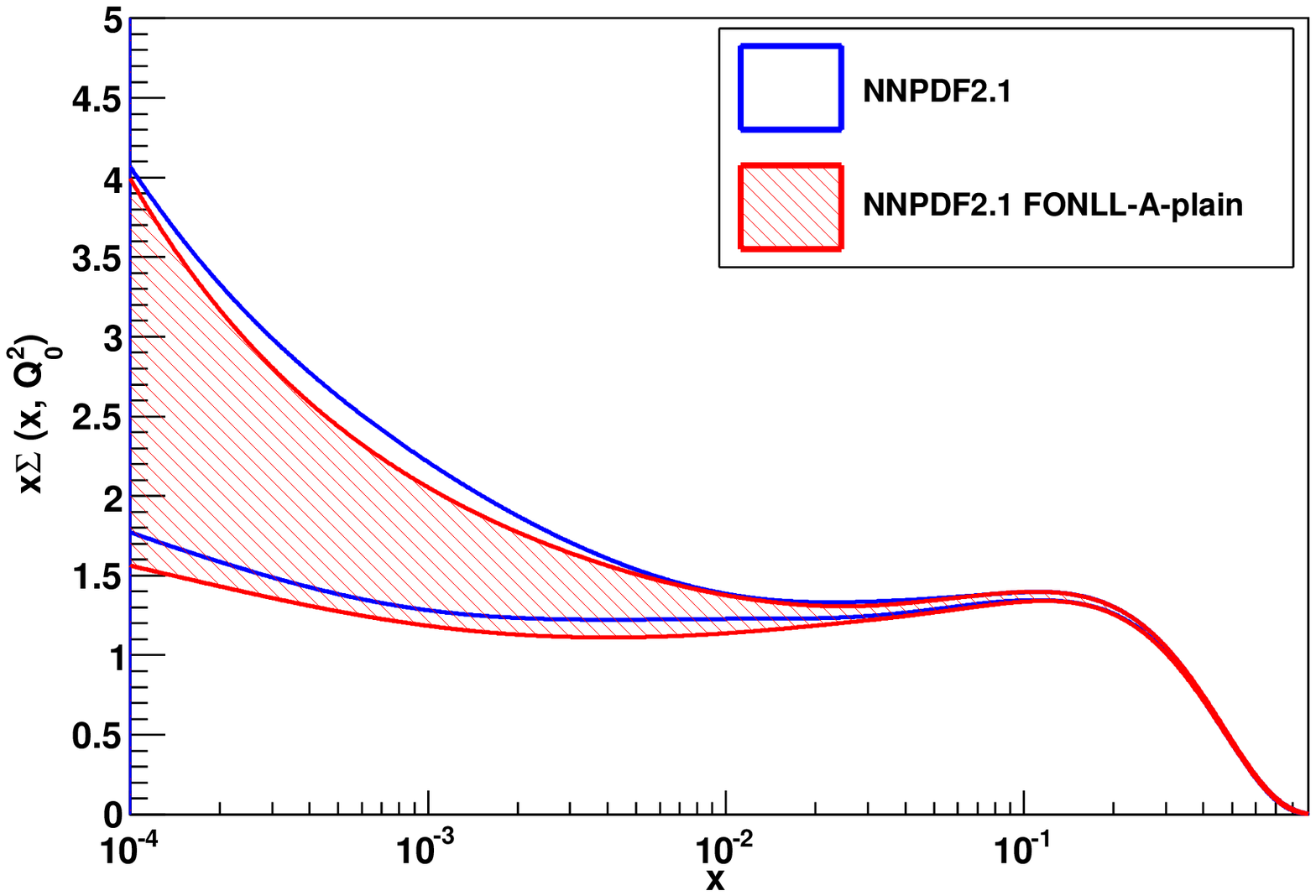}
\epsfig{width=0.49\textwidth,figure=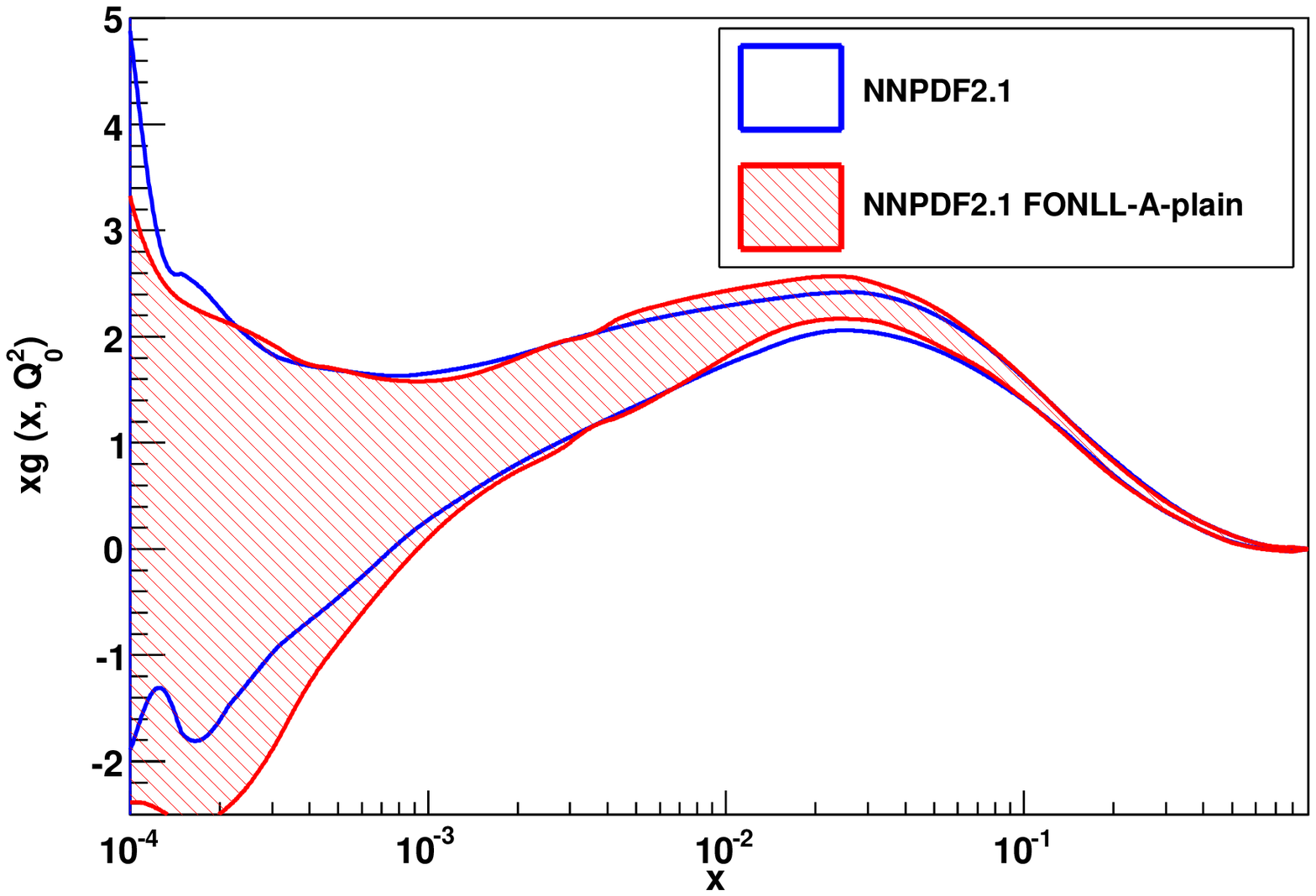}
\caption{\small The small--$x$ singlet
and gluon PDFs, in the reference NNPDF2.1 set and in 
NNPDF2.1 obtained with FONLL-A without threshold
damping factor (distances are shown  in
Fig.~\ref{fig:dist-21-nodamp}).
 \label{fig:PDFs-nodamp}} 
\end{center}
\end{figure}

\end{itemize}

\section{Phenomenological implications}
\label{sec:pheno}

In this Section we discuss the implications of the NNPDF2.1 set for
LHC physics. We begin comparing the prediction for LHC benchmark
cross sections obtained using  
NNPDF2.1 and NNPDF2.0.   This comparison allows us to assess the
impact of heavy quark mass effects.
We also compare to predictions obtained using the CT10 and MSTW08
sets, both using  
their preferred 
values of $\alpha_s$ and with a common value. We then compare 
parton luminosities relevant for LHC processes and determine 
correlations  between PDFs and some observables. Next, we
determine the correlation between PDFs and $\alpha_s$ and 
discuss PDF sets with varying $\alpha_s$, which are
needed to compute the combined PDF+$\alpha_s$ uncertainties, and present 
sets in which PDF and $\alpha_s$ uncertainties 
are pre-combined. Finally, we briefly revisit  implications
of the strangeness asymmetry on the NuTeV anomaly, confirming our
previous result that the anomaly disappears once the strangeness
asymmetry is properly determined.
We conclude with a  comparison of NNPDF2.1 results with 
published HERA $F_L$ and $F_2^c$ data and predictions for 
the upcoming $F_L$ and 
$F_2^c$ HERA combined datasets.

\subsection{LHC benchmark cross-sections}
\label{sec:lhcimplications}

\begin{table}
  \centering
  {\footnotesize
  \begin{tabular}{|c|c|c|c|}
    \hline
        & $\sigma(W^+)B_{l\nu}$ [nb]
         & $\sigma(W^-)B_{l\nu}$ [nb]
         & $\sigma(Z^0)B_{ll}$ [nb]\\
    \hline
    \hline
    NNPDF2.0  &  $5.84\pm 0.14$& $3.97\pm 0.09$ & $0.91 \pm 0.02$ \\
\hline
    NNPDF2.1  & $5.99\pm 0.14$  & $4.09\pm 0.09$ & $0.93 \pm 0.02 $ \\
   \hline
   CT10 - $\alpha_s=0.118$ & $6.00\pm 0.13$ & $4.10\pm 0.09$ & 
$0.94\pm 0.02$\\
  CT10 - $\alpha_s=0.119$ & $6.04\pm 0.13$ & $4.13\pm 0.09$ & $0.95\pm 0.02$ \\
  \hline
   MSTW08 - $\alpha_s=0.119$  & $5.91\pm 0.11$ & $4.16\pm 0.08$ &
$0.94 \pm 0.02$ \\
  MSTW08 - $\alpha_s=0.120$  & $5.95\pm 0.11$ & $4.19\pm 0.08$ &
$0.95 \pm 0.02$ \\
    \hline
  \end{tabular}}\\
\vspace{0.3cm}
 {\footnotesize
  \begin{tabular}{|c|c|c|}
    \hline
        & $\sigma(t\bar{t})$ [pb] 
          & $\sigma(H,m_H=120\,{\rm GeV})$ [pb]
          \\
    \hline
\hline
    NNPDF2.0 & $168 \pm 7$ & $11.59 \pm 0.22$ \\
\hline
    NNPDF2.1 & $170 \pm 5$ & $11.64\pm 0.17$ \\
  \hline
   CT10 - $\alpha_s=0.118$ & $158\pm 7$ & $10.99 \pm 0.21$  \\
  CT10 - $\alpha_s=0.119$ &$161\pm 7$ & $11.17\pm 0.21$ \\
  \hline
   MSTW08 - $\alpha_s=0.119$  & $164 \pm 5$ & $11.48\pm 0.18$ \\ 
  MSTW08 - $\alpha_s=0.120$  & $168 \pm 5$ & $11.69\pm 0.18$ \\
 \hline
  \end{tabular}}
  \caption{\label{tab:LHCobs}  
\small Cross-sections for W, Z, $t\bar{t}$ and Higgs production
at the LHC at $\sqrt{s}=7$ TeV and the associated
PDF uncertainties. All quantities have been computed at NLO using
    MCFM for the NNPDF2.1, NNPDF2.0, CT10 and MSTW08 PDF sets. All 
uncertainties shown are one-$\sigma$.}
\end{table}
\begin{figure}[t]
\begin{center}
\epsfig{width=0.32\textwidth,figure=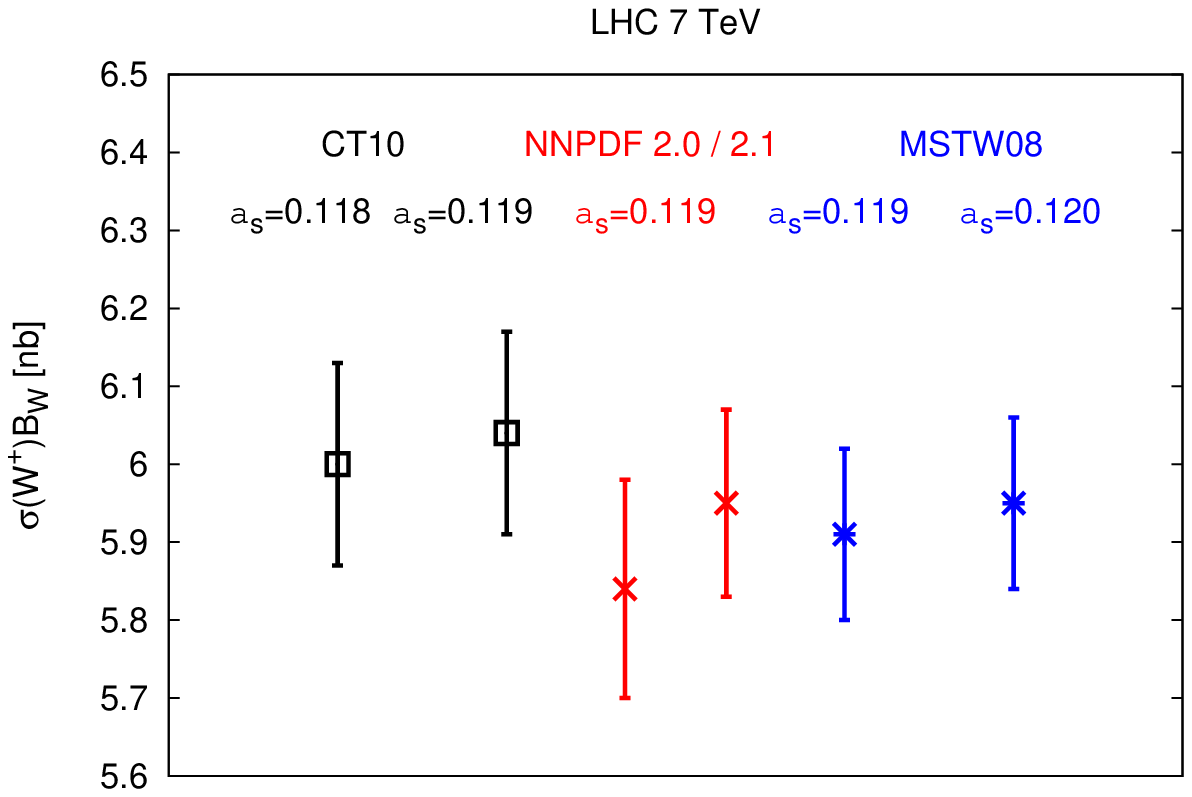}
\epsfig{width=0.32\textwidth,figure=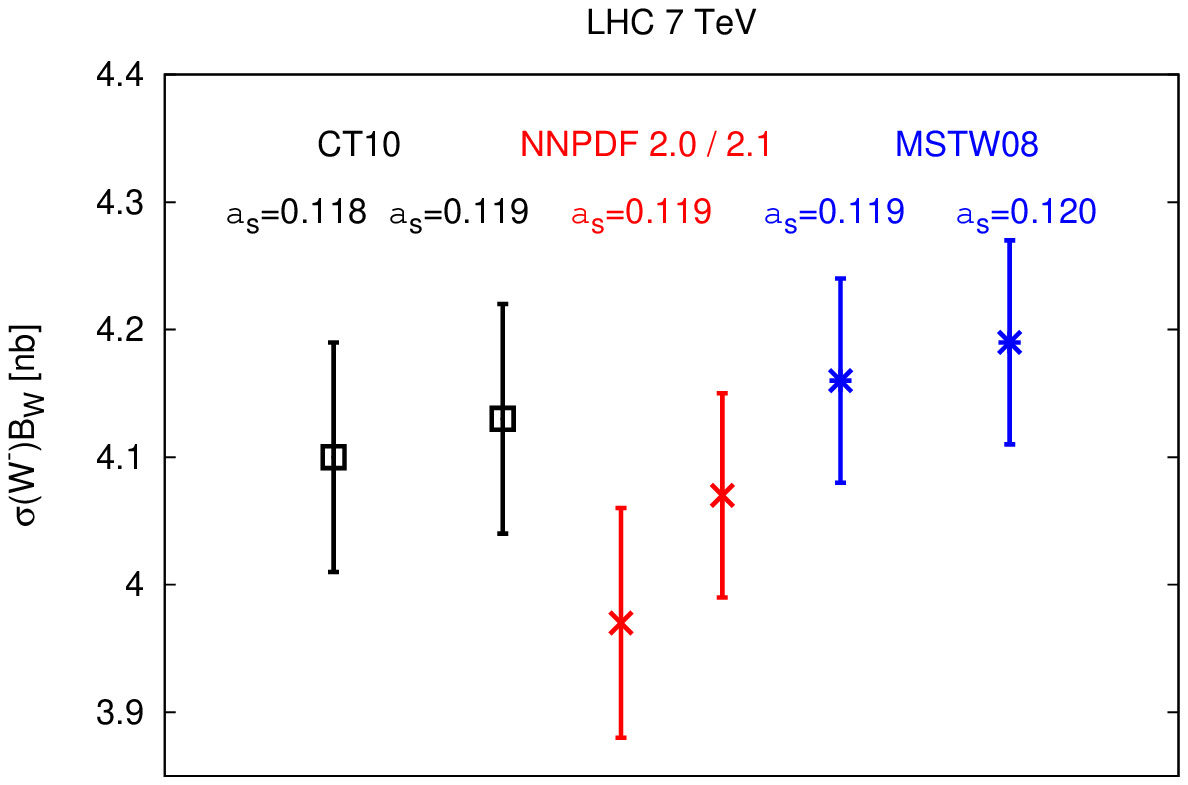}
\epsfig{width=0.32\textwidth,figure=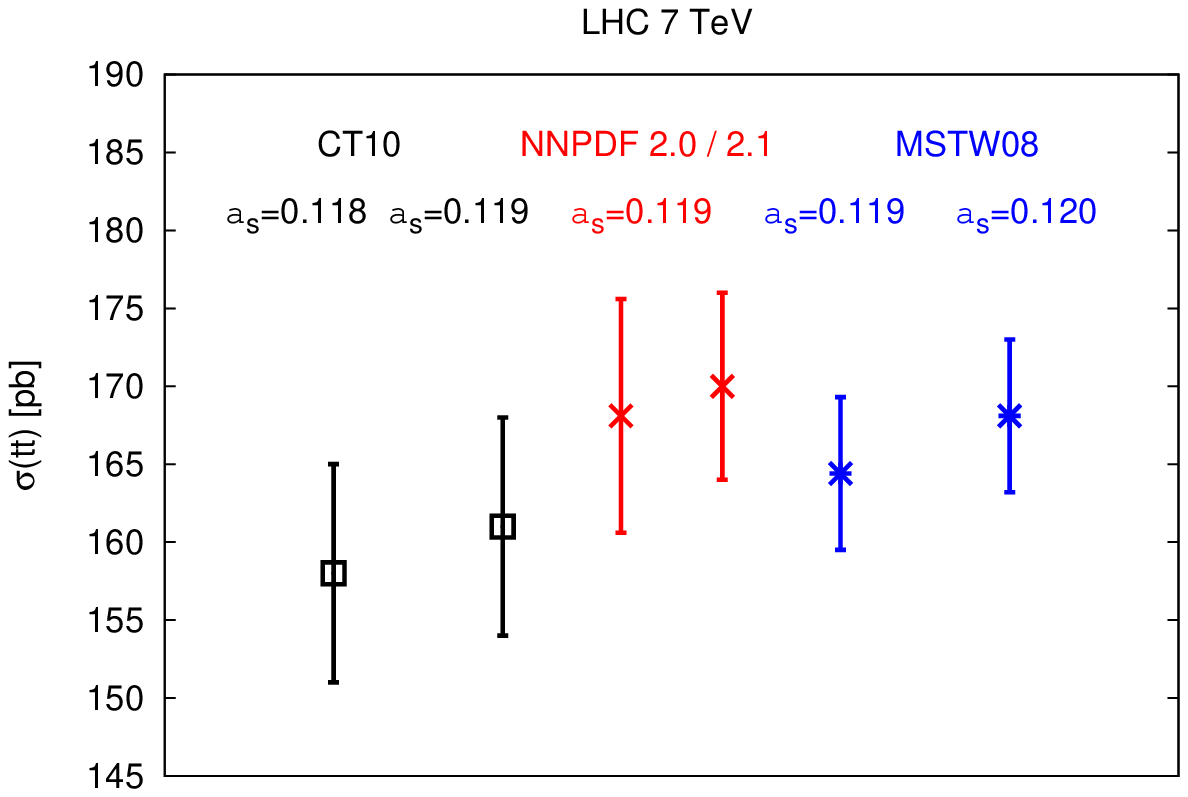}
\epsfig{width=0.32\textwidth,figure=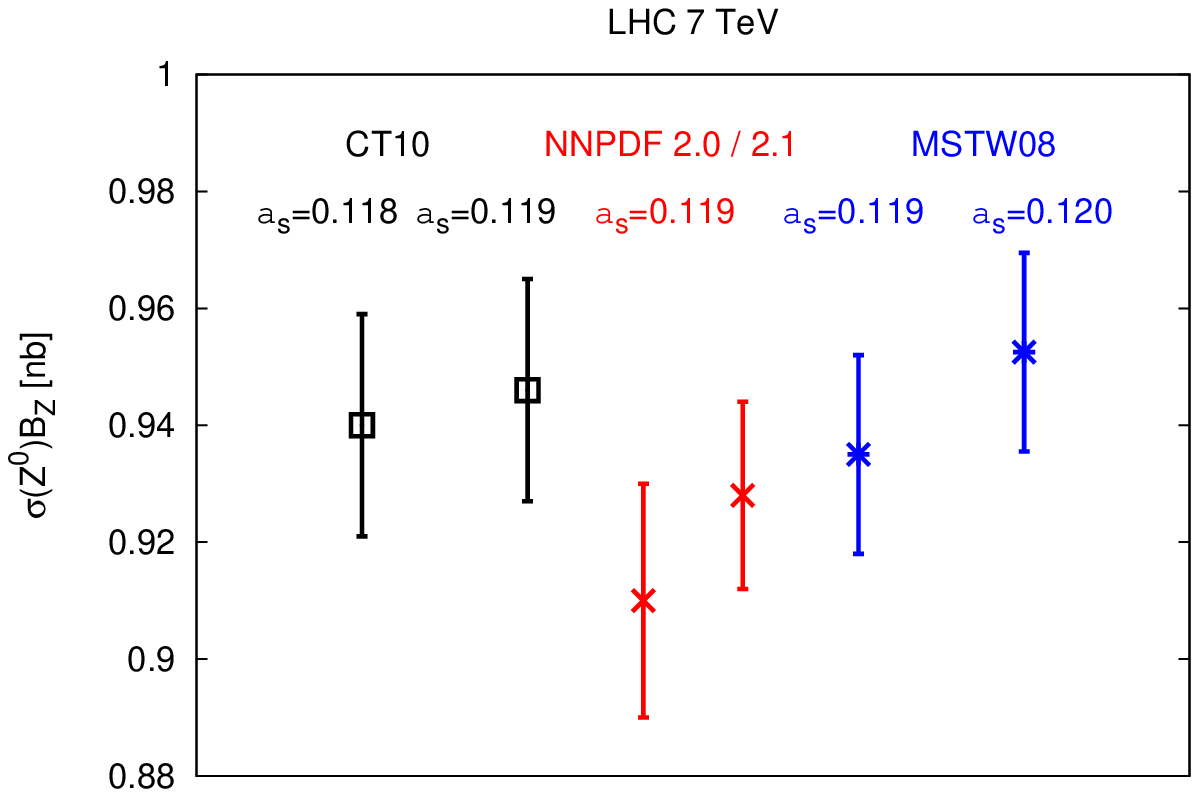}
\epsfig{width=0.32\textwidth,figure=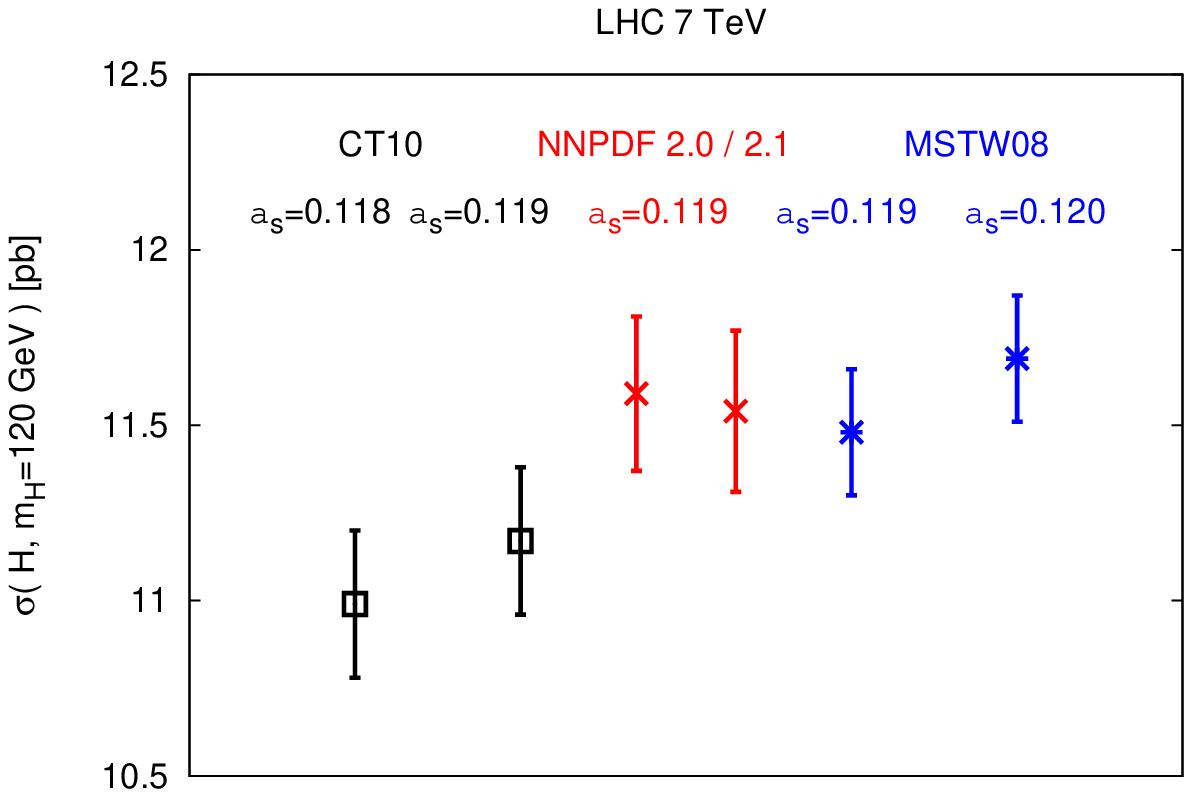}
\caption{\small Graphical representation of the results of
  Table~\ref{tab:LHCobs}. 
 \label{fig:comp7tev}} 
\end{center}
\end{figure}

The assessment of the theoretical uncertainties on LHC standard
candles is especially important now 
that the first 7 TeV LHC results on inclusive cross-sections are 
appearing~\cite{ATLAS:2010yt,Collaboration:2010ez,Collaboration:2010wv,Collaboration:2010ey}.   
In this Section we present results  at  $\sqrt{s}=$7 TeV
and $\sqrt{s}=$14 TeV for $W^{\pm}$,  $Z^{0}$, $t\bar{t}$ 
and Higgs production 
via gluon fusion with $m_H=120$ GeV.
All observables are  computed at NLO QCD using
MCFM~\cite{Campbell:2002tg,MCFMurl}. 

In Tables~\ref{tab:LHCobs} and~\ref{tab:LHCobs2}, and the corresponding Figs.~\ref{fig:comp7tev} 
and~\ref{fig:comp14tev}, we compare the predictions for these cross-sections obtained using the NNPDF2.1, 
NNPDF2.0, CT10 and MSTW08 sets. 
In the case of the last two sets, 
we show results both using the respective default value of 
$\alpha_s(M_Z)$ and at the common value of $\alpha_s(M_Z)=0.119$,
obtained using the PDF sets of Refs.~\cite{Martin:2009bu,Lai:2010nw}.

Predictions obtained using the NNPDF2.0 and NNPDF2.1 PDF sets mostly
differ because of heavy quark mass effects, but
other differences such as different kinematic cuts also play a role.
As can be seen from Fig.~\ref{fig:comp7tev} and Table~\ref{tab:LHCobs}, the differences between NNPDF2.0 
and NNPDF2.1 are at most at the one-$\sigma$ level for  $W^{\pm}$ and
$Z$ production 7 TeV, while predictions for the $t\bar{t}$ and Higgs are 
essentially unchanged: these observables are only minimally affected by the 
heavy quark treatment.

\begin{figure}[t]
\begin{center}
\epsfig{width=0.38\textwidth,figure=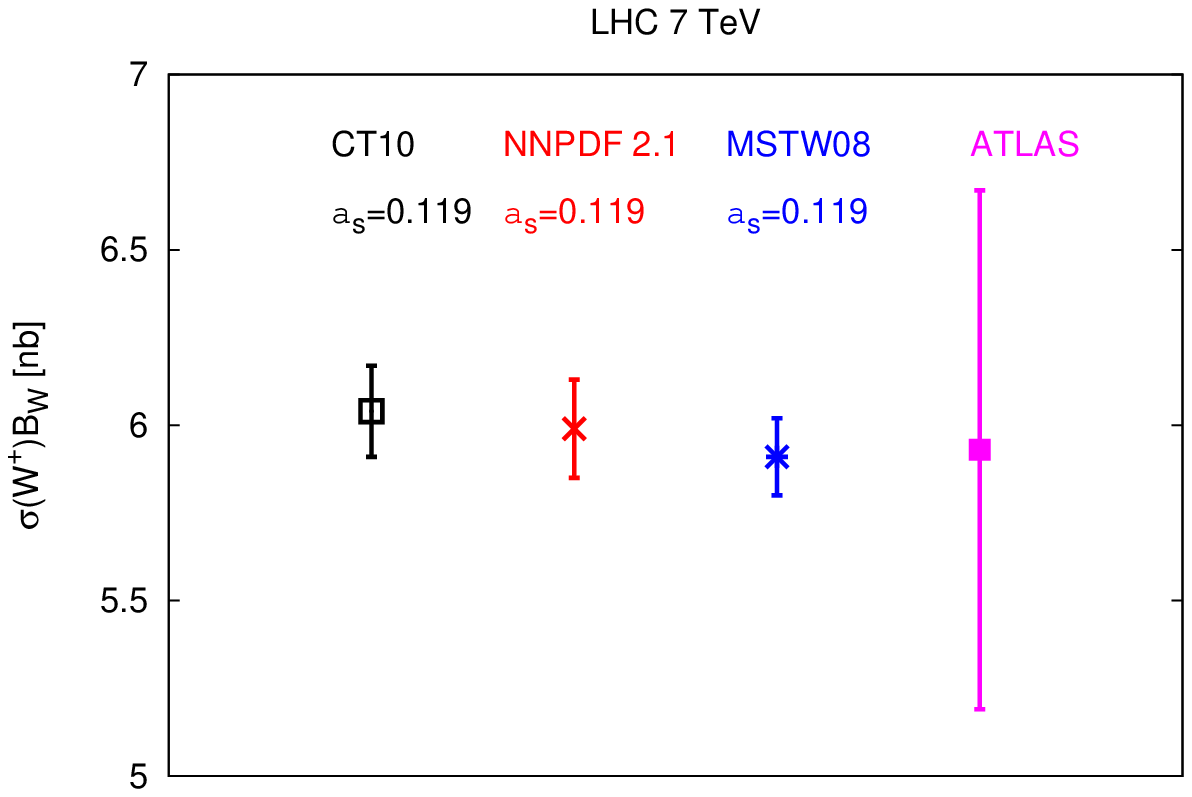}
\epsfig{width=0.38\textwidth,figure=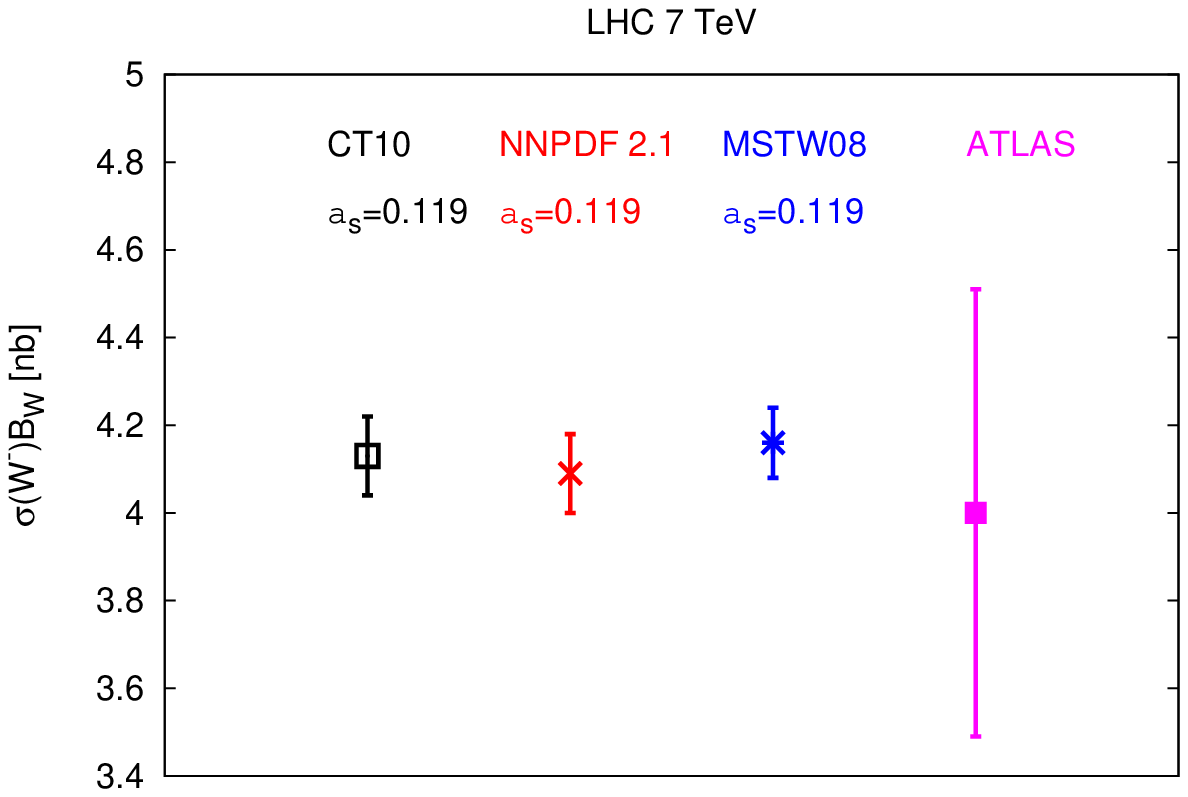}\\
\epsfig{width=0.38\textwidth,figure=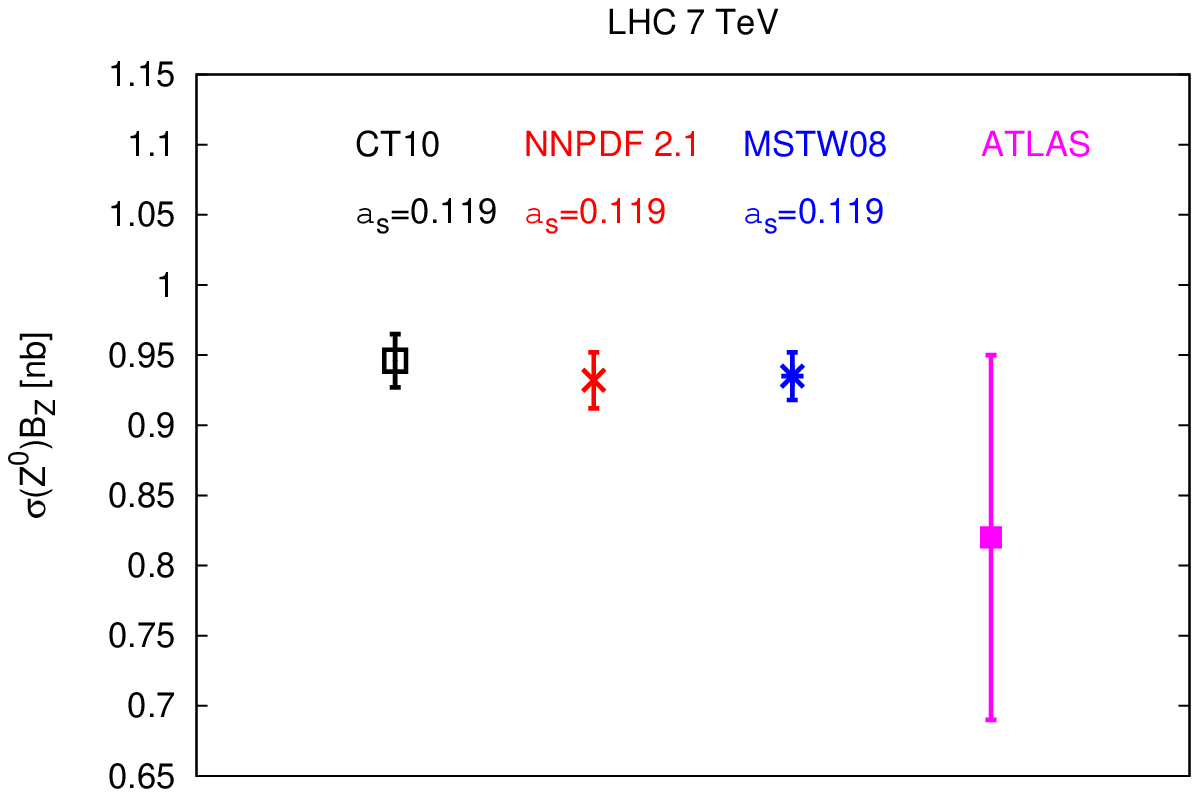}
\epsfig{width=0.38\textwidth,figure=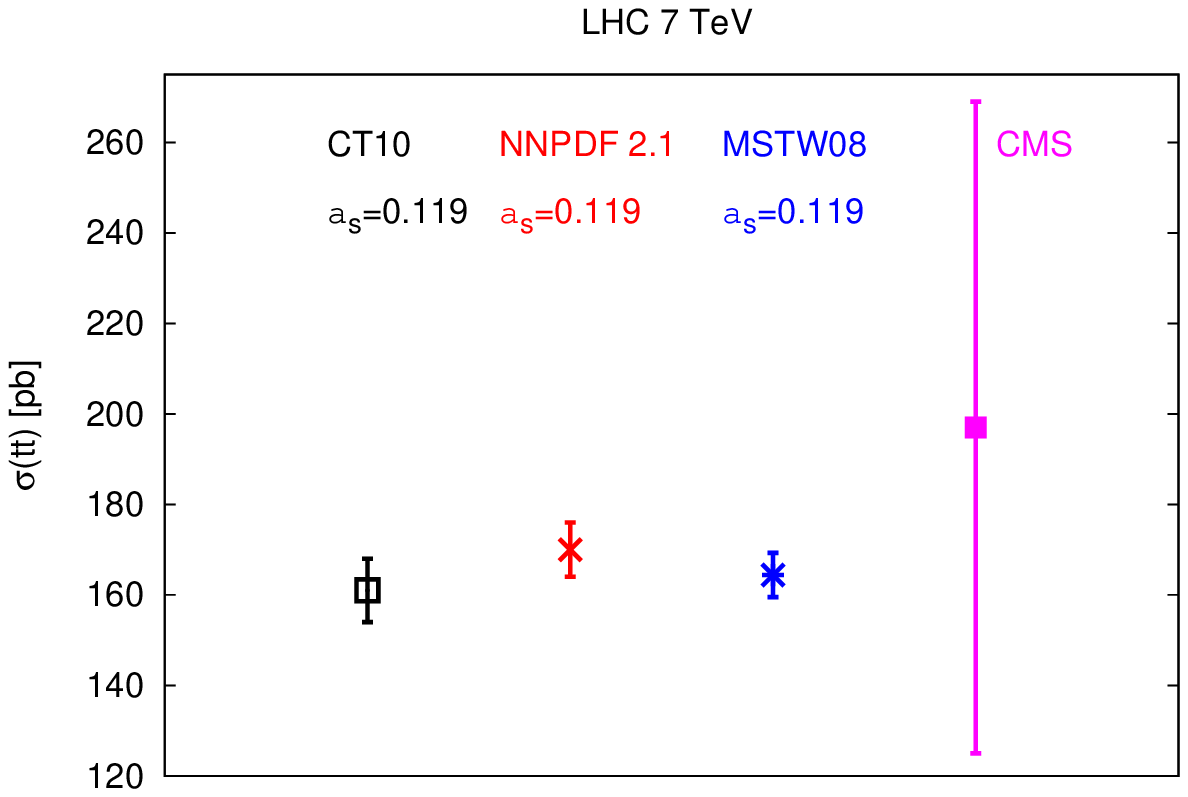}
\caption{\small 
LHC measurements of the $W^{\pm},Z$ and $t\bar{t}$
cross-sections at $\sqrt{s}=$7 TeV from the ATLAS~\cite{ATLAS:2010yt}
and CMS experiments~\cite{Collaboration:2010ez} compared to the
predictions of Fig.~\ref{fig:comp7tev}.
 \label{fig:comp7tev-data}} 
\end{center}
\end{figure}

\begin{figure}[t]
\begin{center}
\epsfig{width=0.32\textwidth,figure=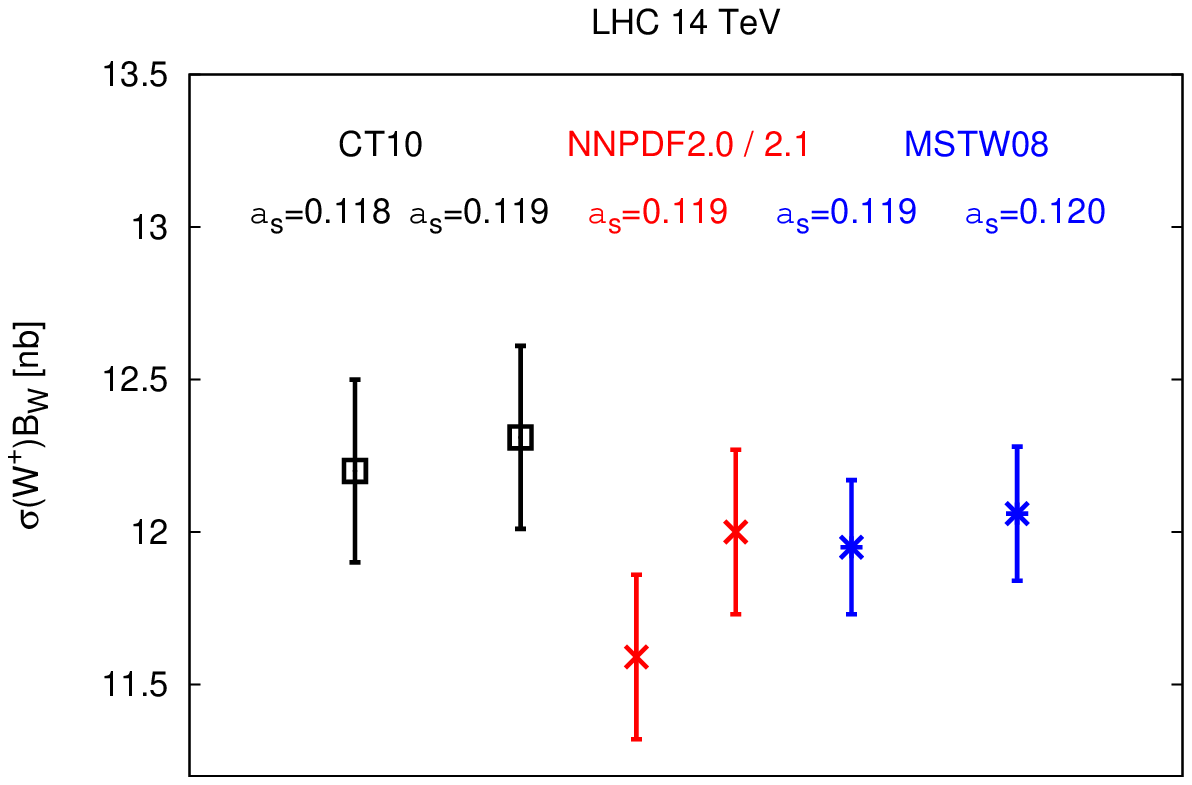}
\epsfig{width=0.32\textwidth,figure=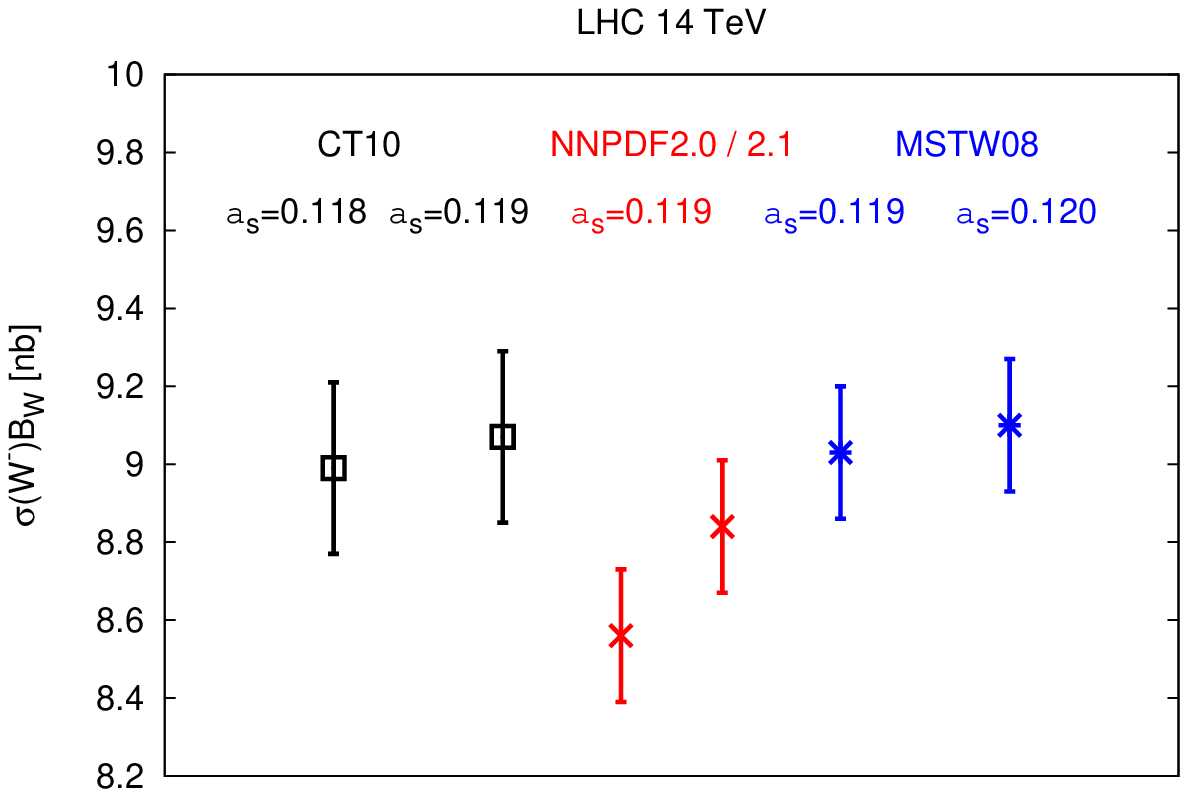}
\epsfig{width=0.32\textwidth,figure=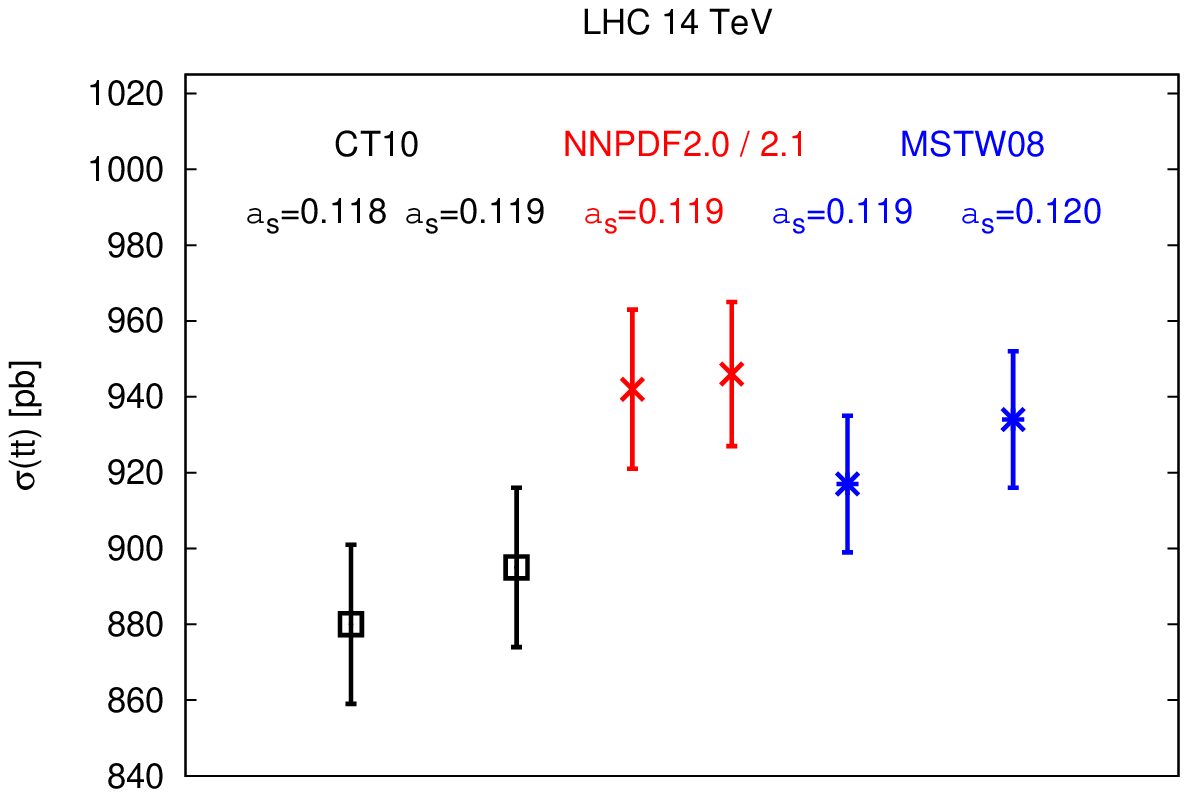}
\epsfig{width=0.32\textwidth,figure=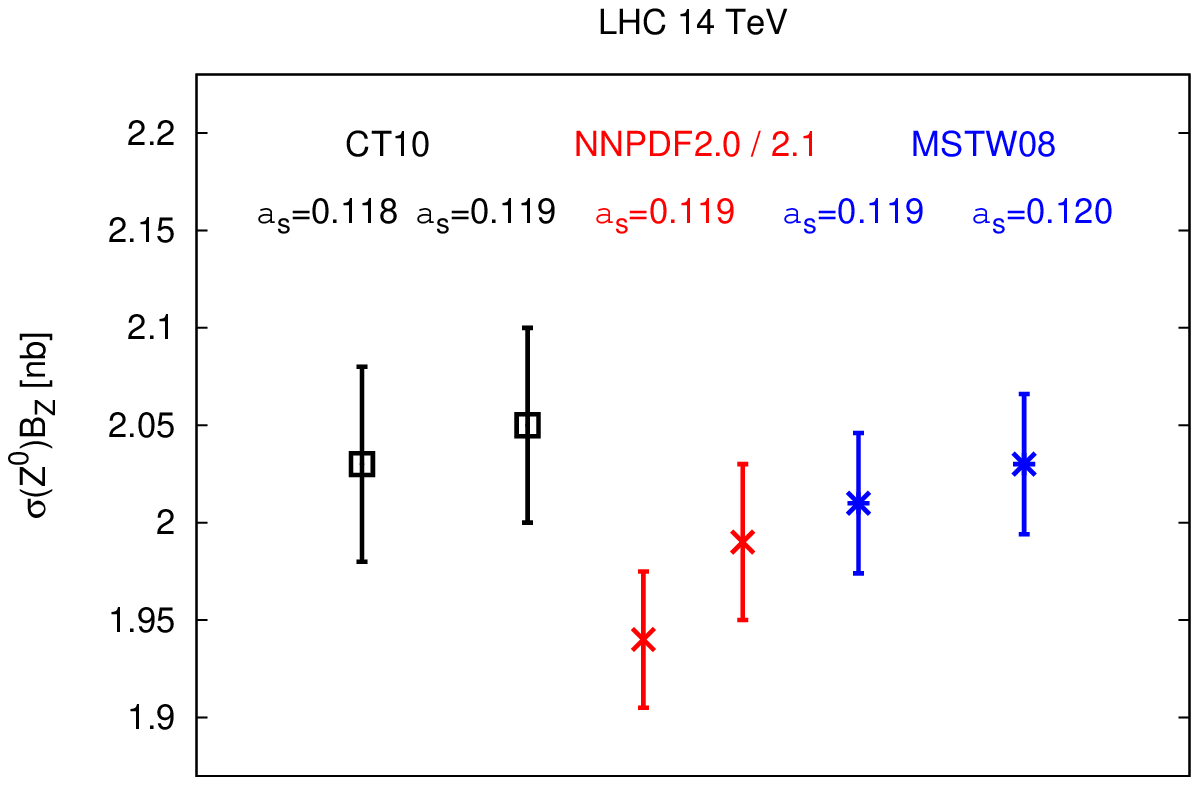}
\epsfig{width=0.32\textwidth,figure=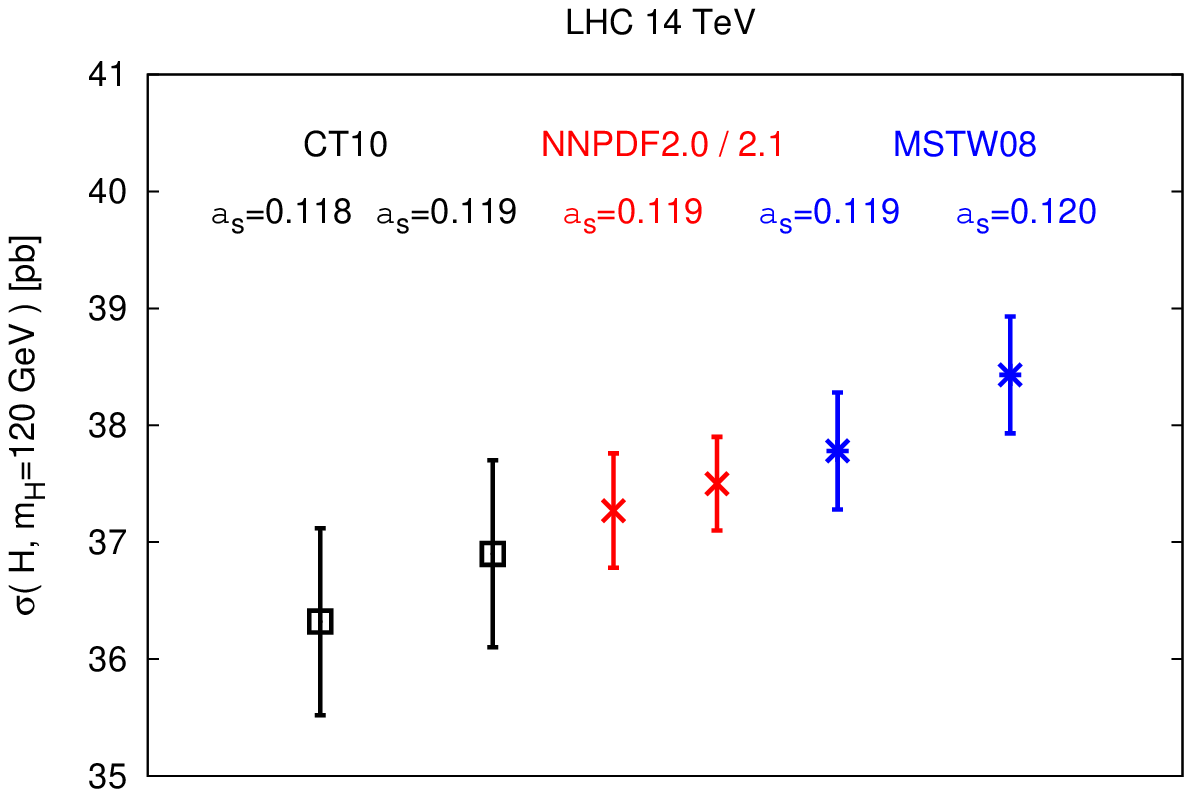}
\caption{\small Graphical representation of the results of
  Table~\ref{tab:LHCobs2}.
 \label{fig:comp14tev}} 
\end{center}
\end{figure}

NNPDF2.1 predictions are in rather good agreement with MSTW08 for all
observables, though  differences with  CT10 are somewhat larger, 
especially for observables which are most sensitive to the gluon distribution, like Higgs 
and $t\bar{t}$ production.
The use of a common value for the strong coupling $\alpha_s$ leads to 
better agreement between predictions, especially for processes 
which depend on $\alpha_s$ already at leading order such as 
Higgs production in gluon 
fusion~\cite{Demartin:2010er}. In Fig.~\ref{fig:comp7tev-data}
first measurements by the ATLAS and CMS
experiments~\cite{Collaboration:2010ez,ATLAS:2010yt} are also shown:
with their large uncertainties, dominated 
by the current large (${\mathcal{O} (11\%)}$) luminosity uncertainty,
they cannot yet provide  constraint on PDFs.

At $\sqrt{s}=$14 TeV we expect the effect of the heavy quark treatment
to be larger: results are collected in Fig.~\ref{fig:comp14tev} and
Table~\ref{tab:LHCobs2}. In this case, the upwards shift in the the
$W^{\pm}$ and $Z$ cross-sections 
from NNPDF2.0 to NNPDF2.1 is at or just above the one-$\sigma$ level,
while as before Higgs 
and top-pair cross-sections are essentially unchanged.
The comparison with CT10 and MSTW08 is similar as before, but with
the agreement somewhat better for the Higgs  and somewhat worse for top.

\begin{table}[t]
  \centering
  {\footnotesize
  \begin{tabular}{|c|c|c|c|}
    \hline
        & $\sigma(W^+)B_{l\nu}$ [nb]
         & $\sigma(W^-)B_{l\nu}$ [nb]
         & $\sigma(Z^0)B_{ll}$ [nb]\\
    \hline
    \hline
    NNPDF2.0  & $11.59\pm 0.27$  & $8.56\pm 0.17$  & $1.94\pm 0.04$ \\
    NNPDF2.1  & $12.00\pm 0.27$  & $8.84 \pm 0.17$ & $1.99\pm 0.04$ \\
   \hline
   CT10 - $\alpha_s=0.118$ & $12.20\pm 0.30$ & $9.00\pm 0.22$ & 
$2.03\pm 0.05$\\
  CT10 - $\alpha_s=0.119$ & $12.31\pm 0.30$ & $9.07\pm 0.22$ & 
$2.05\pm 0.05$\\
  \hline
   MSTW08 - $\alpha_s=0.119$   & $11.95\pm 0.22$ & $9.03\pm 0.17$ &
$2.01\pm 0.04$ \\
  MSTW08 - $\alpha_s=0.120$  & $12.06\pm 0.22$ & $9.10\pm 0.17$ &
$2.03\pm 0.04$ \\
    \hline
  \end{tabular}}\\
\vspace{0.3cm}
 {\footnotesize
  \begin{tabular}{|c|c|c|}
    \hline
        & $\sigma(t\bar{t})$ [pb] 
          & $\sigma(H,m_H=120\,{\rm GeV})$ [pb]
          \\
    \hline
\hline
    NNPDF2.0   & $942\pm 21$& $37.3\pm 0.50$ \\
    NNPDF2.1  & $946\pm 19 $ & $37.5\pm 0.40$  \\
  \hline
   CT10 - $\alpha_s=0.118$& $880\pm 21$& $36.32\pm 0.80$\\
  CT10 - $\alpha_s=0.119$ & $895\pm 21$& $36.90\pm 0.80$\\
  \hline
   MSTW08 - $\alpha_s=0.119$ & $917\pm 18$& $37.78\pm 0.50$\\
  MSTW08 - $\alpha_s=0.120$  & $934\pm 18$& $38.43\pm 0.50$\\
 \hline
  \end{tabular}}
  \caption{\label{tab:LHCobs2}  
\small Same as Table~\ref{tab:LHCobs} for the LHC at
$\sqrt{s}=14$ TeV.  
}
\end{table}

\begin{figure}[t]
\begin{center}
\epsfig{width=0.38\textwidth,figure=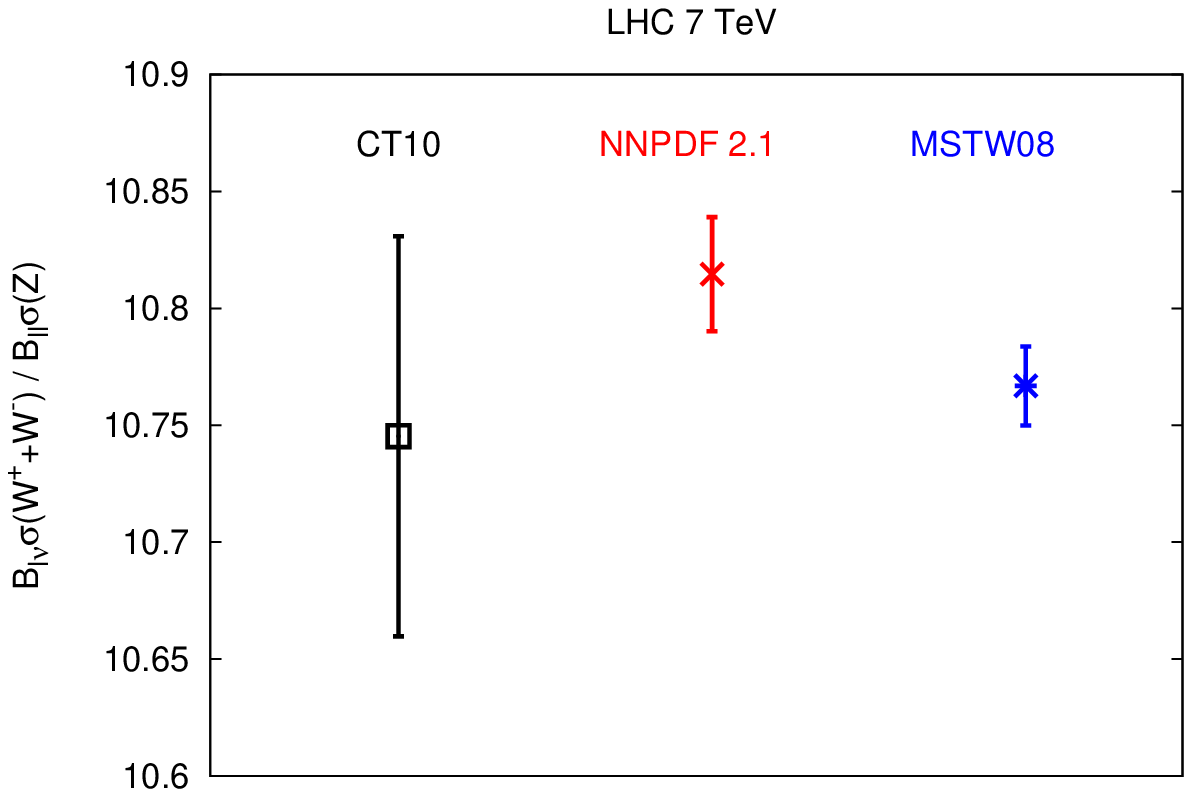}
\epsfig{width=0.38\textwidth,figure=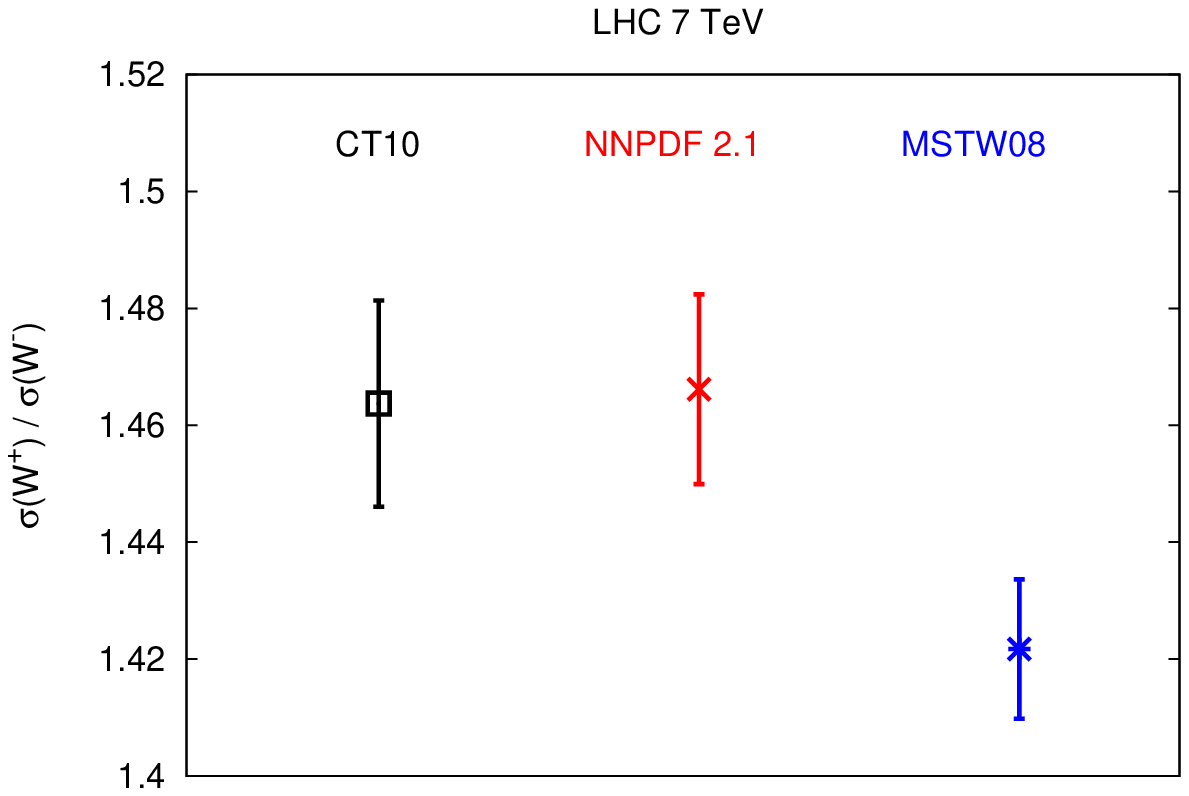}\\
\epsfig{width=0.38\textwidth,figure=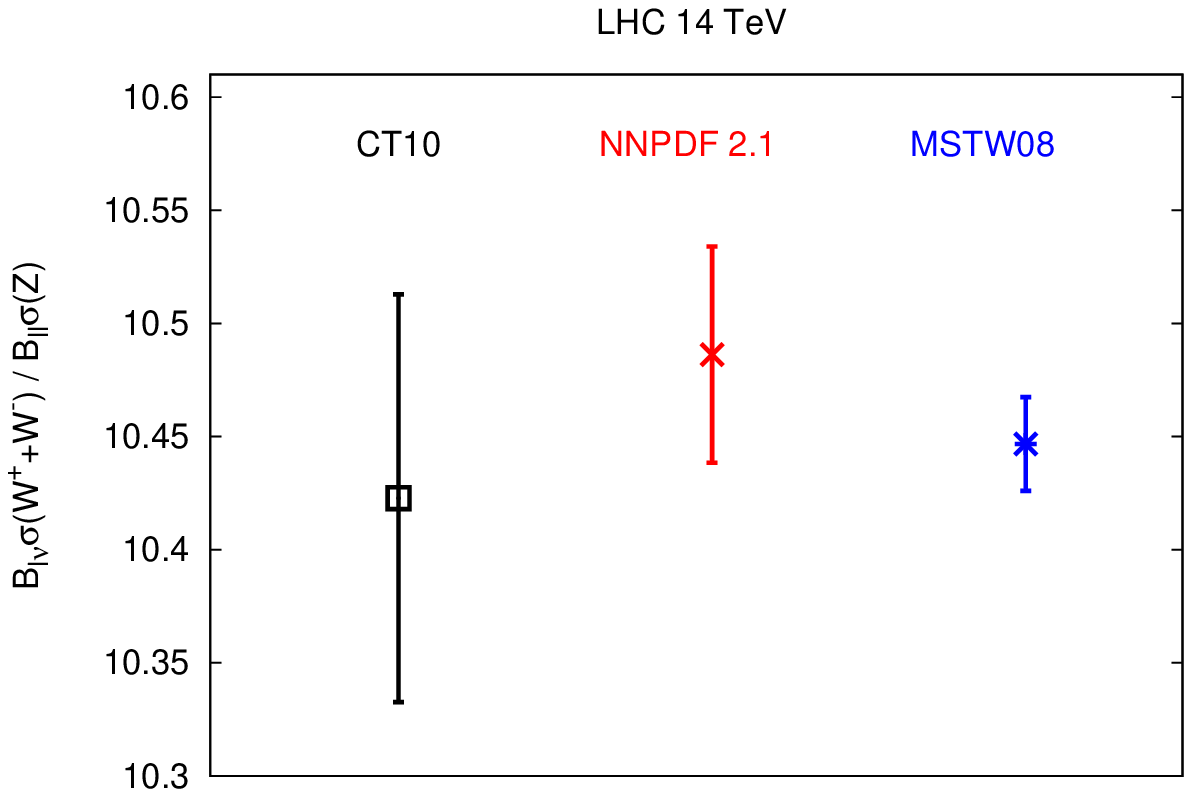}
\epsfig{width=0.38\textwidth,figure=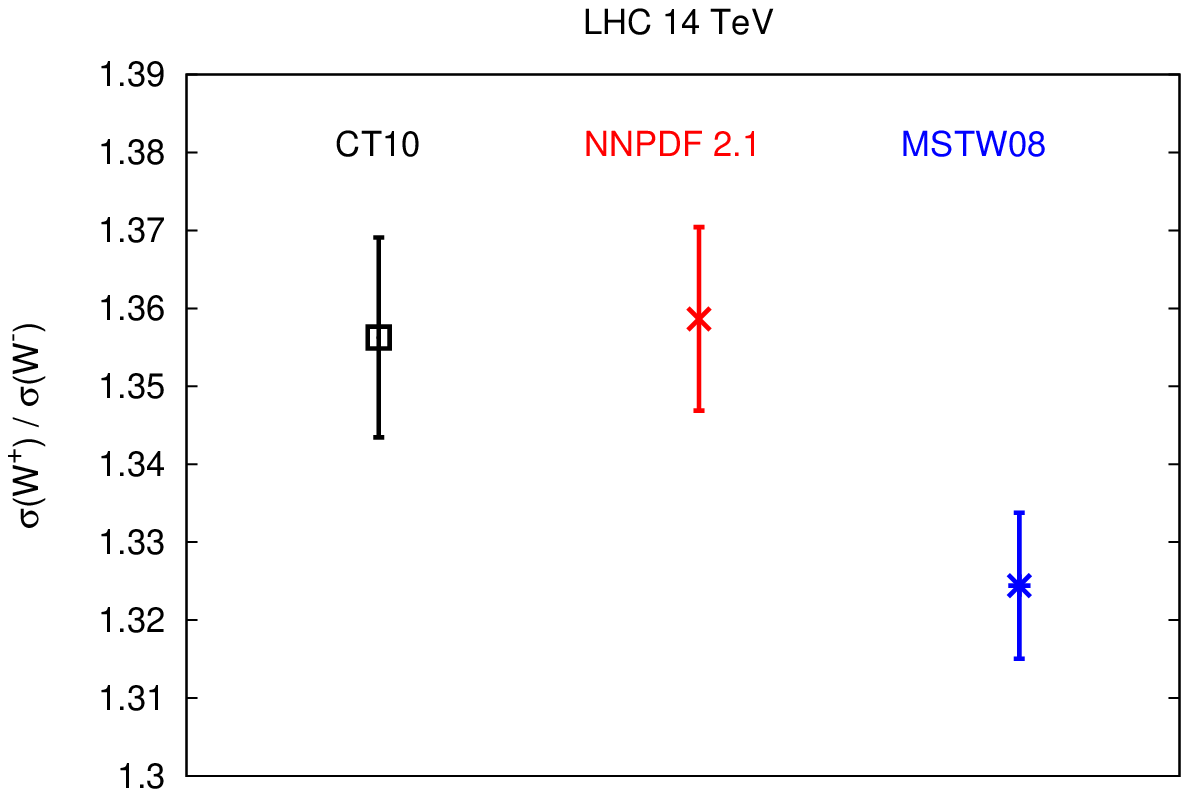}
\caption{\small Comparison between predictions from
different PDF sets for the $W^{+}/W^-$ and
$W/Z$ ratios  at the LHC $\sqrt{s}=$7 TeV (upper plots) and
$\sqrt{s}=$14 TeV (lower plots). 
 \label{fig:WZratios}} 
\end{center}
\end{figure}

Related important observables at the LHC are the $W^+/W^-$ and $W/Z$
cross-section ratios. These have generally 
reduced experimental uncertainties, since e.g.
normalization uncertainties cancel in the ratio.
Predictions at 7 and 14 TeV for NNPDF2.1, CT10 and MSTW08 are
compared in Fig.~\ref{fig:WZratios}. For these observables the dependence
on $\alpha_s$ is negligible. The agreement
for cross-section ratios seems to be worse than for total
cross-sections: for example for the $W^+/W^-$ ratio CT10 and NNPDF2.1
are in good agreement but MSTW08 is lower by more than two-$\sigma$. The
agreement for the $W/Z$ ratio is better but still marginal at 7 TeV.

\begin{figure}[t]
\begin{center}
\epsfig{width=0.38\textwidth,figure=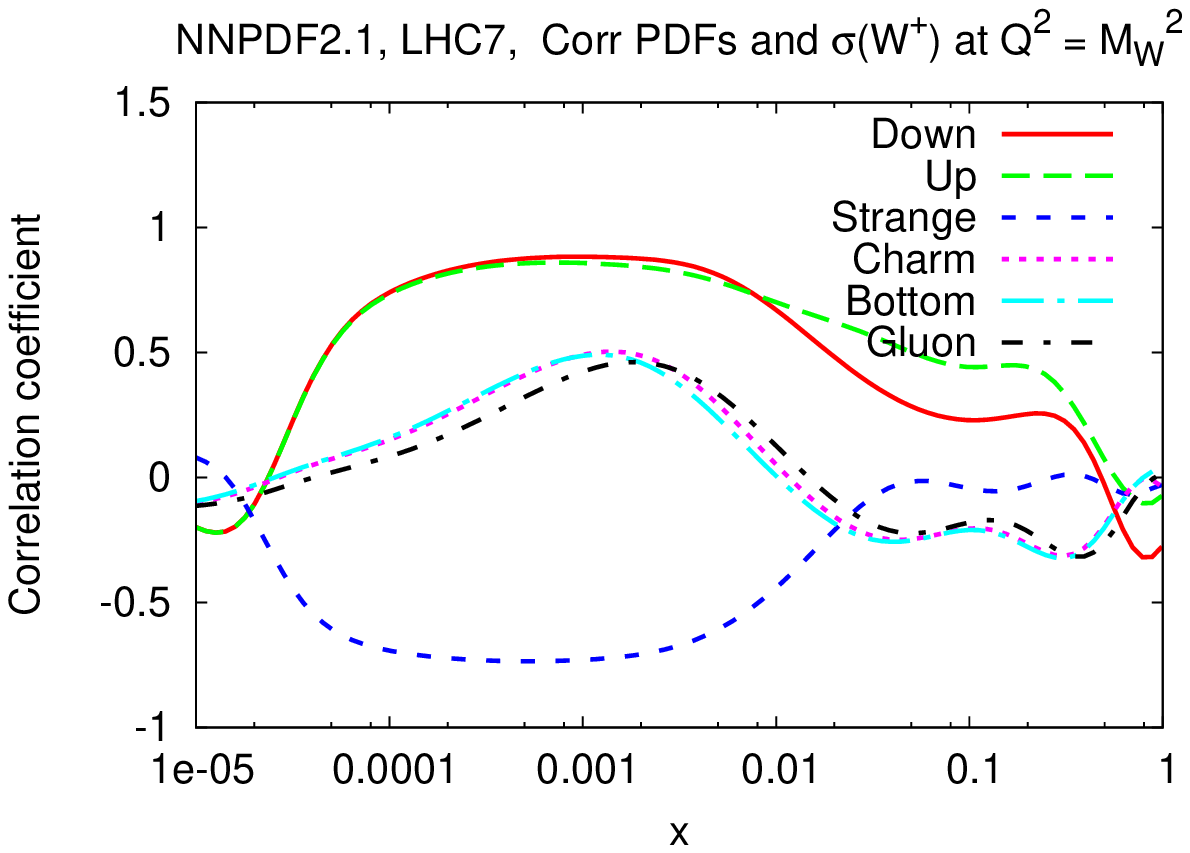}
\epsfig{width=0.38\textwidth,figure=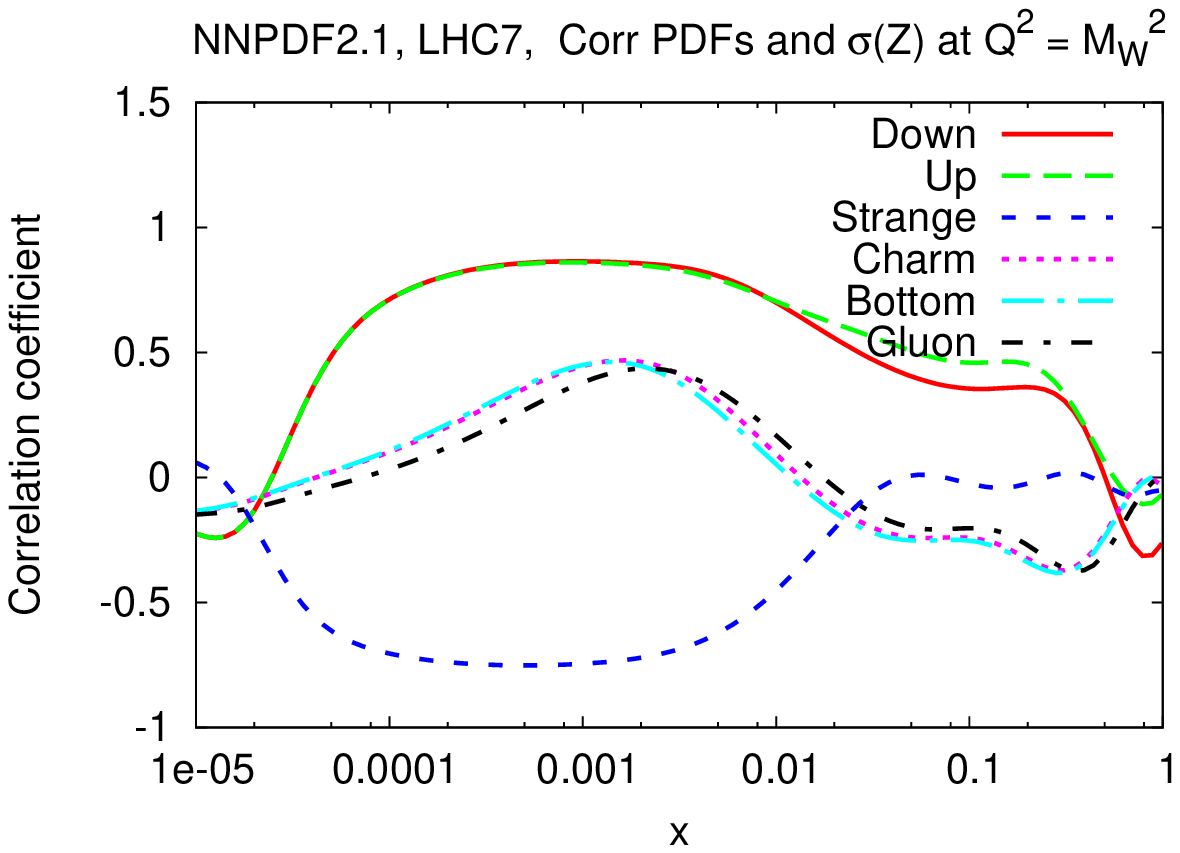}\\
\epsfig{width=0.38\textwidth,figure=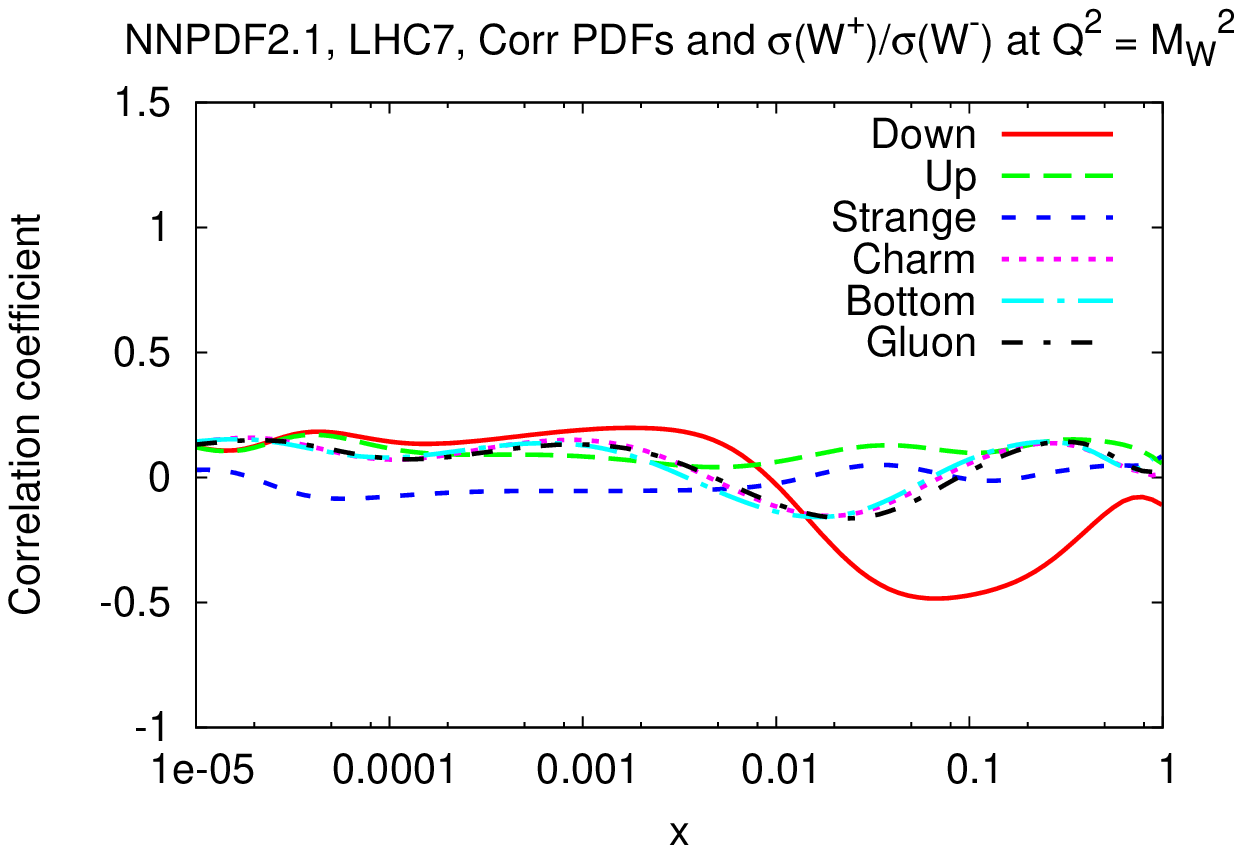}
\epsfig{width=0.38\textwidth,figure=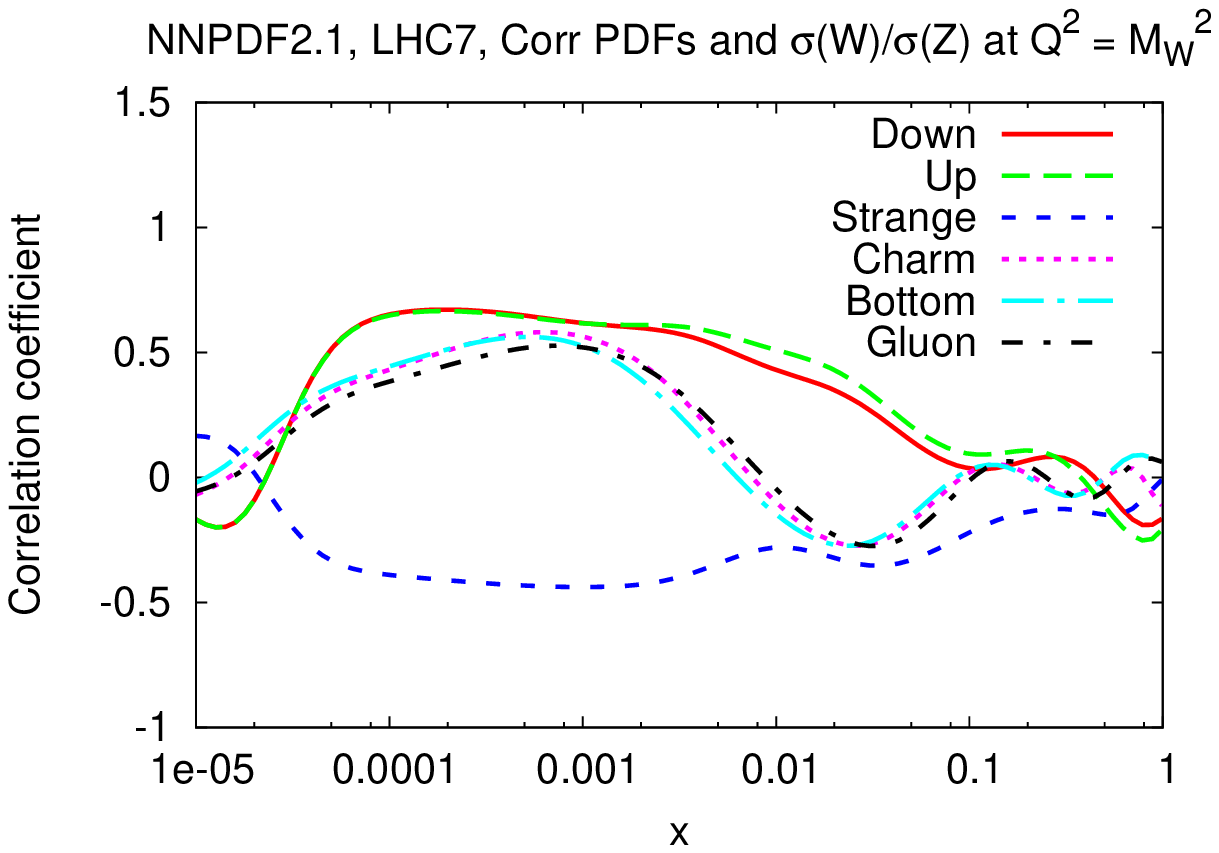}
\caption{\small The correlation between PDFs and vector
boson production total cross-sections (upper plots) and their
ratios (lower plots) for LHC 7 TeV. Correlations for $W^-$ (not
shown) are very similar.
 \label{fig:WZcorrelations}} 
\end{center}
\end{figure}

The correlation between PDFs and physical observables quantifies 
the relevance of each PDF for different
observables (and conversely)
as a function of $x$~\cite{Guffanti:2010yu,Nadolsky:2008zw}.
As an illustration with NNPDF2.1, 
we have computed the correlations between PDFs  and vector boson
production total cross-sections and their ratios at LHC 7 TeV.
Results are shown in Fig.~\ref{fig:WZcorrelations}. The total
$W$ and $Z$ cross-sections are as expected
mostly correlated with the $u$ and $d$ sea quarks and
anticorrelated with the strange quarks, the correlation
with the gluon (and the heavy flavours generated dynamically
from it) being milder. It is also interesting to note that correlations
between PDFs and the corresponding physical observable are only moderately
reduced in the $W/Z$ ratio as compared to the individual cross-sections, and
they are instead almost suppressed in the $W^+/W^-$ ratio. This observation
suggest that the latter ratio should be less sensitive to PDF uncertainties.

As for NNPDF2.0~\cite{Ball:2010de}, we have produced variants of
the NNPDF2.1 fit based on reduced datasets: DIS only, 
DIS and inclusive jet data only. These fits 
are useful to study the impact on PDF of the various observables used
in the global fit. 
Results for  LHC cross-sections  at 7 TeV determined using these fits
are
collected in Table~\ref{tab:LHCobsdatasets}. 
For these very inclusive observables, it turns out that a 
purely DIS fit already provides a rather good approximation, though this
need not be always the case for other observables.

\begin{table}[t]
  \centering
  {\footnotesize
    \begin{tabular}{|c|c|c|c|}
      \hline
      & $\sigma(W^+)B_{l\nu}$ [nb]
      & $\sigma(W^-)B_{l\nu}$ [nb]
      & $\sigma(Z^0)B_{ll}$ [nb]\\
      \hline 
      \hline
      NNPDF2.1 DIS     & $6.03\pm 0.11$ & $4.15\pm 0.08$ & $0.940\pm 0.014$ \\
      NNPDF2.1 DIS+JET & $6.03\pm 0.12$ & $4.14\pm 0.08$ & $0.939\pm 0.015$\\
      NNPDF2.1         & $5.99\pm 0.14$ & $4.09\pm 0.09$ & $0.932\pm 0.020 $ \\
      \hline
    \end{tabular}}\\
  \vspace{0.3cm}
  {\footnotesize
    \begin{tabular}{|c|c|c|}
      \hline
      & $\sigma(t\bar{t})$ [pb] 
      & $\sigma(H,m_H=120\,{\rm GeV})$ [pb]
      \\
      \hline
      \hline
      NNPDF2.1 DIS     & $167\pm  7$ & $11.66\pm 0.21$ \\
      NNPDF2.1 DIS+JET & $170\pm  5$ & $11.66\pm 0.22$ \\
      NNPDF2.1         & $170\pm  5$ & $11.64\pm 0.17$ \\
      \hline
    \end{tabular}}
  \caption{\label{tab:LHCobsdatasets}  
    \small Cross-sections for W, Z, $t\bar{t}$ and Higgs production
    at the LHC at $\sqrt{s}=7$ TeV and the associated
    PDF uncertainties for the reference NNPDF2.1 set compared to those
    obtained using sets determined from reduced datasets:
    DIS only, DIS+JET.}
\end{table}

\begin{figure}[t]
\begin{center}
\epsfig{width=0.32\textwidth,figure=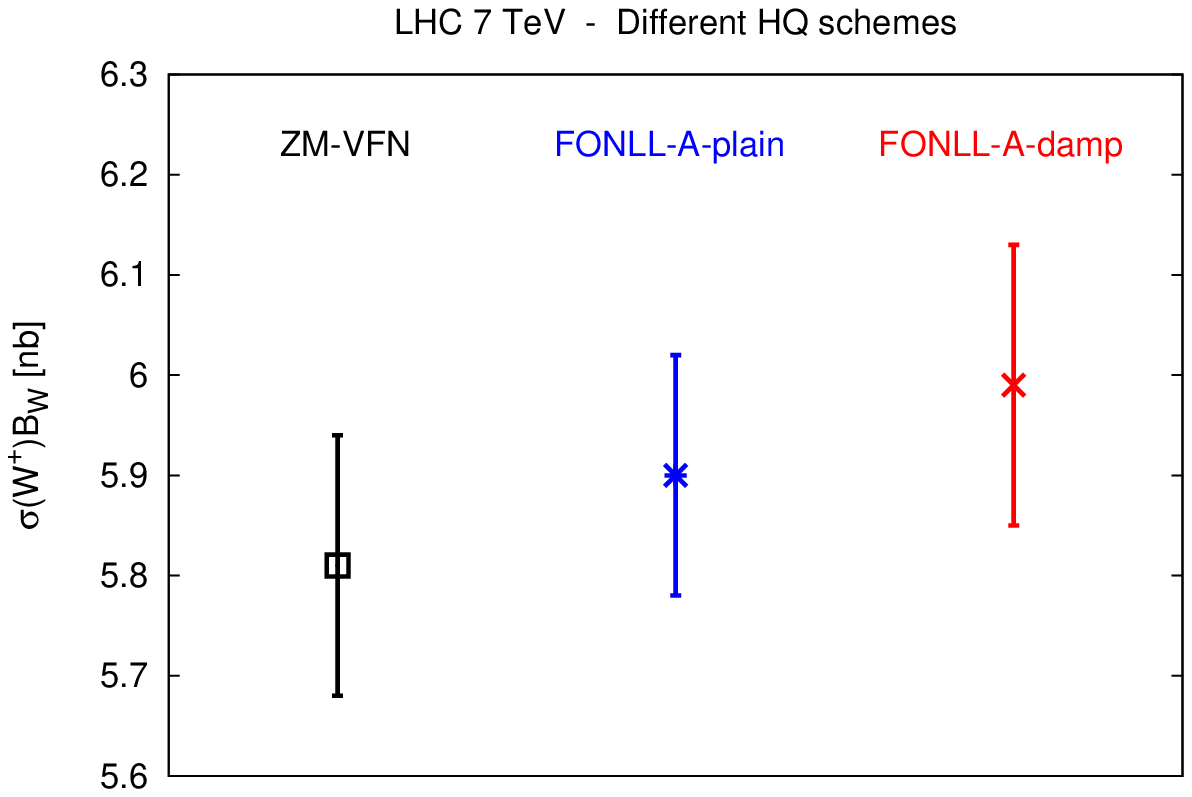}
\epsfig{width=0.32\textwidth,figure=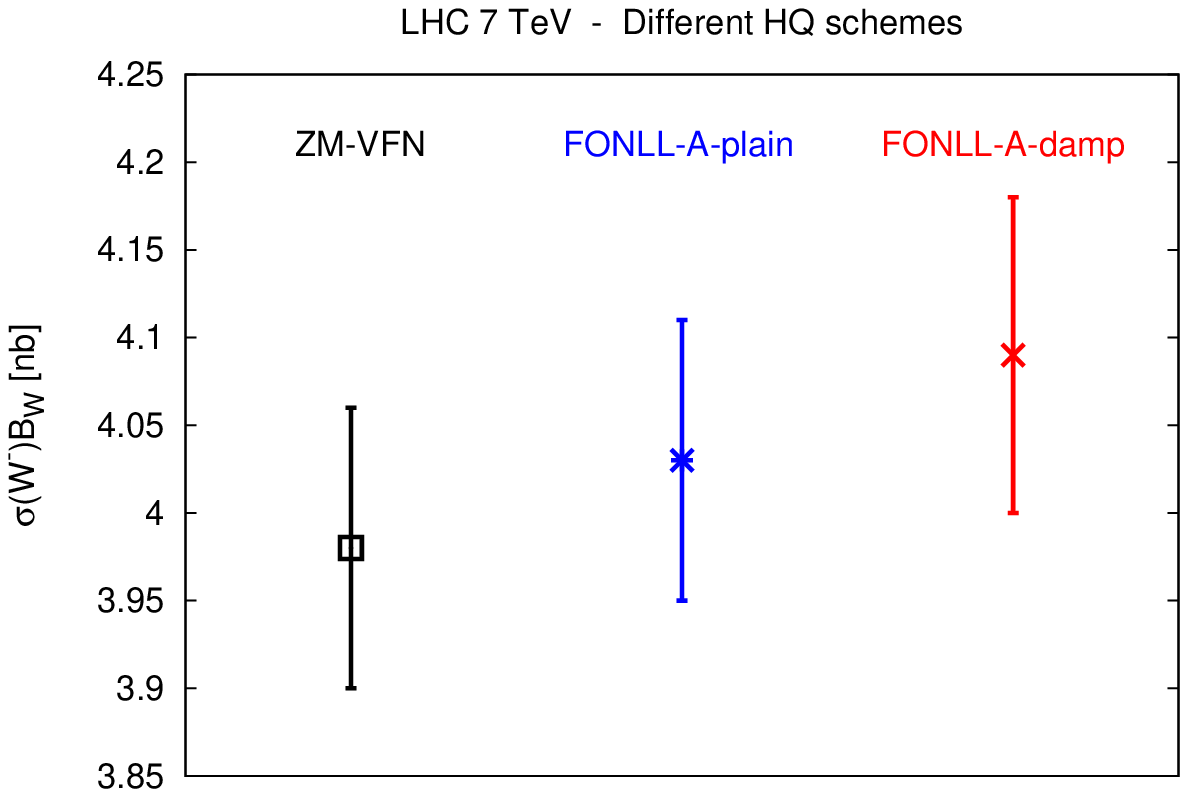}
\epsfig{width=0.32\textwidth,figure=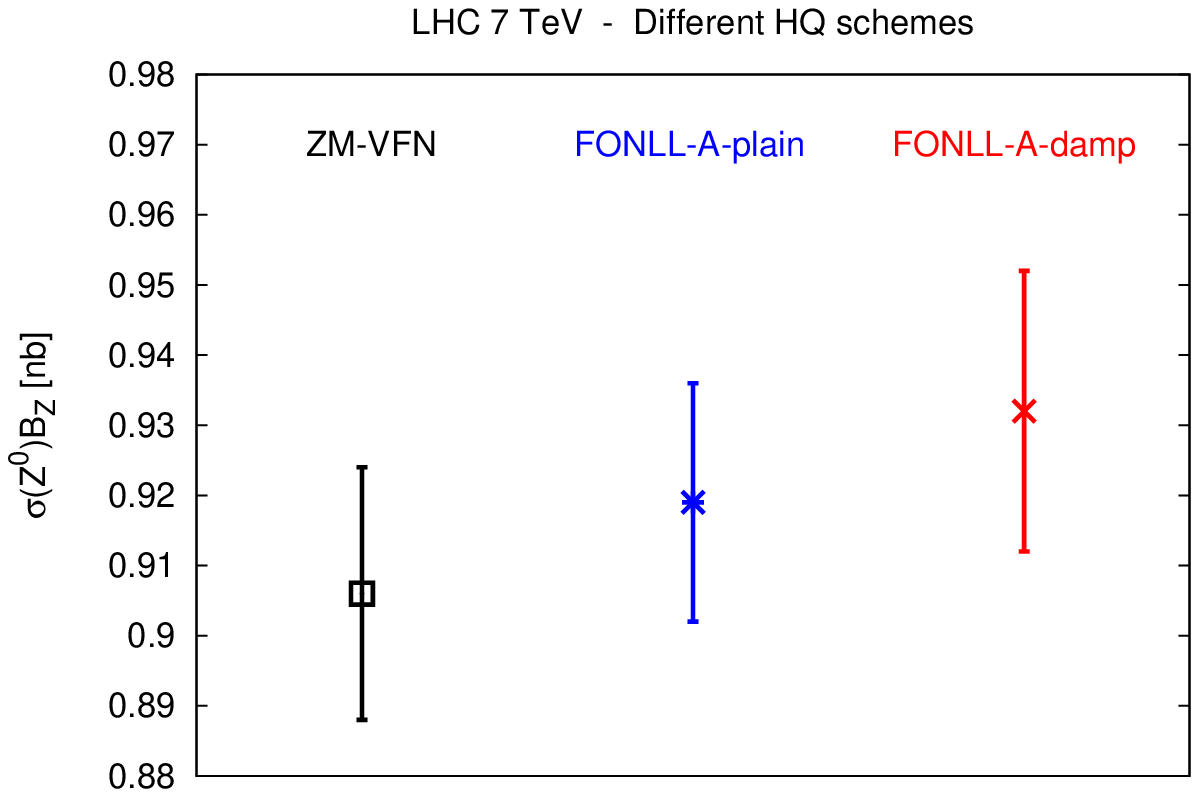}
\epsfig{width=0.32\textwidth,figure=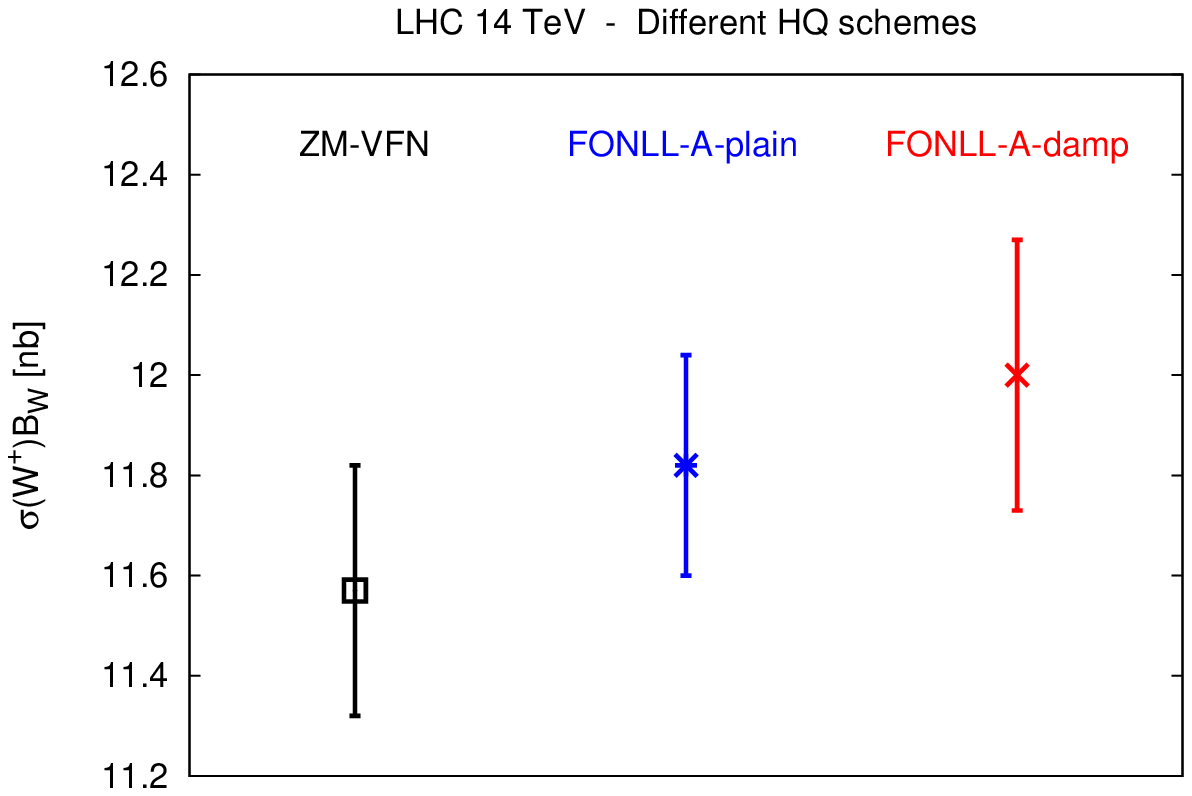}
\epsfig{width=0.32\textwidth,figure=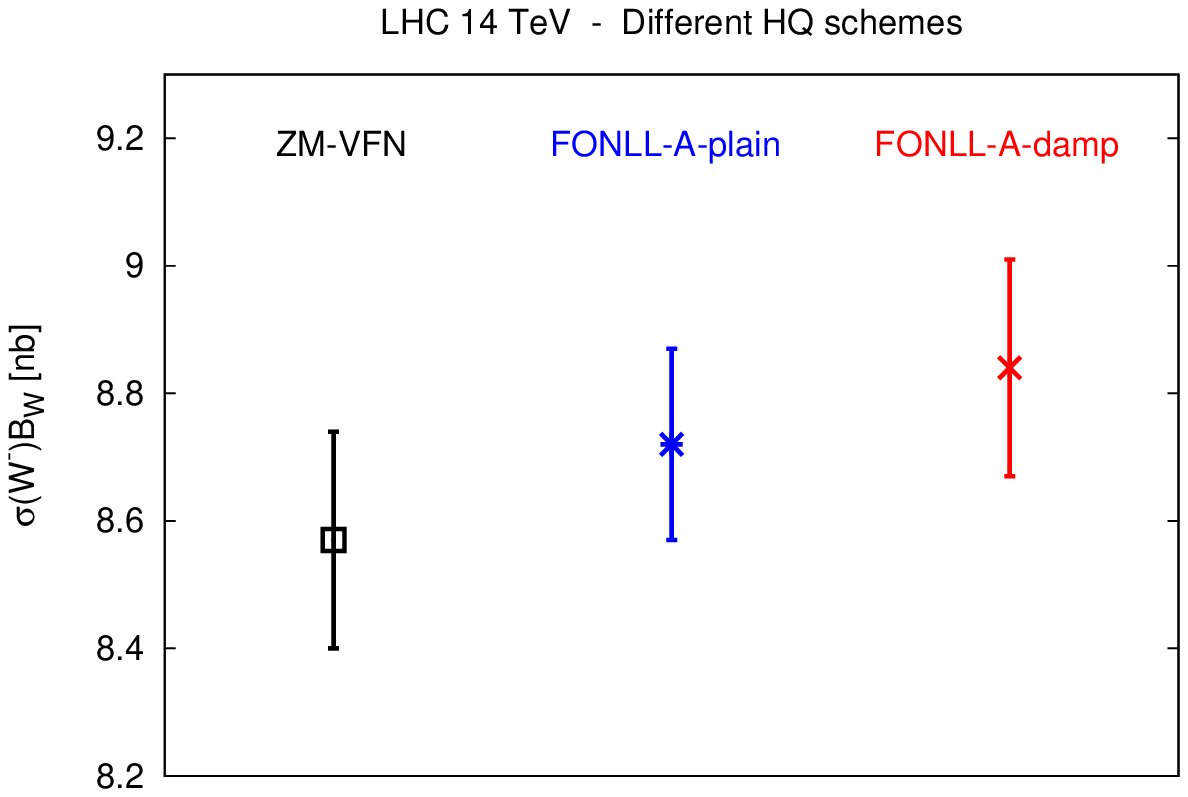}
\epsfig{width=0.32\textwidth,figure=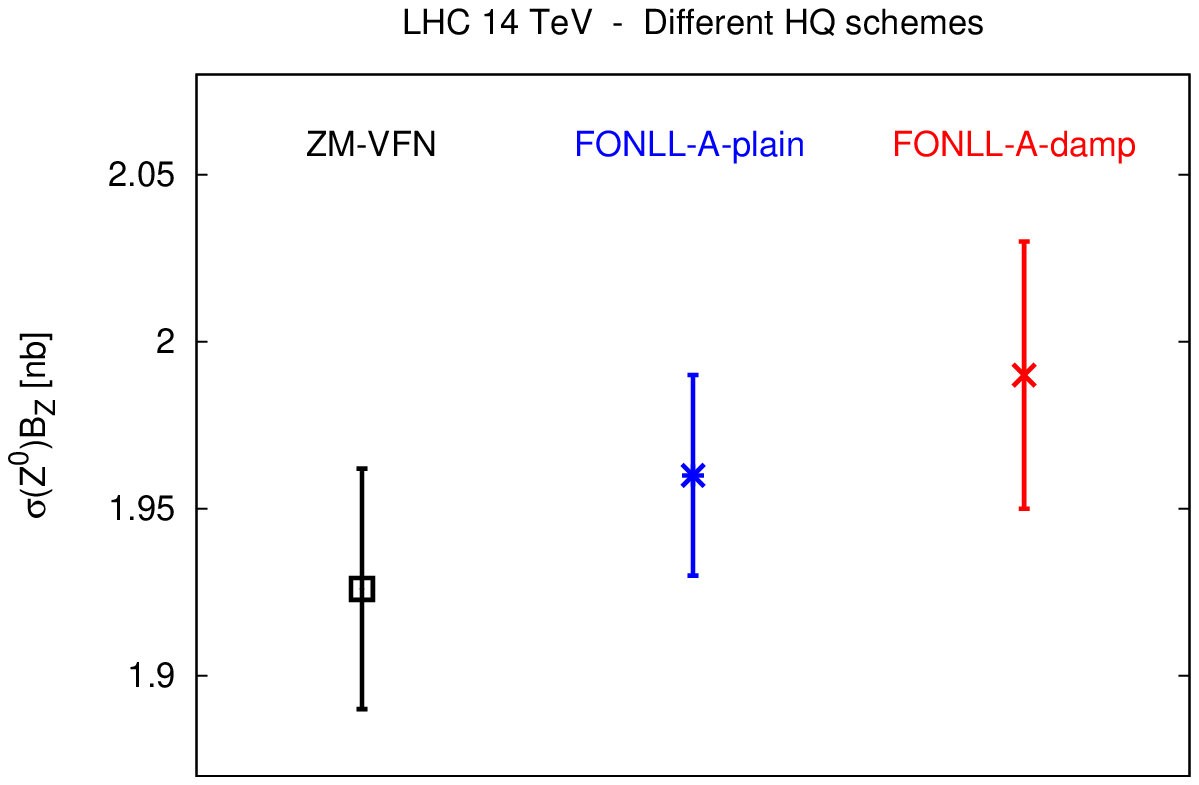}
\caption{\small Graphical representation of the results of 
Table~\ref{tab:LHCgmtreat}.
 \label{fig:comp7tevgm}} 
\end{center}
\end{figure}

The inclusion of heavy quark mass effects 
has a theoretical ambiguity due to subleading terms.
A full study of theoretical uncertainties on PDFs has never been
performed and goes  beyond the scope of this paper. However, we
provide here a first estimate of the uncertainty related to the
inclusion of heavy quark mass effects to ${\mathcal O}(\alpha_s)$ by
comparing results obtained from the three sets discussed in
Sect.~\ref{sec:dataset}:  NNPDF2.0 RED (without heavy quark mass
terms, but the same kinematic cuts NNPDF2.1), the default NNPDF2.1,
and  NNPDF2.1 without damping terms in the FONLL-A method. Results are shown in 
Table~\ref{tab:LHCgmtreat}  and in Fig.~\ref{fig:comp7tevgm}. 
As expected, results obtained without damping prescription sit half
way between NNPDF2.0RED and  NNPDF2.1 at 7 TeV, and closer
to the latter at 14 TeV. As discussed in Sect.~\ref{sec:dataset},
the difference between the NNPDF2.1 results with and without damping terms
can be 
taken as a conservative estimate of the theoretical uncertainty
associated to the uncertainty in the inclusion of heavy quark mass
effects to ${\mathcal O}(\alpha_s)$.

\begin{table}[t]
  \centering
  {\footnotesize
    \begin{tabular}{|c|c|c|c|}
      \hline
      7 TeV     
      & $\sigma(W^+)B_{l\nu}$ [nb]
      & $\sigma(W^-)B_{l\nu}$ [nb]
      & $\sigma(Z^0)B_{ll}$ [nb]\\
      \hline
      \hline
      NNPDF2.1               & $5.99\pm 0.14$ & $4.09\pm 0.09$ & $0.93\pm 0.02$ \\
      \hline
      NNPDF2.0 RED           & $5.81\pm 0.13$ & $3.98\pm 0.08$ & $0.91\pm 0.02$ \\
      NNPDF2.1 FONLL-A plain & $5.90\pm 0.12$ & $4.03\pm 0.08$ & $0.92\pm 0.02$\\
    \hline
  \end{tabular}}\\
\vspace{0.3cm}
{\footnotesize
  \begin{tabular}{|c|c|c|c|}
    \hline
    14 TeV        
    & $\sigma(W^+)B_{l\nu}$ [nb]
    & $\sigma(W^-)B_{l\nu}$ [nb]
    & $\sigma(Z^0)B_{ll}$ [nb]\\
    \hline
    \hline
    NNPDF2.1               & $12.00\pm 0.27$ & $8.84\pm 0.17$ & $1.99\pm 0.04$ \\
    \hline
    NNPDF2.0 RED           & $11.57\pm 0.25$ & $8.57\pm 0.17$ & $1.93\pm 0.04$\\
    NNPDF2.1 FONLL-A plain & $11.82\pm 0.22$ & $8.72\pm 0.15$ & $1.96\pm 0.03$ \\
    \hline
  \end{tabular}}\\
\caption{\label{tab:LHCgmtreat}  
  \small Cross-sections for W, Z, $t\bar{t}$ and Higgs production
    at the LHC at $\sqrt{s}=7$ TeV and the associated
    PDF uncertainties for the reference NNPDF2.1 set compared to those
    obtained using sets with different treatment of heavy quarks:
NNPDF2.0RED, without heavy quark mass effects, and  NNPDF2.1 FONLL-A
plain with heavy quark mass effects but without threshold damping terms.
}
\end{table}

\subsection{Parton luminosities}

The processes discussed in Sect.~\ref{sec:lhcimplications} are a  
small subset of the LHC observables which are sensitive to
PDFs. A detailed, systematic study of these would be quite
interesting; however, a good deal of information can be gathered by
simply studying parton luminosities.
Following Ref.~\cite{Campbell:2006wx}, we define the parton luminosity
\be
\Phi_{ij}\lp M_X^2\rp = \frac{1}{s}\int_{\tau}^1
\frac{dx_1}{x_1} f_i\lp x_1,M_X^2\rp f_j\lp \tau/x_1,M_X^2\rp \ ,
\label{eq:lumdef}
\ee
where $f_i(x,M^2)$ is a PDF and $\tau \equiv M_X^2/s$.
We consider in particular the gluon-gluon luminosity, the various
heavy quark-antiquark luminosity, and the quark-gluon and quark-quark
luminosity respectively defined as
\be
\Phi_{qg}\equiv\sum_{i=1}^{N_f}
\Phi_{q_ig};\quad\Phi_{qq}\equiv\sum_{i=1}^{N_f} \Phi_{q_i\bar q_i}.
\label{eq:qqqgdef}
\ee

\begin{figure}[ht!]
  \centering
\epsfig{width=0.32\textwidth,figure=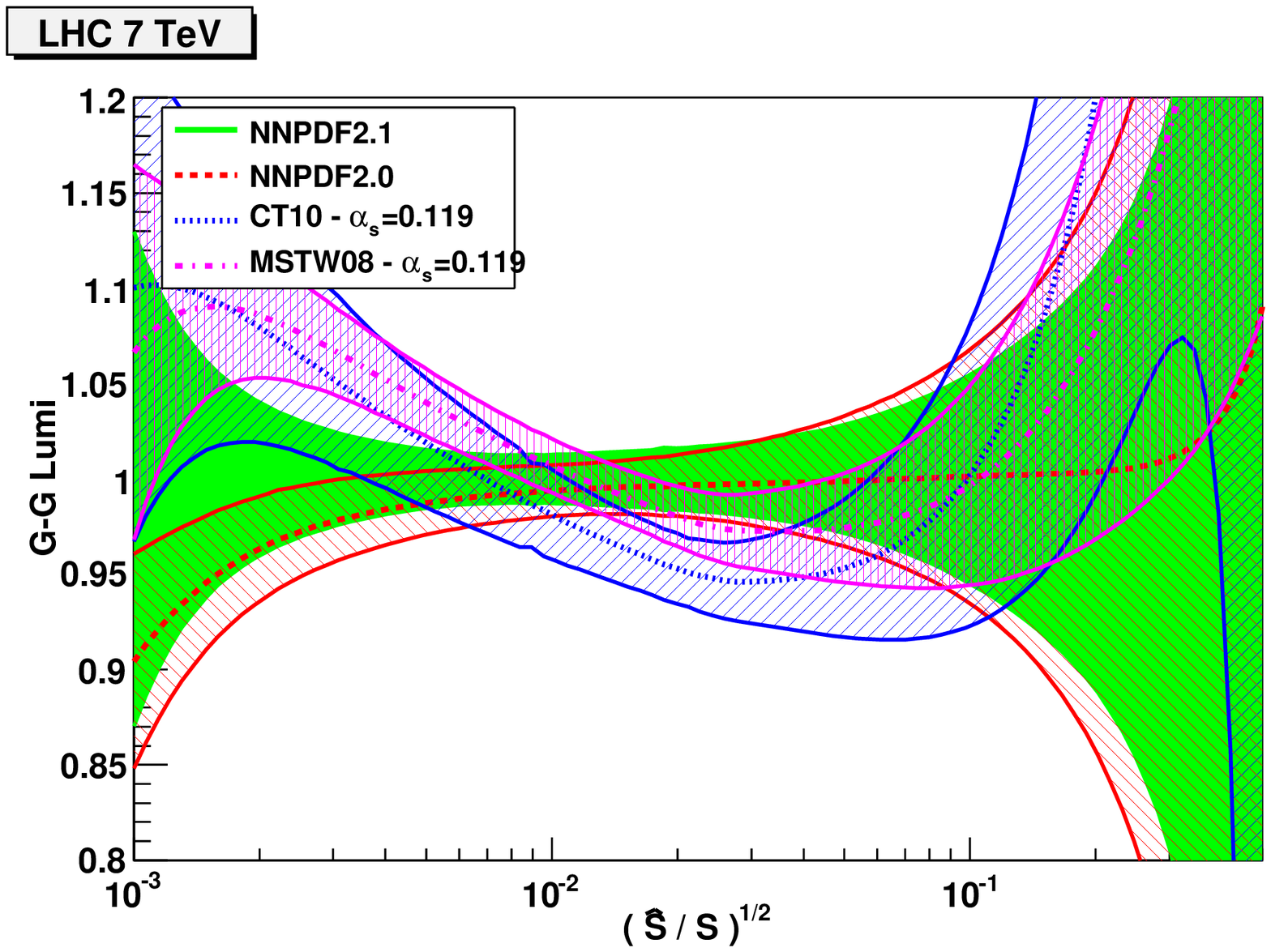}
\epsfig{width=0.32\textwidth,figure=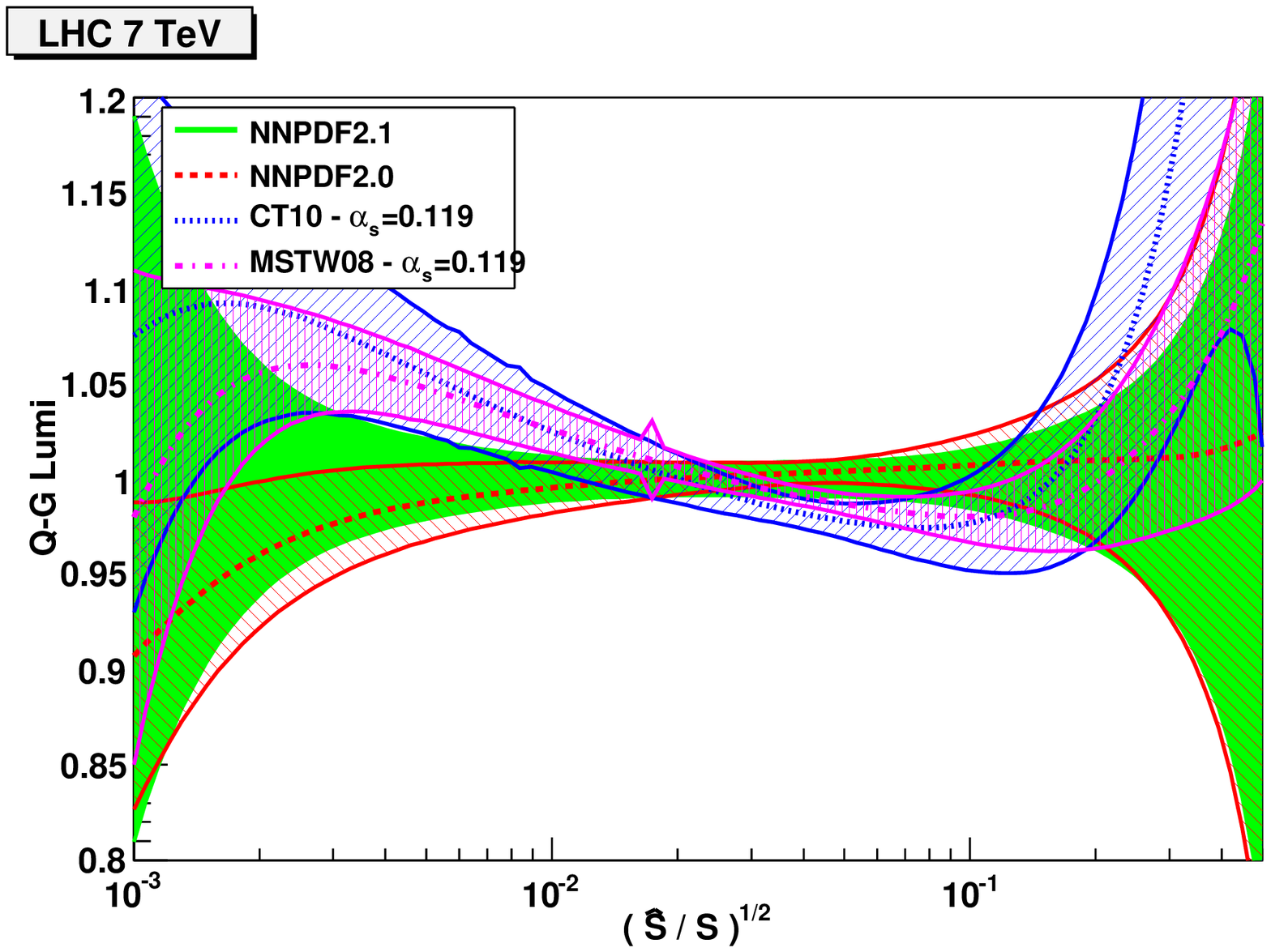}
\epsfig{width=0.32\textwidth,figure=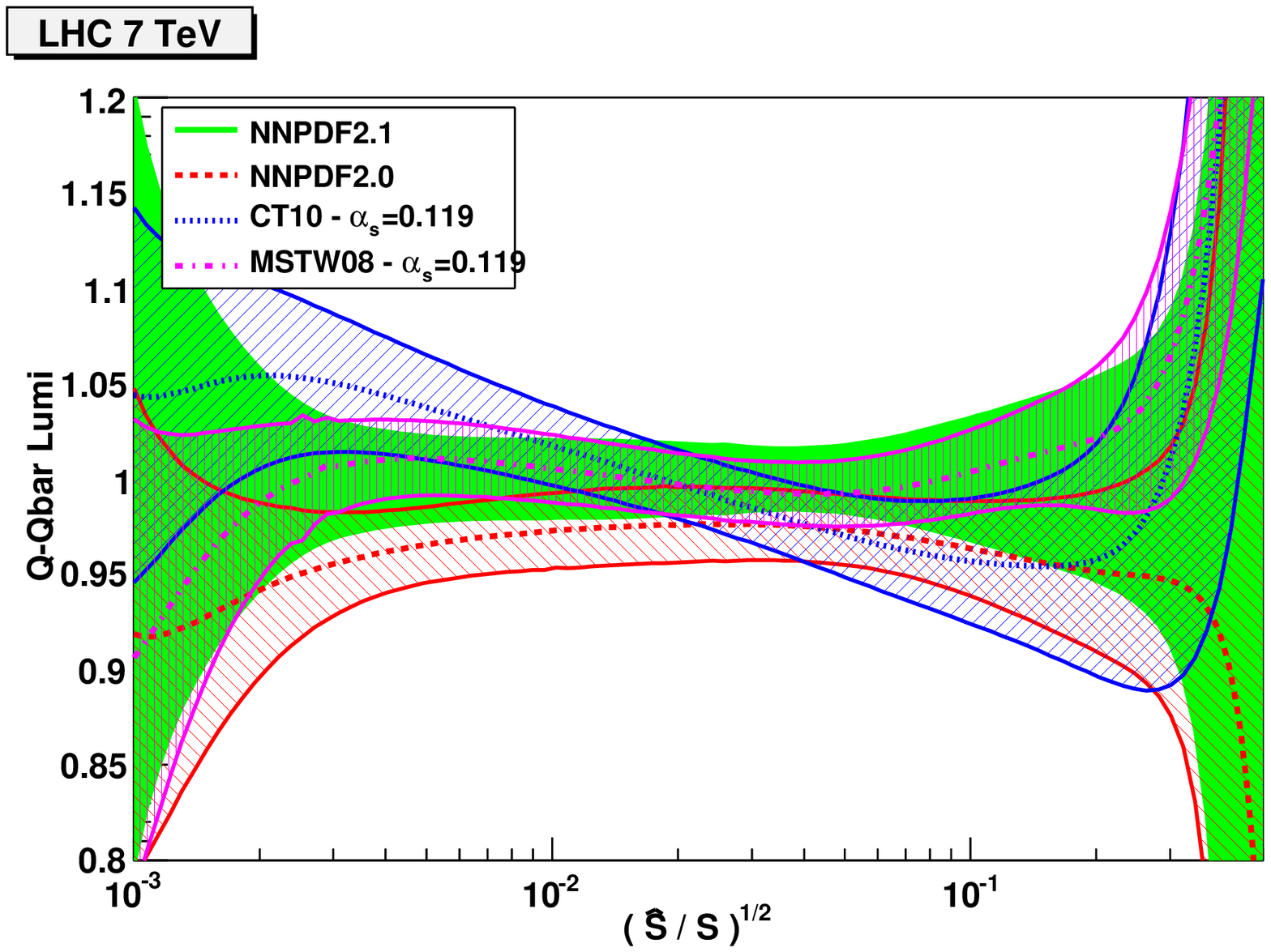}
\epsfig{width=0.32\textwidth,figure=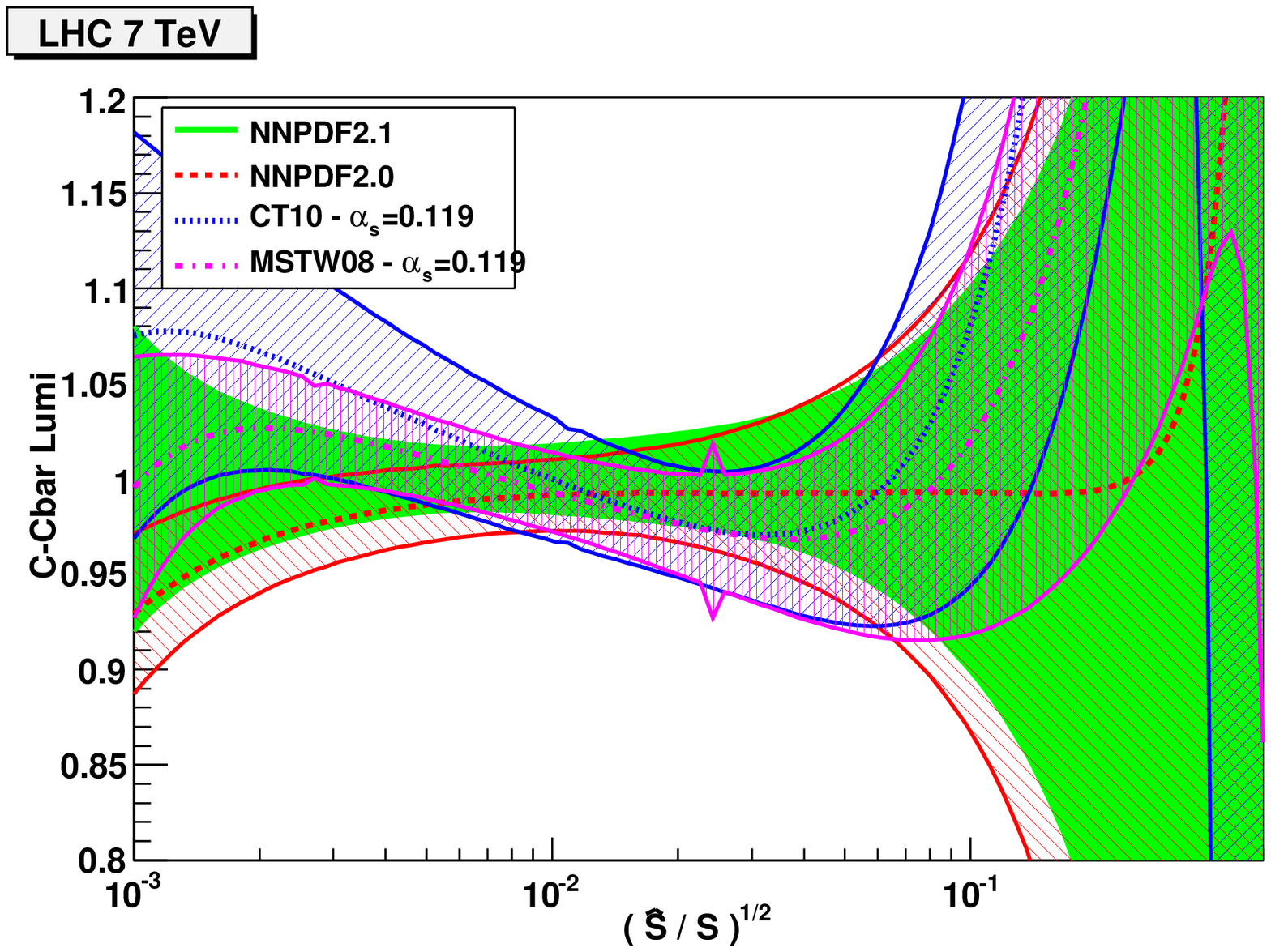}
\epsfig{width=0.32\textwidth,figure=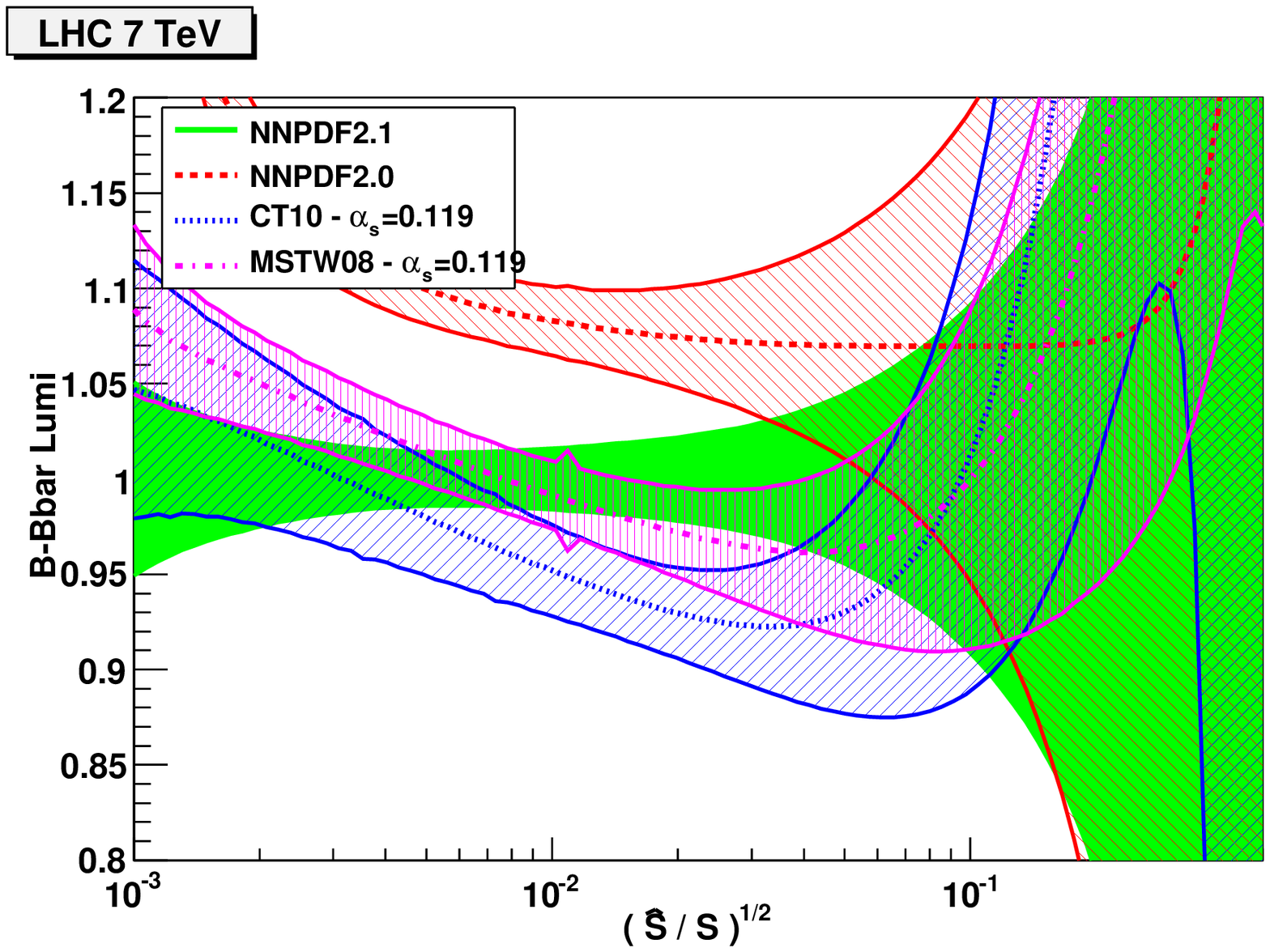}
\epsfig{width=0.32\textwidth,figure=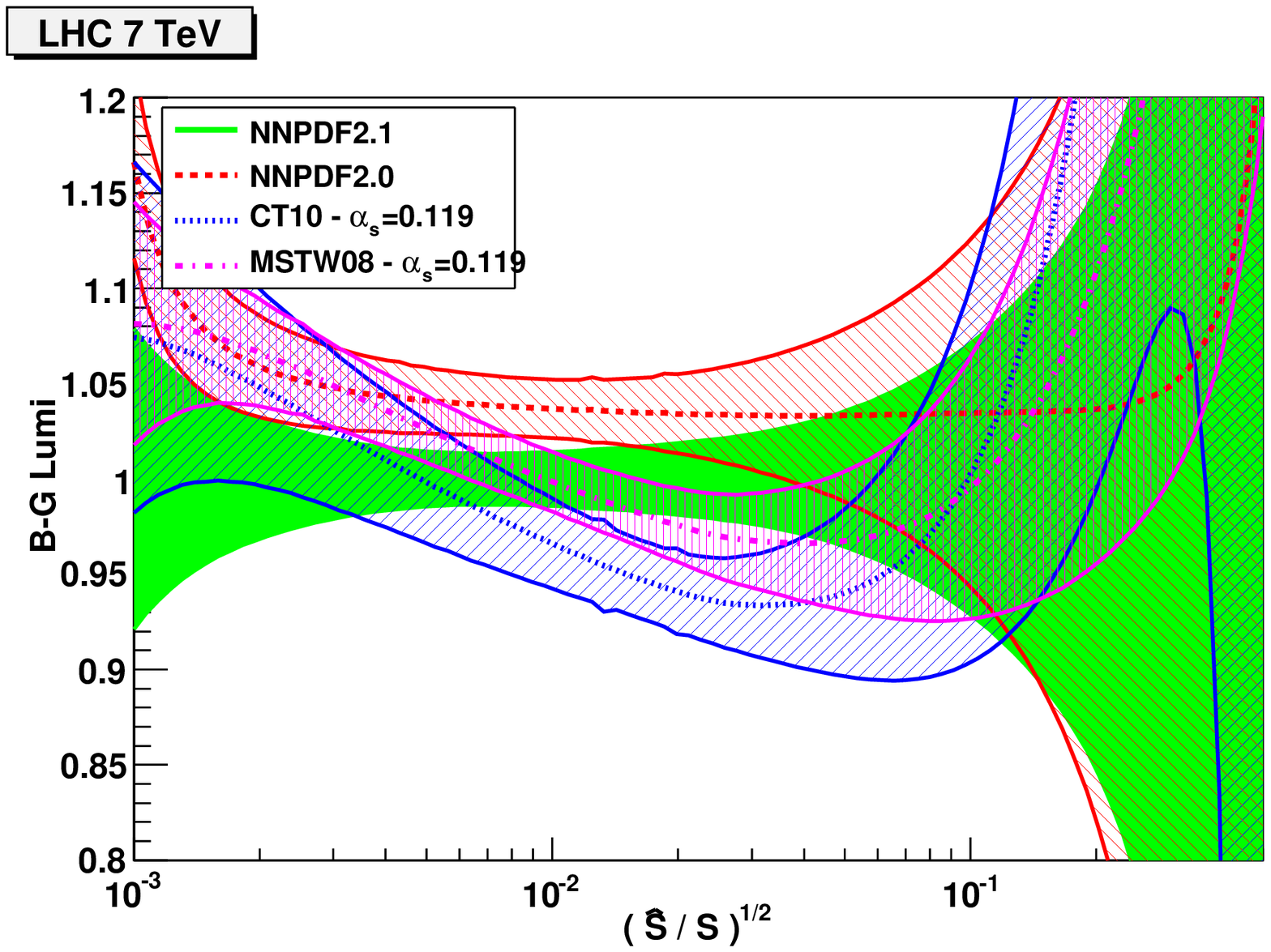}
\caption{\small The parton luminosities Eqs.~(\ref{eq:lumdef}-\ref{eq:qqqgdef})
for NNPDF2.1 compared to NNPDF2.0, CT10 and MSTW2008:
from left to right and from top to bottom $\Phi_{gg}$,  $\Phi_{qg}$,
 $\Phi_{qq}$,  $\Phi_{cc}$,  $\Phi_{bb}$, $\Phi_{bg}$.
 All luminosities
are plotted as ratios to the NNPDF2.1 central value. 
PDF sets with 
$\alpha_s(M_Z)=0.119$ have been used in all cases.
All uncertainties shown are one-$\sigma$.
\label{fig_fluxes}}
\end{figure}

\begin{figure}[ht!]
  \centering
  \epsfig{width=0.32\textwidth,figure=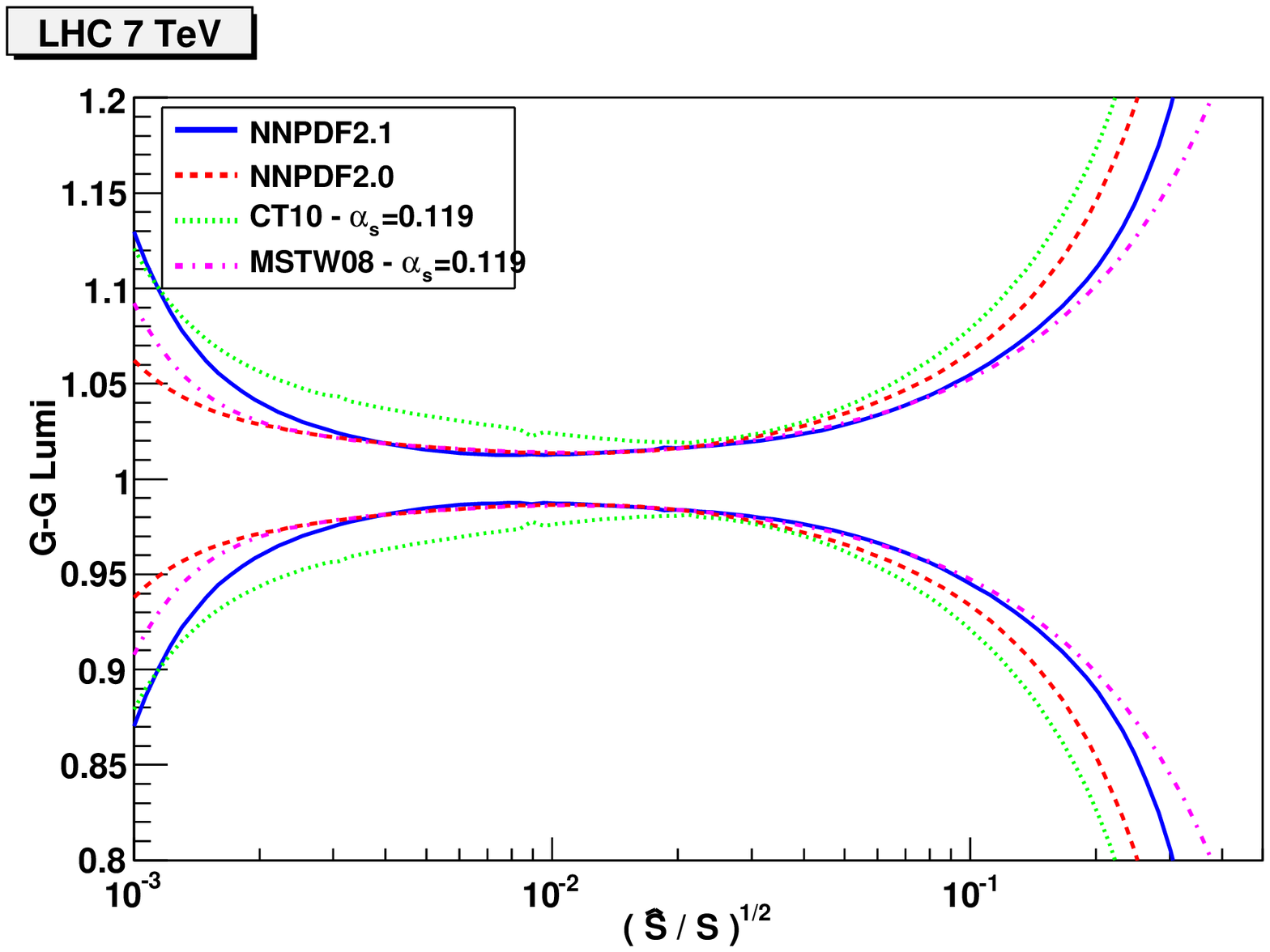}
  \epsfig{width=0.32\textwidth,figure=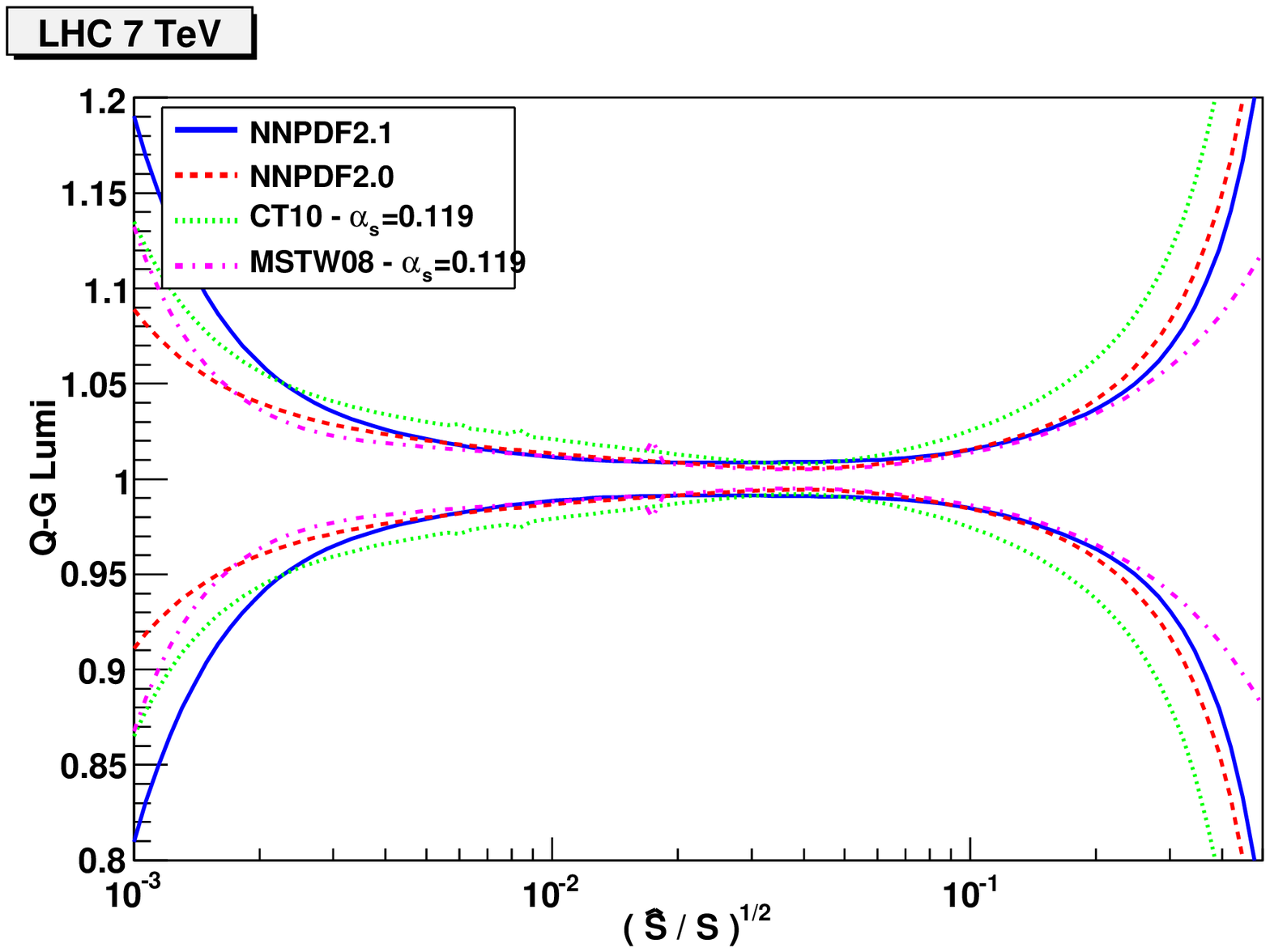}
  \epsfig{width=0.32\textwidth,figure=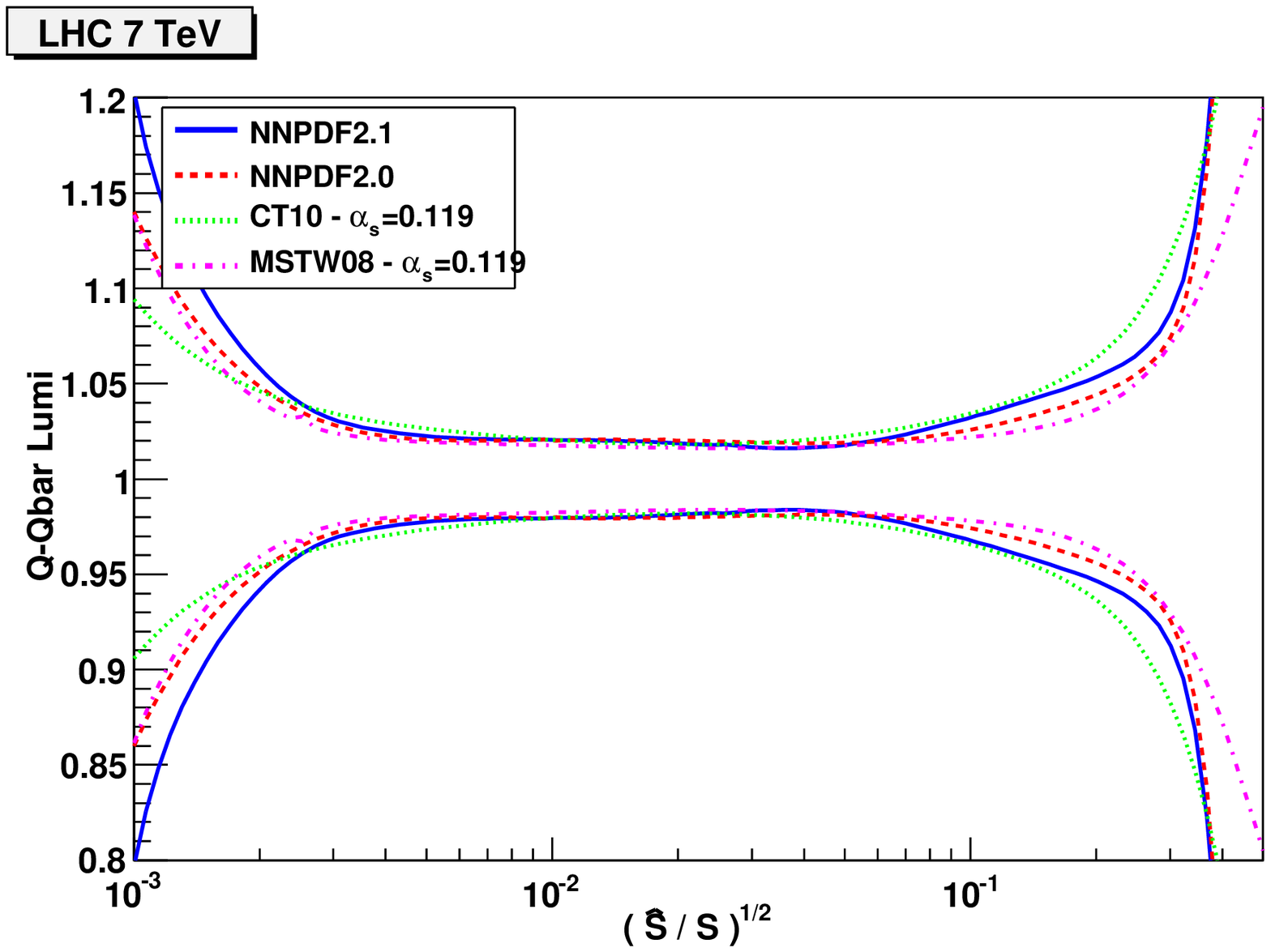}
  \epsfig{width=0.32\textwidth,figure=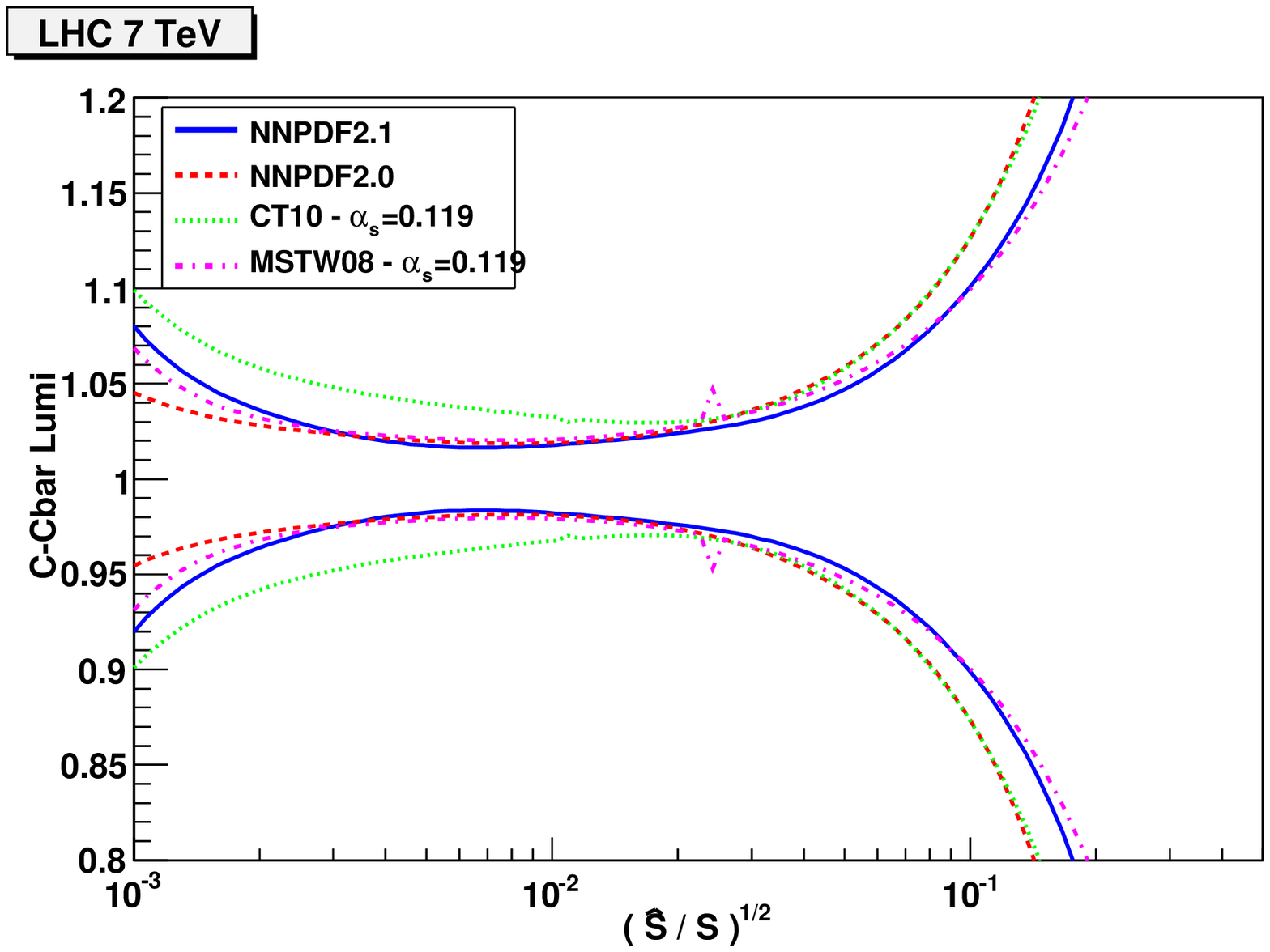}
  \epsfig{width=0.32\textwidth,figure=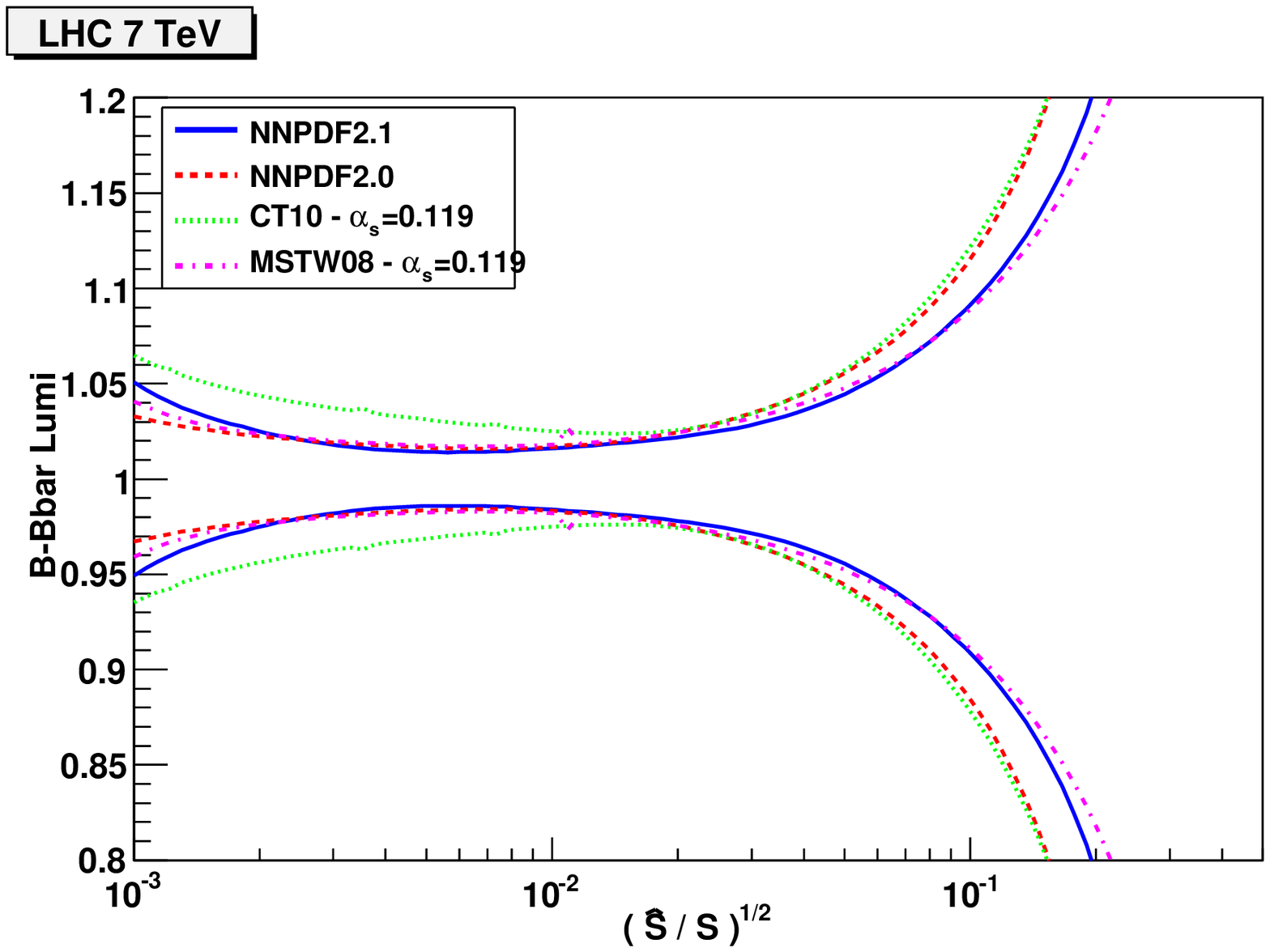}
  \epsfig{width=0.32\textwidth,figure=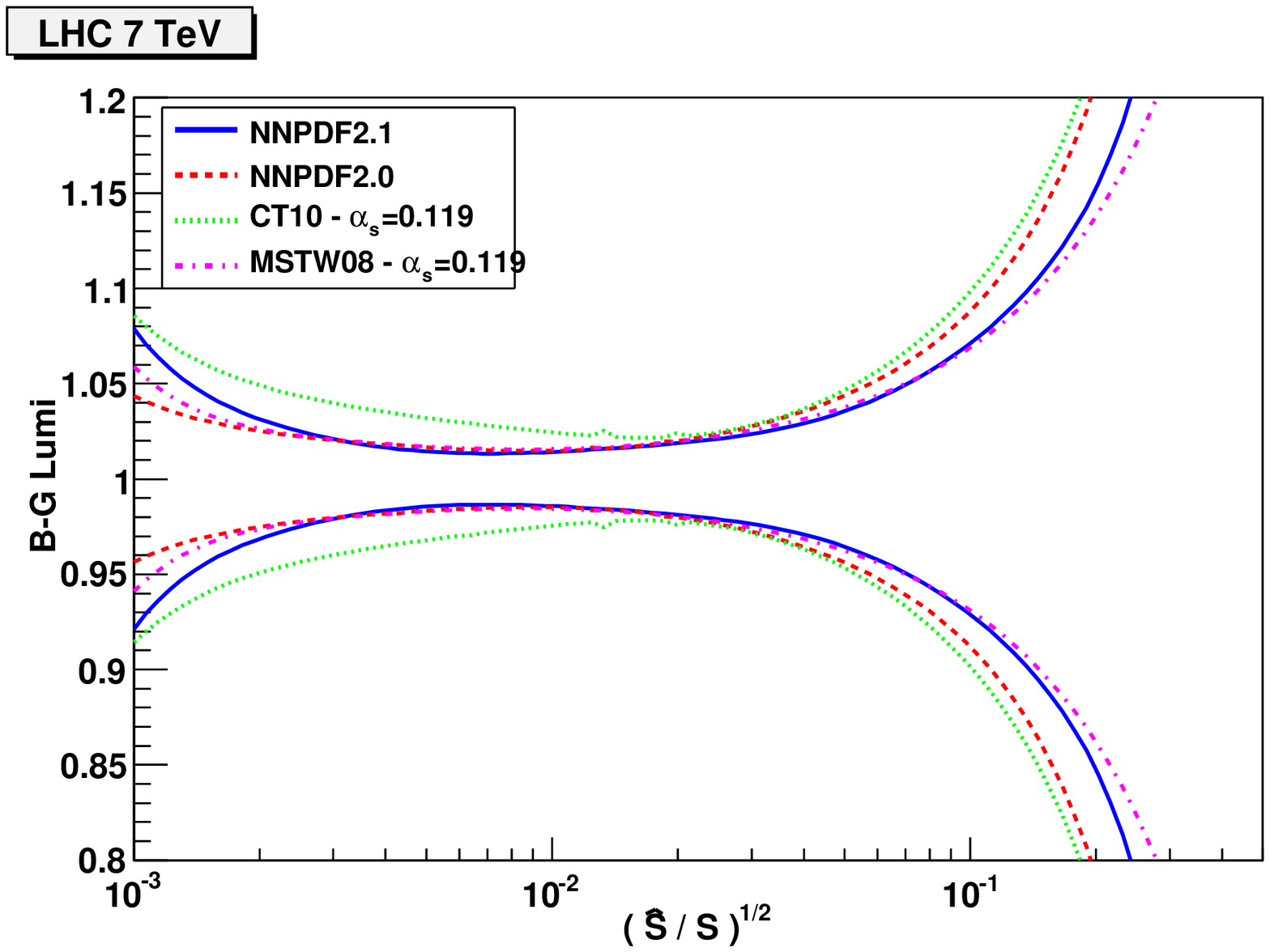}
  \caption{\small Same as Fig.~\ref{fig_fluxes}, but will all
    luminosities normalized to their respective central values.
\label{fig_fluxes-errs}}
\end{figure}

Parton luminosities form NNPDF2.1, NNPDF2.0, 
CT10 and MSTW08 sets, normalized to the NNPDF2.1 central prediction, 
are shown in Fig.~\ref{fig_fluxes} for LHC $\sqrt{s}= 7$ 
TeV, all determined with $\alpha_s\lp M_Z\rp=0.119$.
In  Fig.~\ref{fig_fluxes-errs} we also compare directly the relative
uncertainties on the luminosity for each set. These comparisons show
good agreement between global fits at the one-$\sigma$ level, 
although in some cases, such as the gluon-gluon luminosity at intermediate
invariant masses, the agreement is only marginal. Uncertainties
blow up both at very small and large values 
of $M_X$ for all sets. Differences between different sets are
larger in these regions: for example NNPDF2.1 uncertainties 
at small $M_X$ are rather larger, for luminosities that involve the gluon PDF.
The peculiar behaviour of the bottom luminosity for NNPDF2.0 is due to
the fact that in this set $m_b=4.3$~GeV, while $m_b=4.75$~GeV for all
other sets. The dependence of results on the values of the heavy quark
masses will be discussed in Sect.~\ref{sec:hqmasses} below.

\subsection{The value of $\alpha_s\lp M_Z\rp$}

\begin{figure}[t!]
  \centering
  \epsfig{width=0.48\textwidth,figure=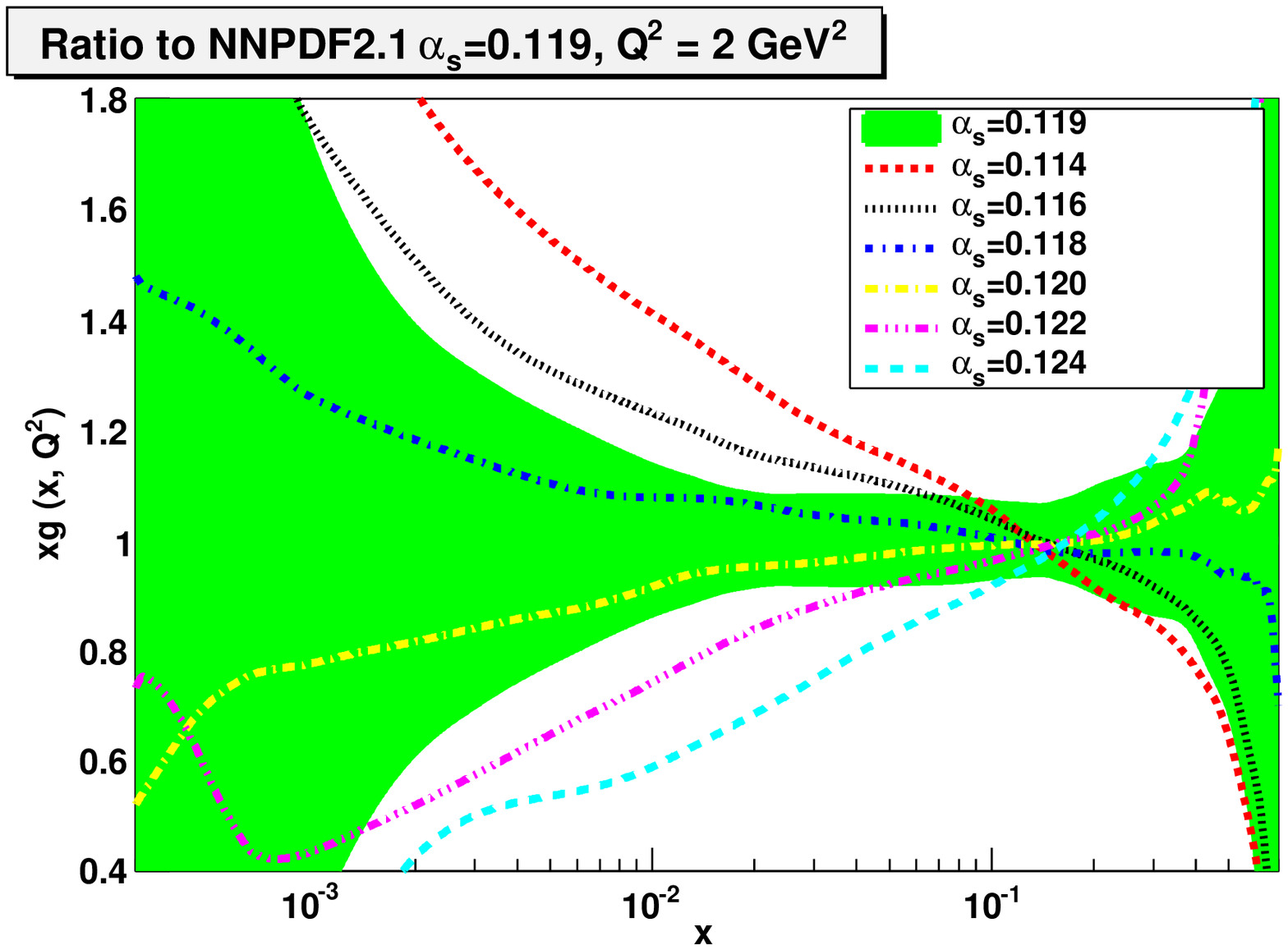}
  \epsfig{width=0.48\textwidth,figure=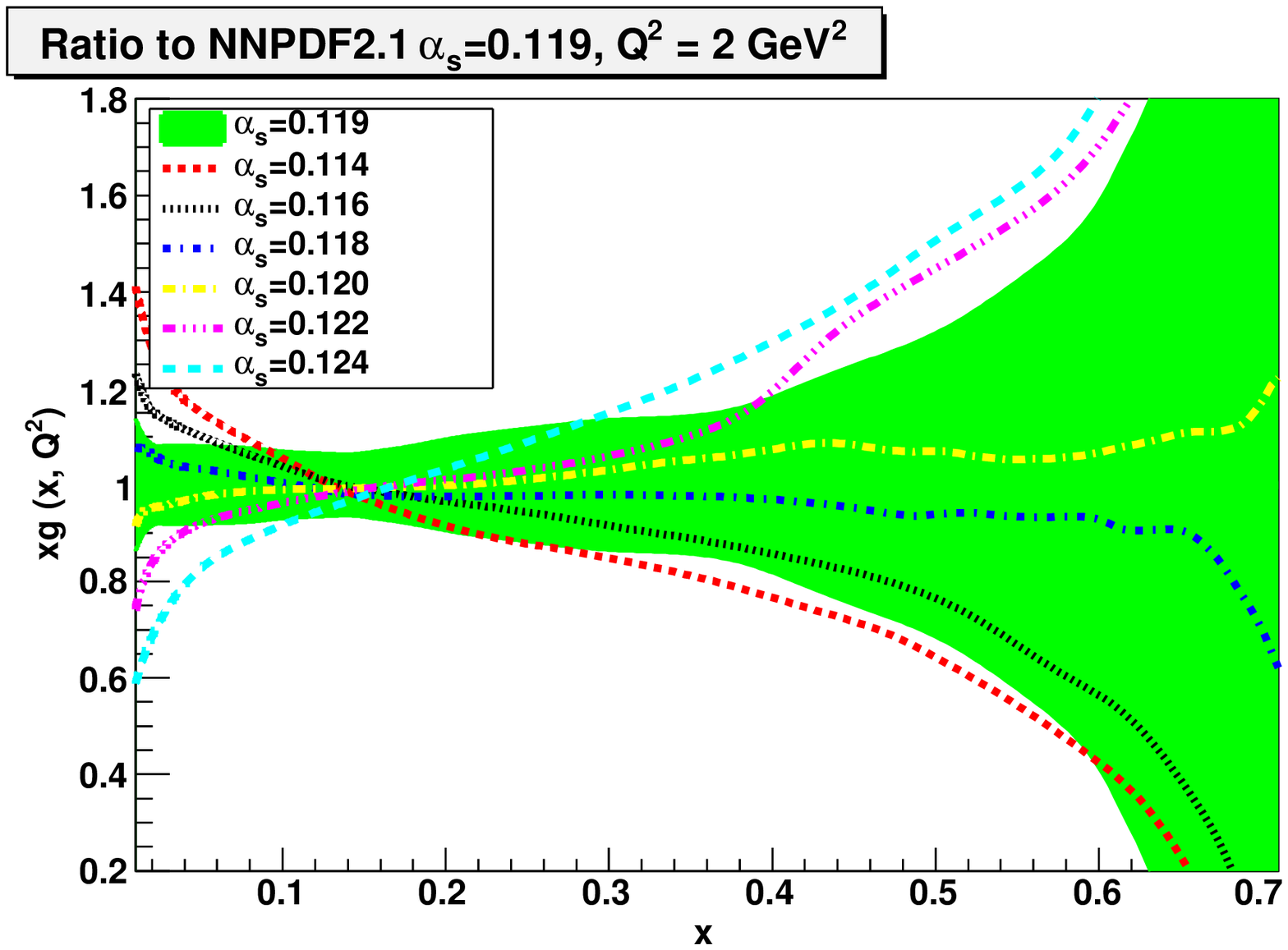}
  \epsfig{width=0.48\textwidth,figure=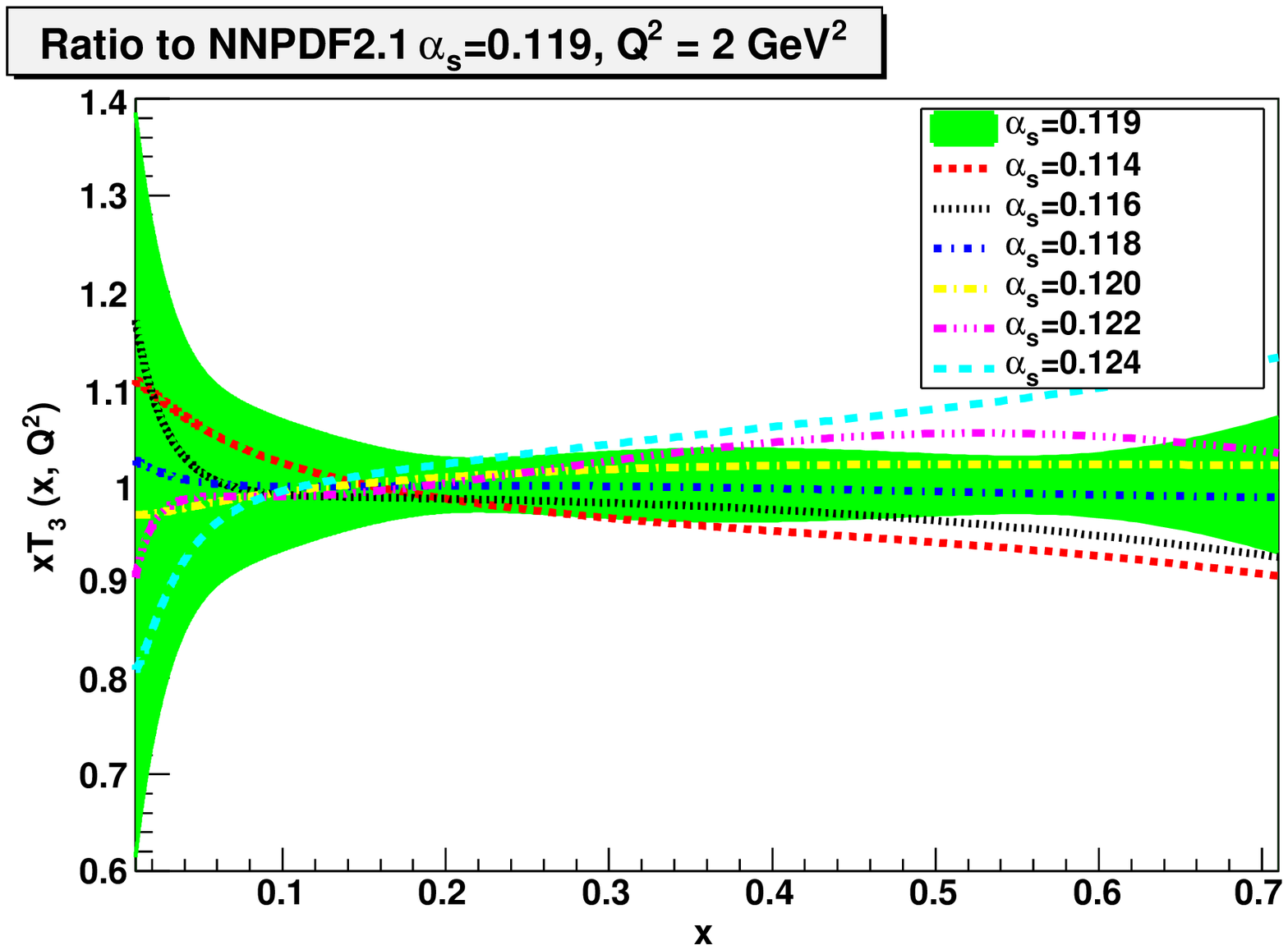}
  \epsfig{width=0.48\textwidth,figure=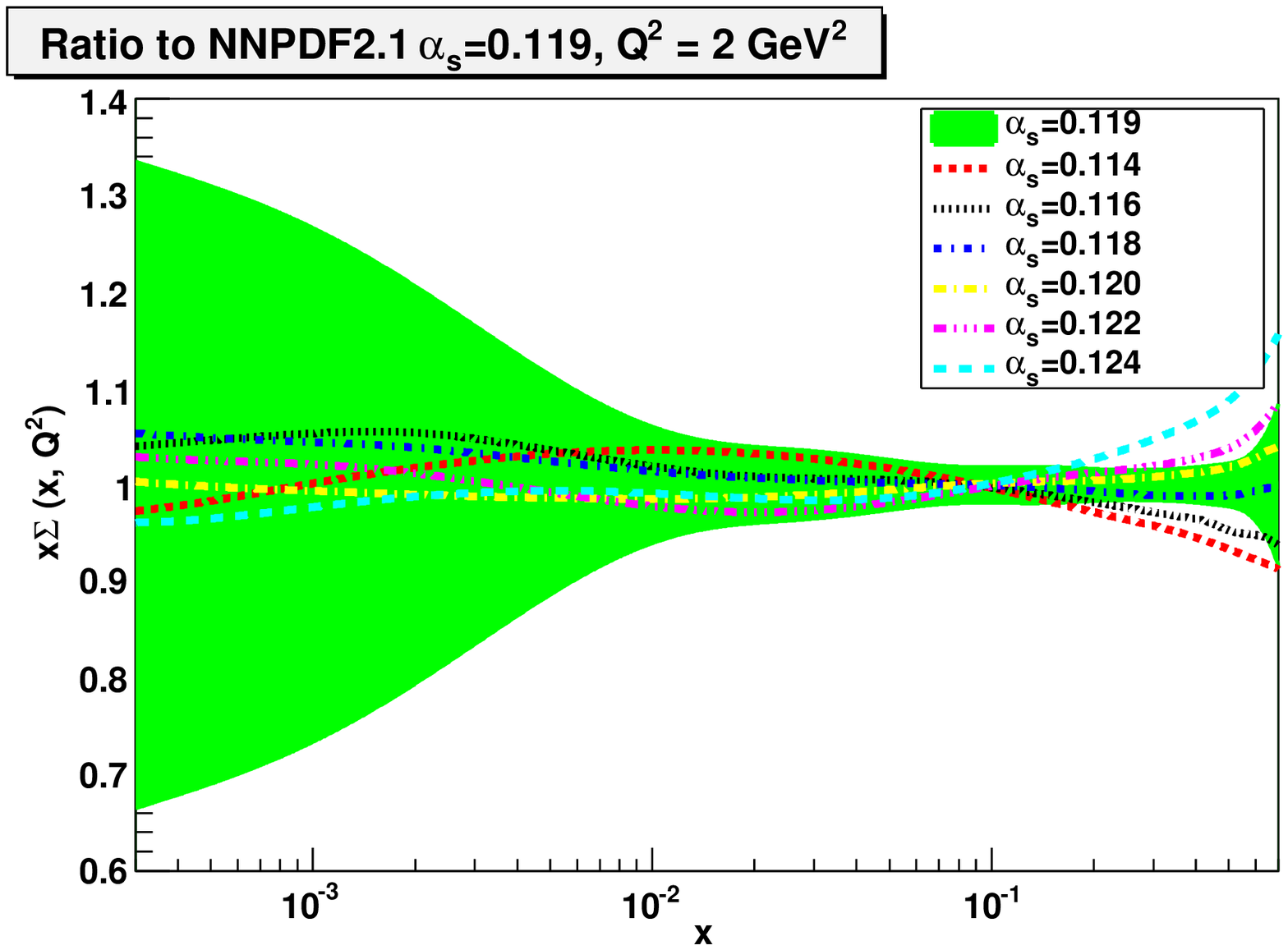}
  \caption{\small Comparison between NNPDF2.1 sets with different values
    of the strong coupling, shown as ratios with respect the reference
    fit with $\alpha_s=0.119$. The PDFs shown are the gluon at small and 
    large-$x$ (upper plots), the triplet at large-$x$ and the singlet at small-$x$
(lower plots).  
    \label{fig:PDFas}}
\end{figure}

\begin{figure}[t]
  \begin{center}
    \epsfig{width=0.49\textwidth,figure=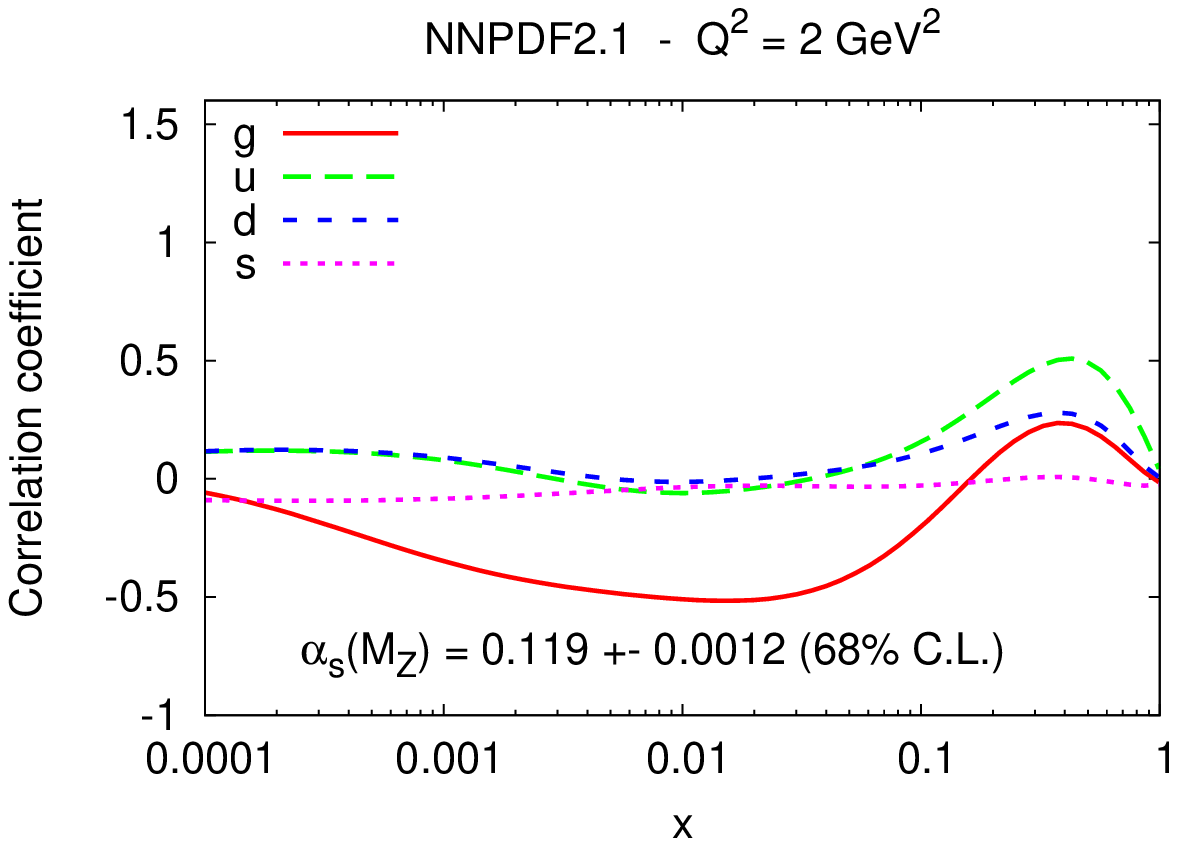}
    \epsfig{width=0.49\textwidth,figure=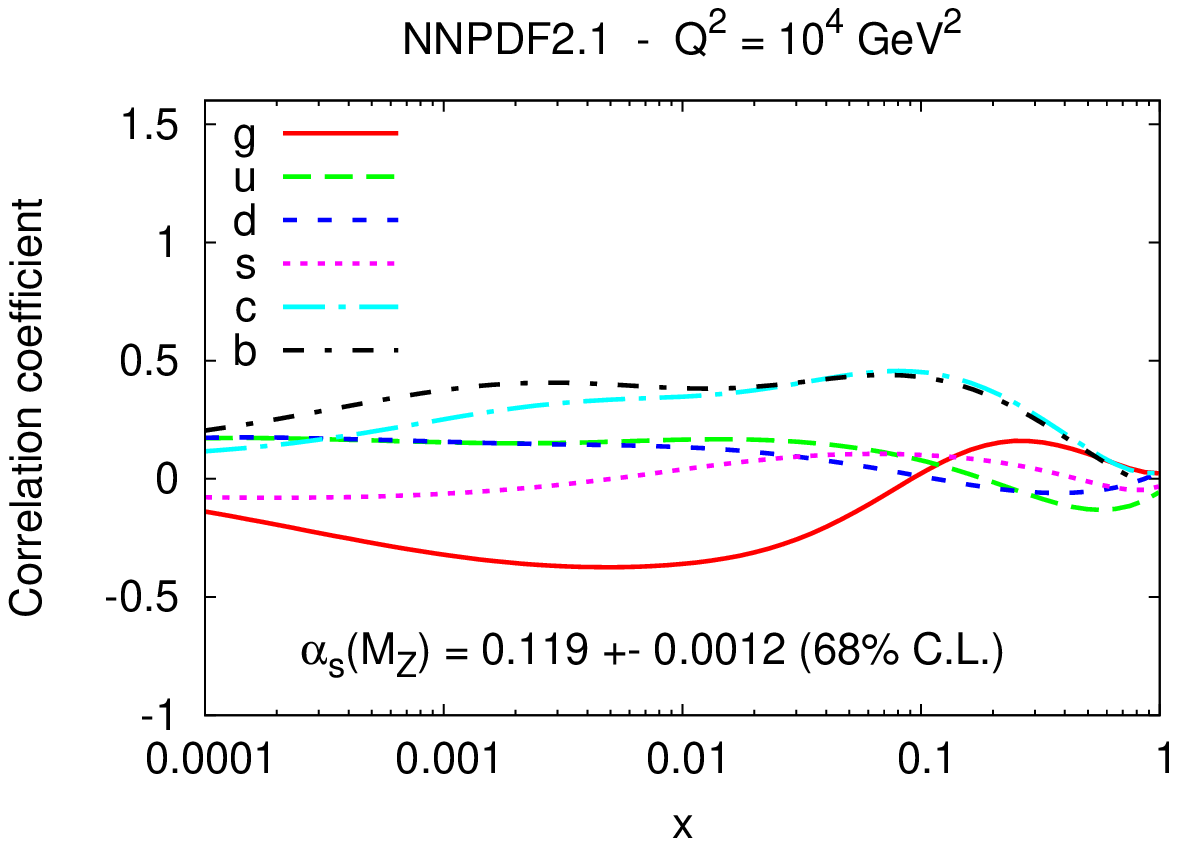}
    \caption{\small Correlation coefficient between PDFs and
      $\alpha_s(M_Z)$ computed assuming $\alpha_s(M_Z)=0.119\pm
      0.0012$ at 68\%~C.L. Results are shown at low scale
      ($Q^2=2$~GeV$^2$, left) and high scale ($Q^2=100$~GeV$^2$,
      right). 
      \label{fig:PDFascorr}} 
\end{center}
\end{figure}

We now consider the correlation between NNPDF2.1 partons and the value 
of $\alpha_s$. To this purpose,
we provide sets with 
 $\alpha_s\lp M_Z\rp$ in the range from 0.114 to 0.124 in
steps of 0.001. PDFs from fits performed using different values
of $\alpha_s$ are show in Fig.~\ref{fig:PDFas}. Results are similar to
those obtained with  NNPDF2.0~\cite{LHas}: as expected, 
the most sensitive PDF is the gluon.
To quantify this 
it is useful to compute the correlation between PDFs and $\alpha_s$
(as defined in Eq.~(82) of Ref.~\cite{LHas}).
We determine it assuming the uncertainty on $\alpha_s$ to be 
$\delta\alpha_s=0.0012$ at the 68\% C.L. Results 
are plotted in Fig.~\ref{fig:PDFascorr} as a function of $x$, both at
$Q^2=2$~GeV$^2$  and $Q^2=10^4$~GeV$^2$. Clearly, because of
asymptotic freedom, correlations are weaker at high scale.

Following the procedure outlined in Ref.~\cite{Demartin:2010er}, it is possible
to combine sets with different values of $\alpha_s$ to compute the
combined PDF+$\alpha_s$ uncertainty on any given observable. 
This procedure has the advantage that both the central value and the uncertainty on the 
strong coupling are not fixed a priori but can be chosen by the PDF user. 
In order to simplify this procedure, we provide prepacked PDF sets with 
combined PDF+$\alpha_s$ uncertainty. Once a central value and
uncertainty for
$\alpha_s(M_Z)$ are assumed, this is done by constructing a set of $N_{\rm
  rep}$ 
replicas, extracted from the original sets with different $\alpha_s$,
in such a way that the prepacked sets contains a number of replicas
for each value of  $\alpha_s$  which corresponds to a gaussian
distribution with given mean and standard
deviation~\cite{Demartin:2010er} (of course, any other distribution
could be used). 
The statistical accuracy of the prediction 
obtained using the prepacked sets scales with the 
number of replicas $N_{\rm rep}$.

We have produced prepacked PDF+$\alpha_s$ uncertainty sets with 
$\alpha_s\lp M_z\rp=0.119$ and uncertainties $\delta_{\alpha_s}=0.0012$ and $\delta_{\alpha_s}=0.002$ 
as one-$\sigma$ errors. These values have been chosen to agree with the
PDF4LHC recommendation~\cite{Botje:2011sn} for the combination of
PDF+$\alpha_s$ uncertainties.  Sets with any other values are easily 
produced and  are available upon request. For completeness, we have produced
the same prepacked sets also for  NNPDF2.0.
We have checked that results for Higgs production 
in gluon fusion at LHC 7 TeV kinematics (which has rather large
$\alpha_s$ uncertainties and correlations)
become essentially independent of the
number of replicas in the prepacked set provided
$N_{\rm rep}\gsim100$. For smaller number of 
replicas there is a certain loss of accuracy, so a minimum of  $N_{\rm
  rep}=$100 is recommended.

An important caveat in the usage of prepacked PDF sets is that  some widely
used codes, such as MCFM, assume the value of $\alpha_s$ is the same
for all PDFs in a given set. Prepacked sets cannot be used with these codes.
Similar prepacked PDF sets could be prepared to include the 
uncertainty on other physical parameters, such as heavy quark masses. 
An important limitation however 
is imposed by the current LHAPDF standard which 
assumes that all physical parameters except the the strong coupling
take the same value for all PDFs in a given set.
For this reason, only PDF+$\alpha_s$ prepacked sets are provided for
the time being.

\subsection{The NuTeV anomaly}
\label{sec:nutev}

\begin{figure}[ht!]
\begin{center}
\epsfig{width=0.65\textwidth,figure=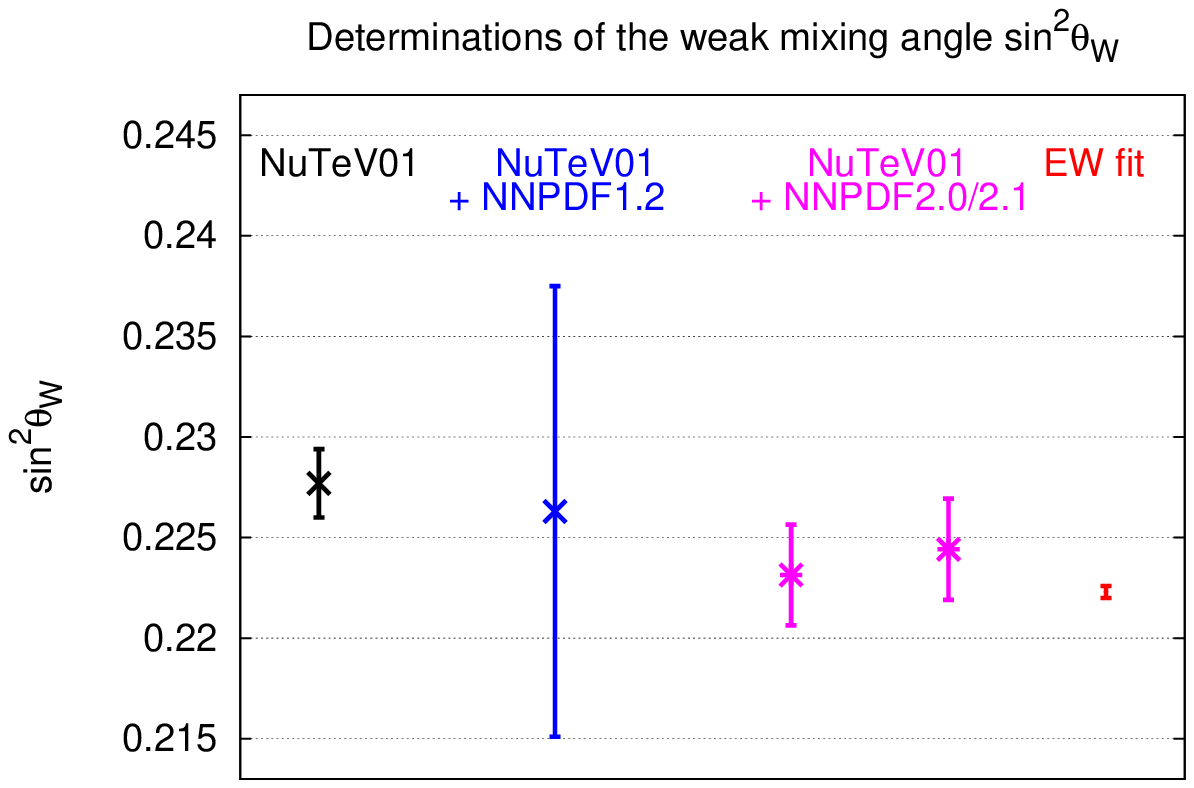}
\caption{\small Determination of the Weinberg angle from the
  uncorrected NuTeV data~\cite{Mason:2007zz}, with 
$\lc S^-\rc$ correction determined from   NNPDF1.2, NNPDF2.0
and NNPDF2.1. The uncertainty shown on NNPDF2.0 and NNPDF2.1 is the
one-$\sigma$ PDF uncertainty only.
\label{fig:nutev}} 
\end{center}
\end{figure}

In previous NNPDF releases~\cite{Ball:2009mk,Ball:2010de} we
studied the implications that the determination of
the strangeness asymmetry $s^-(x,Q^2)$  has on 
the  so--called NuTeV
anomaly~\cite{Davidson:2001ji}. These results are updated here. For 
the first
moment of the strangeness asymmetry with NNPDF2.1 we find
\be
R_S(Q^2)\equiv
2 \frac{\int_0^1 dx xs^-(x,Q^2)}
{\int_0^1 dx x\lp u^-(x,Q^2) +  d^-(x,Q^2) \rp}
=2\frac{\lc S^-\rc}{\lc U^-+D^-\rc}=\lp 1.37 \pm 0.77\rp\, 10^{-2}.
\ee  
In Fig.~\ref{fig:nutev} we show the
NuTeV determination of the Weinberg angle~\cite{Mason:2007zz},
uncorrected and then corrected for the strangeness asymmetry using the
values from previous~\cite{Ball:2009mk,Ball:2010de} and the current
NNPDF sets.   The three corrected values are in excellent 
agreement with the electroweak
fit and with each other, with the NNPDF2.0 and NNPDF2.1 values very
close to each other, thereby showing
that the impact of heavy quark mass effects on the determination of
the strangeness asymmetry is very small.

\subsection{Comparison with present and future HERA $F_2^c$ and $F_L$ data}
\label{sec:combinedpred}

\begin{figure}[t]
  \begin{center}
    \epsfig{width=0.99\textwidth,figure=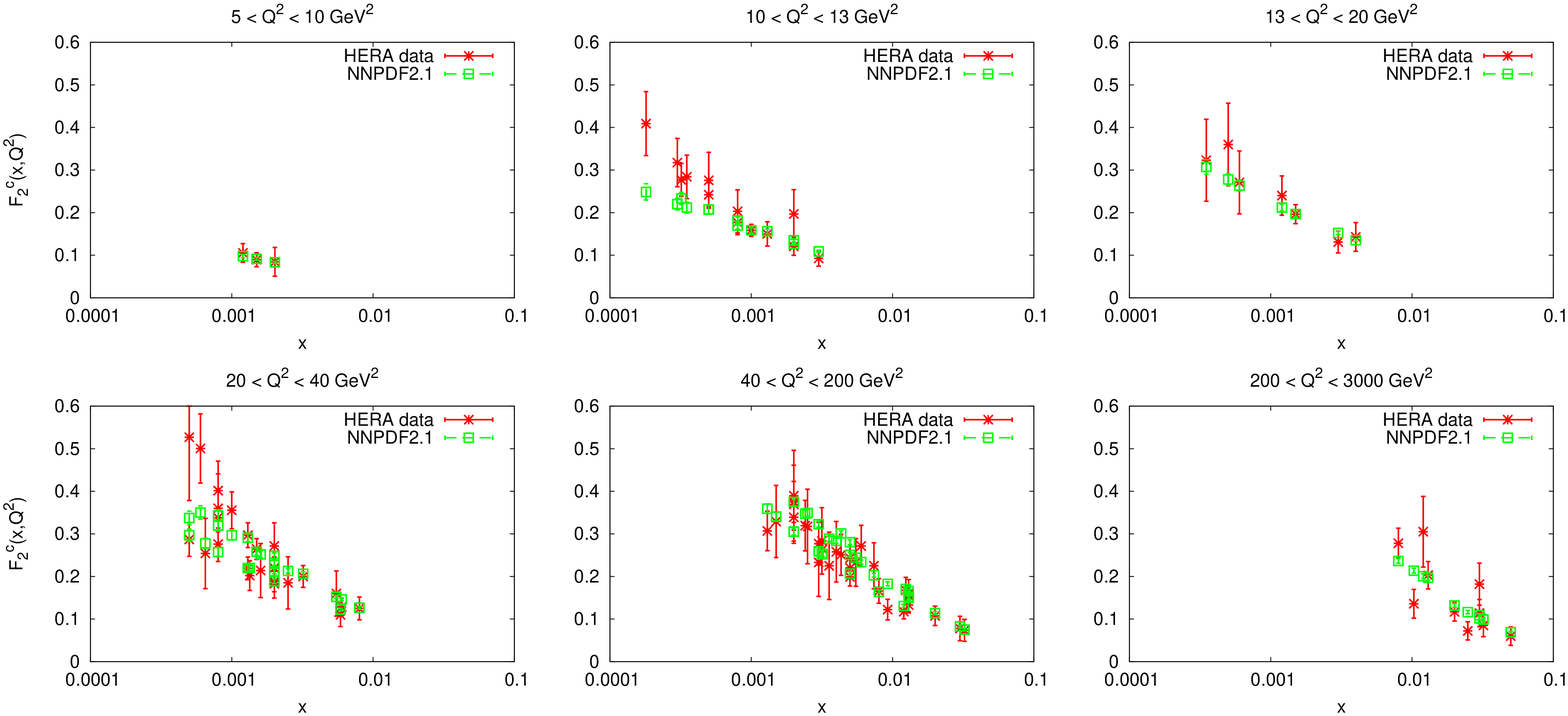}
    \caption{\small \label{fig:f2c} Comparison between the 
H1 and ZEUS $F_2^c$ data included in the present analysis and the
NNPDF2.1 best fit. The data uncertainty includes 
statistical and systematic errors added in quadrature, while the
theoretical uncertainty is the PDF uncertainty only.}
\end{center}
\end{figure}
\begin{figure}[t]
  \begin{center}
    \epsfig{width=0.49\textwidth,figure=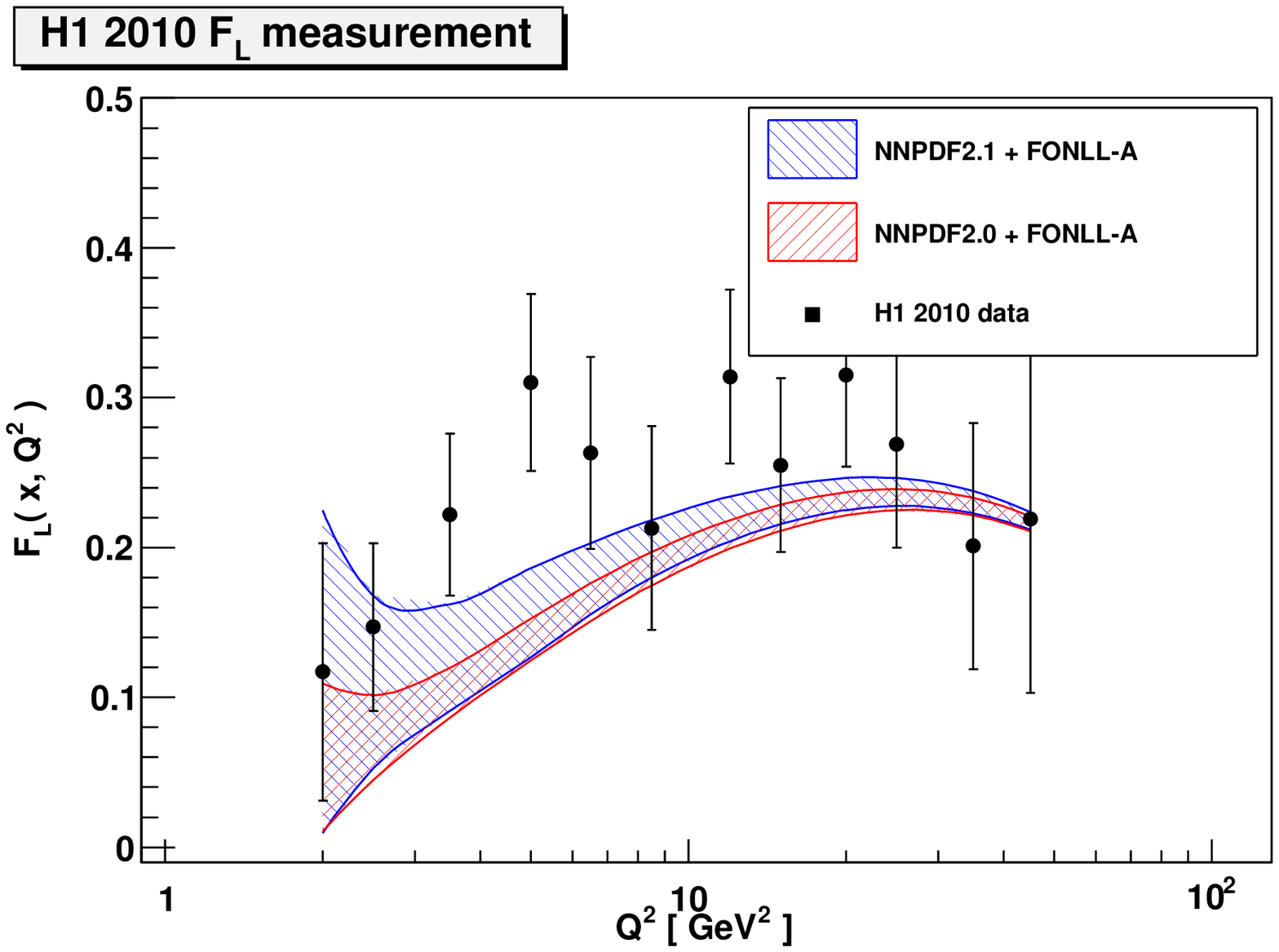}
    \epsfig{width=0.49\textwidth,figure=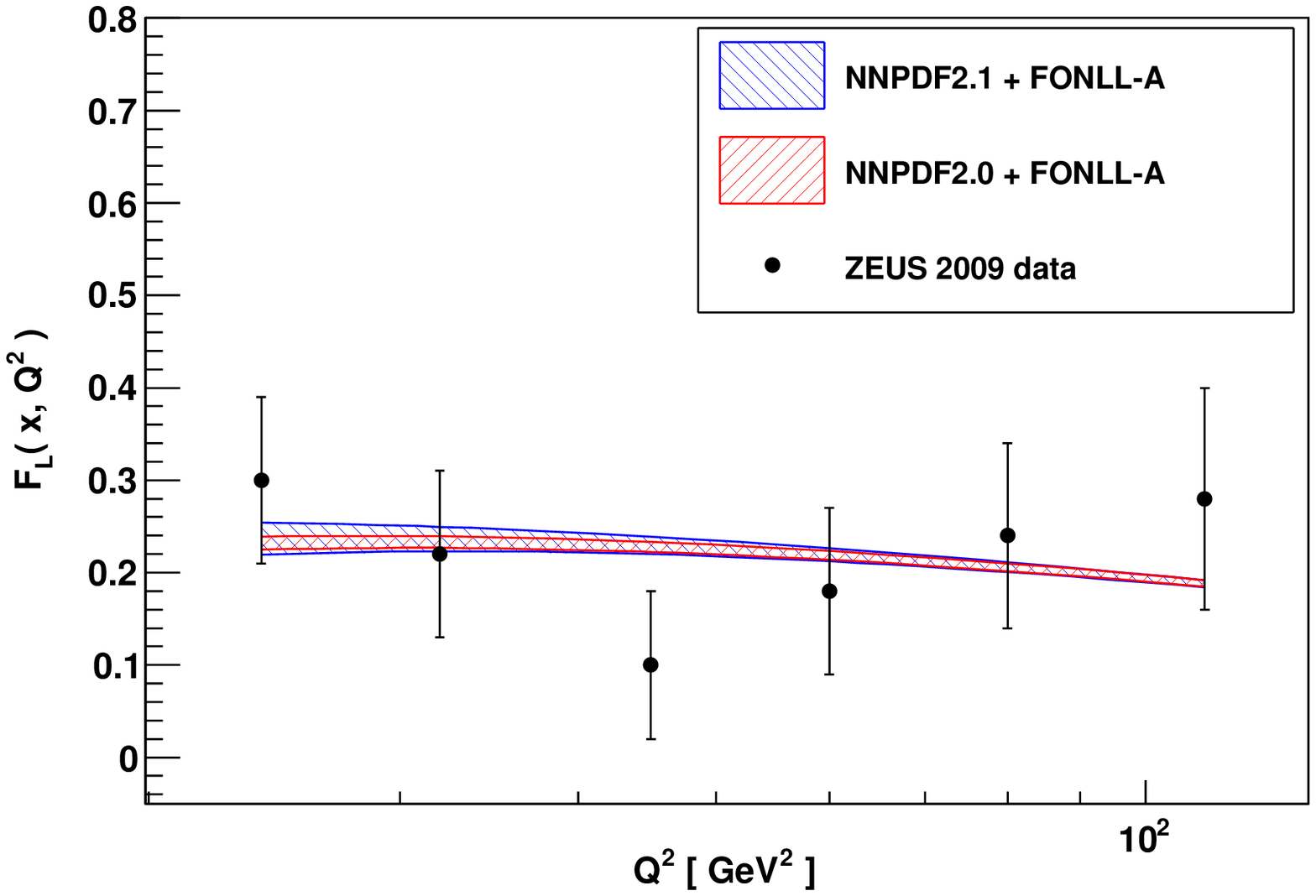}
    \caption{\small Comparison between  
H1~\cite{H1Collaboration:2010ry} (left) and  ZEUS 
~\cite{Chekanov:2009na} (right) $F_L$ data and 
NNPDF2.1 predictions. Predictions using NNPDF2.0 with the FONLL-A are
also shown.
      \label{fig:flhera1}} 
  \end{center}
\end{figure}

In conclusion, we look  at NNPDF2.1 predictions  
for $F_2^c$ and $F_L$, which 
are especially sensitive to the treatment of heavy quark mass effects.
For instance, 
heavy quark mass corrections to $F_L$ for $Q^2 \le 20 $ GeV$^2$
are larger than 30\%, see
Fig.~\ref{fig:f2c_flc_cont}.
In Fig.~\ref{fig:f2c} we compare to the best fit result
the HERA $F_2^c$ data which have been 
included in the present analysis:  in general the agreement is rather
good, though the lowest $Q^2$ and $x$ bins have been removed 
from the fitted dataset, because  $\mathcal{O}\lp \alpha_s^2\rp$ heavy
quark corrections, not included in the present analysis, are large 
there~\cite{Forte:2010ta}.
The NNPDF2.1 predictions for $F_L(x,Q^2)$ is compared to published 
ZEUS~\cite{Chekanov:2009na} and H1~\cite{H1Collaboration:2010ry}  data in Fig.~\ref{fig:flhera1}. Note that while
the  H1  data are included in the fit, they have rather large
uncertainties and thus  carry very little weight in the global fit. 
Predictions obtained
using NNPDF2.0 PDFs, but including heavy quark mass effects in the
computation of the structure function through FONLL-A 
are also shown. This comparison is particularly interesting, because
heavy quark mass effects are quite large especially at low $Q^2$: this
correction is included here with both sets, though NNPDF2.0 PDFs were
determined without it. The good agreement between results found using
the two sets shows that NNPDF2.0 PDFs are quite accurate despite the
lack of heavy quark mass corrections in the fit.

\begin{figure}[t]
  \begin{center}
    \epsfig{width=0.60\textwidth,figure=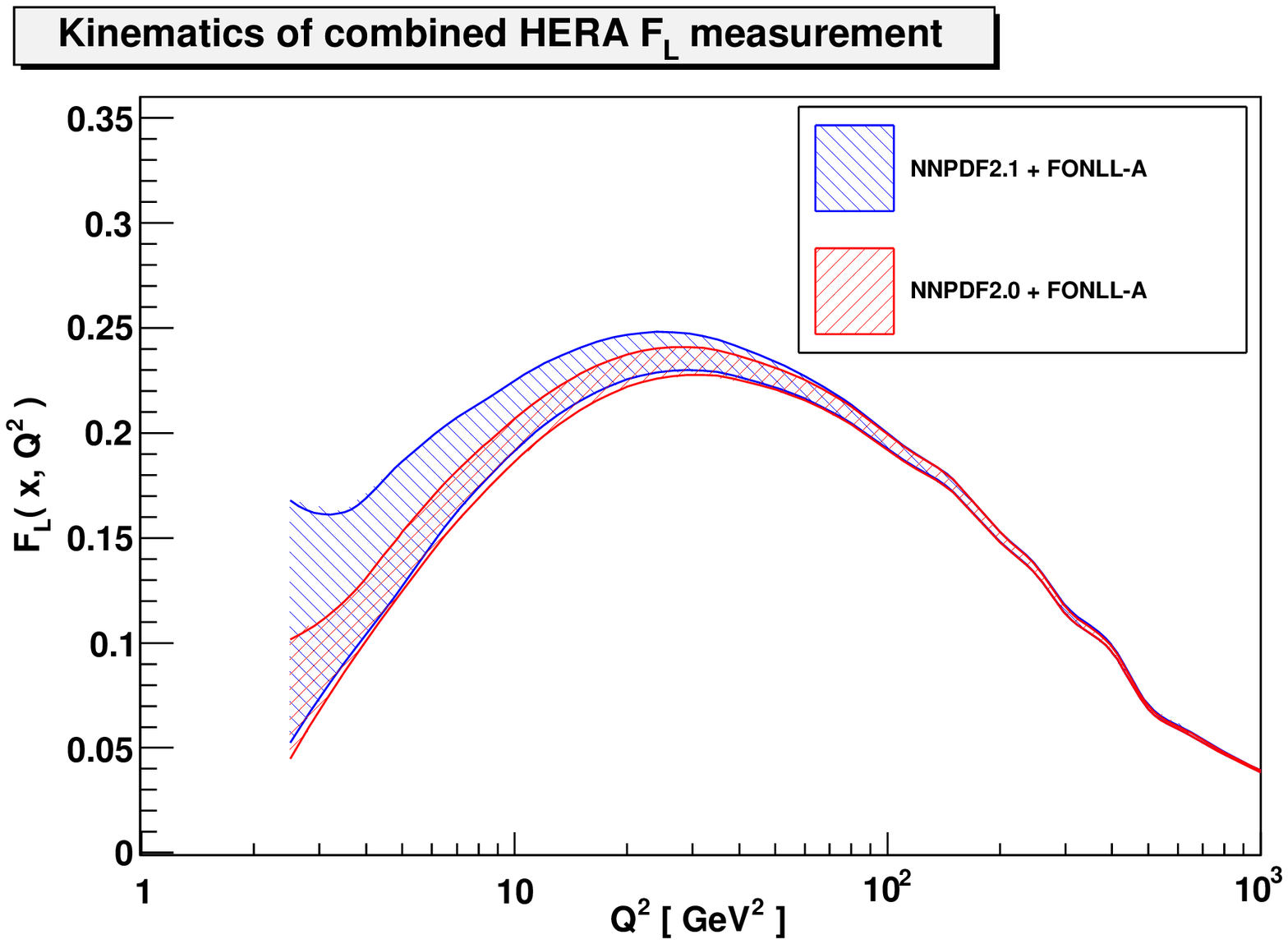}
    \caption{\small The NNPDF2.1 predictions for $F_L$ in the
      kinematics of upcoming combined HERA data.
      Predictions using NNPDF2.0 with the FONLL-A are also shown. 
      \label{fig:flhera2}} 
  \end{center}
\end{figure}

We now turn to the predictions in view of upcoming combined HERA data.
In Fig.~\ref{fig:flhera2} we provide the 
NNPDF2.1 predictions for $F_L$ in the kinematic region of the upcoming
combined HERA 
data~\footnote{We thank S. Glazov for providing us this information.}.
The increase in uncertainty at small-$x$ is driven by the larger uncertainty on the gluon at 
small-$x$, as seen in  Fig.~\ref{fig:singletPDFs}. 
We also show the results using
FONLL-A with both NNPDF2.0 and 2.1 input PDFs.
The NNPDF2.1 results have been compared with 
preliminary combined HERA $F_L$ dataset in 
Ref.~\cite{pdf4lhctalk}.

\begin{figure}[t]
  \begin{center}
    \epsfig{width=0.80\textwidth,figure=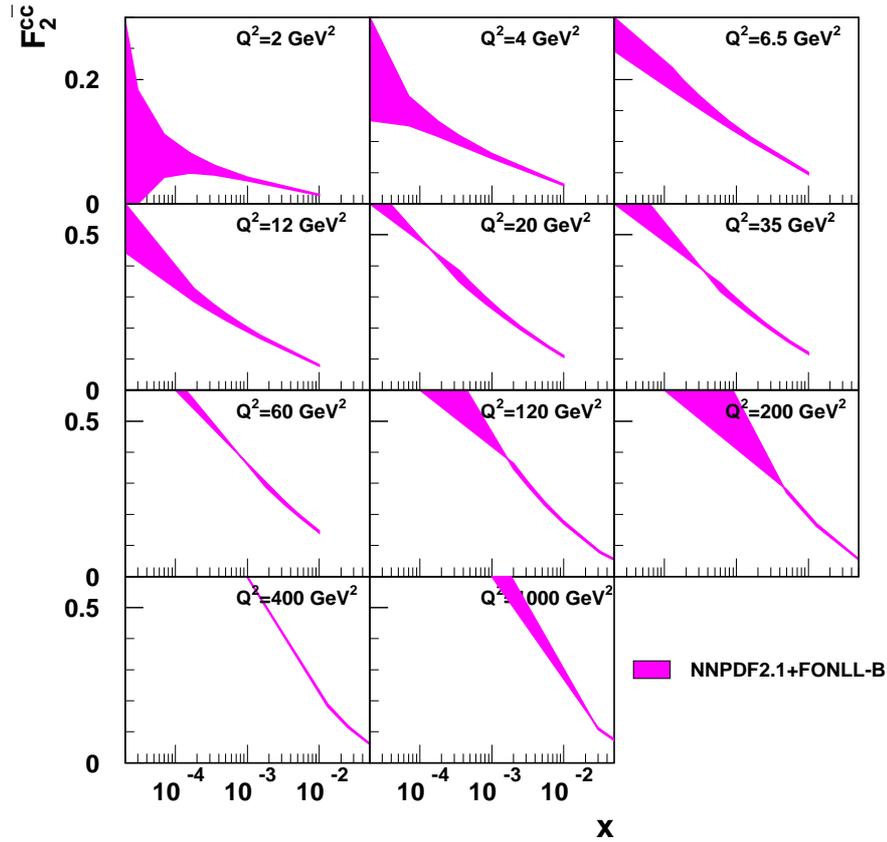}
    \caption{\small NNPDF2.1 predictions for $F_2^c$
with heavy quark mass effects
      included up to
  ${\mathcal O}(\alpha^2)$ through the  FONLL-B scheme    in the kinematic
      region of upcoming combined  HERA data.
      \label{fig:f2c_vs_nnpdf}} 
  \end{center}
\end{figure}

In conclusion, we give predictions for $F_2^c$ in 
Fig.~\ref{fig:f2c_vs_nnpdf} in the range of upcoming 
combined HERA data.\footnote{We thank K. Lipka for providing these plots.} 
These predictions are obtained with heavy quark mass effects included
up to ${\mathcal O}(\alpha^2)$ through the  FONLL-B scheme, but using
input PDFs determined with ${\mathcal
  O}(\alpha)$ heavy quark mass corrections. 
The fact that FONLL-A and B coincide for moderate and large values of $Q^2$, where the $F_2^c$ data 
included in NNPDF2.1 lie, justifies the use of FONLL-B to extrapolate to the low $Q^2$ region with 
the same input PDF set. PDF uncertainties at small-$x$ and $Q^2$ are rather large, suggesting that
the combined HERA $F_2^c$ data will impose severe constraints on the
small-$x$ gluon; comparison with 
preliminary data~\cite{pdf4lhctalk} suggests that the NNPDF2.1 will be
in very good agreement with the HERA data down to the smallest values of $Q^2$.

\section{Heavy quark mass dependence}
\label{sec:hqmasses}

We will now discuss the dependence of PDFs on the
values of heavy quark masses $m_c$ and $m_b$, and present some
preliminary investigations on uncertainties related to heavy quark
masses, along the lines of the recent detailed study by the MSTW
group~\cite{Martin:2010db}; as  an outcome of this analysis we will present
NNPDF2.1 sets with varying $m_c$ and $m_b$ masses.
We will first discuss how different features of the NNPDF2.1 PDFs 
depend on the values of
$m_c$ and $m_b$. We then briefly 
investigate the dependence on $m_c$ of some LHC observables and
summarize how uncertainties on heavy quark masses can be treated in
the Monte Carlo approach. Finally, as an example of a possible application we
evaluate the combined PDF+$m_b$ uncertainty for MSSM Higgs boson production. 

\begin{table}[ht]
  \begin{center}
    \small
    \begin{tabular}{|c|c|c|}
      \hline
      & $m_c$ [GeV] & $m_b$ [GeV] \\
      \hline
      \hline
      NNPDF2.1 &  1.414 &  4.75\\
      \hline
      NNPDF2.0~\cite{Ball:2010de} & $\sqrt{2}$ & $4.3$ \\
      CT10~\cite{Lai:2010vv} & 1.30 & 4.75 \\
      MSTW2008~\cite{Martin:2010db} & 1.40 & $4.75$ \\
      ABKM09~\cite{Alekhin:2009ni} & 1.50 & 4.50 \\
      HERAPDF1.0~\cite{H1:2009wt} & 1.40 & 4.75 \\
      \hline
    \end{tabular}
  \end{center}
  \caption{\small The default values  of the heavy quark masses used
    in NNPDF2.1 and in several
 recent PDF sets.
    \label{tab:hqmasscomp} }
\end{table}

\subsection{Dependence of PDFs on heavy quark masses}
\label{sec:pdfhqmass}

The default value of heavy quark masses used so far are summarized and
compared to those of other PDF sets in Table~\ref{tab:hqmasscomp}.
The dependence of PDFs on the heavy quark masses is studied  by repeating the 
NNPDF2.1 fit with different mass values. 
In particular, we have repeated the reference fit for charm quark masses of 1.5, 1.6 and 1.7 GeV 
as well as for bottom masses of 4.25, 4.5, 5.0 and 5.25 GeV. It is
important to observe that at the order at which we are working,
the perturbative definition of the heavy quark mass is immaterial: indeed
different definitions (such as, for example, the pole and
$\overline{\rm MS}$ mass definitions) differ by terms of
${\mathcal O}(\alpha_s)$. However, we are including heavy quark mass
corrections up to ${\mathcal O}(\alpha_s)$ only, so the difference is
subleading (it becomes relevant once one includes ${\mathcal
  O}(\alpha_s^2)$ heavy quark corrections, for example using the
FONLL-B scheme). Therefore, the value of the quark mass in our PDF
determination  (as well as in other PDF determinations based on an
NLO ACOT treatment of heavy quarks, such as CT/CTEQ) can be
equivalently interpreted as, say, a pole mass or an $\overline{\rm
  MS}$ mass. The $\overline{\rm
  MS}$ mass is better known, and  it has been recently
shown~\cite{Alekhin:2010sv} to lead to perturbatively more stable
results for deep-inelastic structure functions.

\begin{figure}[t]
  \begin{center}
    \epsfig{width=0.49\textwidth,figure=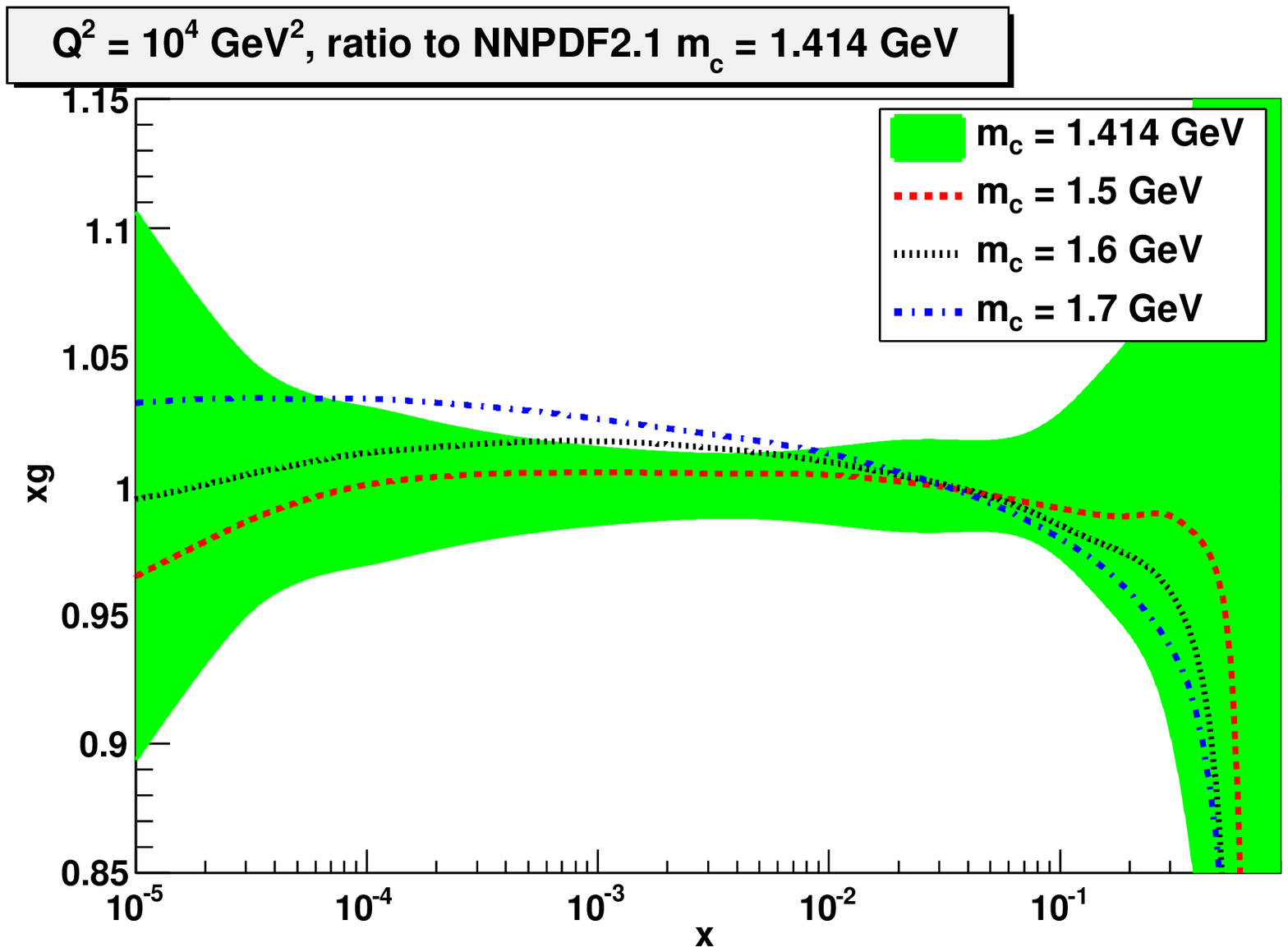}
    \epsfig{width=0.49\textwidth,figure=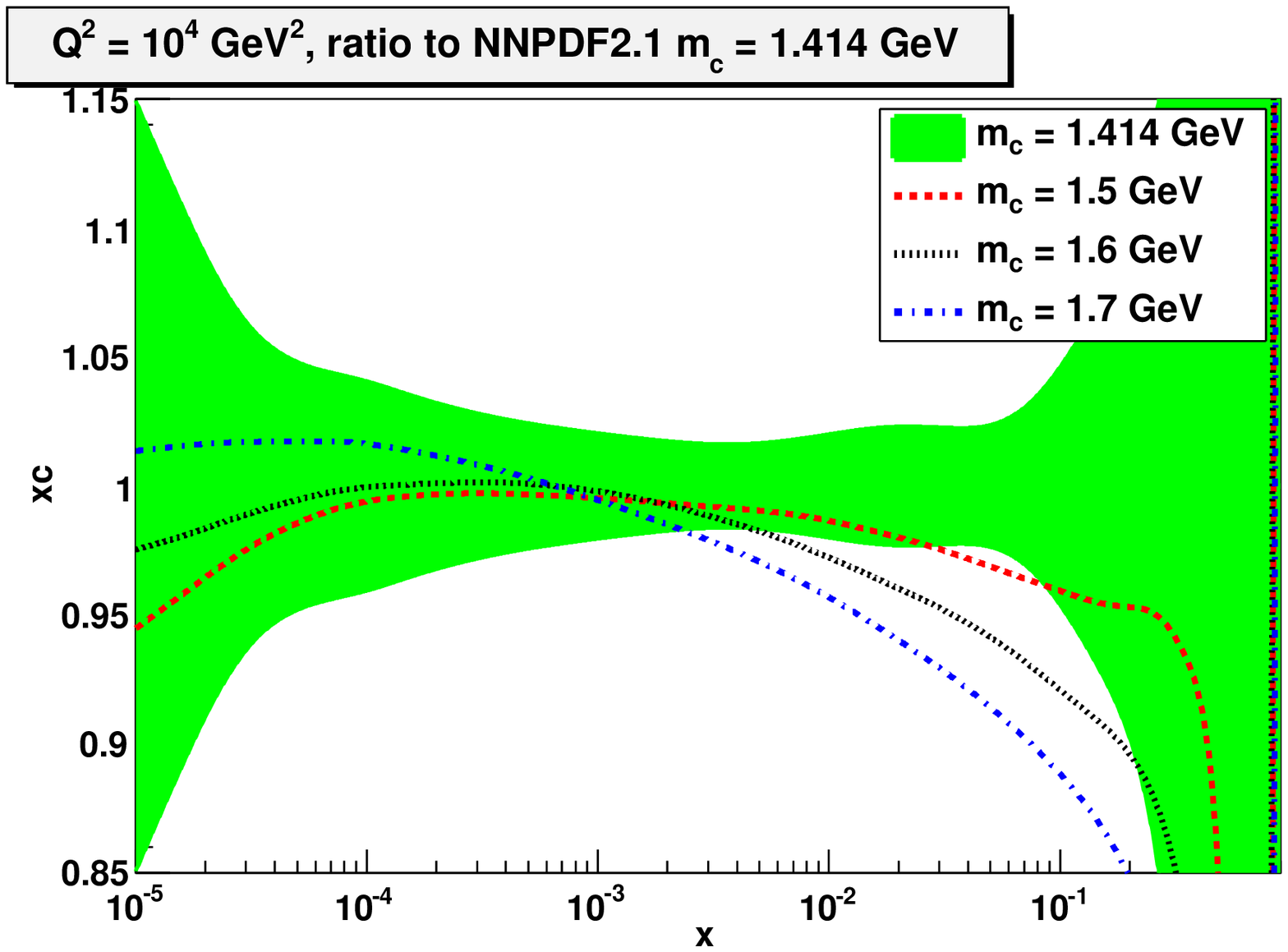}
    \epsfig{width=0.49\textwidth,figure=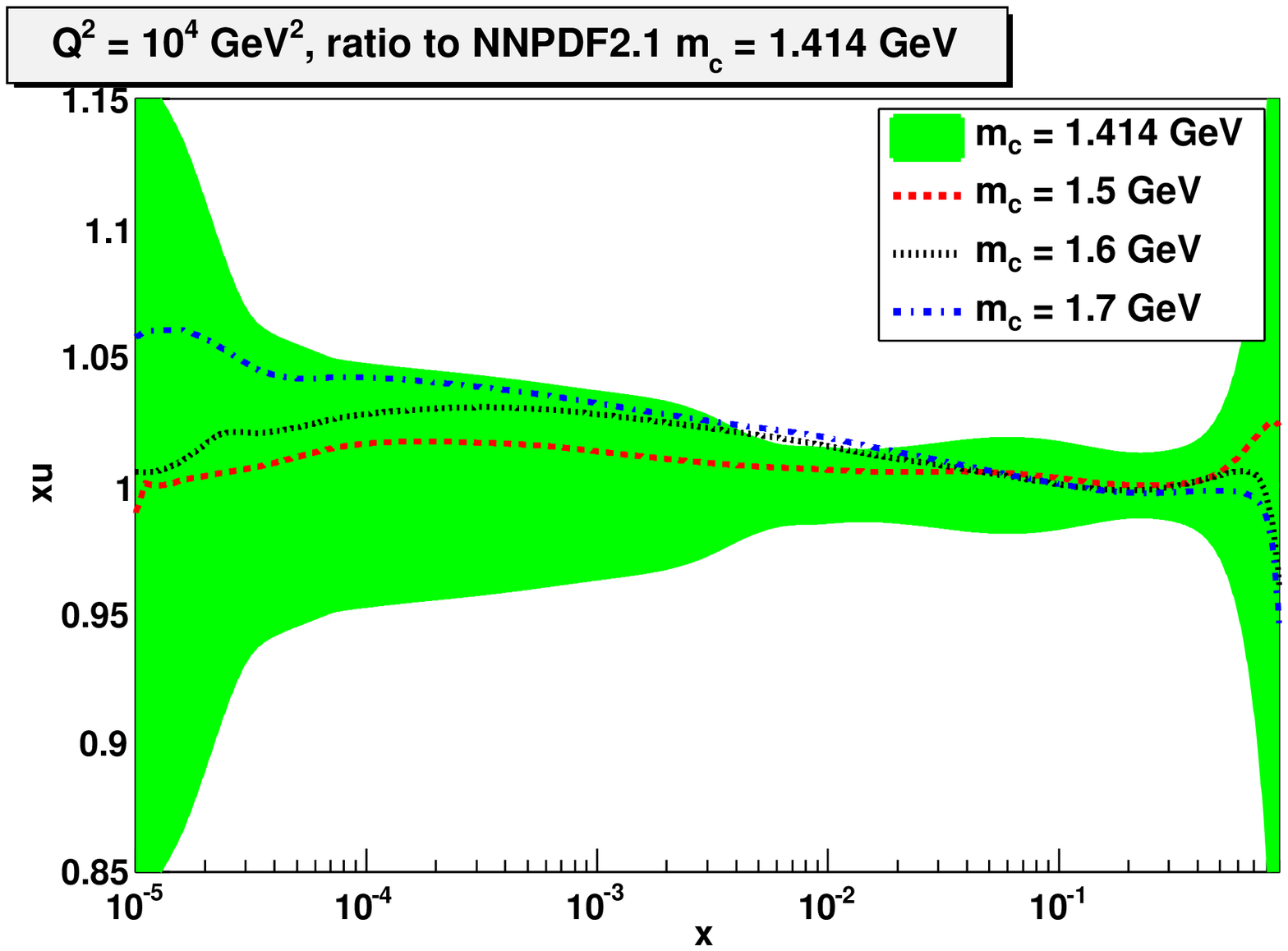}
    \epsfig{width=0.49\textwidth,figure=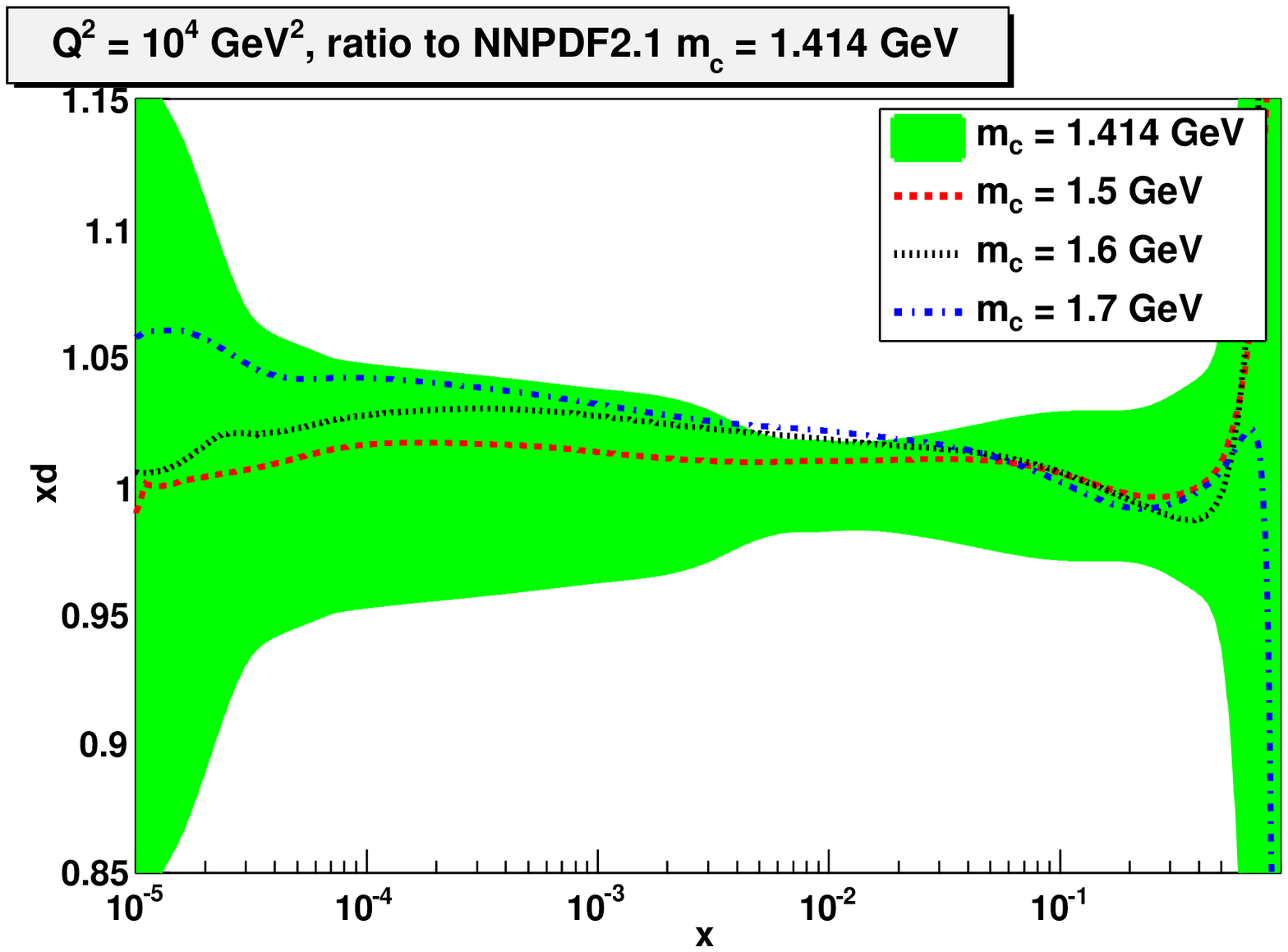}
    \caption{\small Ratio of
NNPDF2.1 PDFs obtained for different values of the charm quark
      mass to the reference NNPDF2.1 set at $Q^2=10^4$ GeV$^2$. Top left:  
 gluon; top right: charm; bottom left: up; bottom right: down.
      \label{fig:PDFmc}} 
  \end{center}
\end{figure}

\begin{figure}[t]
  \begin{center}
    \epsfig{width=0.49\textwidth,figure=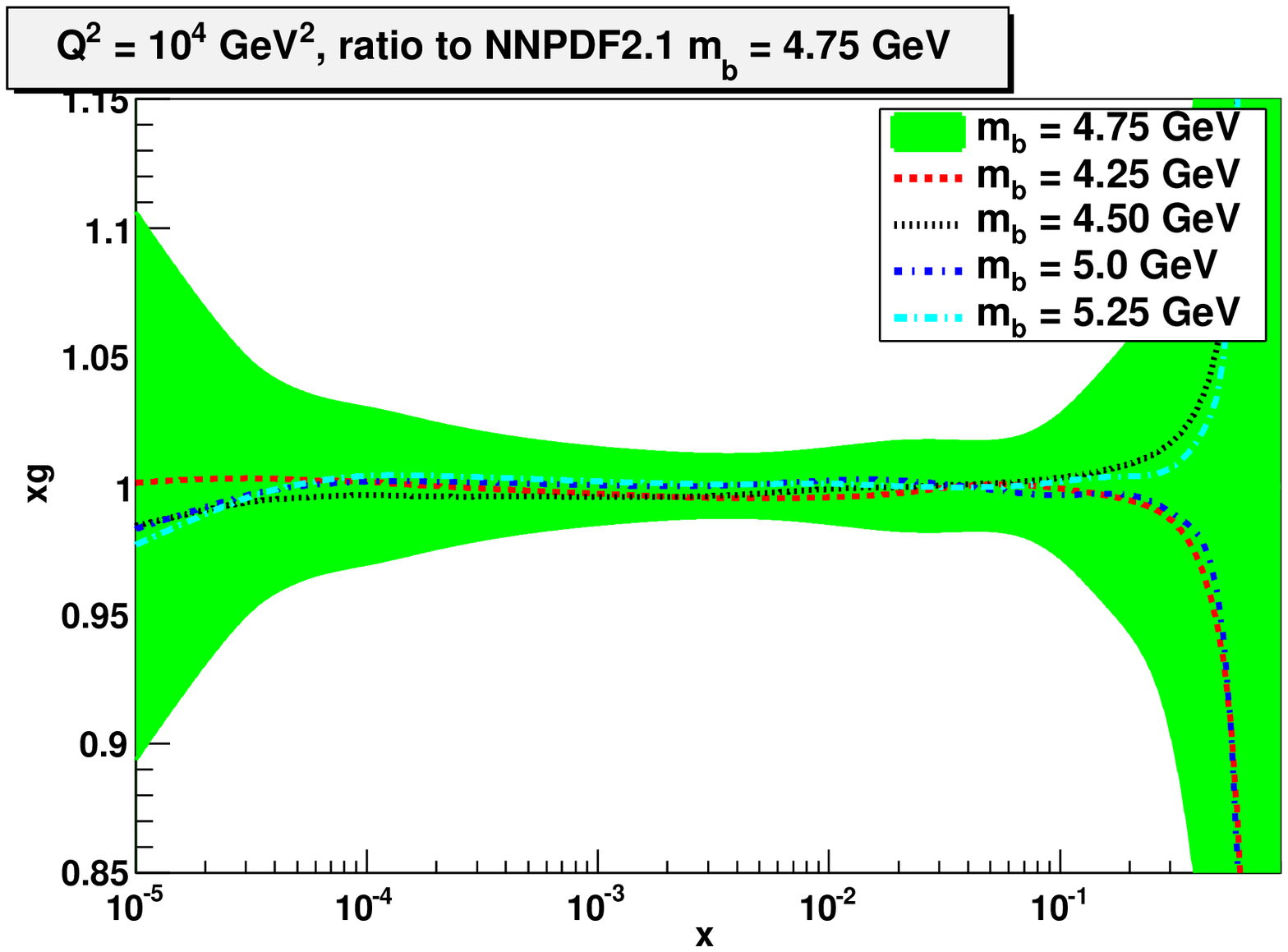}
    \epsfig{width=0.49\textwidth,figure=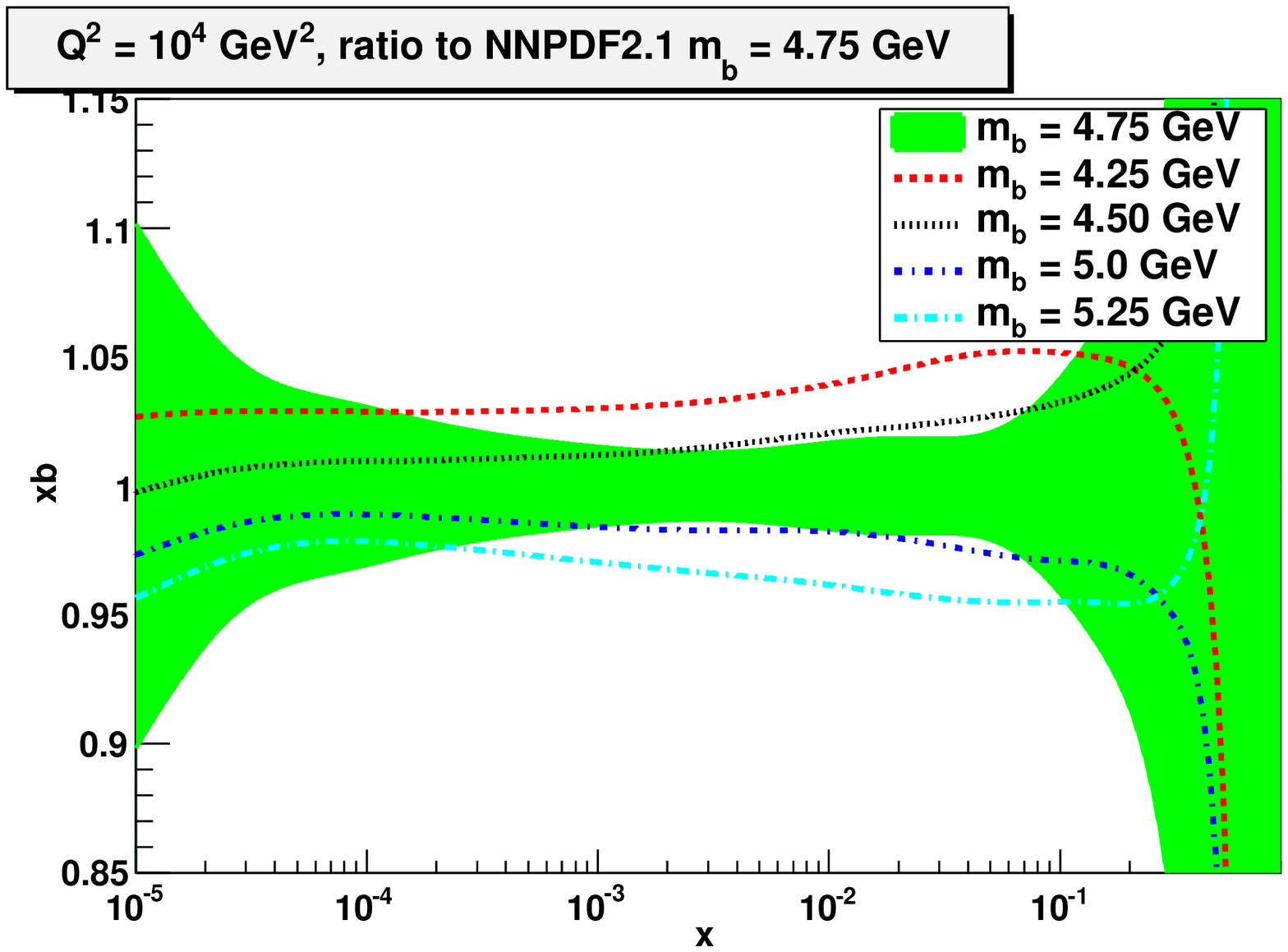}
    \epsfig{width=0.49\textwidth,figure=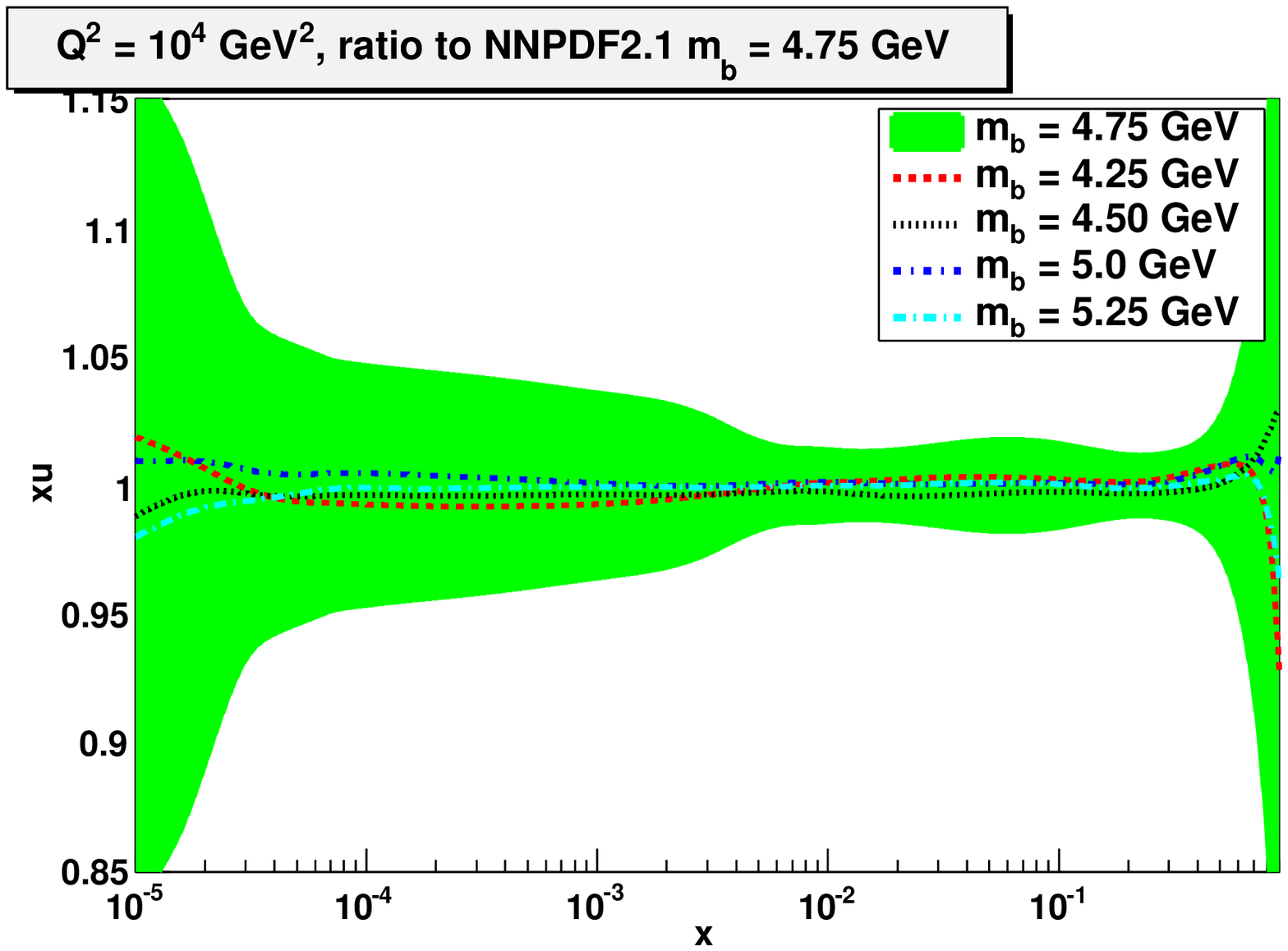}
    \epsfig{width=0.49\textwidth,figure=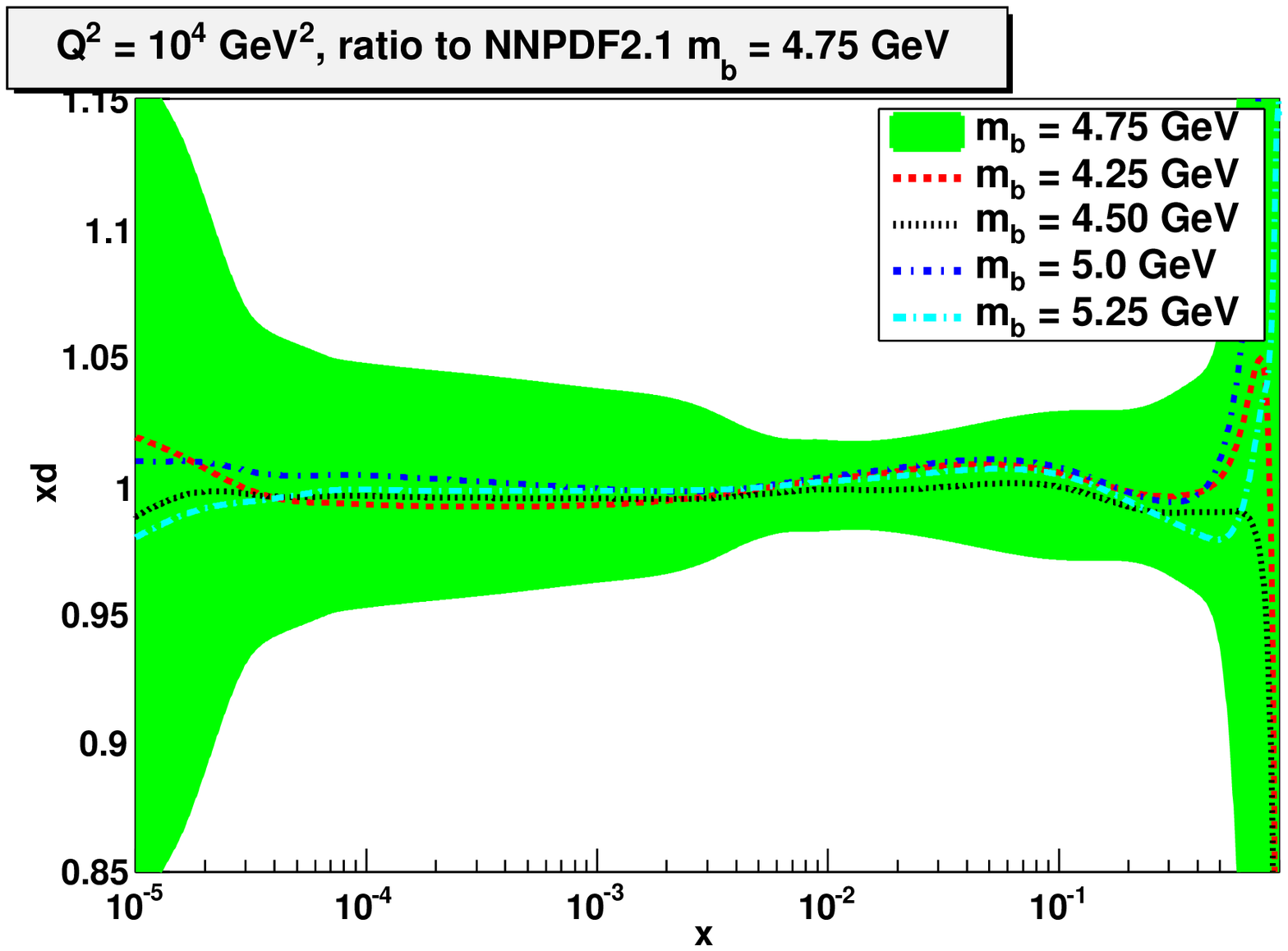}
    \caption{\small Ratio of
NNPDF2.1 PDFs obtained for different values of the bottom quark
      mass to the reference NNPDF2.1 set at $Q^2=10^4$ GeV$^2$. Top left:  
 gluon; top right: bottom; bottom left: up; bottom right: down.
      \label{fig:PDFmb}} 
  \end{center}
\end{figure}
Results are shown in Figs.~\ref{fig:PDFmc},\ref{fig:PDFmb} 
where the ratio of PDFs for different values of $m_c$ and $m_b$ to the
reference NNPDF2.1 fit are plotted as a function  of $x$ for
$Q^2=10^4$~GeV$^2$. 
The dependence of the heavy quark PDFs on the value of the mass is
easily  understood: heavy quark PDFs are generated radiatively, and
assumed to vanish at a scale equal to their mass. Therefore, a lower
mass value corresponds to a longer evolution length and thus to a
larger heavy quark PDF, and conversely. Thus, if one allowed~\cite{Pumplin:2007wg}  for an
``intrinsic''~\cite{Brodsky:1980pb} 
heavy component (i.e. for a nonvanishing initial condition) 
this uncertainty would be absorbed in the initial
intrinsic heavy PDF. Because of the momentum sum rule, if 
the charm PDF becomes larger, other PDFs are
accordingly smaller (and conversely). For bottom in principle the same
mechanism is at work, but in practice the effect on all other PDFs is
negligible. 

\subsection{Mass uncertainties and LHC observables}

The dependence of light quark distributions and the gluon on the charm
mass displayed in Fig.~\ref{fig:PDFmc} is strong enough to affect the
LHC standard candles studied in Sect.~\ref{sec:lhcimplications} at the
percent level or more, as was discovered relatively
recently~\cite{Nadolsky:2008zw}. On the other hand, the dependence on
the bottom mass of all PDFs but $b$ itself is below the percent level,
so only observables which depend on bottom are affected significantly:
an example will be considered in Sect.~\ref{sec:mssmb} below. 

\begin{figure}[t]
  \begin{center}
    \epsfig{width=0.32\textwidth,figure=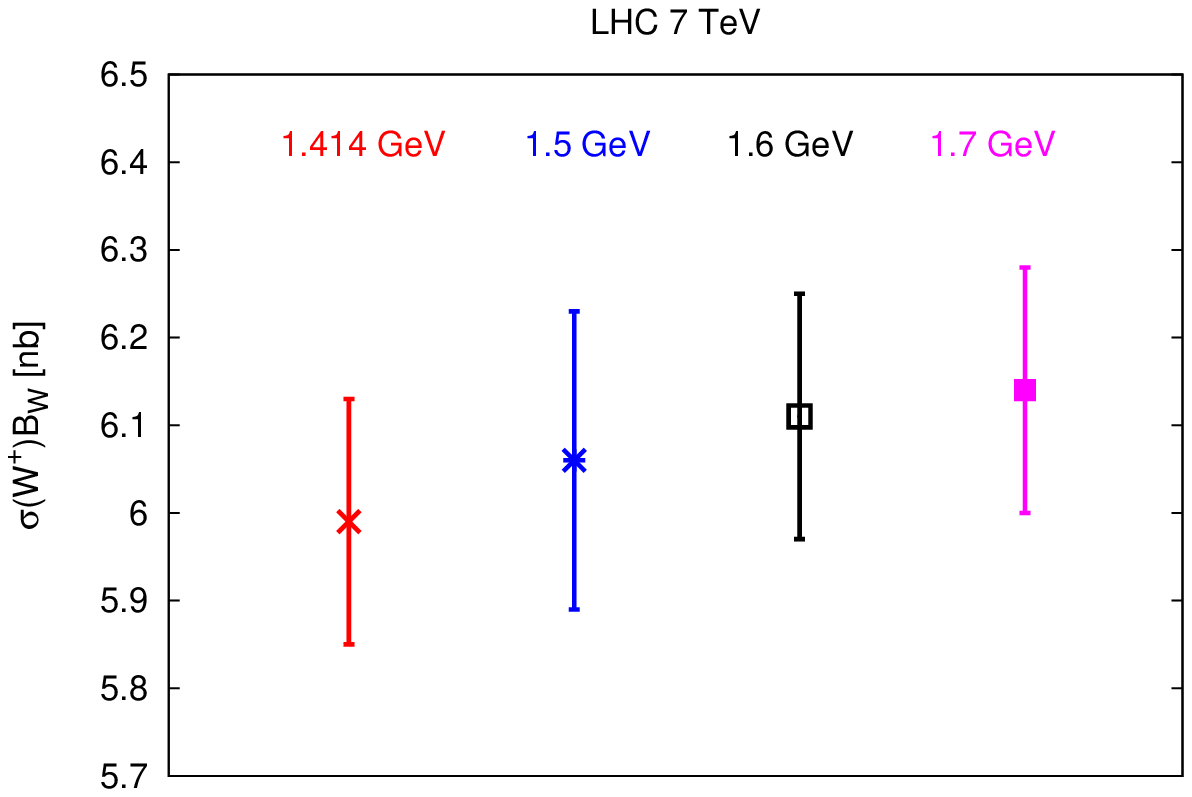}
    \epsfig{width=0.32\textwidth,figure=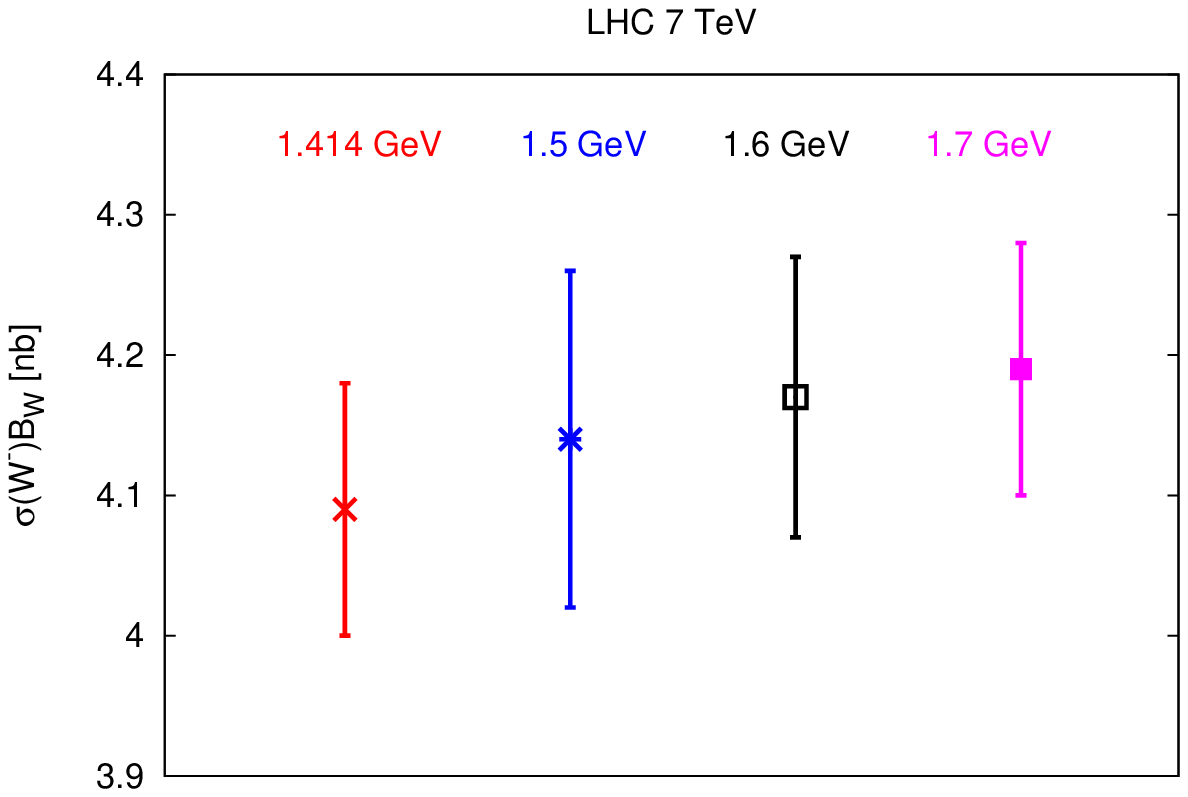}
    \epsfig{width=0.32\textwidth,figure=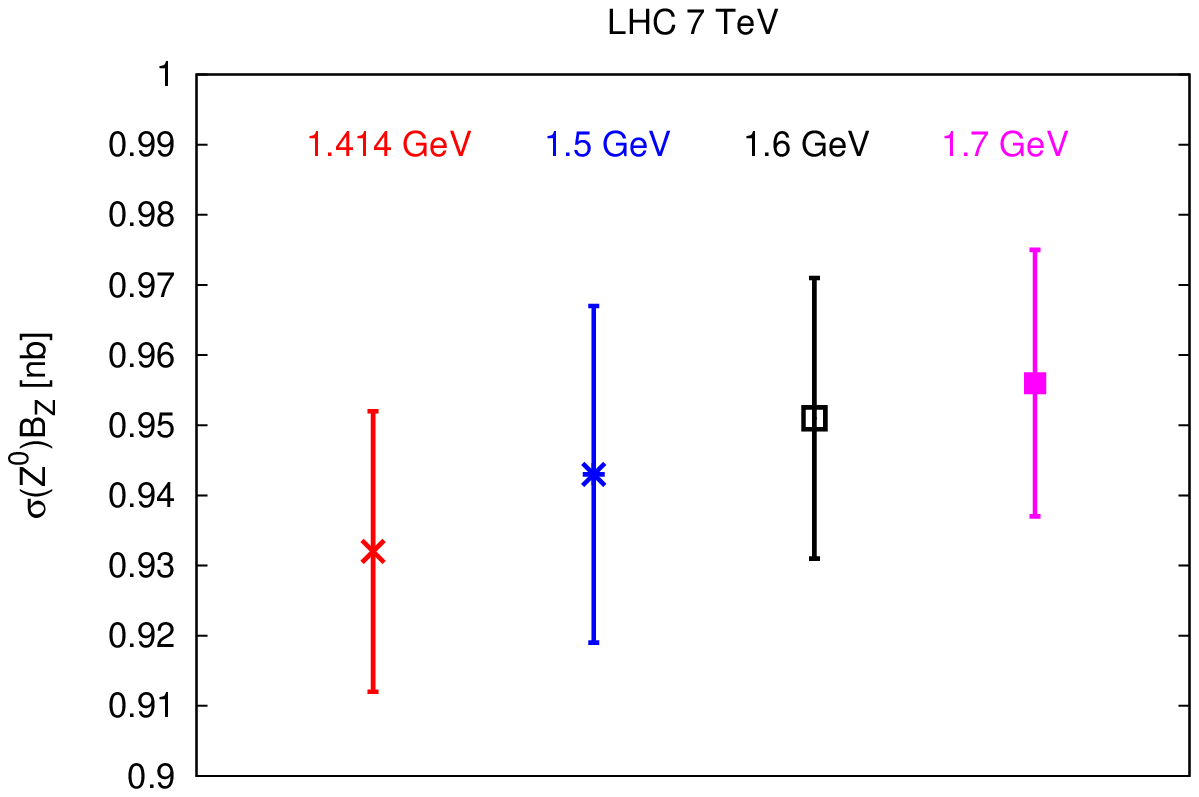}\\
    \epsfig{width=0.32\textwidth,figure=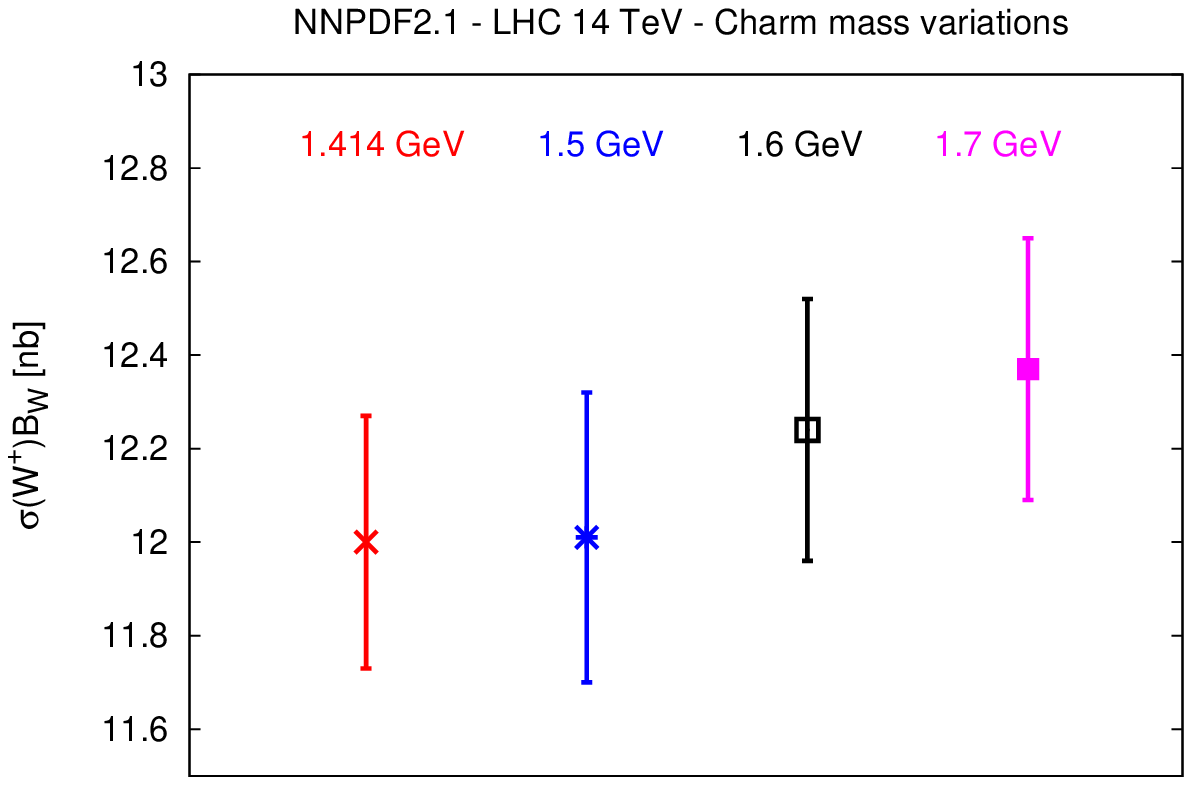}
    \epsfig{width=0.32\textwidth,figure=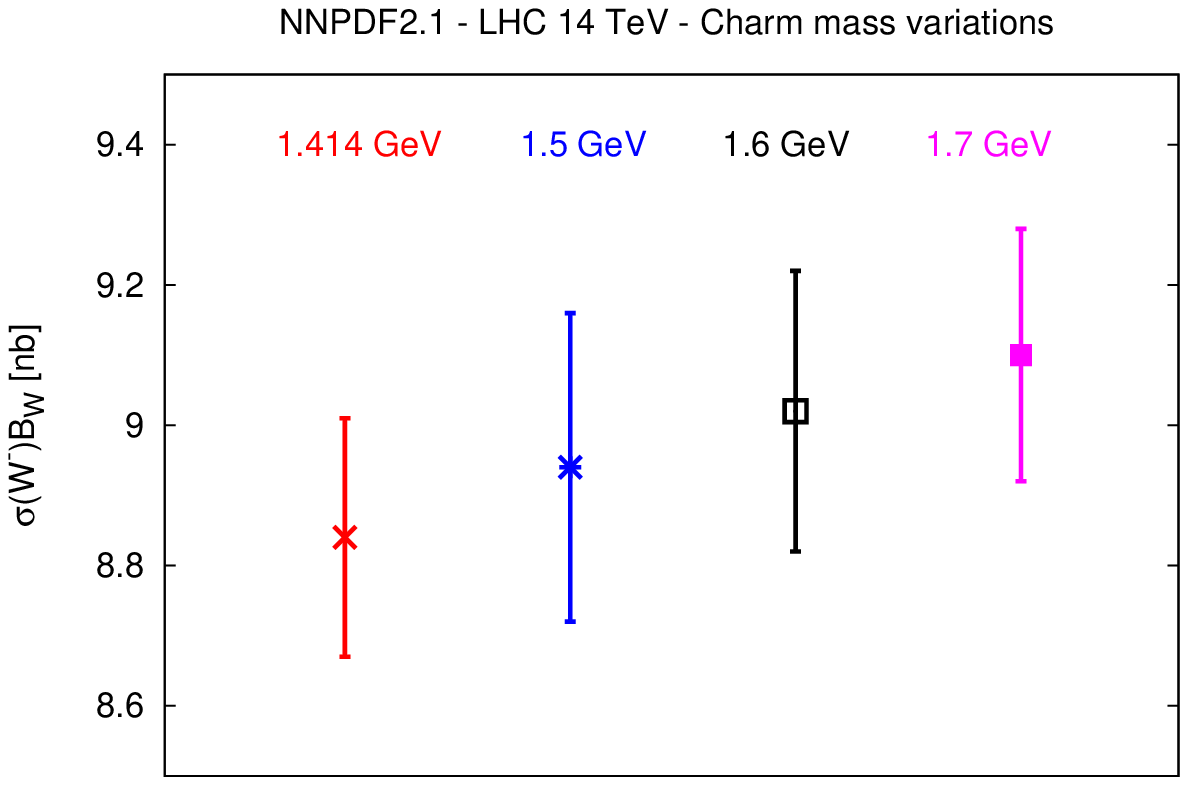}
    \epsfig{width=0.32\textwidth,figure=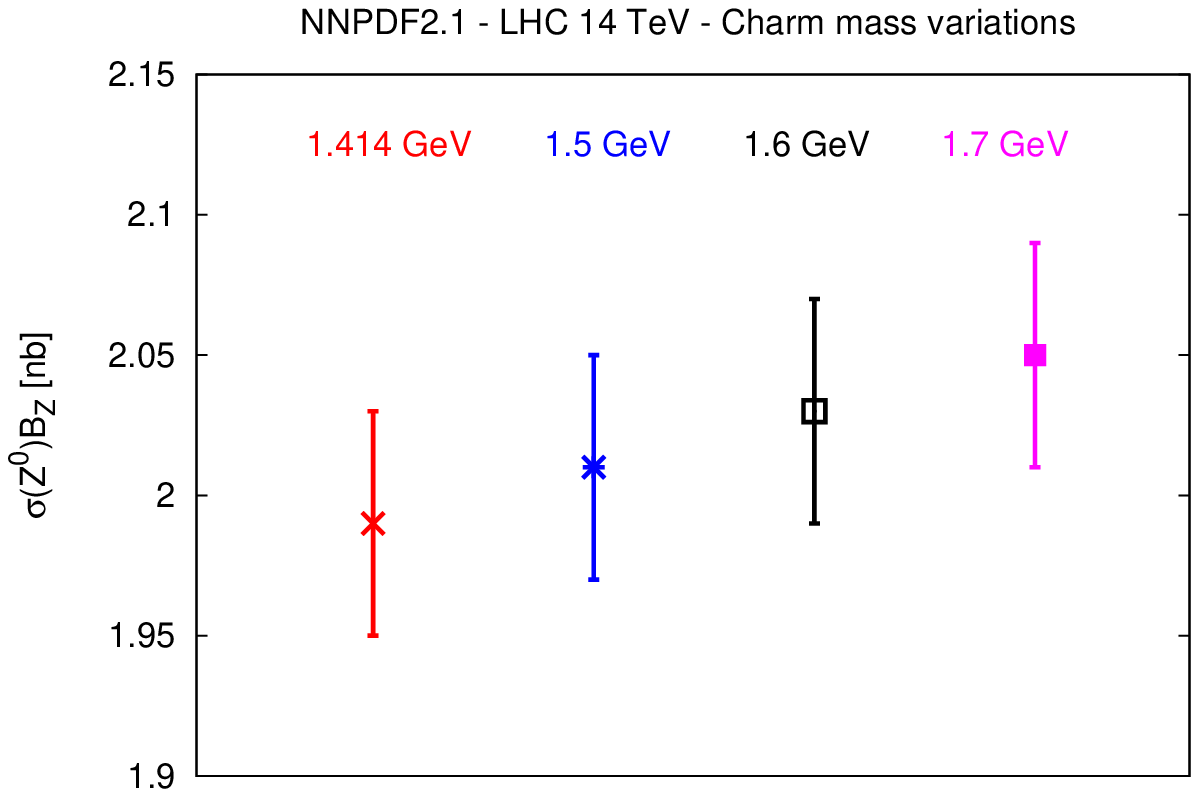}
    \caption{\small Graphical representation of the results of Table~\ref{tab_LHCobs_mc}.
      \label{fig:comp7tev-mc} \label{fig:comp14tev-mc}} 
  \end{center}
\end{figure}

Values of LHC standard candles computed as in
Sect.~\ref{sec:lhcimplications} but using the PDF sets with different
values of the charm mass of Fig.~\ref{fig:PDFmc} are collected in 
Table~\ref{tab_LHCobs_mc}, and shown in
Fig.~\ref{fig:comp7tev-mc}. For completeness, we also give in
Table~\ref{tab_LHCobs_mb} standard candles at 7~TeV for several values
of the $b$ mass. The variation of all standard candles is at the
percent level for charm mass variations of order of 10\%. It is
interesting to observe that the variation seen
when modifying subleading charm mass
terms is (recall Table~\ref{tab:LHCgmtreat}) of the same order of
magnitude and in fact somewhat larger. This suggests that even though
PDF uncertainties on standard candles are still dominant at present, theoretical
uncertainties related to the treatment of charm will become relevant
and possibly dominant
as soon as PDF uncertainties are reduced by a factor of two or three.

\begin{table}[t]
  \centering
  \begin{tabular}{|c|c|c|c|c|c|}
    \hline
    LHC 7 TeV & $W^+B_{l\nu}$ [nb] & $W^-B_{l\nu}$ [nb]
    & $Z^0B_{l\bar{l}}$ [nb] &   $t\bar{t}$ [pb]& $gg\to H$ [pb] \\
    \hline
    \hline
    $ m_c=1.414$ GeV & $5.99\pm 0.14$ & $4.09\pm 0.09$ & $0.932\pm 0.020$ &
    $170\pm 5$ & $11.64\pm 0.17$ \\
    \hline
    $ m_c=1.5$ GeV & $6.06\pm 0.17 $ & $4.14\pm 0.12 $ & $0.943\pm 0.024 $ & $169\pm 6$ & $11.65\pm 0.25$ \\
    $ m_c=1.6$ GeV & $6.11 \pm 0.14 $ & $4.17\pm 0.10$ &
    $0.951 \pm 0.020 $ & $167\pm 6$ & $11.70\pm 0.21$ \\
    $ m_c=1.7$ GeV & $6.14 \pm 0.14 $  & $4.19 \pm 0.09$ & $0.956\pm 0.019$ & $166\pm 5$ & $11.71 \pm 0.22$\\
    \hline
    $\rho\lc \sigma,m_c\rc$ & 0.44 & 0.41 & 0.48 & -0.31 & 0.16 \\
    \hline
  \end{tabular}
  
  \vspace{0.5cm}
  
  \begin{tabular}{|c|c|c|c|c|c|}
    \hline
    LHC 14 TeV & $W^+B_{l\nu}$ [nb] & $W^-B_{l\nu}$ [nb]
    & $Z^0B_{l\bar{l}}$ [nb] &   $t\bar{t}$ [pb ]& $gg\to H$ [pb] \\
    \hline
    \hline
    $ m_c=1.414$ GeV & $12.00\pm 0.27$  & $8.84\pm 0.17$ & $1.99\pm 0.036$ & $946 \pm 19$ & $37.50\pm 0.40$   \\
    \hline
    $ m_c=1.5$ GeV  &  $12.01 \pm 0.31 $ & $8.94\pm 0.22 $ & $2.01\pm 0.04$ & 
    $942 \pm 24$& $37.62 \pm 0.62$ \\
    $ m_c=1.6$ GeV & $12.24\pm 0.28$ & $9.02\pm 0.20$ & $2.03\pm 0.04$ &
    $939 \pm 22$ &  $37.90\pm 0.55$ \\
    $ m_c=1.7$ GeV & $12.37 \pm 0.28 $ & $9.10\pm 0.18$ & $2.05\pm 0.04 $ &
 $935\pm 19$ & $38.15\pm 0.58$  \\
    \hline
    $\rho\lc \sigma,m_c\rc$ & 0.48 & 0.50 & 0.56 & -0.19  & 0.41 \\
    \hline
  \end{tabular}
  
  \caption{\small LHC standard candles  at $\sqrt{s}=7$ TeV 
    (upper table) and 14 TeV (lower table) obtained using NNPDF2.1 fits with different values of 
    the charm mass $m_c$; the values in the top line of each table are the same
    given
in Sect.~\ref{sec:lhcimplications}. The bottom line of each table
gives the correlation coefficient between the observable and the mass.
    \label{tab_LHCobs_mc}}
\end{table}


\begin{table}
  \centering
  \begin{tabular}{|c|c|c|c|c|c|}
    \hline
    & $W^+B_{l\nu}$ [nb] & $W^-B_{l\nu}$ [nb]
    & $Z^0B_{l\bar{l}}$ [nb] &   $t\bar{t}$ [pb ]& $gg\to H$ [pb] \\
    \hline
    \hline
    $ m_b=4.25$ GeV & $5.97\pm 0.12$ & $4.07\pm 0.08$ & $0.930\pm 0.016$ & 
    $170\pm 6$& $11.58\pm 0.26$ \\
    $ m_b=4.5$ GeV & $5.95\pm 0.21$ & $4.07\pm 0.11$ &  $0.928\pm 0.025$&
    $171\pm 7$ & $11.64 \pm 0.18$ \\
    $ m_b=4.75$ GeV  & $5.99\pm 0.14$ & $4.09\pm 0.09$ & $0.932\pm 0.020$ &
    $170\pm 5$ & $11.64\pm 0.17$ \\
    $ m_b=5.0$ GeV & $5.99\pm 0.12 $ & $4.11 \pm 0.07$ & $0.932 \pm 0.016$ &
    $170\pm 5$ & $11.64 \pm 0.17$ \\
    $ m_b=5.25$ GeV & $5.98\pm 0.11$ & $4.10\pm 0.07$ & $0.930\pm 0.015$ & 
    $171\pm 6$&
    $11.66\pm 0.18$ \\
    \hline
  \end{tabular}
  \caption{\small LHC standard candles  at $\sqrt{s}=7$ TeV 
    obtained using NNPDF2.1 fits with different values of 
    the bottom mass $m_b$; the values in the top line of each table are the same
    given
in Sect.~\ref{sec:lhcimplications}.
    \label{tab_LHCobs_mb}}
\end{table}

\subsection{Combined PDF+$m_h$ uncertainties and correlations}
\label{sec:combined_pdf_mh}

Uncertainties which combine  PDF uncertainty and heavy quark mass
dependence
are easily
determined in a Monte Carlo approach, provided that PDF sets for
several values of the quark masses are available. Here we provide
several such sets, and more with a finer mass spacing and wider range
will be made available in the future (such as already available for
the MSTW08 sets~\cite{Martin:2010db}).
Given sets of PDF replicas labelled by heavy quark mass values, such that
${\rm PDF}^{(k_{ij},i,j)}$ is the $k_{ij}$-th replica of the PDF set
with heavy quark mass values $m_c^{(i)}$ and 
$m_b^{(j)}$,
the mean value  of any observable $\mathcal{F}$ is
\be
\label{eq:pdfmhav}
\la \mathcal{F}\ra_{\rep} =\frac{1}{N_{\rep}}\sum_{i=1}^{N_{m_c}}
\sum_{j=1}^{N_{m_b}}
\sum_{k_{ij}=1}^{N_{\rm rep}^{(i,j)}} 
\mathcal{F}\lp  {\rm PDF}^{(k_{ij},i,j)},m_c^{(i)},m_b^{(j)}\rp \ ,
\ee
where 
\be
N_{\rm rep} = \sum_{i=1}^{N_{m_c}}\sum_{j=1}^{N_{m_b}}N^{(i,j)}_{\rm rep} \ ,
\ee
is the total number of replicas, and
$N^{(i,j)}_{\rm rep}$  are
distributed according to a two dimensional gaussian distribution 
with mean  $(m_c^{(0)},m_b^{(0)})$ and width  $(\delta m_c,\delta
m_b)$ (assuming the values of charm and bottom masses are uncorrelated):
\be
N^{(i,j)}_{\rm rep}\propto \exp\lp 
-\frac{\lp m_c^{(i)}- m_c^{(0)}\rp^2}{
  2 \delta_{m_c}^2} -\frac{\lp m_b^{(j)}- m_b^{(0)}\rp^2}{
  2 \delta_{m_b}^2}\rp \,.
\label{eq:bigauss}
\ee
The combined PDF+$m_h$ uncertainty is then 
the standard deviation of the observable over 
the replica sample:
\be
\delta_{\rm PDF+m_h}\mathcal{F}=\sqrt{ \la \mathcal{F}^2\ra -
\la \mathcal{F}\ra^2 } \ ,
\ee
where averages over replicas are to be understood as in
Eq.~(\ref{eq:pdfmhav}).
Of course, a different probability distribution (possibly including a
correlation between heavy quark mass values) could be assumed instead
of Eq.~(\ref{eq:bigauss}).

\begin{figure}[t]
  \begin{center}
    \epsfig{width=0.49\textwidth,figure=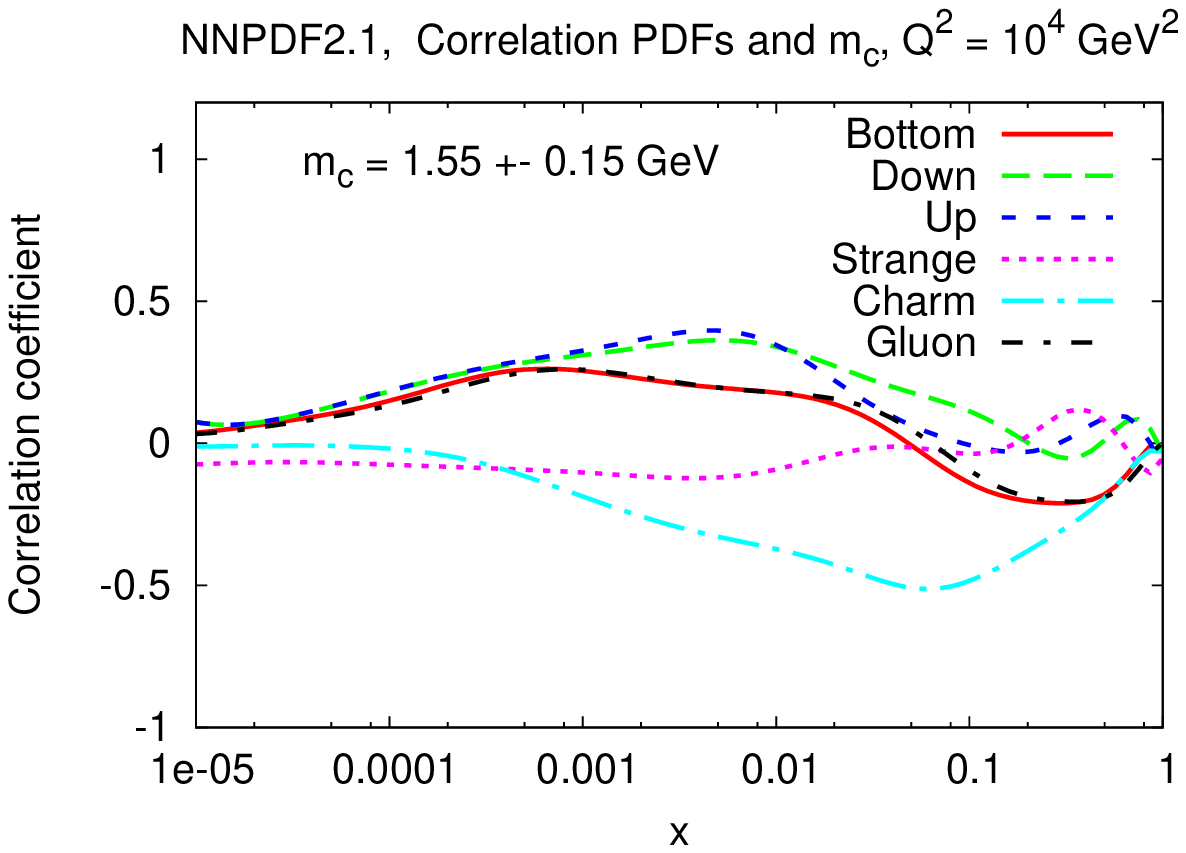}
    \epsfig{width=0.49\textwidth,figure=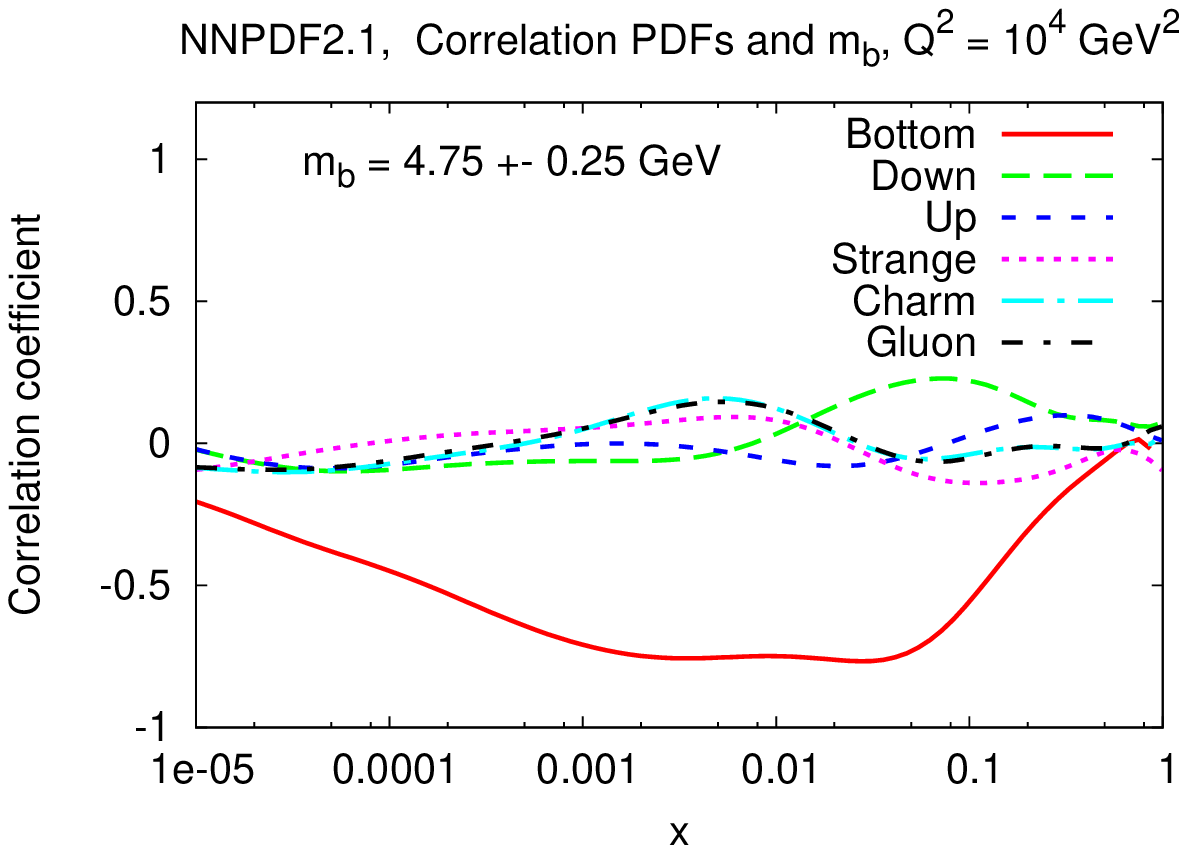}
    \caption{\small Correlation between PDFs and the heavy quark masses
      at a typical LHC scale $Q^2=10^4$ GeV$^2$: charm mass variations (left plot)
      and bottom mass variations (right plot). These correlations
quantify the qualitative behaviour observed in 
Figs.~\ref{fig:PDFmc}-~\ref{fig:PDFmb}. \label{fig:hqcorrplot}} 
  \end{center}
\end{figure}

We can easily compute the correlation between PDFs and 
heavy flavour masses $m_h$:
\be
\label{eq:hqcorr}
\rho \lc  m_h,{\rm PDF}\lp x,Q^2\rp\rc=
\frac{\la m_h {\rm PDF}\lp x,Q^2\rp \ra_{\rep}-
  \la m_h \ra_{\rep}\la  {\rm PDF}\lp x,Q^2\rp \ra_{\rep}
}{\sigma_{m_h} \sigma_{{\rm PDF}\lp x,Q^2\rp}} \ .
\ee
where averages over replicas are to be understood in the sense of Eq.~(\ref{eq:pdfmhav}). 
The correlation Eq.~(\ref{eq:hqcorr}), computed assuming
$m_c=1.55 \pm 0.15 ~{\rm GeV}$ and 
$m_b=4.75 \pm 0.25 ~{\rm GeV}$, is displayed in 
Fig.~\ref{fig:hqcorrplot}, as a function of $x$ for
$Q^2=10^4$ GeV$^2$. As discussed in Sect.~\ref{sec:pdfhqmass}, as the
mass is increased the corresponding heavy quark PDF is reduced,
i.e. the PDF is strongly anticorrelated with its mass. As a
consequence of the heavy quark suppression other PDFs are enhanced and
this appears as a positive correlation, though the effect for bottom is
negligibly small.
Correlations between $m_h$ and physical observables can be computed
analogously. They are given in Tab.~\ref{tab_LHCobs_mc} and can be
used for a quick estimate of the corresponding uncertainty.

\subsection{The combined PDF+$m_b$ uncertainties in MSSM $b\bar{b}\to H$ production}
\label{sec:mssmb}
As an
illustration of the procedure to combine PDF and $m_b$ uncertainties discussed in 
Section~\ref{sec:combined_pdf_mh}, we have evaluated the combined uncertainty for Higgs production in 
association with bottom quarks~\cite{Harlander:2003ai}. Higgs
production via bottom fusion is enhanced in the MSSM in large
$\tan \beta$ scenarios as compared to the SM, so this channel is
important for supersymmetry searches.

\begin{figure}[h!]
  \begin{center}
    \epsfig{width=0.49\textwidth,figure=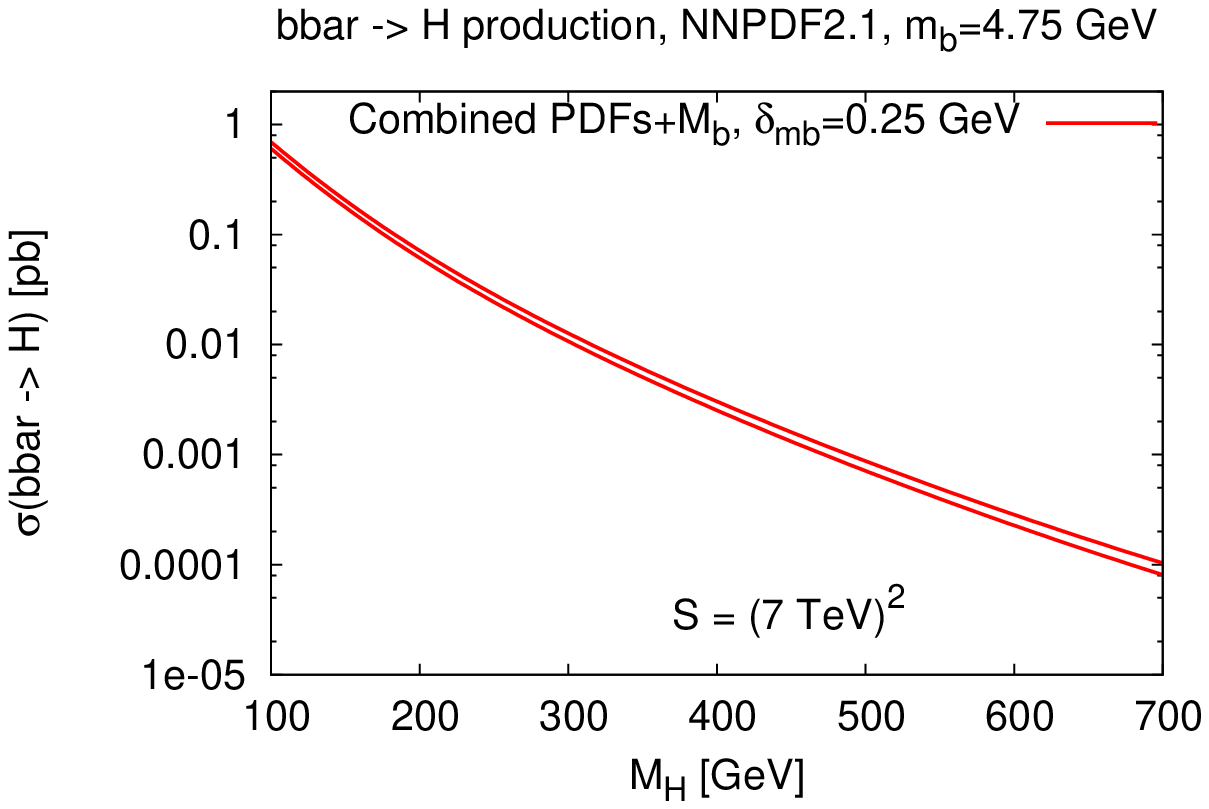}
    \epsfig{width=0.49\textwidth,figure=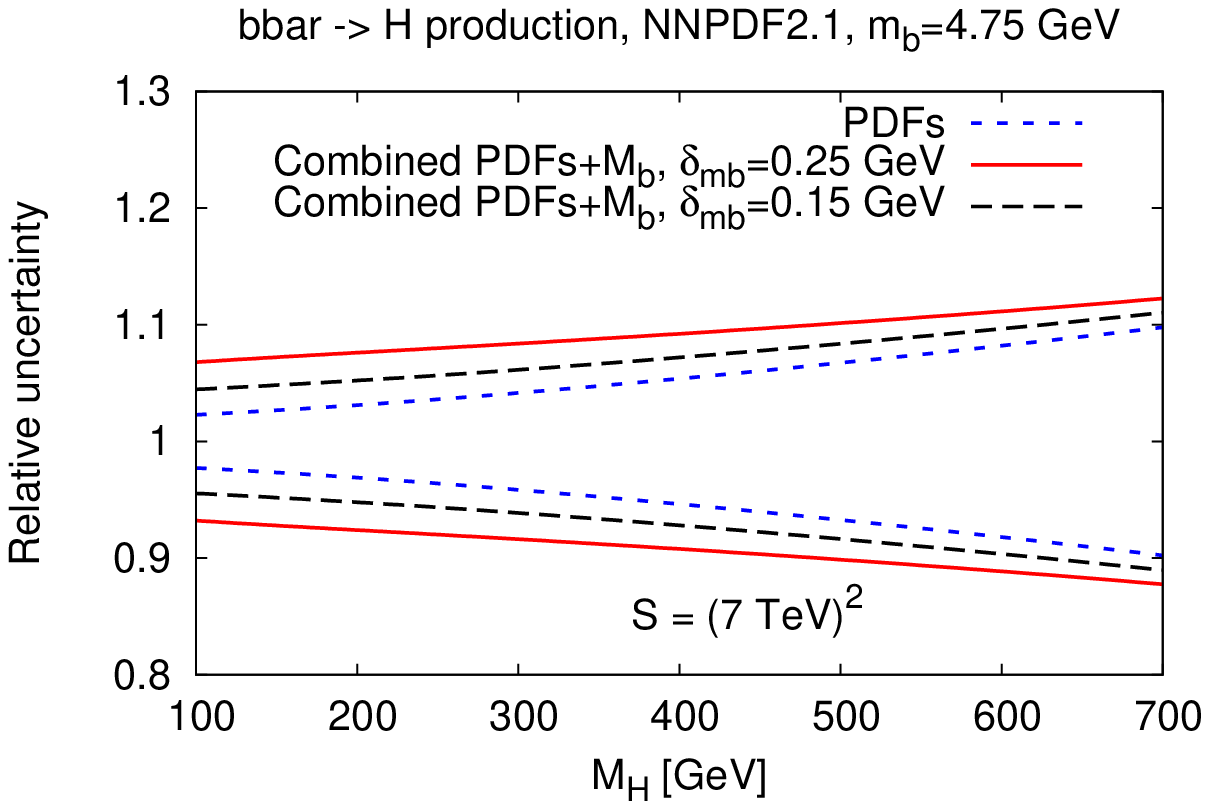}
    \caption{\small Combined PDF+$m_b$ uncertainties on the total cross section for $b\bar{b}\to H$ 
      Higgs production at the LHC 7 TeV with $m_b=4.75$~GeV.  
The absolute cross section
      (left) and relative uncertainty (right) are shown. In the right
      plot, the PDF-only uncertainty is compared to
the combined PDF+$m_b$ uncertainty with  $\delta_{m_b}=0.15$ or
$\delta_{m_b}=0.25$ at one-$\sigma$.
      \label{fig:bbg}} 
  \end{center}
\end{figure}

We have used the code of 
Ref.~\cite{Harlander:2003ai} to computed the $b\bar{b}\to H$
cross-section to NLO in the MSSM, using the NNPDF2.1 sets with
variable $m_b$ of Sect.~\ref{sec:pdfhqmass}. 
For other physical parameters we take  the default values.
Results are shown in Fig.~\ref{fig:bbg} as a function of the Higgs
mass for LHC~7~TeV, with two different uncertainty ranges for the
bottom mass (the current PDG~\cite{Nakamura:2010zzi} quotes
${}^{+0.17}_{-0.07}$ as uncertainty on the $\overline{\rm
  MS}$ $b$ mass).
  Even with the smaller range the mass uncertainty is not negligible
  in comparison to the PDF  uncertainty.

It would be interesting to extend this analysis to several LHC
processes which are expected to depend significantly on heavy quark
masses, such as for instance $t$-channel single-top production.

\section{PDFs with Fixed Flavour Number}
\label{sec:ffnpdfs}
The NNPDF2.1 PDFs discussed so far are determined in a factorization
scheme in which the number of flavours depends on the scale, and in
particular it varies from $N_f=3$ when $Q^2< m_c^2$ to $N_f=6$ when
$Q^2>m_t^2$. Such a scheme is advantageous in that it includes
terms to all orders
in $\alpha_s$ up to the desired logarithmic order (NLO in our case)
both for light and for heavy quarks, while in a general mass scheme
such as the FONLL-A scheme used here heavy quark mass terms are also
included up to some fixed order in $\alpha_s$. Indeed, given that the
charm mass is of the same order as the starting scale of perturbative
evolution at which PDFs are parametrized (in fact, they coincide for
the default NNPDF2.1 set), the resummation 
 to all orders in $\alpha_s$ of 
charm mass logarithms of the form $\ln Q^2/m_c^2$ is as important as that
for any other parton distribution. However, the LO and NLO resummation
of logarithms  related to heavier
flavours  is
usually rather less important at scales relevant for LHC phenomenology,
especially for top, but in practice sometimes also for bottom. In these cases,
PDF sets in which the maximum number of flavors is fixed at some value
lower than $N_f^{\rm max}=6$ may lead to equally accurate
results. Furthermore, use of these sets is  necessary 
in conjunction with matrix elements computed with a number of active
flavour smaller than six, such as single top production~\cite{Campbell:2009ss}, Higgs production in association with 
bottom quarks~\cite{Maltoni:2003pn}, as well as with Monte Carlo codes
based on similar computations, such as the HVQDIS~\cite{Harris:1995tu}
Monte Carlo, widely used for the analysis of $F_2^c$, 
which is based on the $N_f=3$ scheme computation of the observable.

With these motivations, we have constructed PDF sets in which the
maximum number of flavours is $N_f=3$, $N_f=4$ and
$N_f=5$ (see Table~\ref{tab:ffnsets}). 
These are simply constructed by freezing the number of
flavours  at some scale 
 $Q^2_{\rm match}$ which is thus viewed as a matching scale between a
scheme in which the number of flavours depends on scale (for
$Q^2<Q^2_{\rm match}$) and a scheme in which the number of flavours is
fixed (for $Q^2>Q^2_{\rm match}$). We will refer to these as
Fixed-Flavour Number (FFN) PDFs, though this is strictly speaking a 
 misnomer except in the $N_f=3$ case: for instance, below the top
 threshold the default NNPDF2.1 set and the $N_f=5$ set are
identical. Note that if these PDFs are to be used with matrix elements
computed with the given number of flavours, 
the strong coupling must be consistently determined with the same
fixed number of flavour, lest a spurious $N_f$ dependence be
introduced in physical observables (see Ref.~\cite{Martin:2006qz},
where the effect of an incorrect choice in this respect is also
estimated). 

\begin{table}[t]
  \centering
  \scriptsize
  \begin{tabular}{|c|c|c|c|c|c|c|}
    \hline
    LHAPDF name & $N_f^{\rm max}$ & $m_c$ [GeV] & $m_b$ [GeV] & $Q_{\rm match}$ [GeV]
    & $\alpha_s\lp Q_{\rm match}\rp$ & $\alpha_s^{(N_f^{\rm max})}\lp M_Z\rp$ \\
    \hline
    \hline
    NNPDF21\_100.LHgrid  &  6  &  $\sqrt{2}$  & 4.75 &  - & - & 0.11900 \\
    \hline
    NNPDF21\_FFN\_NF3\_100.LHgrid  &  3  & - & - &  $\sqrt{2}$  & 0.359912 & 0.10585  \\
    \hline
    NNPDF21\_FFN\_NF4\_100.LHgrid  &  4  & $\sqrt{2}$ &  -  & 4.75 & 0.218200  & 
    0.11343 \\
\hline
    NNPDF21\_FFN\_NF5\_100.LHgrid  &  5  & $\sqrt{2}$ &  4.75  & 175 & 0.108283 & 
    0.11900 \\
    \hline
  \end{tabular}
  \caption{\small NNPDF2.1 PDF sets with maximum fixed flavour number of active quarks.
    In all cases, the number of flavours of the reference   NNPDF2.1
    PDF set is frozen
at $Q^2_{\rm match}$, and PDFs are then  evolved upwards 
    with a fixed number of flavours. For each set, the values of the
    heavy quark masses, matching scale, and strong coupling at the
    matching scale and at $Q^2=M_z^2$ are shown. 
 \label{tab:ffnsets}}
\end{table}

\begin{figure}[t]
  \begin{center}
    \epsfig{file=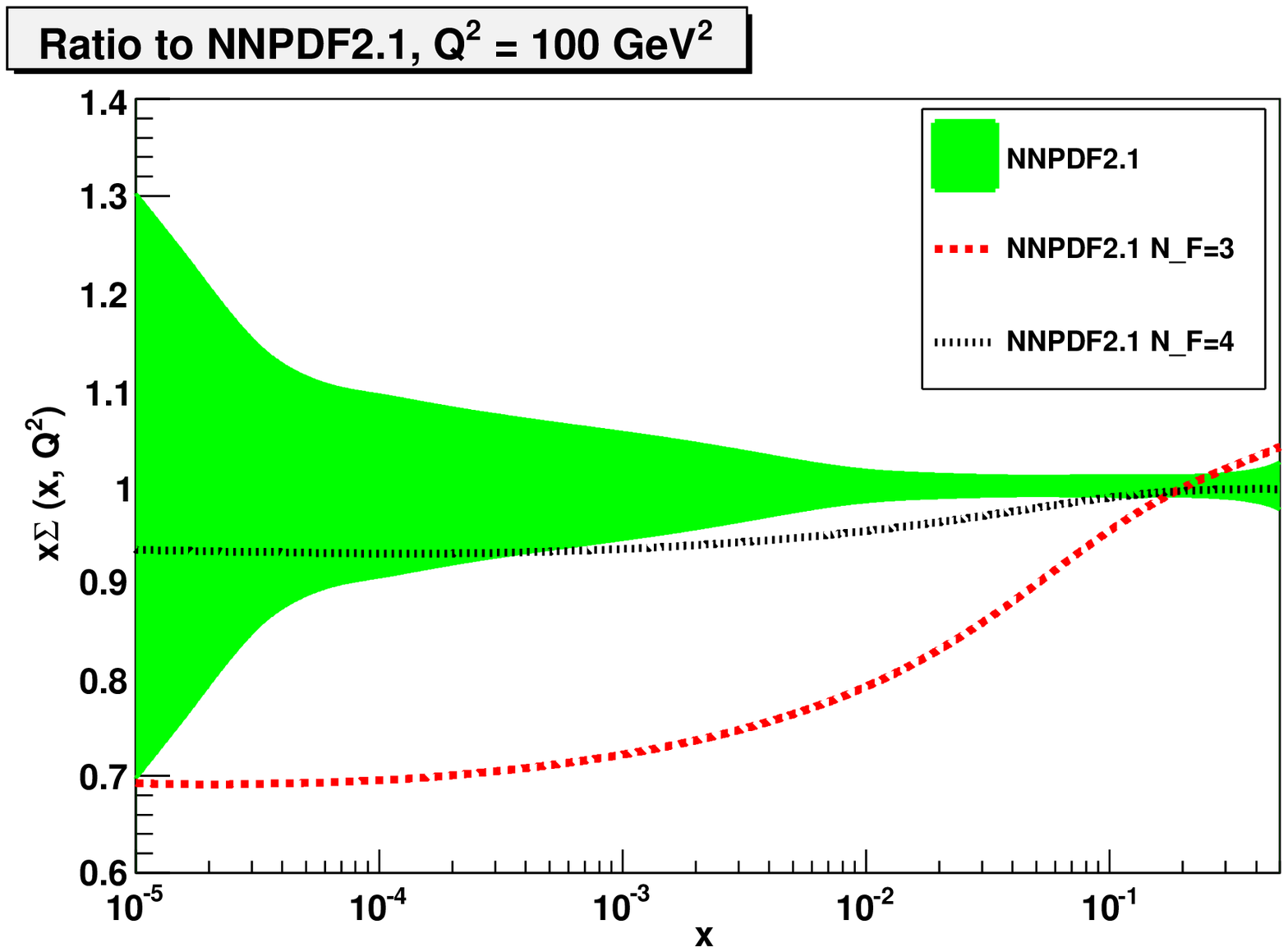,width=0.49\textwidth}
    \epsfig{file=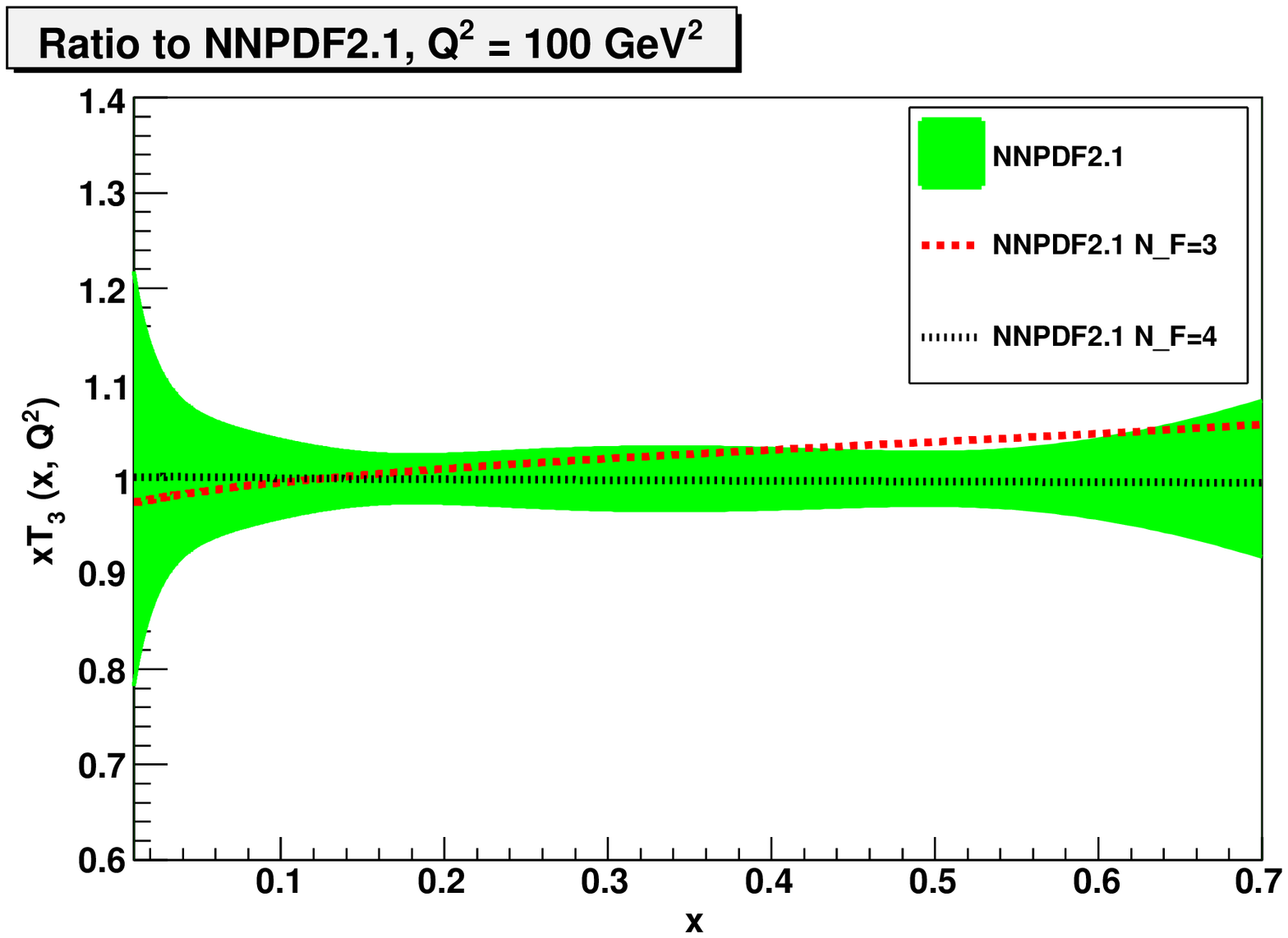,width=0.49\textwidth}
    \epsfig{file=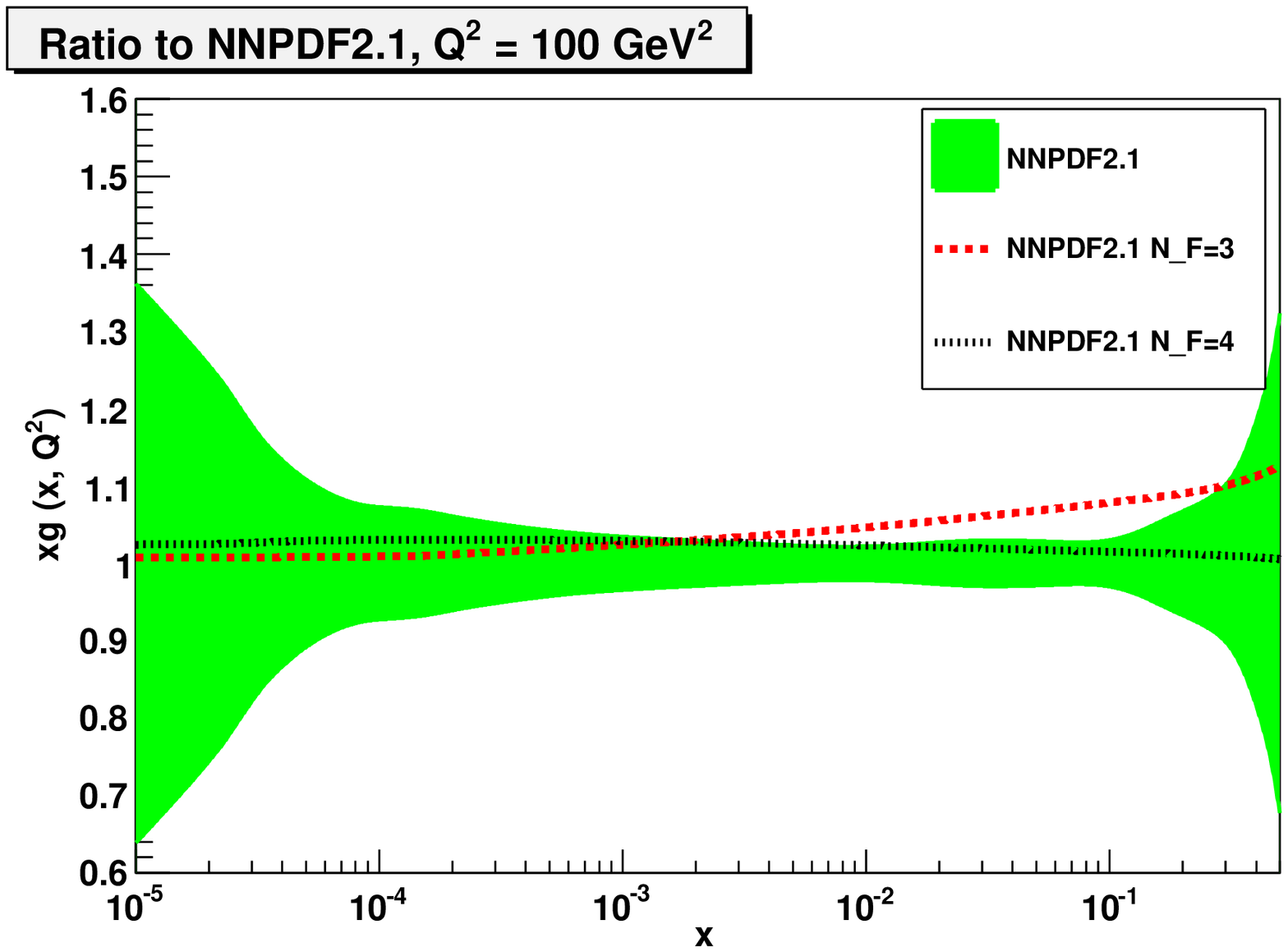,width=0.49\textwidth}
    \epsfig{file=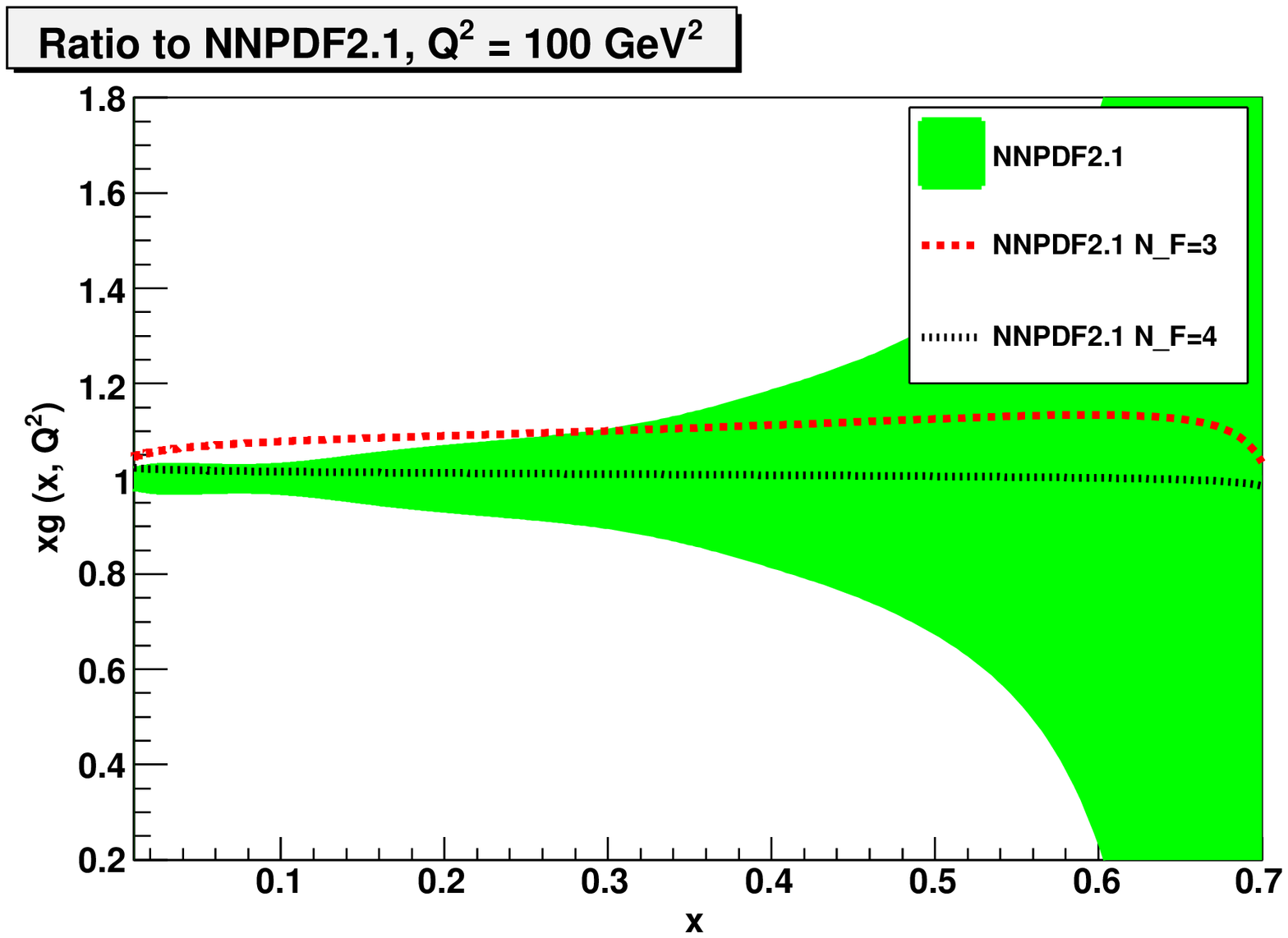,width=0.49\textwidth}
  \end{center}
  \caption{\small Comparison between the default
NNPDF2.1 set and the $N_f=3$ and $N_f=4$ FFN 
    scheme sets at the scale $Q^2=10^2$ GeV$^2$. Results are shown as
    ratios to the default set. The singlet and the triplet (top) and
    the gluon 
 in a linear and 
    logarithmic scale (bottom) are shown.
    \label{ffncompnf3}}
\end{figure}

In order to illustrate the differences between these PDFs in 
Fig.~\ref{ffncompnf3} we compare the FFN PDFs in the $N_f=3$ and $N_f=4$ to the reference 
NNPDF2.1 PDFs at the scale $Q^2=10^2$ GeV$^2$.
Differences stem both from the milder evolution 
due to the reduced number of quark 
flavours and from the correspondingly smaller value of $\alpha_s$. 
This is particularly clear in the case of the singlet at small-$x$ which is
substantially smaller in the $N_f=3$ scheme due to the missing contribution from the charm and 
bottom PDFs.
In the $N_f=4$ case differences are smaller both because we are closer to the heavy quark threshold 
and because now only bottom is not included into the beta function running and the DGLAP evolution equations.

\begin{figure}[t]
  \begin{center}
    \epsfig{file=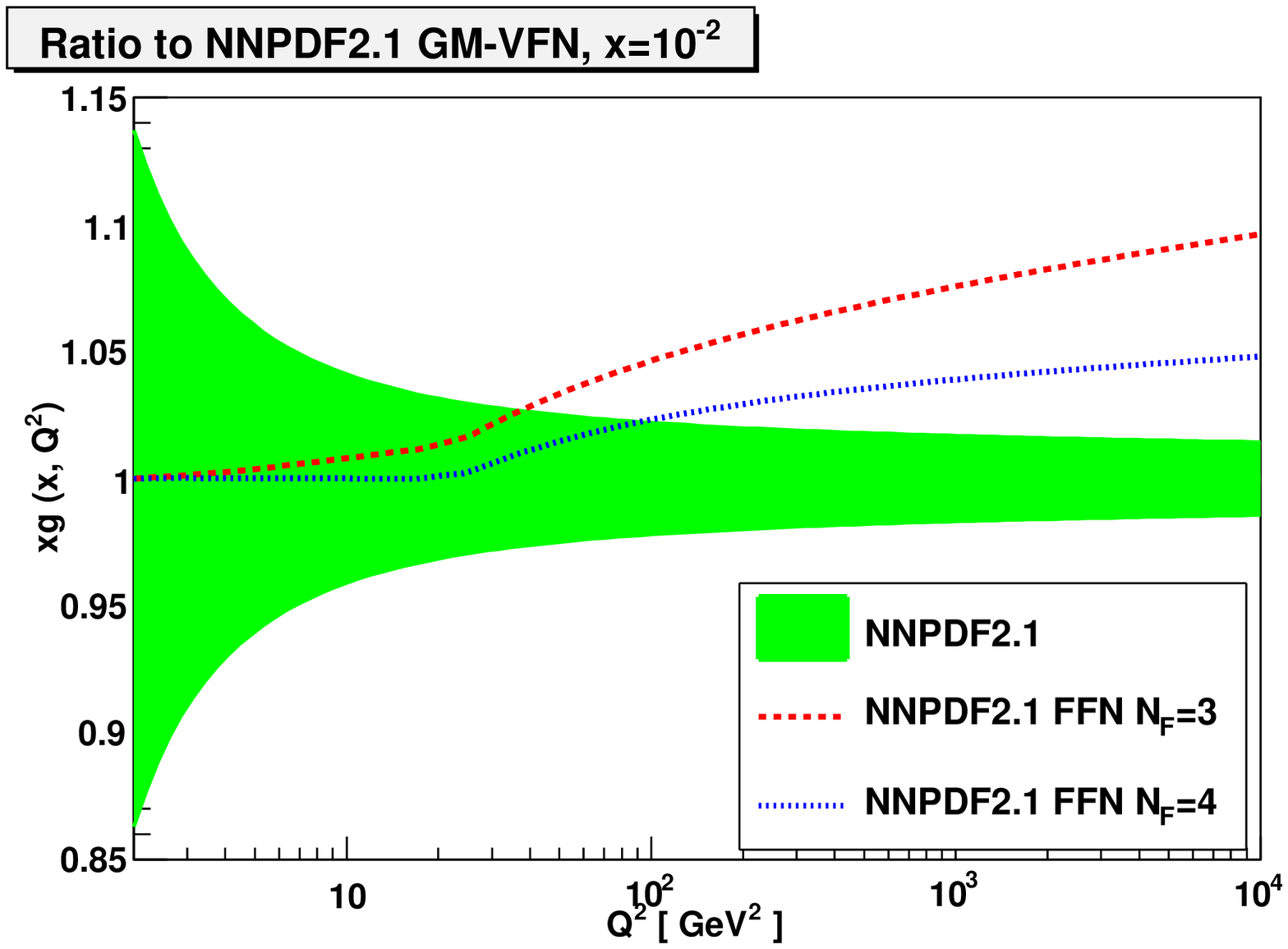,width=0.49\textwidth}
    \epsfig{file=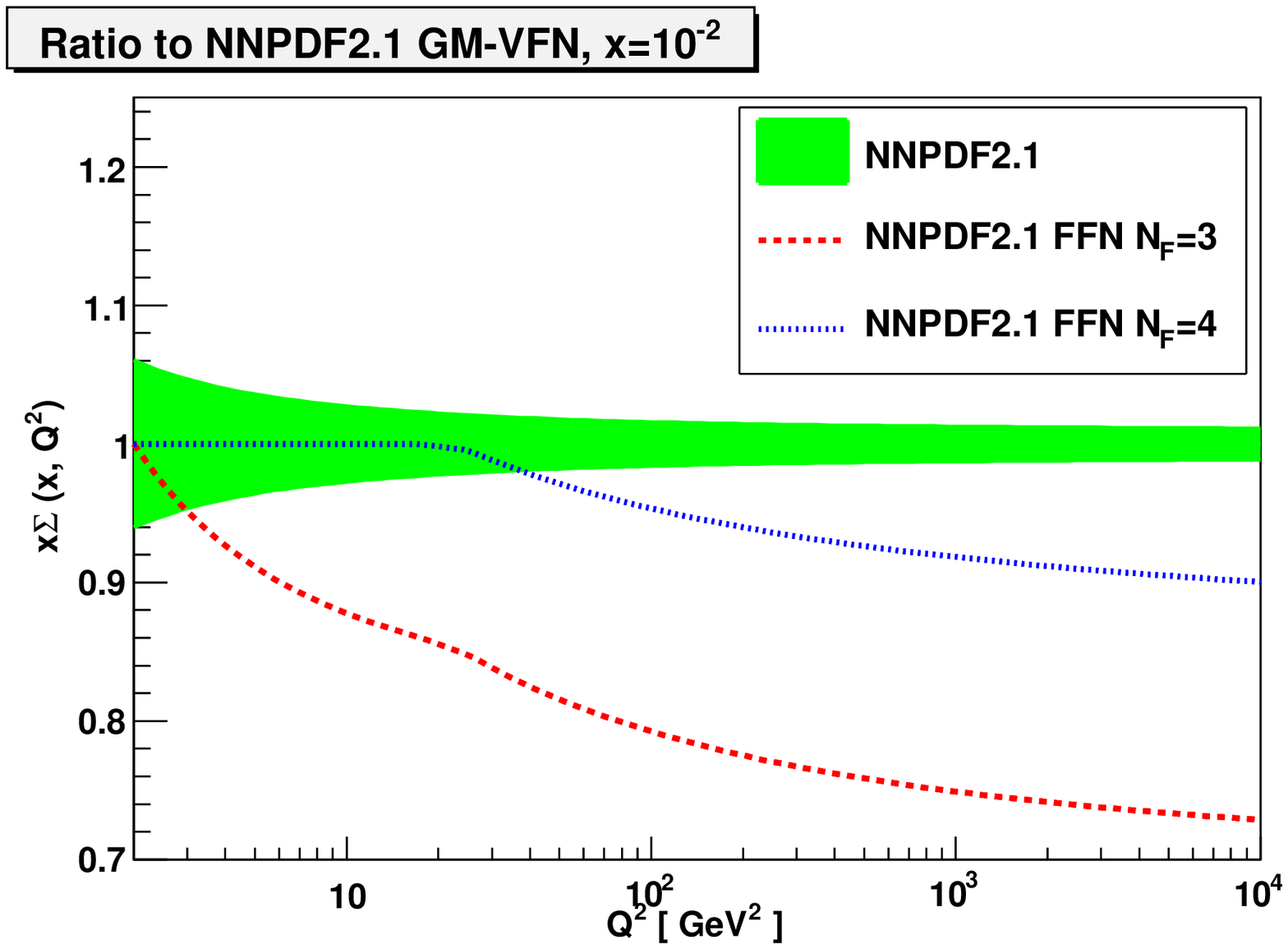,width=0.49\textwidth}
    \epsfig{file=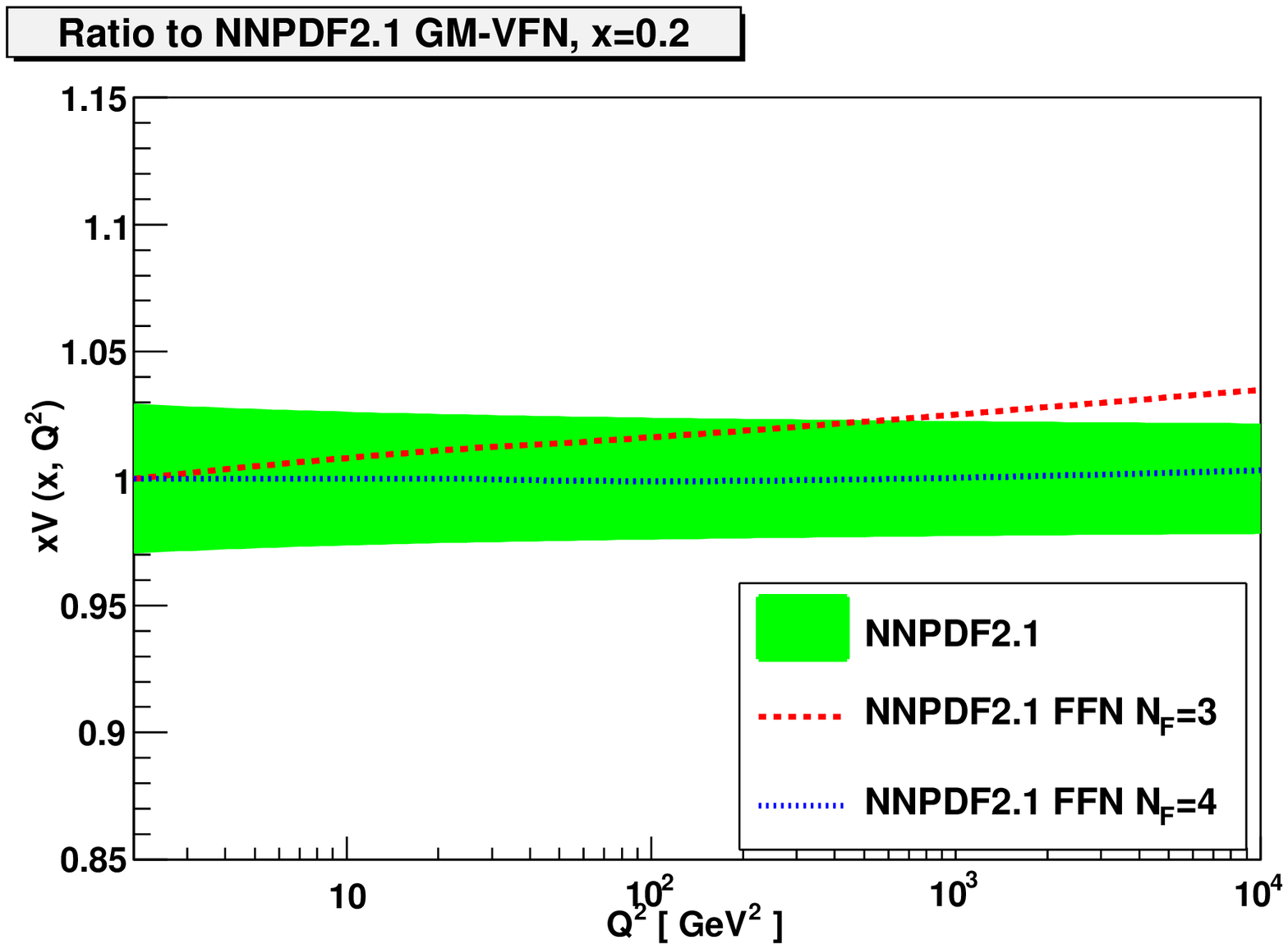,width=0.49\textwidth}
    \epsfig{file=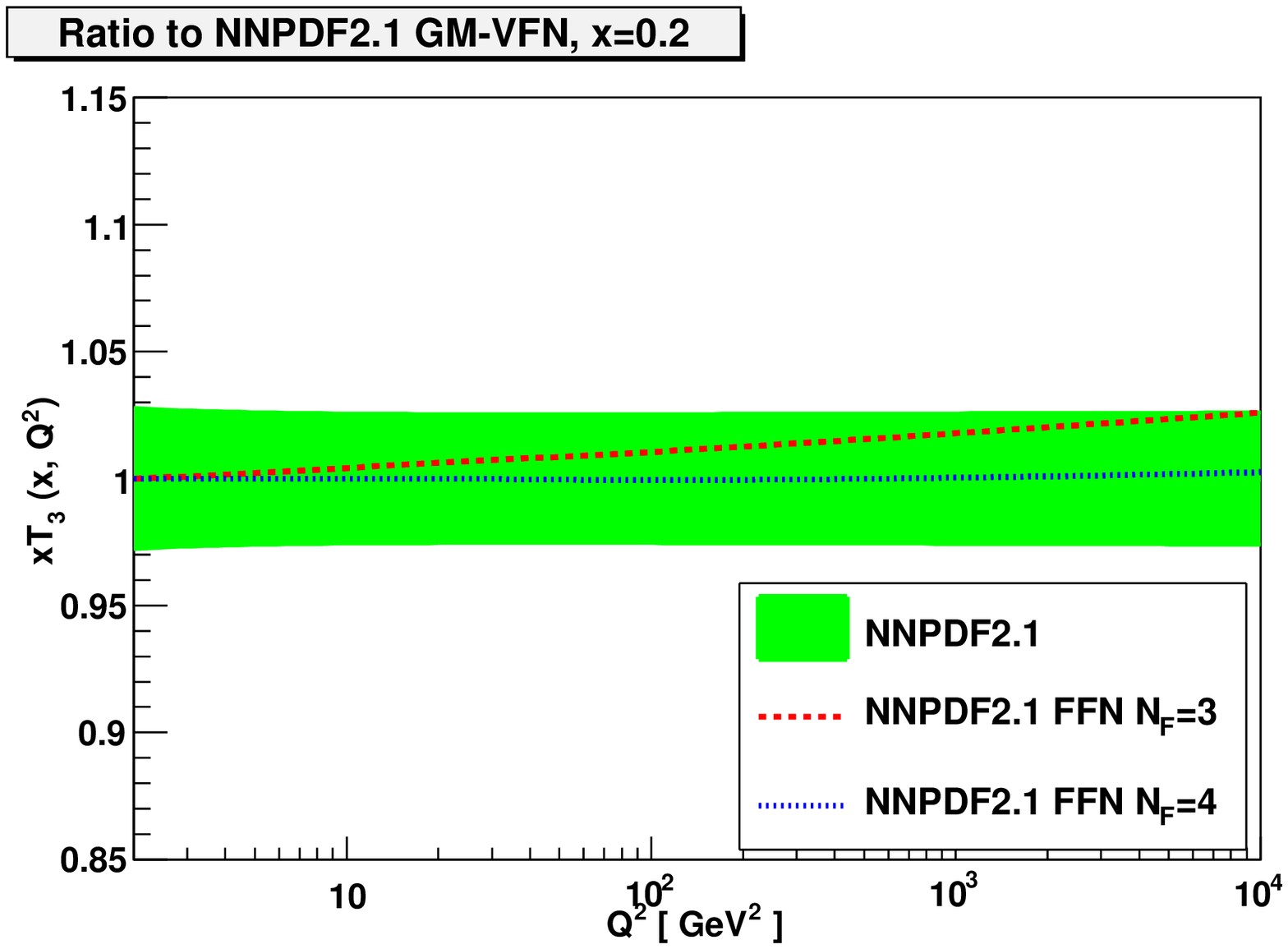,width=0.49\textwidth}
  \end{center}
  \caption{\small Comparison between the default 
NNPDF2.1 set and the $N_f=3$ and  $N_f=4$ FFN 
    sets as a function of scale. Results are shown as
    ratios to the default set.
The gluon and singlet  at $x=0.01$ (top) 
and the triplet  and the total valence at $x=0.2$ (bottom) are shown. 
    \label{ffncompnfQ2}}
\end{figure}

A similar comparison, but now as a function of scale for fixed $x$ is
performed
in Fig.~\ref{ffncompnfQ2}. The default and FFN PDFs coincide
below the matching scale,
and become increasingly different as $Q^2$ grows. 
Differences are larger for the gluon, which is coupled by evolution to
the singlet, which depends on $N_f$ already at leading order.

\begin{figure}[b!]
  \begin{center}
    \epsfig{file=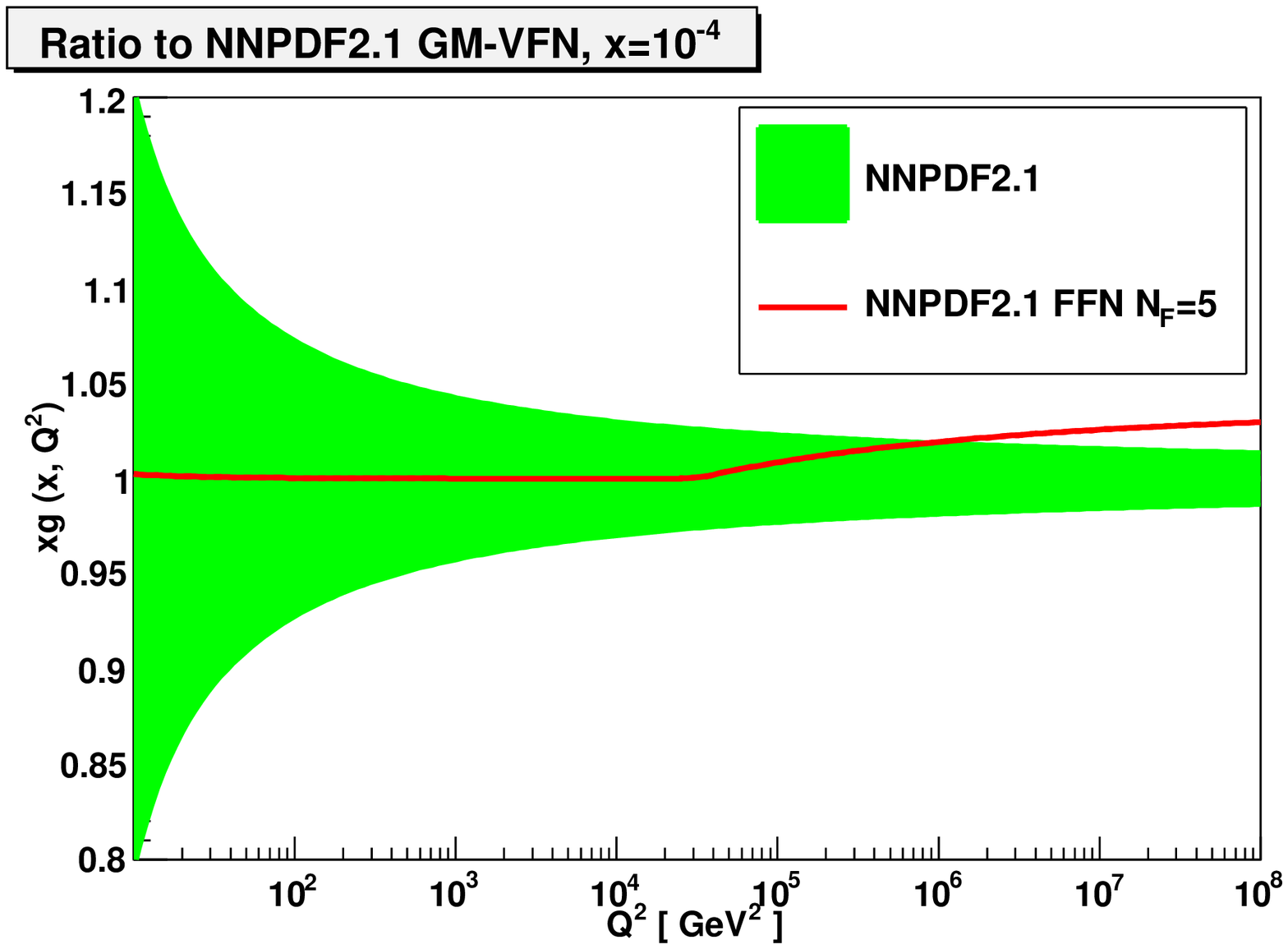,width=0.49\textwidth}
    \epsfig{file=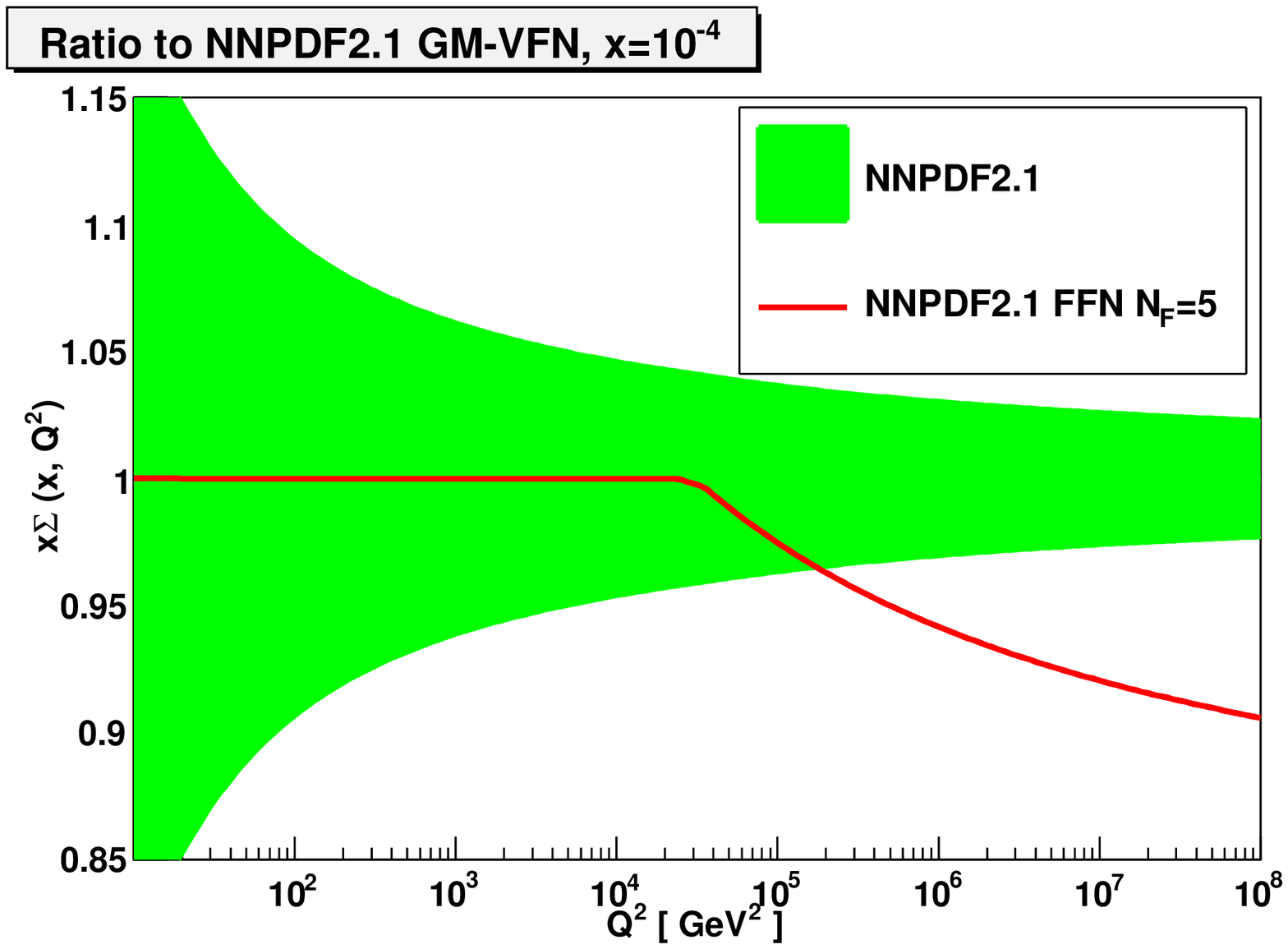,width=0.49\textwidth}
  \end{center}
  \caption{\small Comparison between the default 
NNPDF2.1 set and the $N_f=5$ FFN set as a function of scale,  shown as 
ratios to the default set. The gluon (left) and singlet (right) are shown.
    \label{ffncompnfQ2nf5}}
\end{figure}

Several commonly used PDF sets, such as for instance MSTW08~\cite{Martin:2009iq}
and CT10~\cite{Lai:2010vv} use $N_f=5$ as a maximum number of
flavours. In most cases this makes very little difference at LHC
energies, but, again, 
care must be taken that a scheme in which $N_f=6$ above top threshold 
is sometimes
used matrix element calculations such as  for example, the NLO Higgs
production cross section of Ref.~\cite{Spira:1995rr}. To illustrate
the size of the effects involved, 
in Fig.~\ref{ffncompnfQ2nf5} we compare the gluon and the singlet PDFs
   for the reference NNPDF2.1
set and the FFN set with $N_f=5$  as a function of scale at $x=10^{-4}$. 

It is beyond the scope of this work to study the phenomenological implications of these FFNS PDFs. 
However several applications can be envisaged, such as helping to determine the optimal scales in the 
massless computations when comparing with the massive result, where in each case the corresponding PDF 
set in the same scheme has been used. We will discuss these applications in future work.


\section{Conclusions and outlook}
\label{sec:conclusions}
The NNPDF2.1 PDFs presented here fulfill most of the requirements of an
ideal~\cite{Forte:2010dt} NLO parton set: they are based on
a global dataset which includes most if not all relevant 
deep inelastic and hadronic data, they are free of
parametrization bias, they are provided with reliable and statistically
meaningful uncertainty estimates, 
they include NLO contributions without resorting to $K$-factors, 
they include a consistent treatment of heavy quark mass effects, 
and they are available for a variety of values of the strong coupling 
and heavy quark masses, which allows for the estimate of the 
associated uncertainties. We find that the heavy quark mass effects, 
while substantial for observables which probe directly the heavy quark 
distributions, are rather small for more inclusive observables. In 
particular the benchmark LHC cross-sections (inclusive vector 
boson, top and Higgs production) change by little more than one standard 
deviation when going from NNPDF2.0 to NNPDF2.1. 

An important missing feature of NNPDF2.1 is a reliable estimate of 
theoretical uncertainties related to missing higher order corrections 
(or equivalently, to choices of renormalization and factorization scale). 
The obvious direction for future improvement is thus the
determination of NNLO corrections to PDFs, the inclusion of mass efects 
at $O(\alpha_s^2)$, and the inclusion of resummation corrections 
at large and small $x$. We expect all these corrections to be generally rather 
less than current PDF uncertainties: nonetheless, they need to be computed.

To the extent that the criteria fulfilled by NNPDF2.1 are the 
dominant ones for accurate phenomenology, NNPDF2.1 is perhaps the most reliable
parton set currently available: no other set fulfills all of these criteria. 
We believe it to be adequate for precise phenomenology at the LHC, at least 
for the levels of experimental precision to be expected in the near future.

\bigskip
\bigskip
\begin{center}
\rule{5cm}{.1pt}
\end{center}
\bigskip
\bigskip

All the NNPDF2.1 PDF sets that have been discussed in this work
are available from the NNPDF web site,
\begin{center}
{\bf \url{http://sophia.ecm.ub.es/nnpdf}~}
\end{center}
and will be also available
through the LHAPDF interface~\cite{Bourilkov:2006cj}.

The PDF sets that have been produced in the present analysis and that 
will be available in LHAPDF are the following:

\begin{itemize}

\item The reference NNPDF2.1 sets, sets of $N_{\rm rep}=100$ and 
1000 replicas:\\
{\tt NNPDF21\_100.LHgrid} and {\tt NNPDF21\_1000.LHgrid}.

\item NNPDF2.1 sets of $N_{\rm rep}=100$ replicas  with $\alpha_s$ varied 
from 0.114 to 0.124 with steps of $\delta\alpha_s=0.001$:\\ 
{\tt NNPDF21\_as\_0114\_100.LHgrid},
$\ldots$,  {\tt NNPDF21\_as\_0124\_100.LHgrid}

\item NNPDF2.1 sets with combined PDF+$\alpha_s$ uncertainty:\\ 
{\tt NNPDF21\_as\_0119\_pm\_00012\_100.LHgrid},\\
{\tt NNPDF21\_as\_0119\_pm\_00020\_100.LHgrid},\\ 
{\tt NNPDF21\_as\_0119\_pm\_00012\_50.LHgrid} and 
{\tt NNPDF21\_as\_0119\_pm\_00020\_50.LHgrid}\\
Note than in this case the corresponding NNPDF2.0 sets have also been produced.

\item NNPDF2.1 sets based on reduced datasets: DIS only, DIS+DY, DIS+jets:\\ 
{\tt NNPDF21\_dis\_100.LHgrid}, {\tt NNPDF21\_dis+dy\_100.LHgrid},\\ 
and {\tt NNPDF21\_dis+jet\_100.LHgrid}, as well as  
{\tt NNPDF21\_dis\_1000.LHgrid}

\item NNPDF2.1 sets of $N_{\rm rep}=100$ replicas with varying charm mass:\\ 
 {\tt NNPDF21\_mc\_150\_100.LHgrid},
 {\tt NNPDF21\_mc\_160\_100.LHgrid},\\ 
and  {\tt NNPDF21\_mc\_170\_100.LHgrid}.

\item NNPDF2.1 sets of $N_{\rm rep}=100$ replicas with varying bottom mass:\\
 {\tt NNPDF21\_mb\_425\_100.LHgrid},  {\tt NNPDF21\_mb\_450\_100.LHgrid},\\
 {\tt NNPDF21\_mb\_500\_100.LHgrid} and  {\tt NNPDF21\_mb\_525\_100.LHgrid}.

\item NNPDF2.1 sets in the $N_f=3$, $N_f=4$ and $N_f=5$ FFN schemes:\\ 
 {\tt NNPDF21\_FFN\_NF3\_100.LHgrid}, {\tt NNPDF21\_FFN\_NF4\_100.LHgrid},\\
and {\tt NNPDF21\_FFN\_NF5\_100.LHgrid}.

\end{itemize}

\bigskip
\bigskip
\begin{center}
\rule{5cm}{.1pt}
\end{center}
\bigskip
\bigskip

{\bf\noindent  Acknowledgments \\}

We are especially grateful to G. Watt for providing us the
MSTW08 benchmark computations for heavy quark charged current
structure functions and for discussions. We are grateful to
R. Harlander for providing us with his bbg@nnlo code. 
We would like to thank M. Corradi, K. Lipka, B. List and
P. Thompson 
for help with HERA $F_2^c$ data, S. Glazov for assistance
with the HERA $F_L$ data, P. Nason for illuminating
discussions on heavy quarks, and 
F. Maltoni, P. Nadolsky and R. Thorne for various discussions. 
L.D.D. is funded by an STFC Advanced Fellowship and
M.U. is supported by the Bundesministerium f\"ur Bildung and Forschung (BmBF) of the Federal 
Republic of Germany (project code 05H09PAE).
This work was 
partly supported by the Spanish
MEC FIS2007-60350 grant
and by the European network HEPTOOLS under contract
MRTN-CT-2006-035505. 
We would like to acknowledge the use of the computing resources provided 
by the Black Forest Grid Initiative in Freiburg and by the Edinburgh Compute 
and Data Facility (ECDF) (http://www.ecdf.ed.ac.uk/). The ECDF is partially 
supported by the eDIKT initiative (http://www.edikt.org.uk).



\appendix

\section{Mellin space implementation of neutral current structure functions}
\label{sec:massive-nc}

In this Appendix we compute the analytic Mellin transform
of the $x$--space $\mathcal{O}\lp \alpha_s\rp$ heavy quark neutral
current coefficient
functions and we discuss the implementation and benchmarking of
FONLL neutral current structure functions in the FastKernel framework.
The corresponding results for charged current structure functions
are collected in Appendix~\ref{sec:massive-cc}.

The $x$--space
 gluon $\mathcal{O}\lp \alpha_s\rp$
heavy quark coefficient function is given by Eq.~(\ref{eq:cgnl1}).
Its Mellin transform is defined in the standard way as
\be
C_{2,g}^{(n_l),1} \left( N, \frac{Q^2}{m_h^2} \right)
=\int_0^{(1+4m_h^2/Q^2)^{-1}} dz z^{N-1}C_{2,g}^{(n_l),1} \left( z, \frac{Q^2}{m_h^2} \right) \ .
\label{eq:C0nlcmellin}
\ee
It is easy to see that the integral Eq.~(\ref{eq:C0nlcmellin})
can be written in the following way
\begin{eqnarray}
C_{2,g}^{(n_l),1} \left( N, \epsilon \right) &=& T_R a^N \int_0^1 dt\, t^{N-1} \Big\{ 
\big[1 + 2a(2\epsilon-1)t+2a^2(1-6\epsilon - 4\epsilon^2)t^2\big]
\ln \frac{1+v}{1-v}\nn\\
&&\qquad\qquad - \big[1+4a(\epsilon-2)t-4a^2(\epsilon-2)t^2\big]v\Big\}\\
&=& T_R a^N \int_0^1 dt\, t^{N-1} \Big\{ 
\big[1 +(1-3a)t-\half(1+4a-9a^2)t^2\big]
\ln \frac{1+v}{1-v}\nn\\
&&\qquad\qquad - \big[1+(1-9a)t-a(1-9a)t^2\big]v\Big\},
\end{eqnarray}
where we have defined
$a\lp \epsilon\rp\equiv \lp 1+4\epsilon\rp^{-1}$
 to simplify the coefficients. The integrals we need are thus
\beq 
J_1(N)\equiv \int_0^1 dt\, t^{N-1}\ln\frac{1+v}{1-v} \ , \qquad
J_2(N) \equiv \int_0^1 dt\, t^{N-1}v,
\eeq
since extra powers of $t$ can be accommodated by a shift in $N$ by an integer. Here 
as usual $v = (1-t)^{1/2}/(1-at)^{1/2}$. 

The two integrals that we need are related by an integration by parts.
To show this, we need
\begin{eqnarray}
\frac{d}{dt}\ln\frac{1+v}{1-v} &=& 
\frac{dv}{dt}\frac{d}{dv}\ln\frac{1+v}{1-v}\nn\\
&=&\Big(-\half\frac{1-a}{(1-t)^{1/2}(1-at)^{3/2}}\Big)\Big(\frac{2(1-at)}{(1-a)t}\Big)
\nn\\
&=& -\frac{1}{t}\frac{1}{(1-t)^{1/2}(1-at)^{3/2}}.
\end{eqnarray}
Thus
\beq
 J_1(N) = -\frac{1}{N}\int_0^1 dt\, t^{N}\frac{d}{dt}\ln\frac{1+v}{1-v}
= \frac{1}{N} I(N),
\eeq
where we have defined
\beq
 I(N)\equiv \int_0^1 dt\, t^{N-1}(1-t)^{-1/2}(1-at)^{-1/2}.
\eeq
Note that the boundary term in the integration by parts vanishes for all 
$\mathrm{Re}\,N > 0$, and thus its analytic continuation vanishes for
all $N$, so it  can be safely ignored.
Trivially $J_2(N) = I(N) - I(N+1)$. The integral $I(N)$ may be evaluated in the usual way in terms of a standard hypergeometric function:
\beq 
I(N) = \frac{\Gamma(N)\Gamma(\half)}{\Gamma\lp N+\half\rp}~_2F_1
(\half,N,N+\half;a) .
\eeq
Note that when $a=0$, this reduces to $B(N,\half)$, as it should.

We thus get
\begin{eqnarray}
&&C_{2,g}^{(n_l),1} \left( N, \epsilon \right)
= T_R a^N \Big\{ 
\big[\smallfrac{1}{N}I(N)+\smallfrac{1-3a}{N+1}I(N+1)-\half
\smallfrac{1+6a-9a^2}{N+2}I(N+2)\big]
\nn\\
&&-\big[I(N)-I(N+1)+(1-9a)(I(N+1)-I(N+2))\nn
\\&&-a(1-9a)(I(N+2)-I(N+3))\big]\Big\}\nn\\
&&= T_R a^N \Big\{(\smallfrac{1}{N}-1)I(N)
+(\smallfrac{1-3a}{N+1}+9a)I(N+1)\nn\\
&&-(\smallfrac{1}{2}\smallfrac{1+4a-9a^2}{N+2}-(1+a)(1-9a))I(N+2)
-a(1-9a)I(N+3)
\Big\}  \ .
\label{eq:cnrdb}
\end{eqnarray}
This result is the required ingredient to
implement the FONLL-A neutral current structure functions
in the FastKernel framework.

A cross-check of the Mellin transform of the massive coefficient
function
Eq.~(\ref{eq:cnrdb}) is provided by the fact that its massless limit
coincides with the Mellin transform of the $x$--space massive
asymptotic
$(n_l,0)$ coefficient function, Eq.~(\ref{eq:mznloexp}).
To this purposes, we need to expand Eq.~(\ref{eq:cnrdb})
near $a=1$.
Near $a=1$, i.e. $\epsilon=0$, we need to use the 
asymptotic expansion 
\begin{eqnarray}
&&F(\half,N,N+\half;a) = \frac{\Gamma(N+\half)}{\Gamma(\half)^2\Gamma(N)^2}
\sum_{n=0}^\infty \frac{\Gamma(n+\half)\Gamma(N+n)}{(n!)^2}\times \nonumber \\
&&\quad [2\psi(n+1)-\psi(n+\half)-\psi(N+n)-\ln(1-a)](1-a)^n,
\label{eq:hypgeoexpansion}
\end{eqnarray}
so that
\begin{eqnarray}
I(N) &=& \frac{1}{\Gamma(\half)\Gamma(N)}
\sum_{n=0}^\infty \frac{\Gamma(n+\half)\Gamma(N+n)}{(n!)^2}\times \nonumber \\
&&\quad [2\psi(n+1)-\psi(n+\half)-\psi(N+n)-\ln(1-a)](1-a)^n.
\end{eqnarray}
The $n=0$ term then gives the $\ln\epsilon$ collinear divergence, which is
subtracted by the massless coefficient function: as $\epsilon\to 0$
\beq
I(N) = -\ln(4\epsilon)-2\gamma_E-\psi(\half)-
\psi(N)  + O(\epsilon).
\eeq
Substituting in Eq.~(\ref{eq:cnrdb}) we obtain
\beq
C_{g}^{(n_l),1} \left( N, \epsilon \right) = T_R
\frac{\lc -N^3+3N^2-N(2+N+N^2)\lp \ln \epsilon +\gamma_E+\psi\lp N\rp\rp\rc}{N^2(1+N)(2+N)} + \mathcal{O}(\epsilon) \ ,
\eeq
as expected: the coefficient of the singularity is precisely the 
LO anomalous dimension $\gamma^{(0)}_{qg}(N)$.
Therefore we have checked that the massless limit is
properly reproduced,
\beq
C_{g}^{(n_l),1} \left( N, \epsilon \right) + \mathcal{O}(\epsilon)
= C_{g}^{(n_l,0),1} \left( N, \epsilon \right) ,
\eeq
with the massive asymptotic coefficient function given 
by the Mellin transform of
Eq.~(\ref{eq:mznloexp}), as desired.

For completeness we also provide the correponding expressions
for the $\mathcal{O}\lp \alpha_s\rp$ heavy quark coefficient
function for the longitudinal structure function $F_{L,c}$, which is
implicitely contained in  Eq.~(\ref{eq:cnrdb}) since $F_{2,c}=F_{T,c}+F_{L,c}$.
The $x$--space expression for the longitudinal heavy quark 
coefficient function is
\begin{equation}
\label{eq:cgnl1l}
  C_{L,g}^{(n_l),1} \left( z, \frac{Q^2}{m^2} \right) = \theta \left( W^2 - 4
  m^2 \right) \times T_R \lc -8\epsilon z^2\log \frac{1 + v}{1 - v}+4vz(1-z) \rc \ .
\end{equation}
Its Mellin transform can be computed using the integrals
discussed above, with the result
\beq
 C_{L,g}^{(n_l),1} \left( N, \frac{Q^2}{m^2} \right)  =  T_Ra^{N+1}\Big[
-8\epsilon a \frac{I(N+2)}{N+2} \nonumber
+ 4\lp I(N+1) -  I(N+2)(1+a) +I(N+2)\rp\Big] \ .
\eeq
The massless limits of the $x$- and $N$-space results are
straightforwardly computed and checked to be related by Mellin
transformation as they ought to.

Now we turn to the implementation and benchmarking of these
results in the FastKernel framework.
The major improvement in the FastKernel framework as compared
to Ref.~\cite{Ball:2010de} 
is the inclusion of heavy quark mass effects in deep--inelastic
scattering structure functions, following
the FONLL-A general-mass scheme~\cite{Forte:2010ta}. 
As discussed in Ref.~\cite{Ball:2010de}, FastKernel requires
to write down all the DIS observables in Mellin space and
precomputing all the associated x--space Green's functions.
Therefore, to extend FastKernel with FONLL structure functions
we need to formulate FONLL in Mellin space.

The $x$--space expression for the
FONLL-A  heavy quark structure functions, Eq.~(\ref{eq:FONLLnlo}),
can be easily written down in $N$--space
as follows:
\bea
\label{eq:f2c_nnpdf}
F_{2,h}^{\rm FONLL}(N,Q^2) &=& 
F_{2,h}^{(n_l)}(N,Q^2) \\
&-&\theta\lp Q^2 - m^2\rp\lp 1-\frac{m^2}{Q^2}\rp^2 \lc
F_{2,h}^{(n_l,0)}(N,Q^2)-F_{2,h}^{(n_l+1)}(N,Q^2) \rc \nonumber \ ,
\eea
with the default damping factor as threshold prescription.
In order to implement Eq.~(\ref{eq:f2c_nnpdf}) in the FastKernel
framework, we need the Mellin space expressions of the
heavy quark coefficient function in the
$(n_l)$, $(n_l,0)$ and $(n_l+1)$ schemes. While the last two
are known, the former was not available in a closed form
suitable for analytical continuation. 
The details of the computation have
been presented above, and the desired
result is Eq.~(\ref{eq:cnrdb}).

With all the Mellin space heavy quark coefficient functions
available, it becomes possible to
implement the FONLL-A heavy quark
structure functions, Eq.~(\ref{eq:f2c_nnpdf}) 
into the FastKernel framework. To show that
the $N$--space implementation has the required accuracy, in 
Table~\ref{tab:tablebench} for $F_{2,c}$ and in Table~\ref{tab:tablebench2} for $F_{L,c}$,
we compare the results for the 
Les Houches Heavy Quark benchmarks~\cite{LHhq}
for FONLL-A obtained with the FONLLdis code~\cite{Forte:2010ta,FONLLdis} 
and with the
FastKernel framework for various relevant values of $Q^2$. The
benchmark settings for the PDFs and $\alpha_s$ are used for
this comparison. What we can see is that the accuracy on
the FONLL heavy quark structure functions
 is essentially always below the percent level, enough
for precision phenomenological studies.  For completeness
we also show the analogous results for the case of
the massive scheme results, where similar accuracies are obtained.

\begin{table}[ht]
\begin{center}
\small
 \begin{tabular}{|c|c|c|c|c|c|c|}
\hline
& \multicolumn{3}{|c|}{FONLL-A} & \multicolumn{3}{|c|}{FFN} \\
 \hline
 $x$ & FONLLdis & FastKernel  
& Accuracy & FONLLdis & FastKernel  
& Accuracy\\
\hline
 \hline
\multicolumn{7}{|c|}{$Q^2=4$ GeV$^2$} \\
 \hline
$10^{-5}$ &      0.1507  & 0.1501   & $0.4\%$  & 0.1088& 0.1091& $0.3\%$\\
 $10^{-4}$  &    0.0936  & 0.0931   & $0.5\%$  & 0.0697& 0.0698& $0.1\%$ \\
 $10^{-3}$  &    0.0506   & 0.0504  & $0.4\%$  & 0.0392& 0.391&$0.2\%$ \\
  $10^{-2}$  &   0.0174   & 0.0177 & $1.5\%$  & 0.0136& 0.0137& $0.7\%$\\
\hline
\hline
\multicolumn{7}{|c|}{$Q^2=10$ GeV$^2$} \\
 \hline
$10^{-5}$ &    0.563 & 0.561 & $0.4\%$  & 0.3598 & 0.3602& $0.1\%$\\
 $10^{-4}$  &  0.312 & 0.311 & $0.3\%$  & 0.2007&0.2011& $0.2\%$\\
 $10^{-3}$  &   0.1499 & 0.1495 & $0.3\%$  & 0.0981& 0.0982& $0.1\%$\\
  $10^{-2}$  &  0.05056 & 0.05052 & $0.1\%$  & 0.0328 & 0.0327& $0.3\%$ \\
\hline
\hline
\multicolumn{7}{|c|}{$Q^2=100$ GeV$^2$} \\
 \hline
$10^{-5}$ &    2.28636 & 2.28577 & $0.02\%$  & 1.9779 & 1.9877& $0.5\%$ \\
 $10^{-4}$  &  1.12186 & 1.12082 & $0.1\%$  & 0.9161& 0.9184& $0.3\%$\\
 $10^{-3}$  &  0.48008 & 0.47919 & $0.2\%$  & 0.3644& 0.3647& $0.1\%$\\
  $10^{-2}$  & 0.15207 & 0.15200 & $0.04\%$  & 0.1037 & 0.1038& $0.1\%$\\
 \hline
 \end{tabular}
\end{center}
\caption{\small Results of the benchmark comparison for the
$F_{2c}(x,Q^2)$ structure function in the
 FONLL-A scheme for the FONLLdis code\cite{Forte:2010ta}
and for the FastKernel code. Results are provided at the
benchmark kinematical points in $x,Q^2$. Results  for the 
massive (FFN) scheme are also given for completeness.
\label{tab:tablebench}}
\end{table}

\begin{table}[h!]
\begin{center}
\small
 \begin{tabular}{|c|c|c|c|c|c|c|}
\hline
& \multicolumn{3}{|c|}{FONLL-A} & \multicolumn{3}{|c|}{FFN} \\
 \hline
 $x$ & FONLLdis & FastKernel  
& Accuracy & FONLLdis & FastKernel  
& Accuracy\\
\hline
 \hline
\multicolumn{7}{|c|}{$Q^2=4$ GeV$^2$} \\
 \hline
$10^{-5}$ &    0.0130174  & 0.013094  & $0.6\%$ & 0.009077 & 0.009081 & $0.04\%$ \\
 $10^{-4}$  &  0.008347  & 0.008316 & $0.4\%$ & 0.005913 & 0.005910 & $0.05\%$ \\
 $10^{-3}$  &  0.004795  & 0.004778 & $0.3\%$ & 0.003511 & 0.003509 & $0.06\%$ \\
  $10^{-2}$  & 0.001910  & 0.001907 & $0.2\%$  & 0.001403 & 0.001406 & $0.2\%$ \\
\hline
\hline
\multicolumn{7}{|c|}{$Q^2=10$ GeV$^2$} \\
 \hline
$10^{-5}$ & 0.073235   & 0.073022   & $0.3\%$ & 0.049856 & 0.049982& $0.2\%$ \\
 $10^{-4}$  & 0.041392 &  0.041251 & $0.3\%$ & 0.028402 & 0.028423 & $0.07\%$ \\
 $10^{-3}$  & 0.020754  & 0.020707 & $0.2\%$ & 0.014463 & 0.014456 & $0.05\%$ \\
  $10^{-2}$  & 0.007616  & 0.007595  & $0.3\%$ & 0.005350 & 0.005346 & $0.07\%$\\
\hline
\hline
\multicolumn{7}{|c|}{$Q^2=100$ GeV$^2$} \\
 \hline
$10^{-5}$ & 0.471889   & 0.4729 & $0.2\%$ & 0.3955& 0.397855 & $0.6\%$ \\
 $10^{-4}$  & 0.2236  & 0.2235 & $0.1\%$ & 0.18656 & 0.186914 & $0.2\%$ \\
 $10^{-3}$  &  0.0920 & 0.09188 & $0.1\%$ & 0.0765 & 0.076393 & $0.1\%$ \\
  $10^{-2}$  & 0.027822  &  0.02782 & $0.1\%$& 0.023079 & 0.023100 & $0.1\%$ \\
 \hline
 \end{tabular}
\end{center}
\caption{\small Same as Table~\ref{tab:tablebench} for the
$F_{Lc}(x,Q^2)$ structure function.
\label{tab:tablebench2} 
}
\end{table}

\section{Mellin space implementation of charged current structure functions}
\label{sec:massive-cc}

In this Appendix the analysis of Appendix A is repeated for
charged current structure functions.

The $F_2^c$ charm structure functions in charged
current DIS is given by Eq.~(\ref{eq:f2cffns}). The 
expression for structure functions in neutrino-induced
 charged current scattering in the
FFN scheme is
\begin{multline}\label{FC}
\mathcal{F}_i^c(x,Q^2)=\frac12s'(\xi,\mu^2)+
\frac12\frac{\alpha_s(\mu^2)}{2\pi}
\bigg\{\int\limits_{\xi}^{1}\frac{d\xi'}{\xi'}\bigg[C^{(n_l)}_{i,q}(\xi',\mu^2,\lambda)s'\left(\frac{\xi}{\xi'},\mu^2\right)+\\
+C^{(n_l)}_{i,g}(\xi',\mu^2,\lambda)g\left(\frac{\xi}{\xi'},\mu^2\right)
\bigg]\bigg\},
\end{multline}
with $i=1,2,3$. In Eq.~(\ref{FC})  have used the following definitions:
\begin{equation}
\begin{array}{l}
\displaystyle s'= 2|V_{cs}|^2s+2|V_{cd}|^2[f\,d+(1-f)u];\quad\quad f=\frac{N_p}{N_p+N_n};\\
\\
\displaystyle \xi=x\bigg(1+\frac{m_c^2}{Q^2}\bigg);\quad\quad\lambda = \frac{Q^2}{Q^2+m_c^2}\,.
\end{array}
\end{equation}
The explicit $x$-space expressions of the  $\mathcal{O}\lp
\alpha_s\rp$ contributions $C^{(n_l)}_{i,q(g)}$ to the coefficient
functions are given in Refs.~\cite{Gottschalk:1980rv,Gluck:1996ve}.
The standard
structure functions are related to those defined in Eq.~(\ref{FC}) through
\begin{equation}\label{convSFdef}
F_1^c\equiv\mathcal{F}_1^c;\qquad F_2^c\equiv2\xi\mathcal{F}_2^c=x\frac{2}{\lambda}\mathcal{F}_2^c;\qquad F_3^c\equiv2\mathcal{F}_3^c\,,
\end{equation}
so that
\begin{equation}\label{convFLdef}
F_L^c\equiv
F_2^c-2xF_1^c=2\xi\left(\mathcal{F}_2^c-\lambda\mathcal{F}_1^c\right). 
\end{equation}

Before Mellin- transforming the $x$-space quark coefficient functions
of Refs.~\cite{Gottschalk:1980rv,Gluck:1996ve} 
we rewrite them all in the form
$$
C^{(n_l)}_{i,q}(x) = K\delta(1-x)+f(x)+[g(x)]_+,
$$
where $K$ is a constant and $f(x)$ is  regular function in $x\in
[0,1]$ (so in general
$g(x)$ is not  regular in $x=1$).
We get
 \begin{equation}\label{xc1}
\begin{array}{rcl}
C^{(n_l)}_{1,q}(z)&=&\displaystyle -C_F\bigg(4+\frac{1}{2\lambda}+\frac{\pi^2}{3}+\frac{1+3\lambda}{2\lambda}K_A\bigg)\delta(1-z)\\
\\
&+&\displaystyle C_F\left[-\frac{(1+z^2)\ln z}{1-z}-2(1+z)\ln(1-z)+(1+z)\ln(1-\lambda z)+(3-z)+\frac{z-z^2}{1-\lambda z}\right]\\
\\
&+&\displaystyle C_F\left[4\frac{\ln(1-z)}{1-z}-2\frac{\ln(1-\lambda
    z)}{1-z}-2\frac1{1-z}+\frac12\frac{1-z}{(1-\lambda
    z)^2}-\frac{1+z^2}{1-z}\ln{\lambda}\right]_+ ;
\end{array}
\end{equation}
\begin{equation}
\begin{array}{rcl}
C^{(n_l)}_{2,q}(z)&=&\displaystyle- C_F\bigg(4+\frac{1}{2\lambda}+\frac{\pi^2}{3}+\frac{1+\lambda}{2\lambda}K_A\bigg)\delta(1-z)\\
\\
&+&\displaystyle C_F\left[-\frac{(1+z^2)\ln z}{1-z}-2(1+z)\ln(1-z)+(1+z)\ln(1-\lambda z)+\left(2z+2-\frac2z\right)+\frac{\tfrac{2}{z}-1-z}{1-\lambda z}\right]\\
\\
&+&\displaystyle C_F\left[4\frac{\ln(1-z)}{1-z}-2\frac{\ln(1-\lambda z)}{1-z}-2\frac1{1-z}+\frac12\frac{1-z}{(1-\lambda z)^2}-\frac{1+z^2}{1-z}\ln{\lambda}\right]_+;
\end{array}
\end{equation}
\begin{equation}\label{xc3}
\begin{array}{rcl}
C^{(n_l)}_{3,q}(z)&=&\displaystyle -C_F\bigg(4+\frac{1}{2\lambda}+\frac{\pi^2}{3}+\frac{1+3\lambda}{2\lambda}K_A\bigg)\delta(1-z)\\
\\
&+&\displaystyle C_F\left[-\frac{(1+z^2)\ln z}{1-z}-2(1+z)\ln(1-z)+(1+z)\ln(1-\lambda z)+(1+z)+\frac{1-z}{1-\lambda z}\right]\\
\\
&+&\displaystyle C_F\left[4\frac{\ln(1-z)}{1-z}-2\frac{\ln(1-\lambda z)}{1-z}-2\frac1{1-z}+\frac12\frac{1-z}{(1-\lambda z)^2}-\frac{1+z^2}{1-z}\ln{\lambda}\right]_+;
\end{array}
\end{equation}
with $K_A=(1-\lambda)\ln(1-\lambda)/\lambda$.

The gluon coefficient functions do not need any further work and are
given by
\begin{eqnarray}
C^{(n_l)}_{1,g}(z)&=&T_f(2z^2-2z+1)\left\{2\ln(1-z)-2\ln{z}-\ln[\lambda(1-\lambda)]\right\}+\nonumber\\
\nonumber\\
&&[4-4(1-\lambda)]z(1-z)+(1-\lambda)\frac{z}{1-\lambda z}+\\
\nonumber \\
&&2(1-\lambda)\left[z\ln\frac{1-\lambda z}{(1-\lambda)z}-2\lambda z^2\ln\frac{1-\lambda z}{(1-\lambda)z}\right]-1;\nonumber
\end{eqnarray}
\begin{eqnarray}
C^{(n_l)}_{2,g}(z)&=&T_f(2z^2-2z+1)\left\{2\ln(1-z)-2\ln{z}-\ln[\lambda(1-\lambda)]\right\}+\nonumber\\
\nonumber\\
&&[8-18(1-\lambda)+12(1-\lambda)^2]z(1-z)+(1-\lambda)\frac{1}{1-\lambda z}+\\
\nonumber\\
&&6\lambda(1-\lambda)\left[z\ln\frac{1-\lambda z}{(1-\lambda)z}-2\lambda z^2\ln\frac{1-\lambda z}{(1-\lambda)z}\right]-1;\nonumber
\end{eqnarray}
\begin{eqnarray}
C^{(n_l)}_{3,g}(z)&=&T_f(2z^2-2z+1)\left\{2\ln(1-z)-2\ln(1-\lambda z)+\ln\left(\frac{1-\lambda}{\lambda}\right)\right\}+\nonumber\\
\\
&&2(1-\lambda)z(1-z)+2(1-\lambda)\left[(1+\lambda )z^2\ln\frac{1-\lambda z}{(1-\lambda)z}-z\ln\frac{1-\lambda z}{(1-\lambda)z}\right].\nonumber
\end{eqnarray}

\begin{table}
\begin{center}
\begin{tabular}{|c|c|}
\hline
\vphantom{\Bigg|}$f(z)$                           &     $\mathbf{M}[f](N)$  \\
\hline\hline
\vphantom{\Bigg|}$\delta(1-z)$                    &     $1$                 \\
\hline
\vphantom{\Bigg|}$z^l$                            &     $\displaystyle\frac{1}{N+l}$     \\
\hline
\vphantom{\Bigg|}$\displaystyle\frac{(1+z^2)\ln z}{1-z}$ & $\displaystyle 2(S_2-\zeta_2)-\frac{1}{N^2}+\frac{1}{(N+1)^2}$\\
\hline
\vphantom{\Bigg|}$(1+z)\ln (1-z)$                 &     $\displaystyle-\frac{S_1}{N}-\frac{S_1}{N+1}-\frac{1}{(N+1)^2}$\\
\hline
\vphantom{\Bigg|}$(1+z)\ln (1-\lambda z)$         &     $\displaystyle\lambda \frac{_2F_1(1,N+1,N+2;\lambda)}{N(N+1)}+\frac{\ln(1-\lambda)}{N}+$\\
\vphantom{\Bigg|}                                 &     $\displaystyle\lambda \frac{_2F_1(1,N+2,N+3;\lambda)}{(N+1)(N+2)}+
                                                           \frac{\ln(1-\lambda)}{N+1}$\\
\hline
\vphantom{\Bigg|}$\displaystyle\frac{z-z^2}{1-\lambda z}$&$\displaystyle \frac{_2F_1(1,N+1,N+2,\lambda)}{N+1}-\frac{_2F_1(1,N+2,N+3,\lambda)}{N+2}$\\
\hline
\vphantom{\Bigg|}$\displaystyle\frac{\frac{2}{z}-1-z}{1-\lambda z}$&$\displaystyle 2\frac{_2F_1(1,N-1,N,\lambda)}{N-1}-\frac{_2F_1(1,N,N+1,\lambda)}{N}$\\
\vphantom{\Bigg|}                                                  &$\displaystyle-\frac{_2F_1(1,N+1,N+2,\lambda)}{N+1}$\\
\hline
\vphantom{\Bigg|}$\displaystyle\frac{1-z}{1-\lambda z}$&$\displaystyle\frac{_2F_1(1,N,N+1,\lambda)}{N}-\frac{_2F_1(1,N+1,N+2,\lambda)}{N+1}$\\
\hline
\vphantom{\Bigg|}$\displaystyle\left[\frac{\ln(1-z)}{1-z}\right]_+$&$\displaystyle\frac{1}{2}\bigg(S_1^2+S_2-2\frac{S_1}{N}\bigg)$\\
\hline
\vphantom{\Bigg|}$\displaystyle\left[\frac{\ln(1-\lambda z)}{1-z}\right]_+$& $\displaystyle J_\lambda(N)=\sum_{k=1}^{\infty}\frac{\lambda^k}{k}\left[S_1(N+k)-S_1(k)-\frac1{N+k}\right]$         \\
\hline
\vphantom{\Bigg|}$\displaystyle\left[\frac{1}{1-z}\right]_+$&$\displaystyle\frac{1}{N}-S_1$\\
\hline
\vphantom{\Bigg|}$\displaystyle\left[\frac{1-z}{(1-\lambda z)^2}\right]_+$&$\displaystyle\frac{_2F_1(2,N,N+2,\lambda)}{N(N+1)}+\frac{\lambda+\ln(1-\lambda)}{\lambda^2}$\\
\hline
\vphantom{\Bigg|}$\displaystyle\left[\frac{1+z^2}{1-z}\right]_+$&$\displaystyle\frac{1}{N}-\frac{1}{N+1}-2S_1+\frac 3 2$ \\
\hline
\end{tabular}
\end{center}
\caption{Mellin transforms of the terms involved in the NLO charged current quark coefficient functions.\label{quarkCFterms}}
\end{table}

\begin{table}[t]
\begin{center}
\begin{tabular}{|c|c|}
\hline
\vphantom{\Bigg|}$f(z)$                           &     $\mathbf{M}[f](N)$  \\
\hline\hline
\vphantom{\Bigg|}$[z^2+(1-z)^2]\times$&$\displaystyle\frac{4-2N(N-3)-N(N^2+N+2)\{2S_1+\ln[\lambda(1-\lambda)]\}}{N^2(N+1)(N+2)}$\\
\vphantom{\Bigg|}$\displaystyle\left\{2\ln\left(\frac{1-z}{z}\right)-\ln[\lambda(1-\lambda)]\right\}$& \\
\hline
\vphantom{\Bigg|}$[z^2+(1-z)^2)]\times$& $\displaystyle-\frac{2}{\lambda}\left(\frac{\lambda^2}{N}-\frac{2\lambda}{N+1}+\frac{2}{N+2}\right)\frac{_2F_1(1,N+1,N+2,\lambda)}{N+1}$\\
\vphantom{\Bigg|}$\displaystyle\left\{2\ln\left(\frac{1-z}{1-\lambda z}\right)+\ln\left(\frac{1-\lambda}{\lambda}\right)\right\}$&$\displaystyle-\frac{4(\lambda-1)}{\lambda(N+1)(N+2)}-\frac{(N^2+N+2)\left\{2S_1-\ln\left(\frac{1-\lambda}{\lambda}\right)\right\}}{N(N+1)(N+2)}$\\
\hline
\vphantom{\Bigg|}$\displaystyle z\ln\frac{1-\lambda z}{(1-\lambda)z}$&$\displaystyle\frac{_2F_1(1,N+1,N+2;\lambda)}{(N+1)^2}$\\
\hline
\vphantom{\Bigg|}$\displaystyle z^2\ln\frac{1-\lambda z}{(1-\lambda)z}$&$\displaystyle\frac{_2F_1(1,N+1,N+2;\lambda)-1}{\lambda(N+1)(N+2)}$\\
\hline
\vphantom{\Bigg|}$\displaystyle\frac{1}{1-\lambda z}$&$\displaystyle\frac{_2F_1(1,N,N+1;\lambda)}{N}$\\
\hline
\vphantom{\Bigg|}$\displaystyle\frac{z}{1-\lambda z}$&$\displaystyle\frac{_2F_1(1,N+1,N+2;\lambda)}{N+1}$\\
\hline
\end{tabular}
\end{center}
\caption{Mellin transforms of the terms involved in the NLO charged current gluon coefficient functions.\label{gluonCFterms}}
\end{table}
In order to transform to the $N$-space the above $x$-space expressions, in Tables~\ref{quarkCFterms} and~\ref{gluonCFterms} we tabulate the Mellin transforms of all terms involved.
In these  tables we use the analytic continuation of the harmonic sum
$$
S_l\equiv S_l(N) = \sum_{k=1}^{N}\frac1{k^l} = \zeta(l)- \frac{(-1)^l}{(l-1)!}\psi^{(l-1)}(N+1),
$$
where $\zeta(l)$ is the Riemann 
$\zeta$-function, with $\zeta(1)=\gamma_{EM}$,  $\psi((l-1))$ is the
polygamma,  and $_2F_1(a,b,c;N)$ is the Gauss hypergeometric function .

As an example of use of Tables~\ref{quarkCFterms}-\ref{gluonCFterms}, we present here the complete $N$-space quark and gluon coefficient functions for $F_2^c$
\begin{eqnarray}\label{xc22}
C^{(n_l)}_{2,q}(N)&=&C_F\bigg[-\left(4+\frac{1}{2\lambda}+\frac{\pi^2}{3}+\frac{1+\lambda}{2\lambda}K_A\right) -2(S_2-\zeta_2)+\frac{1}{N^2}\nonumber\\
\nonumber\\
&-&\frac{1}{(N+1)^2}+2\left(\frac{S_1}{N}+\frac{S_1}{N+1}+\frac{1}{(N+1)^2}\right)+\lambda \frac{_2F_1(1,N+1,N+2;\lambda)}{N(N+1)}\nonumber\\
\nonumber\\
&+&\frac{\ln(1-\lambda)}{N}+\lambda\frac{_2F_1(1,N+2,N+3;\lambda)}{(N+1)(N+2)}+\frac{\ln(1-\lambda)}{N+1}+\frac{2}{N+1}+\frac{2}{N}\nonumber\\
\nonumber\\
&-&\frac{2}{N-1}+2\frac{_2F_1(1,N-1,N,\lambda)}{N-1}-\frac{_2F_1(1,N,N+1,\lambda)}{N}\\
\nonumber\\
&-&\frac{_2F_1(1,N+1,N+2,\lambda)}{N+1}+ 2\left(S_1^2+S_2-2\frac{S_1}{N}\right)-2J_\lambda(N)-2\left(\frac1N-S_1\right)\nonumber\nonumber\\
\nonumber\\
&+&\frac12\left(\frac{_2F_1(2,N,N+2,\lambda)}{N(N+1)}+\frac{\lambda+\ln(1-\lambda)}{\lambda^2}\right)-\left(\frac{1}{N}-\frac{1}{N+1}-2S_1+\frac32\right)\ln{\lambda}\bigg];\nonumber
\end{eqnarray}

\begin{eqnarray}
C^{(n_l)}_{2,g}(N)&=&T_f\frac{4-2N(N-3)-N(N^2+N+2)\{2S_1+\ln[\lambda(1-\lambda)]\}}{2N^2(N+1)(N+2)}+\nonumber\\
\nonumber\\
&&\frac{8-18(1-\lambda)+12(1-\lambda)^2}{(N+1)(N+2)}+\frac{(1-\lambda)_2F_1(1,N,N+1;\lambda)}{N}+\\
\nonumber\\
&&6\lambda(1-\lambda)\left[\frac{_2F_1(1,N+1,N+2;\lambda)}{(N+1)^2}-\frac{2_2F_1(1,N+1,N+2;\lambda)-1}{(N+1)(N+2)}\right]-\frac{1}{N}.\nonumber
\end{eqnarray}

As a cross-check of the Mellin space results, it is possible
to compute the asymptotic limit $\lambda \to 1$ of these
expressions. We need
 the asymptotic expansion of the hypergeometric
functions, Eq.~(\ref{eq:hypgeoexpansion}) up to $\mathcal{O}\lp \lambda-1\rp$ terms. In particular,
\be
_2F_1(1,N+1,N+2;\lambda) =-(1+N)\lp \ln\lp 1-\lambda \rp +\gamma_E +
\psi^{(0)}(N+1)\rp + \mathcal{O}\lp (\lambda-1)\rp,
\ee
\be
_2F_1(2,N,N+2;\lambda) =-N(1+N)\lp \ln\lp 1-\lambda \rp +\gamma_E +
\psi^{(0)}(N)\rp + \mathcal{O}\lp (\lambda-1)\rp.
\ee
Substituting in Eq.~(\ref{xc22}), one can see that
all collinear heavy quark logarithms and that the massless limit
of the massive charged current heavy quark coefficient functions reduces
to the usual ZM-VFN result, as we know from  $x$--space.

Now we turn to discuss the implementation and benchmarking
of the above results into the FastKernel framework.
Analogously to the neutral current sector, the FONLL-A
charged current structure functions
in Mellin space can be written as
\bea
\label{eq:f2c_cc_nnpdf}
F_{i,h}^{\rm CC,FONLL}(N,Q^2) &=& 
F_{i,h}^{CC(n_l)}(N,Q^2) \\
&-&\theta\lp Q^2 - m^2\rp\lp 1-\frac{m^2}{Q^2}\rp^2 \lc
F_{i,h}^{CC(n_l,0)}(N,Q^2)-F_{2,h}^{CC(n_l+1)}(N,Q^2) \rc \nonumber \ .
\eea
with $i=1,2,3$. The Mellin space expressions of the massive
heavy quark coefficient functions have been computed above,
and the other ingredients of Eq.~(\ref{eq:f2c_cc_nnpdf}) are
their massless limits and the standard Mellin transform
of the ZM-VFN coefficient functions.

With these results, we have implemented the
FONLL-A charged current structure functions
Eq.~(\ref{eq:f2c_cc_nnpdf}) into
the FastKernel framework. As it has been done in the case of neutral
current observables, here we benchmark the
accuracy of this FONLL scheme implementation. 
We use again the same settings of the
Les Houches heavy quarks benchmark study.
The benchmarking of the FONLL-A CC structure function
implementation in FastKernel is performed for
the charm production cross section in neutrino
induced DIS, defined by Eq.~(\ref{eq:nuxsecdimuon}),
that combines all three charged current structure functions.
We have checked that the comparison of individual
structure functions has a similar level of accuracy.

Results for the benchmark
comparison are shown in Table~\ref{tab:tablebench-ccdimuons}.
As discussed above, the FONLL-A calculation
 of charged current structure functions has
been implemented in a $x$--space code, FONLLdisCC, that we will use
for the  benchmarking with the FastKernel
implementation. Results are shown
for various values of $Q^2$ relevant for the analysis of
experimental data.
The accuracy is similar to the one achieved for neutral current
structure functions (see Tables~\ref{tab:tablebench}-\ref{tab:tablebench2}),
at the per mil level, suitable for precision PDF determinations.

\begin{table}[ht]
\begin{center}
\small
 \begin{tabular}{|c|c|c|c|c|c|c|}
\hline
& \multicolumn{3}{|c|}{FONLL-A} & \multicolumn{3}{|c|}{FFN} \\
 \hline
 $x$ & FONLLdisCC & FastKernel  
& Accuracy & FONLLdisCC & FastKernel  
& Accuracy\\
\hline
 \hline
\multicolumn{7}{|c|}{$Q^2=4$ GeV$^2$} \\
 \hline
$10^{-5}$    &  163.14   & 164.06 & 0.6\% & 158.70 & 158.15 & 0.3\% \\
 $10^{-4}$   & 109.48  &  109.55 & 0.1\% & 106.81 & 106.64 & 0.2\% \\
 $10^{-3}$   & 69.24  & 69.35 & 0.2\% & 67.86 & 67.88 & 0.1\% \\
  $10^{-2}$  & 37.75 & 37.87 & 0.3\% & 37.27 & 37.30 & 0.1\% \\
  $10^{-1}$  & 13.56 & 13.57  & 0.1\% & 13.53  & 13.51 & 0.1\% \\
\hline
\hline
\multicolumn{7}{|c|}{$Q^2=10$ GeV$^2$} \\
 \hline
$10^{-5}$ &  279.31  & 278.71  & 0.2\%  & 261.49 & 261.55 & 0.02\%\\
 $10^{-4}$  &  167.02 & 166.85 & 0.1\% & 157.27 & 157.11 & 0.1\% \\
 $10^{-3}$  &  92.90 & 92.87  & 0.03\% & 88.33 & 88.12  & 0.2\% \\
  $10^{-2}$  & 44.92  & 44.93  & 0.02\% & 43.36 & 43.23 & 0.3\% \\
  $10^{-1}$  & 14.50  & 14.48 & 0.1\% & 14.26 & 14.28 & 0.1\% \\
\hline
\hline
\multicolumn{7}{|c|}{$Q^2=100$ GeV$^2$} \\
 \hline
$10^{-5}$ &     674.55 & 674.53 & 0.02\% & 651.21 & 645.94 & 0.1\% \\
 $10^{-4}$  & 345.73  & 345.81 & 0.02\% & 331.17 & 329.14 & 0.5\% \\
 $10^{-3}$  & 161.70 &  161.78 & 0.05\% & 153.94 & 152.36  & 0.1\% \\
  $10^{-2}$  & 64.20  & 64.26  & 0.1\% & 61.11 & 61.06  & 0.1\% \\
  $10^{-1}$  & 15.79  & 15.83  & 0.2\% & 15.33 & 15.42 & 0.1\%\\
 \hline
 \end{tabular}
\end{center}
\caption{\small Results of the benchmark comparison for the
dimuon charm production cross section Eq.~(\ref{eq:nuxsecdimuon}), 
 in the
 FONLL-A scheme for the FONLLdisCC charged current code
and for the FastKernel framework. Results are provided at the
benchmark kinematical points in $x,Q^2$. Results  for the 
massive (FFN) scheme are also given for completeness.
 The inelasticity variable
in the dimuon cross section for this benchmark table has been
taken to be $y=0.5$. The Les Houches Heavy Quark benchmark 
settings~\cite{LHhq}
have been used for the comparison.
\label{tab:tablebench-ccdimuons}}
\end{table}

\bibliography{nnpdf21}

\begin{thebibliography}{10}

\bibitem{Forte:2010dt}
S. Forte,
\newblock (2010), 1011.5247.

\bibitem{Stump:2003yu}
D. Stump et~al.,
\newblock JHEP 10 (2003) 046, hep-ph/0303013.

\bibitem{Tung:2006tb}
W.K. Tung et~al.,
\newblock JHEP 02 (2007) 053, hep-ph/0611254.

\bibitem{Guffanti:2010yu}
A. Guffanti and J. Rojo,
\newblock (2010), 1008.4671.

\bibitem{Forte:2002fg}
S. Forte et~al.,
\newblock JHEP 05 (2002) 062, hep-ph/0204232.

\bibitem{DelDebbio:2004qj}
The NNPDF collaboration, L. Del~Debbio et~al.,
\newblock JHEP 03 (2005) 080, hep-ph/0501067.

\bibitem{DelDebbio:2007ee}
The NNPDF collaboration, L. Del~Debbio et~al.,
\newblock JHEP 03 (2007) 039, hep-ph/0701127.

\bibitem{Ball:2008by}
The NNPDF Collaboration, R.D. Ball et~al.,
\newblock Nucl. Phys. B809 (2009) 1, 0808.1231.

\bibitem{Ball:2009mk}
The NNPDF collaboration, R.D. Ball et~al.,
\newblock Nucl. Phys. B823 (2009) 195, 0906.1958.

\bibitem{Ball:2010de}
{The NNPDF collaboration}, R.D. Ball et~al.,
\newblock Nucl. Phys. B838 (2010) 136, 1002.4407.

\bibitem{Forte:2010ta}
S. Forte et~al.,
\newblock Nucl. Phys. B834 (2010) 116, 1001.2312.

\bibitem{Cacciari:1998it}
M. Cacciari, M. Greco and P. Nason,
\newblock JHEP 05 (1998) 007, hep-ph/9803400.

\bibitem{Kramer:2000hn}
M. Kramer, 1, F.I. Olness and D.E. Soper,
\newblock Phys. Rev. D62 (2000) 096007, hep-ph/0003035.

\bibitem{acot2}
M.A.G. Aivazis et~al.,
\newblock Phys. Rev. D50 (1994) 3102, hep-ph/9312319.

\bibitem{Nadolsky:2008zw}
P.M. Nadolsky et~al.,
\newblock Phys. Rev. D78 (2008) 013004, 0802.0007.

\bibitem{Lai:2010vv}
H.L. Lai et~al.,
\newblock Phys. Rev. D82 (2010) 074024, 1007.2241.

\bibitem{LHas}
R.D. Ball et~al.,
\newblock {Chapter 21 in: J.~R.~Andersen et al., "The SM and NLO multileg
  working group: Summary report"},
\newblock arXiv:1003.1241, 2010.

\bibitem{Demartin:2010er}
F. Demartin et~al.,
\newblock Phys. Rev. D82 (2010) 014002, 1004.0962.

\bibitem{Caola:2010cy}
F. Caola, S. Forte and J. Rojo,
\newblock (2010), 1007.5405,
\newblock accepted for publication in Nucl. Phys. A.

\bibitem{Caola:2009iy}
F. Caola, S. Forte and J. Rojo,
\newblock Phys. Lett. B686 (2010) 127, 0910.3143.

\bibitem{Martin:2009iq}
A.D. Martin et~al.,
\newblock Eur. Phys. J. C63 (2009) 189, 0901.0002.

\bibitem{Alekhin:2009ni}
S. Alekhin et~al.,
\newblock Phys. Rev. D81 (2010) 014032, 0908.2766.

\bibitem{H1:2009wt}
H1 and ZEUS collaborations, A. F. et~al.,
\newblock (2009), 0911.0884.

\bibitem{reweighting}
The NNPDF Collaboration, R.D. Ball et~al.,
\newblock (2010), 1012.0836.

\bibitem{Breitweg:1999ad}
ZEUS, J. Breitweg et~al.,
\newblock Eur. Phys. J. C12 (2000) 35, hep-ex/9908012.

\bibitem{Chekanov:2003rb}
ZEUS, S. Chekanov et~al.,
\newblock Phys. Rev. D69 (2004) 012004, hep-ex/0308068.

\bibitem{Chekanov:2008yd}
ZEUS, S. Chekanov et~al.,
\newblock Eur. Phys. J. C63 (2009) 171, 0812.3775.

\bibitem{Chekanov:2009kj}
ZEUS, S. Chekanov et~al.,
\newblock Eur. Phys. J. C65 (2010) 65, 0904.3487.

\bibitem{Adloff:2001zj}
H1, C. Adloff et~al.,
\newblock Phys. Lett. B528 (2002) 199, hep-ex/0108039.

\bibitem{Collaboration:2009jy}
H1, F.D. Aaron et~al.,
\newblock Phys. Lett. B686 (2010) 91, 0911.3989.

\bibitem{H1F2c10:2009ut}
H1, F.D. Aaron et~al.,
\newblock Eur. Phys. J. C65 (2010) 89, 0907.2643.

\bibitem{Arneodo:1996kd}
New Muon Collaboration, M. Arneodo et~al.,
\newblock Nucl. Phys. B487 (1997) 3, hep-ex/9611022.

\bibitem{Arneodo:1996qe}
New Muon Collaboration, M. Arneodo et~al.,
\newblock Nucl. Phys. B483 (1997) 3, hep-ph/9610231.

\bibitem{Whitlow:1991uw}
L.W. Whitlow et~al.,
\newblock Phys. Lett. B282 (1992) 475.

\bibitem{bcdms1}
BCDMS, A.C. Benvenuti et~al.,
\newblock Phys. Lett. B223 (1989) 485.

\bibitem{bcdms2}
BCDMS, A.C. Benvenuti et~al.,
\newblock Phys. Lett. B237 (1990) 592.

\bibitem{Onengut:2005kv}
CHORUS, G. Onengut et~al.,
\newblock Phys. Lett. B632 (2006) 65.

\bibitem{h1fl}
H1, F.D. Aaron et~al.,
\newblock Phys. Lett. B665 (2008) 139, 0805.2809.

\bibitem{Goncharov:2001qe}
NuTeV, M. Goncharov et~al.,
\newblock Phys. Rev. D64 (2001) 112006, hep-ex/0102049.

\bibitem{MasonPhD}
D.A. Mason,
\newblock FERMILAB-THESIS-2006-01.

\bibitem{Chekanov:2009gm}
ZEUS, S. Chekanov et~al.,
\newblock Eur. Phys. J. C62 (2009) 625, 0901.2385.

\bibitem{Chekanov:2008aa}
ZEUS, S. Chekanov et~al.,
\newblock Eur. Phys. J. C61 (2009) 223, 0812.4620.

\bibitem{Moreno:1990sf}
G. Moreno et~al.,
\newblock Phys. Rev. D43 (1991) 2815.

\bibitem{Webb:2003ps}
NuSea, J.C. Webb et~al.,
\newblock (2003), hep-ex/0302019.

\bibitem{Webb:2003bj}
J.C. Webb,
\newblock (2003), hep-ex/0301031.

\bibitem{Towell:2001nh}
FNAL E866/NuSea, R.S. Towell et~al.,
\newblock Phys. Rev. D64 (2001) 052002, hep-ex/0103030.

\bibitem{Aaltonen:2009ta}
CDF, T. Aaltonen et~al.,
\newblock Phys. Rev. Lett. 102 (2009) 181801, 0901.2169.

\bibitem{Abazov:2007jy}
D0, V.M. Abazov et~al.,
\newblock Phys. Rev. D76 (2007) 012003, hep-ex/0702025.

\bibitem{Aaltonen:2009pc}
CDF, T. Aaltonen et~al.,
\newblock (2009), 0908.3914.

\bibitem{Abulencia:2007ez}
CDF - Run II, A. Abulencia et~al.,
\newblock Phys. Rev. D75 (2007) 092006, hep-ex/0701051.

\bibitem{D0:2008hua}
D0, V.M. Abazov et~al.,
\newblock Phys. Rev. Lett. 101 (2008) 062001, 0802.2400.

\bibitem{Altarelli:1998gn}
G. Altarelli, S. Forte and G. Ridolfi,
\newblock Nucl. Phys. B534 (1998) 277, hep-ph/9806345.

\bibitem{LHhq}
J. Rojo et~al.,
\newblock {Chapter 22 in: J.~R.~Andersen et al., "The SM and NLO multileg
  working group: Summary report"},
\newblock arXiv:1003.1241, 2010.

\bibitem{Nadolsky:2009ge}
P.M. Nadolsky and W.K. Tung,
\newblock Phys. Rev. D79 (2009) 113014, 0903.2667.

\bibitem{Witten:1975bh}
E. Witten,
\newblock Nucl. Phys. B104 (1976) 445.

\bibitem{Shifman:1977yb}
M.A. Shifman, A.I. Vainshtein and V.I. Zakharov,
\newblock Nucl. Phys. B136 (1978) 157.

\bibitem{Leveille:1978px}
J.P. Leveille and T.J. Weiler,
\newblock Nucl. Phys. B147 (1979) 147.

\bibitem{Alekhin:2003ev}
S.I. Alekhin and J. Blumlein,
\newblock Phys. Lett. B594 (2004) 299, hep-ph/0404034.

\bibitem{Buza:1997mg}
M. Buza and W.L. van Neerven,
\newblock Nucl. Phys. B500 (1997) 301, hep-ph/9702242.

\bibitem{Corcella:2003ib}
G. Corcella and A.D. Mitov,
\newblock Nucl. Phys. B676 (2004) 346, hep-ph/0308105.

\bibitem{Gluck:1996ve}
M. Gluck, S. Kretzer and E. Reya,
\newblock Phys. Lett. B380 (1996) 171, hep-ph/9603304.

\bibitem{Gottschalk:1980rv}
T. Gottschalk,
\newblock Phys. Rev. D23 (1981) 56.

\bibitem{Zijlstra:1992qd}
E.B. Zijlstra and W.L. van Neerven,
\newblock Nucl. Phys. B383 (1992) 525.

\bibitem{Nakamura:2010zzi}
Particle Data Group, K. Nakamura,
\newblock J. Phys. G37 (2010) 075021.

\bibitem{Mason:2007zz}
D. Mason et~al.,
\newblock Phys. Rev. Lett. 99 (2007) 192001.

\bibitem{FONLLdis}
J. Rojo et~al.,
\newblock Fonlldis: {\url{http://wwwteor.mi.infn.it/~rojo/fonlldis.html}},
\newblock 2010.

\bibitem{wattpriv}
G. Watt,
\newblock {Private communication},
\newblock 2010.

\bibitem{Ball:2009qv}
The NNPDF collaboration, R.D. Ball et~al.,
\newblock JHEP 05 (2010) 075, 0912.2276.

\bibitem{ATLAS:2010yt}
The ATLAS collaboration,
\newblock (2010), 1010.2130.

\bibitem{Collaboration:2010ez}
The CMS collaboration,
\newblock (2010), 1010.5994.

\bibitem{Collaboration:2010wv}
The ATLAS Collaboration,
\newblock (2010), 1009.5908.

\bibitem{Collaboration:2010ey}
The ATLAS Collaboration,
\newblock (2010), 1012.1792.

\bibitem{Campbell:2002tg}
J. Campbell and R.K. Ellis,
\newblock Phys. Rev. D65 (2002) 113007, hep-ph/0202176.

\bibitem{MCFMurl}
MCFM, http://mcfm.fnal.gov.

\bibitem{Martin:2009bu}
A.D. Martin et~al.,
\newblock Eur. Phys. J. C64 (2009) 653, 0905.3531.

\bibitem{Lai:2010nw}
H.L. Lai et~al.,
\newblock Phys. Rev. D82 (2010) 054021, 1004.4624.

\bibitem{Campbell:2006wx}
J.M. Campbell, J.W. Huston and W.J. Stirling,
\newblock Rept. Prog. Phys. 70 (2007) 89, hep-ph/0611148.

\bibitem{Botje:2011sn}
M. Botje et~al.,
\newblock (2011), 1101.0538.

\bibitem{Davidson:2001ji}
S. Davidson et~al.,
\newblock JHEP 02 (2002) 037, hep-ph/0112302.

\bibitem{Chekanov:2009na}
ZEUS, S. Chekanov et~al.,
\newblock Phys. Lett. B682 (2009) 8, 0904.1092.

\bibitem{H1Collaboration:2010ry}
The H1 Collaboration,
\newblock (2010), 1012.4355.

\bibitem{pdf4lhctalk}
J. Rojo,
\newblock {Talk at the PDF4LHC workshop, DESY, Hamburg,
  \url{http://indico.cern.ch/materialDisplay.py?contribId=7\&sessionId=2\&mate%
rialId=slides\&confId=103872}},
\newblock 2010.

\bibitem{Martin:2010db}
A.D. Martin et~al.,
\newblock Eur. Phys. J. C70 (2010) 51, 1007.2624.

\bibitem{Alekhin:2010sv}
S. Alekhin and S. Moch,
\newblock (2010), 1011.5790.

\bibitem{Pumplin:2007wg}
J. Pumplin, H.L. Lai and W.K. Tung,
\newblock Phys. Rev. D75 (2007) 054029, hep-ph/0701220.

\bibitem{Brodsky:1980pb}
S.J. Brodsky et~al.,
\newblock Phys. Lett. B93 (1980) 451.

\bibitem{Harlander:2003ai}
R.V. Harlander and W.B. Kilgore,
\newblock Phys. Rev. D68 (2003) 013001, hep-ph/0304035.

\bibitem{Campbell:2009ss}
J.M. Campbell et~al.,
\newblock Phys. Rev. Lett. 102 (2009) 182003, 0903.0005.

\bibitem{Maltoni:2003pn}
F. Maltoni, Z. Sullivan and S. Willenbrock,
\newblock Phys. Rev. D67 (2003) 093005, hep-ph/0301033.

\bibitem{Harris:1995tu}
B.W. Harris and J. Smith,
\newblock Nucl. Phys. B452 (1995) 109, hep-ph/9503484.

\bibitem{Martin:2006qz}
A.D. Martin, W.J. Stirling and R.S. Thorne,
\newblock Phys. Lett. B636 (2006) 259, hep-ph/0603143.

\bibitem{Spira:1995rr}
M. Spira et~al.,
\newblock Nucl.Phys. B453 (1995) 17, hep-ph/9504378.

\bibitem{Bourilkov:2006cj}
D. Bourilkov, R.C. Group and M.R. Whalley,
\newblock (2006), hep-ph/0605240.

\end{thebibliography}

\end{document}